%% file: report_thesis.tex
\documentclass[12pt]{report}
\usepackage{suthesis}
\usepackage{graphicx}
\usepackage{makeidx}
\usepackage{url}

\makeindex

\begin{document}
\widowpenalty=10000
\clubpenalty=10000
\input{definitions}

\input{nonastrodef}

\title{A Search for Gamma-Ray Bursts and Pulsars, \\
and the Application of Kalman Filters to Gamma-Ray Reconstruction}
\author{Brian Jones}
\dept{Physics}
\principaladviser{Peter Michelson}
\firstreader{Vah\eacc\ Petrosian}
\secondreader{Robert Wagoner}
 
\submitdate{October 1998}
\copyrightyear{1999}

\beforepreface
\prefacesection{Abstract}
\input{Abstract}

\prefacesection{Acknowledgements}
\input{Ack}
\clearpage
\input{Dedication}

\afterpreface

\part{Methods for Time-Series Analysis}
\label{part1}

\input{Introduction}
\input{Stats}
\input{GRB}

\input{Timevar}

\input{Timevar2}

\input{Conclusions}

\part{The October 1997 \glast-prototype Beam Test}
\label{part2}
\input{Beamtest}

\input{Beamtest2}

\input{Beamresults}

\appendix
\input{timedelayapp}
\input{part2appen}


\begin{singlespace}
\bibliography{Bibliography,glast,pulsars,grb,kalman}
\bibliographystyle{bbjthesis}
\end{singlespace}

\end{document}

%% file: definitions.tex
\newcommand{\cosb}{{\em COS~B}\/}
\newcommand{\sas}{{\em SAS~2}\/}
\newcommand{\cgro}{{\em CGRO}\/}
\newcommand{\egret}{{\em EGRET}\/}
\newcommand{\glast}{{\em GLAST}\/}
\newcommand{\batse}{{\em BATSE}\/}
\newcommand{\comptel}{{\em COMPTEL}\/}
\newcommand{\osse}{{\em OSSE}\/}
\newcommand{\rosat}{{\em ROSAT}\/}
\newcommand{\sax}{{\em BeppoSAX}\/}
\newcommand{\xte}{{\em RXTE}\/}
\newcommand{\ulysses}{{\em Ulysses}\/}
\newcommand{\asca}{{\em ASCA}\/}

\newcommand{\sage}{{\tt SAGE}\/}
\newcommand{\likeprog}{{\tt LIKE}\/}
\newcommand{\timevar}{{\tt timevar}\/}
\newcommand{\tjrecon}{{\tt tjrecon}\/}
\newcommand{\gismo}{{\tt gismo}\/}
\newcommand{\glastsim}{{\tt glastsim}\/}
\newcommand{\pulsar}{{\tt PULSAR}\/}
\newcommand{\spectral}{{\tt SPECTRAL}\/}

\newcommand{\gammaray}{$\gamma$-ray}
\newcommand{\gammarays}{$\gamma$-rays}
\newcommand{\Gammaray}{$\gamma$-Ray}
\newcommand{\Gammarays}{$\gamma$-Rays}
\newcommand{\perareasec}{~cm$^{-2}$~s$^{-1}$}
\newcommand{\perareasecmev}{~cm$^{-2}$~s$^{-1}$~MeV$^{-1}$}
\newcommand{\perareasecsr}{~cm$^{-2}$~s$^{-1}$~sr$^{-1}$}
\newcommand{\eg}{e.g.}
\newcommand{\ie}{i.e.}
\newcommand{\epm}{$e^\pm$}
\newcommand{\etal}{et~al.}
\newcommand{\sub}[1]{_{\rm #1}}
\newcommand{\expct}[1]{\left\langle #1 \right\rangle}
\newcommand{\radlen}{$X_0$}
\newcommand{\BO}{B$\emptyset$\index{$\emptyset$}}
\newcommand{\pos}{$e^+$}
\newcommand{\el}{$e^-$}
\newcommand{\beamtest}{beam test}
\newcommand{\Beamtest}{Beam test}
\newcommand{\BeamTest}{Beam Test}
\newcommand{\eacc}{\'{e}\index{\'{e}}}
\newcommand{\aeind}{\ae \index{\ae}}

\newcommand{\dbyd}[1]{\frac{\partial}{\partial #1}\hbox{\index{$\partial$}}}
\newcommand{\deriv}[2]{\frac{\partial #1}{\partial #2}\index{$\partial$}}
\newcommand{\secderiv}[2]{\frac{\partial^2 #1}{\partial #2^2}\index{$\partial$}}
\newcommand{\myvec}[1]{ {\bf #1}}

\def\overbo{\O verb\o\index{\O}\index{\o}}
\def\pacz{Pa\-czy\'{n}\-ski\index{\'{n}}}
\newcommand{\meszaros}{M\'{e}sz\'{a}ros\index{\'{e}}\index{\'{a}}}
\def\cerenkov{\v{C}erenkov\index{\v{C}}}
\def\stos{PSR~1706-44\/}
\def\tff{PSR~1055-52\/}
\def\ntfo{PSR~B1951+32\/}

\def\accang{\hbox{$\theta_{67}$}}
\def\chisq{\hbox{$\chi^2$}}
\def\expsr{\hbox{$\cal E$}}
\def\htest{\hbox{$H$-test}}
\def\ipred{\hbox{$\dot E/D^2$}}
\def\ij{\hbox{\/$_{ij}$}}
\def\kl{\hbox{\/$_{kl}$}}
\def\psf{\hbox{\rm PSF}}
\def\psd{\hbox{\rm PSD}}
\def\sa{\hbox{\rm SA}}
\def\ed{\hbox{\rm ED}}
\def\TS{\hbox{\rm TS}}
\def\sn{\hbox{\rm S/N}}
\def\sqrtTS{\hbox{${\rm TS}^{1/2}$}}
\def\zsqr{\hbox{$Z_m^2$}}
\def\like{\hbox{${\cal L}$\index{$\cal L$}}}
\def\by{\hbox{$\times$\index{$\times$}}}
\def\mysim{\hbox{$\sim$\index{$\sim$}}}
\def\mystar{\hbox{$^\star$\index{$^\star$}}}
\def\infinity{\infty \hbox{\index{$\infty$}}}
\def\ranal{\hbox{$R_{anal}$}}
\def\myell{\ell\hbox{\index{$\ell$}}}
\def\mypm{\pm\hbox{\index{$\pm$}}}
\def\gtsim{\stackrel{\gt}{\sim}}
\def\utc{\hbox{\scriptsize UTC}}

\newcommand{\chapt}[1]{Chapter~\ref{#1}}
\newcommand{\eq}[1]{equation~(\ref{#1})}
\newcommand{\Eq}[1]{Equation~(\ref{#1})}
\newcommand{\fig}[1]{Figure~\ref{#1}}
\newcommand{\sect}[1]{\S\ref{#1}\index{\S}}
\newcommand{\tbl}[1]{Table~\ref{#1}}
\newcommand{\app}[1]{Appendix~\ref{#1}}
\newcommand{\pt}[1]{Part~\ref{#1}}

\def\aj{AJ}			
\def\araa{ARA\&A}		
\def\apj{ApJ}			
\def\apjl{ApJ}		
\def\apjs{ApJS}		
\def\ao{Appl.Optics}		
\def\apss{Ap\&SS}		
\def\aap{A\&A}		
\def\aapr{A\&A~Rev.}		
\def\aaps{A\&AS}		
\def\azh{AZh}			
\def\baas{BAAS}		
\def\jrasc{JRASC}		
\def\memras{MmRAS}		
\def\mnras{MNRAS}		
\def\nature{Nature}		
\def\physrev{Phys.Rev.}		
\def\pra{Phys.Rev.A}		
\def\prb{Phys.Rev.B}		
\def\prc{Phys.Rev.C}		
\def\prd{Phys.Rev.D}		
\def\prl{Phys.Rev.Lett}	
\def\pasp{PASP}		
\def\pasj{PASJ}		
\def\qjras{QJRAS}		
\def\science{Science}		
\def\skytel{S\&T}		
\def\solphys{Solar~Phys.}	
\def\sovast{Soviet~Ast.}	
\def\ssr{Space~Sci.Rev.}
\def\thesis{Ph.D. Thesis}	
\def\zap{ZAp}			
\def\GROworkshop{in Proc. of the Gamma-Ray Observatory Science Workshop,
ed. W. N. Johnson (Greenbelt, MD: NASA)}
\def\EGRETsymp{in The Energetic Gamma-Ray Experiment Telescope (EGRET)
Science Symposium, ed. C. Fichtel \etal\ (Greenbelt, MD: NASA)}

\def\gt{\hbox{$>$}}
\def\lt{\hbox{$<$}}
\newcommand{\cross}{{\boldmath \times}}
\newcommand{\vdot}{{\bf \cdot}}

\def\deg{\hbox{$^\circ$}}
\def\hr{\hbox{$^{\rm h}$}\index{$^{\rm h}$}}
\def\mn{\hbox{$^{\rm m}$}\index{$^{\rm m}$}}
\def\mysc{\hbox{$^{\rm s}$}\index{$^{\rm s}$}}
\def\sun{\hbox{\scriptsize $\odot$}}
\def\earth{\hbox{\scriptsize $\oplus$}}
\def\sq{\hbox{\rlap{$\sqcap$}$\sqcup$}}
\def\arcmin{\hbox{$^\prime$}}
\def\arcsec{\hbox{$^{\prime\prime}$}}
\def\fd{\hbox{$.\!\!^{\rm d}$}}
\def\fh{\hbox{$.\!\!^{\rm h}$}}
\def\fm{\hbox{$.\!\!^{\rm m}$}}
\def\fs{\hbox{$.\!\!^{\rm s}$}}
\def\fdg{\hbox{$.\!\!^\circ$}\index{$.\!\!^\circ$}}
\def\farcm{\hbox{$.\mkern-4mu^\prime$}}
\def\farcs{\hbox{$.\!\!^{\prime\prime}$}}
\def\fp{\hbox{$.\!\!^{\scriptscriptstyle\rm p}$}}
\def\us{\hbox{$\mu$s}}

\def\onehalf{1/2}
\def\onethird{1/3}
\def\twothirds{2/3}
\def\onequarter{1/4}
\def\threequarters{3/4}
\def\oneeighth{1/8}
\def\ubvr{\hbox{$U\!BV\!R$}}		
\def\ub{\hbox{$U\!-\!B$}}		
\def\bv{\hbox{$B\!-\!V$}}		
\def\vr{\hbox{$V\!-\!R$}}		
\def\ur{\hbox{$U\!-\!R$}}

%% file: nonastrodef.tex
\newcommand{\micron}{$\mu$m}

%% file: Abstract.tex
High-energy \gammaray\ astronomy was revolutionized in 1991 with the
launch of the Energetic Gamma-Ray Experiment Telescope (\egret) on board
the {\em Compton Gamma-Ray Observatory}.  In addition to unprecedented
instrument effective area and a narrow point-spread function, \egret\
provided photon time-tagging to an absolute accuracy of 100~\us.
The opportunity to analyze high-quality \gammaray\ data requires
sophisticated statistical and analytic tools.
\pt{part1} describes
the analysis of periodic and transient signals in \egret\ data.  
A method to search for the transient flux from \gammaray\ bursts 
independent of triggers from other \gammaray\ instruments is
developed.  Several known \gammaray\ bursts were independently detected,
and there is evidence for a previously unknown \gammaray\ burst candidate.
Statistical methods
using maximum likelihood and Bayesian inference are developed and implemented
to extract periodic signals from \gammaray\ sources
in the presence of significant astrophysical background radiation.  The methods
allow searches for periodic modulation without {\em a priori} knowledge of the
period or period derivative.  The analysis was performed on six
pulsars and three pulsar candidates.  The three brightest pulsars,
Crab, Vela, and Geminga, were readily identified, and would have been
detected independently in the \egret\ data without knowledge of the
pulse period.  No significant pulsation was detected in the three
pulsar candidates.  Furthermore, the method allows the
analysis of sources with periods on the same order as the time scales 
associated with changes in the
instrumental sensitivity, such as the orbital time scale of \cgro\ 
around the Earth.   Eighteen X-ray binaries were examined.
None showed any evidence of periodicity.  In addition, methods for 
calculating the detection threshold of periodic flux modulation
were developed.

The future hopes of \gammaray\ astronomy lie in the development of
the {\em Gamma-ray Large Area Space Telescope}, or \glast.  
\pt{part2} describes the development and results of the particle track 
reconstruction software for a \glast\ science prototype instrument beamtest.  
The Kalman filtering method of track reconstruction is introduced
and implemented.  
Monte Carlo simulations, very similar to those used for the full
\glast\ instrument, were performed to predict the instrumental response
of the prototype.  The prototype was tested
in a \gammaray\ beam at SLAC.  
The reconstruction software was
used to determine the incident \gammaray\ direction. 
It was found that the simulations did an excellent
job of representing the actual instrument response.

%% file: Ack.tex
The acknowledgements page of the modern physics thesis has become a 
stage upon which seasoned and world-weary graduate students perform
a tragicomic stand-up routine to amuse their family and friends.  It
is altogether fitting and proper that they should do so, since it is
unlikely that those family and friends will read any of the rest of
the thesis.  Besides, they hope they will be mentioned.  Nevertheless,
it is a good medium, and a useful way to point out the people and 
institutions without whom, for better or worse, this work would never
have been performed.

The number of people who have been a part of this cause is enormous.
There is no way I can mention all of the friends and colleagues whom I 
appreciate, so I will just touch on a few. Beginning with the most general,
 I must thank the taxpayers of the United States who, whether they
knew it or not, spent a vast amount of money to advance the frontiers
of \gammaray\ astrophysics.  I can only hope to have done them proud.
I was fortunate enough to profit from data taken by the \egret\ instrument,
on board the Compton Gamma-Ray Observatory.  I have deep appreciation for
the work done to get \egret\ into orbit and taking data by the scientists
and staff of Goddard Space Flight Center, especially Dave Thompson, who
was always particularly supportive of my work,
and the Max Planck Institut f\"{u}\index{\"{u}}r Extraterrestriche Physik.
This collaboration has done an outstanding job in designing, building, and
operating an instrument which has radically changed our understanding of
the \gammaray\ sky.  I'm proud to have been a small part of that group.

I'm also proud to be a part of a second international collaboration,
this one responsible for \glast.  I have benefitted from the chance
to work with scientists from Goddard, the Naval Research Laboratory,
UC Santa Cruz, and Columbia University in the United States, as well
as the \'Ecole Polytechnique and CEA in France, the University of Trieste
in Italy and the University of Tokyo, as well as dozens of other 
institutions worldwide too numerous to mention.  Being on the inside
of a developing Big Science project has been eye-opening, to say the
least.  Specifically, I'd like to thank Steve Ritz, Neil Johnson, and
Eric Grove for listening to the brash opinions of a grad student as
if he were a Real Scientist.

The fact that no Stanford (or SLAC) people are included above 
reflects not their absence, but their proximity.  A number of people
in and around Stanford have made an indelible impression on me.
My advisor Peter Michelson has given me tremendous freedom to pursue
my own interests, even when they were not what most astrophysicists
would call mainstream.  Bill Atwood has been a tremendous motivator,
and taught me a great deal.  His methods are generally quite harsh, 
by his own admission, and he is never satisfied.  For some reason
which I still do not understand, this had the effect of spurring me
on to better, more careful, and more thorough work, completed more rapidly,
rather than to hatred and bitterness.  Joe Fierro's Pucklike sensibilities
drew me to the group in the first place, in the belief that it would be 
a fun place to work.  He was a great help once I pestered him enough.
Tom Willis graciously put up with a newcomer in his office, and helped 
me through the steep learning curve in \egret\ minuti\aeind .  
Although he's not at Stanford, Prof. Dave Wilkinson along with the
rest of the Princeton faculty showed me what I am really capable of,
and gave me the best undergraduate physics education available.
After that, grad school was a breeze.  And of 
course, without Marcia Keating in the Varian office, I would never have
filed a study list, gotten paid, or managed to file the appropriate
graduation forms.  It doesn't matter how many Nobel
Laureates we have wandering around; without Marcia, this place would fall apart.
She will be greatly missed as she moves on to new challenges.

Pat Nolan, my {\em de facto} advisor, deserves a paragraph to himself.
His patience, advice, and knowledge have been a prime factor in whatever
success I may have had.  He also deserves rich thanks for maintaining 
the computer systems and knowing arcane details about UNIX so that I 
didn't have to.  It is a great privilege to have had the opportunity
to learn from him.

Also critical have been my fellow graduate students in the Physics
Department, especially Doug Natelson and Bill Tompkins.  Doug, as my
erstwhile roommate, had to put up with a lot and endured it well.
Bill, as roommate and officemate, found he could never 
avoid me.  He therefore sought to educate me, so that I would be
more bearable.  If I have given half as good as I've gotten from him,
I consider myself a success.

On a more personal note, I would like to thank all of the members
of TMBS for their support and prayers.  I know the Lord would have
given me strength without them, but they knocked at the door persistently,
and it has paid off.

Primarily, however, support for an undertaking such as a Ph.D. has to
come from the family.  Fortunately, I have been blessed with the best
family I could imagine.  My parents have supported me in my quest 
``to be a scientist'' from the very beginning, and always believed I 
could do it, even if they weren't sure why I wanted to.  Plus, they paid for 
college.  My brother Bill, aside from the non-trivial accomplishment of
being a fantastically supportive brother, decided to study physics just
so I wouldn't feel alone.  And last, but by no means least, my beautiful
wife Kim met me while I was yet a graduate student, and saw some kernel
of value in me even then.  Her love, support, and confidence in me 
have given me the strength to finish this work.  My great blessing is
that although this thesis is now part of the past, 
we have many wonderful years together in our future.

%% file: Dedication.tex
\begin{verse}
\vspace*{1 in}
{\em
{\sc O LORD}, our Lord, how majestic is your name in all the earth! \\
\hspace*{.25 in}You have set your glory above the heavens. \\

When I consider your heavens, the work of your fingers, \\
\hspace*{.25 in}the moon and the stars, which you have set in place, \\

what is man that you are mindful of him, \\
\hspace*{.25 in}the son of man that you care for him? \\
}
{\small Psalm 8 {\sc (NIV)}} \\
\vspace{.5 in}
{\em
The heavens declare the glory of God; \\
\hspace*{.25 in}the skies proclaim the work of his hands.  \\

Day after day they pour forth speech; \\
\hspace*{.25 in}night after night they display knowledge. \\

There is no speech or language where their voice is not heard. \\
\hspace*{.25 in}Their voice goes out into all the earth, \\
\hspace*{.25 in}their words to the ends of the world. \\

In the heavens he has pitched a tent for the sun, \\
\hspace*{.25 in}which is like a bridegroom coming forth from his pavilion, \\
\hspace*{.25 in}like a champion rejoicing to run his course. \\
}

{\small Psalm 19 {\sc (NIV)}}

\end{verse}

%% file: Introduction.tex
\chapter{Introduction \label{Intro}}

\begin{figure}
\centering
\includegraphics[width = 4in]{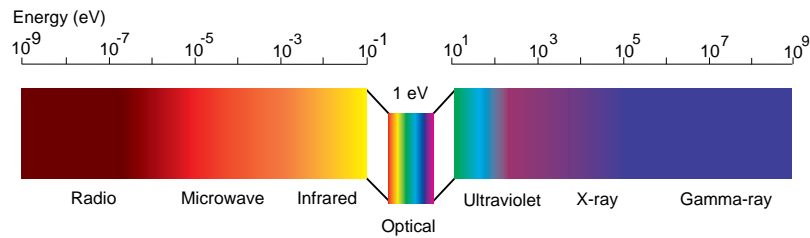}
\caption[The electromagnetic spectrum]{\label{intro:spectrum}
The Electromagnetic Spectrum.  \Gammarays\ occupy the highest energy extreme
of the electromagnetic spectrum, from 10~MeV to over 300~GeV.}
\end{figure}

The desire to experiment with the extremes of nature is innate.  While 
it is almost always amusing, it can be informative as well.  Such is 
the case in the study of some of the most energetic photons produced 
in space: the \gammarays\ (\fig{intro:spectrum}).  With at least ten 
million times the energy of ordinary optical photons, \gammarays\ 
represent a unique window into the most energetic processes in 
astrophysics---the (electromagnetically) roaring jets from active 
galactic nuclei, the arcing plasmas in the intense gravitational 
fields of pulsars, and the enigmatic and inordinately powerful 
explosions known as \gammaray\ bursts.

The immense value of \gammarays\ for astrophysics lies both in their 
role as telltale markers of large energy-generation processes and in 
their likelihood of passing unperturbed through vast reaches of 
intergalactic space.  Measuring the \gammaray\ spectra of 
astrophysical objects sharply constrains estimates of their total 
energy output.  Since \gammarays\ are so energetic, many astrophysical 
sources emit the bulk of their total power output at these high 
energies.  In addition, \gammarays\ travel relatively unimpeded 
through space.  Since they carry no charge, they are nearly unaffected 
by galactic and intergalactic magnetic fields.  Their small 
interaction cross section means that they are relatively unaffected by 
dust and gas in the intervening space between the source and the 
detector.  A high-energy \gammaray\ can travel through the central 
plane of the Galaxy with only a 1\% chance of being absorbed 
\cite{joethesis}.  \Gammarays\ may be observed from Earth essentially 
unchanged since they left the distant violence in which they were 
created.

However, every silver lining has its cloud.  The 
Earth's atmosphere is very good at absorbing \gammarays. 
Unfortunately, it means that precise astrophysical \gammaray\ observations 
must be done in space.  The second difficulty in \gammaray\ astronomy 
is intrinsic to the energy production mechanisms that produce 
\gammarays\ in the first place.  Because \gammarays\ are so energetic, 
most sources produce very few of them.  Detectors must be very 
efficient to collect these rare photons, and at the same time be able 
to discriminate against the sea of undesirable charged particles trapped
in the Earth's magnetic field and 
albedo \gammarays\ which are generated in the Earth's atmosphere.
This discrimination
requires background rejection on the order of one part in \mysim$10^5$ 
or better.

\section{The \Gammaray\ Observatories}
\begin{figure}[t]
\centering
\includegraphics[width = 5 in]{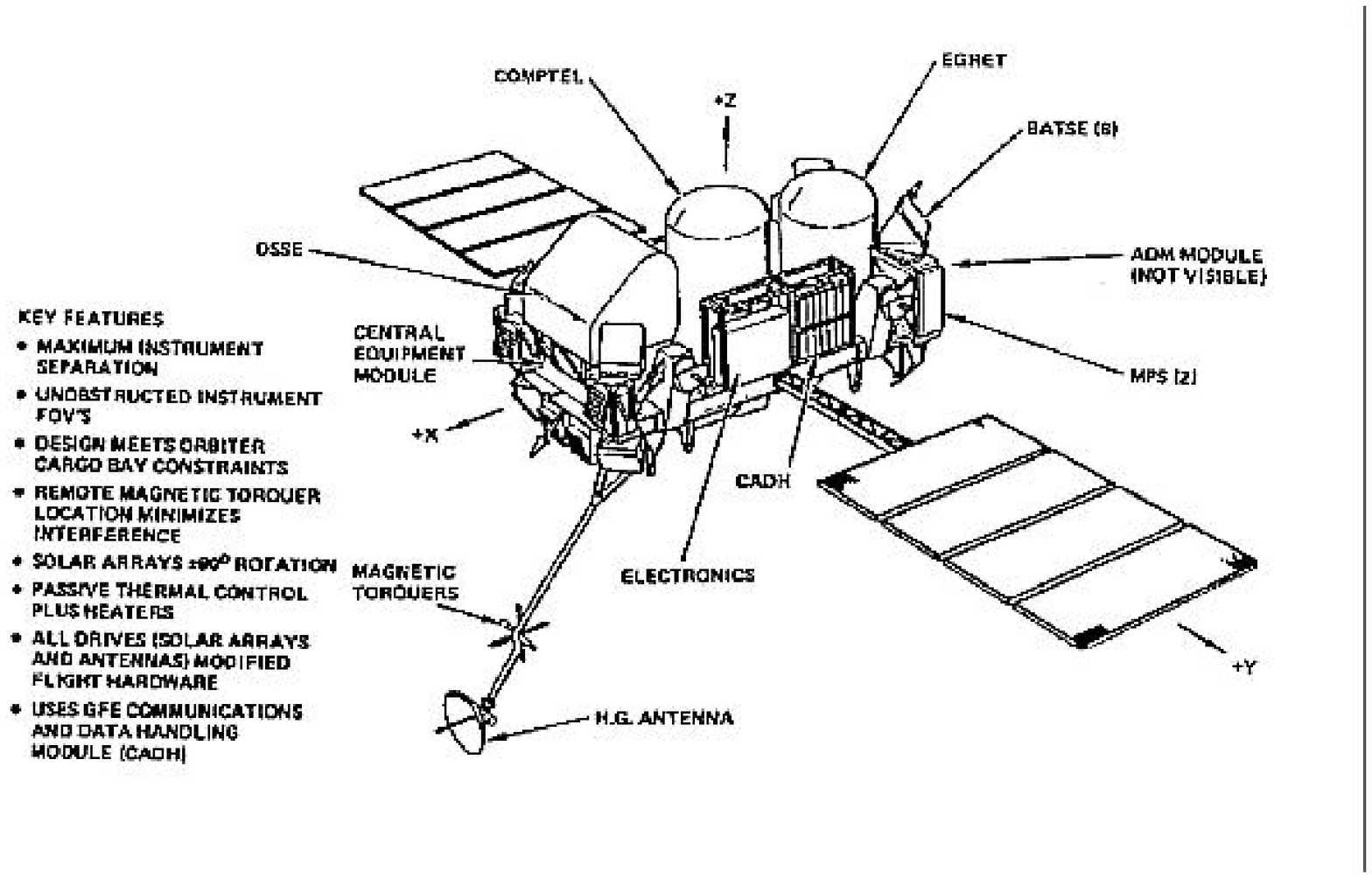}
\caption[The Compton Gamma-Ray Observatory]{\label{intro:cgro}
The Compton Gamma-Ray Observatory.  \egret\ is the dome on the right
end of the spacecraft.  It is coaligned with \comptel.  Rounding out
the instruments aboard \cgro\ are \osse, which is very
sensitive to \el\pos\ annihilation lines, and the omni-directional \batse\ 
\gammaray\ burst detector.}
\end{figure}

Experience with terrestrial accelerator-based \gammaray\ detectors suggested
that a spark chamber might be an effective astrophysical \gammaray\ detector.
In the mid-1970s, \sas\ \cite{sas2} and \cosb\ \cite{cosb} proved the 
concept of a \gammaray\ satellite telescope, while discovering several of
the brightest \gammaray\ sources.  Simultaneously, NASA envisioned the
{\em Great Observatories for Space Astrophysics} program:  a series of
satellite telescopes designed to give unprecedented insights into 
electromagnetic emission from infrared to \gammarays.  Under the auspices
of this program, the Space InfraRed Telescope Facility ({\em SIRTF\/})
infrared telescope has been designed, the Advanced X-ray Astrophysics
Facility (AXAF) has been designed and built,
the Hubble Space Telescope has revolutionized optical astronomy, and 
the Compton Gamma-Ray Observatory (\cgro) has explored the \gammaray\ sky
from less than 0.1~MeV to more than 10~GeV (\fig{intro:cgro}).

The five orders of magnitude in energy of the electromagnetic spectrum 
observed by \cgro\ require four different instruments on the 
satellite.  The lowest energy \gammarays\ interact primarily through 
the photo-electric effect.  The Oriented Scintillation Spectrometer 
Experiment (\osse) covers the energy range from 0.05--10~MeV with a 
field of view of $3\fdg 8 \by 11\fdg4$ \cite{osse}.  The Compton 
Telescope (\comptel) detects Compton scattered electrons, the most 
significant \gammaray\ interaction in the energy range between 1~MeV and almost 
30~MeV, to image the \gammaray\ sky with a field of view of \mysim1~sr 
\cite{comptel}.  The Energetic Gamma-Ray Telescope Experiment (\egret) 
measures pair-conversion events in a spark chamber, like \sas\ and 
\cosb.  It is sensitive to energies between 20~MeV and 30~GeV, with a 
field of view of \mysim1~sr~\cite{egret,egretcalibrate88}.  In 
addition, a fourth instrument aboard \cgro\ is optimized to detect 
\gammaray\ bursts.  The Burst and Transient Source Experiment (\batse) 
consists of eight uncollimated detectors, one on each corner of the 
\cgro\ spacecraft, sensitive to 25~keV--2~MeV \gammarays\ with nearly 
uniform coverage of the sky~\cite{batse}.

The \egret\ instrument is the focus of \pt{part1} of this work.  The instrument
was built and operated by a collaboration of scientists at Stanford University,
Goddard Space Flight Center (Greenbelt, Maryland), 
the Max Planck Institut f\"{u}r\index{\"{u}} 
Extraterrestrische Physik (Garching, Germany),
 and the Grumman Aerospace Corporation (Bethpage, New York).  It was 
launched aboard \cgro\ on the Space Shuttle Atlantis (STS-37) on April 5,
1991 and was deployed two days later.  It was activated on April 15, and
began taking data on April 20.  The instrument and its characteristics 
have been extensively documented
 \cite{egret,egretcalibrate88,egretcalibrate89,egretcalibrate92,egretcalibrate93};
 we will briefly touch on the highlights relevant to data analysis in 
\sect{stats:insteffects}.

Future \gammaray\ telescopes will further extend our understanding of 
astrophysical \gammaray\ processes.  The \glast\ instrument, scheduled for
launch in 2005 \cite{glast:proposal95,glast:atwood94,glast:bloom,glast:doe_proposal}, 
will improve upon the successes of \egret\ as it brings
\gammaray\ astronomy to the 21st century.  The test of a \glast\ science
prototype will be the focus of \pt{part2} of this work.  Details of the
current instrument baseline design are given in \sect{bt:baselinesect}.

\section{The \egret\ instrument}
\label{intro:egret}

Spark chambers detect \gammarays\ {\em via} pair production.  Pair
production refers to the process whereby a \gammaray\ converts to 
an electron--positron pair in the presence of matter.
The detection process is more fully described in \sect{bt:pairproduction}.  
The resulting electrons and positrons are easily detected because they are 
charged particles.

\subsection{Instrumental Design}

\begin{figure}[t]
\centering
\includegraphics[width = 4 in]{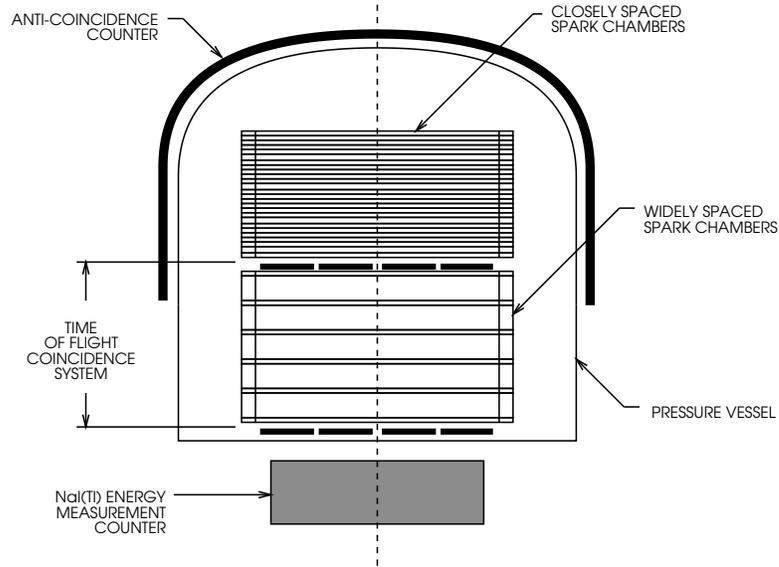}
\caption[The \egret\ Instrument]{\label{intro:egret_schem}
The \egret\ instrument.  The total height of the spark chambers
is approximately 1 m.}
\end{figure}

To optimize the detection and resolution of \gammarays, the \egret\ 
instrument consists of a series of thin tantalum (Ta) sheets 
interleaved with planes of conducting wires spaced by 0.8~mm.  Below this 
multilayer spark chamber is a NaI(Tl) calorimeter known as the TASC 
(Total Absorption Shower Counter).  Surrounding the spark chamber is a 
monolithic plastic scintillator to reject charged cosmic ray 
particles.  The schematic design is shown in \fig{intro:egret_schem}.

99.5\% of \gammarays\ will pass undetected 
through the anticoincidence scintillator into the 28 closely-spaced 
spark chamber modules, where it has a \mysim33\% chance of converting 
to an \el\pos\ pair.  If it does so, the pair will ionize the (mostly) 
neon gas in the spark chamber along their trajectories.  Below the 
closely-spaced modules is a time-of-flight system designed to measure 
whether the particles are upward- or downward-moving.  The system 
consists of two layers of $4 \by 4$ arrays of plastic scintillator 
tiles, spaced 60~cm apart.  By combining measurements from the two 
layers, the time-of-flight delay can be measured to within 
\mysim1.5~ns, and the general direction of the particles can be 
estimated.  At any given time, a limited number of general directions 
are considered valid for an instrument trigger.  The valid directions 
depend on orbital parameters, and are designed to exclude the Earth's 
limb, as well as limiting the field-of-view when desired.

If the time-of-flight system registers the passage of a downward 
particle from a valid region of the sky, and the anticoincidence 
system has not been triggered by a charged particle, a high voltage 
pulse is applied to the wires in the spark chamber modules.  Ionized 
paths short the wires to ground, and the affected wires are recorded 
digitally in ferrite cores.

The Total Absorption Spectrometer Calorimeter (TASC), located below the 
spark chamber, measures the total energy of the \gammaray\ event.  In 
consists of 8 radiation lengths of NaI(Tl), and has an energy 
resolution of \mysim20\% FWHM from a few tens of MeV to several tens 
of GeV. Events are tagged with an arrival time by the \cgro\ on-board 
clock to an absolute accuracy of 100~\us\ and a relative accuracy of 8 
\us.  The energy measurements made by the TASC are corrected on the 
ground for energy lost in the spark chamber and shower leakage.

\subsection{Instrumental Calibration and Performance}

\begin{table}[t]
\begin{minipage}{5.5 in}
\centering
\begin{tabular}{l c c c c}
& \sas & \cosb & \egret & \glast \\ \hline
Field of View & 0.25 sr & 0.25 sr & 1.0 sr & 2.6 sr \\
Effective Area $\gt$100 MeV & 100 cm$^2$ & 70 cm$^2$ & 1200 cm$^2$ & \mysim7000 cm$^2$ \\
Angular Resolution\footnote{\scriptsize \em RMS at 500 MeV} & 1\fdg5 & 1\fdg5 & 0\fdg6 & \mysim0\fdg1 \\
Energy Resolution\footnote{\scriptsize \em full-width, half maximum at 100 MeV} & \mysim100\% & 42\% & 18\% & 10\% \\
Point Source Sensitivity\footnote{\scriptsize {\em $\gt$100 MeV, 
10$^{\mbox{\scriptsize{\it 6}}}$ s exposure, unless noted}}
 & 10$^{-6}$
 & 10$^{-6}$ & 10$^{-7}$ 
& $\lt 4$ \by 10$^{-9}$\footnote{\scriptsize \em 1 year, high-latitude, $\gt$100 MeV, $5\sigma$} \\
(photons \perareasec) \\
\end{tabular}
\end{minipage}
\caption[\Gammaray\ telescope performance]{\label{intro:compare}
Performance of four high-energy \gammaray\ telescopes.  Continually 
improving technology is reflected in improving performance from the 
earliest instruments, \sas\ and \cosb, through \egret\ and on to the 
proposed \glast\ instrument.}
\end{table}

Good calibration of the \egret\ instrument was critical to the proper 
understanding of its data.  The calibration was as extensive as its 
literature 
\cite{egretcalibrate93,egretcalibrate88,egretcalibrate89,egretcalibrate87,egretcalibrate92}; 
only the results will be stated here.  There are three areas in which 
we will need to know the instrument performance: the point-spread 
function, or the distribution of the measured \gammaray\ incident 
angles as a function of the true incident angle; the sensitive (or 
effective) area, or the physical area for collecting \gammarays\ 
multiplied by the efficiency, as a function of position on the sky at 
any given time; and the energy dispersion, or the distribution of 
measured energy as a function of the true energy.

These three functions were measured and recorded in tabular form as a 
function of aspect angle and energy.  
Their use in data analysis will be discussed in \sect{stats:insteffects}.
A reasonable approximation to the point-spread width assumes a relatively
simple functional form.  The half-angle which
defines a cone containing \mysim68\% of the \gammarays\ from a point on 
the sky may be taken as
\begin{equation}
\label{intro:containment}
\theta_{68} = 5\fdg85 (E/\hbox{100 MeV})^{-0.534}
\end{equation}
where $E$ is the energy in MeV \cite{egretcalibrate93}.

The sensitive area and energy dispersion are not easily expressible 
in functional form.  Tables of their values were created in machine readable
form, and analysis programs access them directly.

The performance of the \egret\ instrument compares very favorably with 
its predecessors, \sas\ and \cosb.  The order of magnitude increase in 
effective area and improved point-spread function lead to the order of 
magnitude improvement in the point source sensitivity.  A comparison 
of the telescopes, along with the proposed \glast\ telescope, is shown 
in \tbl{intro:compare}.

\section{Successes with \egret}
The {\em Compton Gamma-Ray Observatory} has been very fortunate in successfully
achieving and exceeding its design goals.  Still operating some 7 years after 
launch, it has almost quadrupled its planned lifetime.  While the entire
observatory has been critical in advancing our understanding of astrophysics
from 15~keV to 30~GeV (most notably the shocking revelation from \batse\ that
the mysterious \gammaray\ bursts are isotropically distributed on the sky), 
we will concentrate here on the contributions made by \egret, with a view
toward future advancements to be made by \glast.

\egret\ has significantly improved our understanding of 
pulsars~\cite{joethesis,thompsonreview}.  Six \gammaray\ pulsars have 
been identified by \egret, and their pulse periods have been measured.  
\egret\ observations of pulsars will be discussed in 
\chapt{timevarchap}.  Significant advancements have been made through 
observations of the Galactic~\cite{diffuse4compt} and 
extragalactic~\cite{tomthesis} diffuse background emission.  The 
largest number of identified \egret\ sources are the \gammaray\ 
blazars.  Roughly 60 blazars have been identified above 100~MeV, 
leading to new insights into blazar emission 
mechanisms~\cite{blazarreview}.  While the \batse\ \gammaray\ burst 
measurements were the most revolutionary discovery, \egret\ has made 
significant contributions to our understanding of \gammaray\ bursts at 
the highest energies~\cite{grbreview}.  \Gammaray\ bursts will be 
discussed in \chapt{grbchap}.

\cgro\ has yielded a wealth of information about the \gammaray\ sky.  Much of that 
information has directly increased our understanding of astrophysical systems.
In particular, \egret\ has given us an unprecedented view of the high-energy 
\gammaray\ sky.  
\egret\ has identified a great number of new sources;
the launch of \glast\ will give us a tool to understand what exactly it is that 
we have found.

%% file: Stats.tex
\chapter{Statistical Methods in \Gammaray\ Astronomy} 
Optical astronomy has long been famous for breathtaking images of 
distant galaxies, star-forming clouds, and beautiful nebul\aeind.  
While \gammaray\ astronomy can produce equally beautiful results, the 
nature of the photons and the instrument yield data which must be 
analyzed very differently than data from other wavelengths.  The first 
major difference is the sparsity of the photons; integration times of 
days are usually required to observe all but the brightest sources.  A 
multitude of astrophysical conditions also affect the nature of the 
data.  Cosmic ray interactions with Galactic dust and gas produce 
diffuse background \gammarays.  Different energy generation mechanisms 
produce different spectral profiles across the energy range observed 
by \egret.  Several specific instrumental responses also shape the 
data analysis process.  A finite point-spread function means that 
\gammaray\ directions will be determined to within a statistical 
distribution around the true source direction.  Instrumental 
sensitivity to three orders of magnitude in energy across the field of 
view of more than a steradian is not uniform as the platform orbits at 
17,000 miles per hour.  Meanwhile, our estimates of all these 
parameters depend on the \gammaray\ energy, which is only known to 
limited accuracy.

Despite these impediments to observation, \egret\ has been very successful
in making \gammaray\ observations.  This chapter will examine the nature
of the \egret\ data, and the statistical methods used to analyze it.

\section{The Nature of the \Gammarays}
\begin{figure}
\centering
\includegraphics[width = 5.5 in]{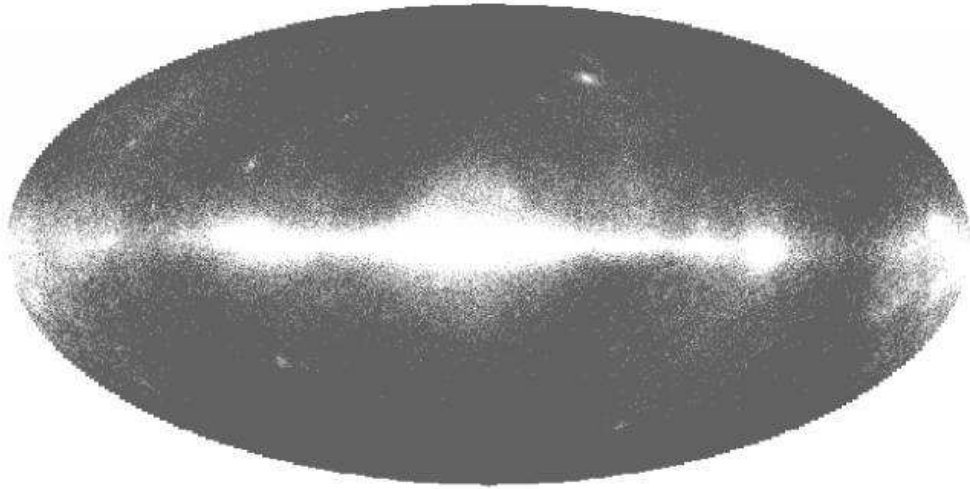}
\caption[Full sky photon counts]{\label{stats:allsky_counts}
All \gammarays\ above 100~MeV measured by \egret\ in phase I, II, and III,
binned at a scale of 0\fdg5.}
\end{figure}

Several major features of the high-energy \gammaray\ sky are evident 
from the simplest possible examination.  \fig{stats:allsky_counts} 
shows the raw photon counts for the whole sky in 0\fdg5 \by 0\fdg5 bins 
in Galactic coordinates.  The Galactic center is evident at the center 
of the map, as is the Galactic disk.  There are several bright spots 
in the plane, as well as some evident high-latitude sources.  We will 
first examine the diffuse background, then see how statistical methods 
along with understanding of the background will help us identify point 
sources and estimate their locations and fluxes.

\subsection{Diffuse Background}
Almost two decades before the launch of \sas, Hayakawa \cite{hayakawa} 
predicted that high-energy \gammarays\ would be produced as cosmic 
rays interacted with interstellar gas yielding pions, which would 
decay, directly or indirectly, into \gammarays.  Indeed, experience 
with \sas, \cosb, and \egret\ has shown strong correlations of the 
diffuse \gammaray\ background at low Galactic latitudes with known 
Galactic structural features such as the spiral arms 
\cite{sas2,hartman_diffuse,hans_diffuse}.  Based on indications that 
the diffuse \gammaray\ flux observed by \sas\ and \cosb\ was 
approximately the intensity and shape expected from cosmic ray 
interactions with interstellar matter, a model of the diffuse flux was 
made for the purposes of \egret\ data analysis 
\cite{bertsch_diffuse,hunter_diffuse}.

There are three main processes by which diffuse \gammarays\ are generated.
The dominant process above \mysim70~MeV is the decay of pions.  Pions are
produced when cosmic-ray protons interact with dust particles or gas.
These pions then decay to high-energy \gammarays.  Below \mysim70~MeV,
\gammarays\ can be produced by either bremsstrahlung of cosmic rays in
interstellar clouds or by inverse-Compton upscattering of low-energy 
photons. 

A good model of the Galactic diffuse \gammaray\ intensity thus 
requires a good model of the distribution of interstellar matter in 
the Galaxy, and a good model of the cosmic ray flux throughout the 
Galaxy.  The first may be well approximated using maps of the 
distribution of hydrogen, which comprises most of the interstellar 
matter in the Galaxy.  Atomic hydrogen has been carefully mapped with 
observations of the 21~cm hyperfine transition line.  Molecular 
hydrogen is more difficult to map, but may be approximated by assuming 
that CO is a good tracer, easily identified by its 2.6~mm emission 
line.  A constant ratio of CO to molecular hydrogen is typically assumed 
throughout the galaxy.

\begin{figure}
\centering
\includegraphics[width = 5.5 in]{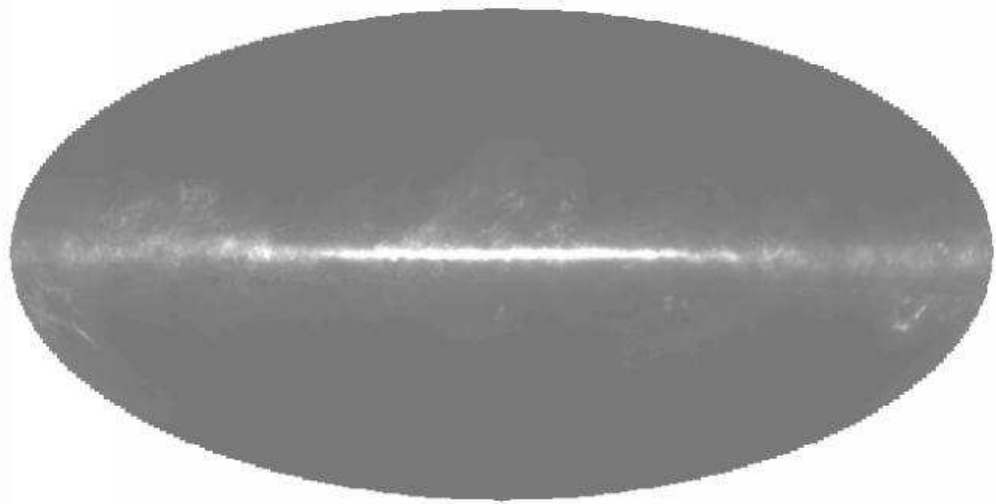}
\caption[Full sky galactic diffuse map]{\label{stats:allsky_gas}
The assumed Galactic contribution to the diffuse \gammaray\ background, 
taken from measurements of atomic and molecular hydrogen, and convolved
with the \egret\ point-spread function.  This is a map of $G_{ij}$, as
defined in \eq{stats:gasmapeq}.}
\end{figure}

Much more difficult is the estimation of the Galactic cosmic ray flux.  
Since the flux cannot be directly measured, assumptions about the 
distribution of cosmic rays must be made.  For the purposes of \egret\ 
analysis, Bertsch \cite{bertsch_diffuse} and Hunter 
\cite{hunter_diffuse} assume that cosmic rays are in dynamic 
equilibrium with the interstellar magnetic pressure and the 
gravitational pressure of the galactic matter.  These assumptions,
convolved with the instrument point-spread function (\sect{stats:egretlike}),
result in the map of diffuse Galactic \gammarays\ shown in 
\fig{stats:allsky_gas}, which is in good agreement with the observed 
intensity map shown in \fig{stats:allsky_intens}.  Of course, at high 
galactic latitudes the diffuse background is primarily 
extragalactic~\cite{kumar_extragalactic}, although there is 
significant galactic diffuse background at high latitude.  Some or all of 
this extragalactic background is due to a large number of weak sources, while some 
of it may be due to a truly diffuse background~\cite{tomthesis}.

\subsection{Spectral Differences}
\egret's broad energy range allows the spectra of sources and the 
diffuse background to be measured.  The spectrum of almost every 
\egret\ source is well fit by a power law.  Since \gammarays\ are so 
sparse, the power laws are usually quoted as a function of the number 
of photons, instead of a function of energy as is sometimes used for 
other wavelengths.  The form is then
\begin{equation}
I(E) = I_0 E^{-\alpha}\:\: \hbox{photons \perareasecmev}
\end{equation}
where $I(E)$ is the differential photon flux.  The spectral index 
$\alpha$ is close to 2.0 for most sources, though it can be as low as 
1.42 for pulsars.  The Galactic diffuse background has a bit softer 
spectrum---about 2.1.

\section{Instrumental Effects}
\label{stats:insteffects}
The quality and nature of the data depend equally on the photons and the 
instrument which observes them.  Any attempt to analyze the data must 
consider the instrumental response as the basis for an analysis 
method.  The three main aspects of the instrumental response that we 
must consider are the point-spread function, the sensitive area, and 
the energy dispersion.  There is no reason, {\em a priori}, to expect 
that the instrument response function should separate cleanly into 
these three functions.  However, it offers great simplification to the 
data analysis, and in practice seems to be a good approximation.

\subsection{Point-Spread Function}
It is important to be able to quantify the ability of a \gammaray\ 
telescope to correctly reconstruct the true incident direction of a 
\gammaray.  To precisely define this, we will distinguish between the 
``point-spread density'' and the ``point-spread function.''  The 
point-spread density, or \psd, refers to the probability density 
distribution of incident \gammaray\ directions measured by the 
instrument from a point source.  This distribution may in general be a 
function the true position of the point source (the inclination and 
azimuth relative to the centerline of the telescope) and the energy of 
the \gammaray:
\begin{equation}
\psd = \psd(\theta, \phi; \theta_0, \phi_0, E_0)
\end{equation}
where $\theta_0, \phi_0$ represents the true source position and  $E_0$ is the true
\gammaray\ energy.  The apparent \gammaray\ direction is ($\theta, \phi$).
Often, we will require a probability, as opposed to a probability 
density.  The differential probability of measuring a photon in a 
differential element $d\theta\,d\phi$ is given by
\begin{equation}
	dP = \psd(\theta, \phi; \theta_{0}, \phi_{0}, E_{0}) \sin \theta 
	\, d\theta\, d\phi
\end{equation}
The point-spread function is the differential probability, often 
integrated over azimuth and energy to yield a function of inclination, 
as in \eq{stats:psfeq}.

The \egret\ PSD was measured at the Stanford Linear Accelerator Center in 1986.
A beam of electrons with tunable energies between 650~MeV and 30~GeV was 
back-scattered off pulsed laser photons.  \Gammarays\ were produced between
15~MeV and 10~GeV by inverse-Compton scattering \cite{egretcalibrate87}.  The
point-spread density was measured as a function of apparent \gammaray\ position
for 10 discrete energies, 5 inclination angles ($\theta_0$) and 3 azimuthal 
angles ($\phi_0$).  The resulting tables yield the relative probability of detecting
a photon at ($\theta, \phi$) assuming values of the other three parameters.  In 
addition, \egret\ operates in up to 87 different ``modes,'' corresponding to 
different triggering criteria.\footnote{These different modes 
correspond to allowing only photons from certain broad regions of the 
sky as defined by coincidence of different combinations of 
time-of-flight tiles.}  These modes are designed to maximize the
operating field of view, even when part of the geometric field of view is
obscured by the Earth or its limb.

\subsection{Sensitive Area}
A second function which is clearly critical for data analysis is the sensitive
area of instrument.  The sensitive area (or effective area) is the projected
 area of the detector multiplied by its efficiency.  It too is a function of
incident \gammaray\ parameters:
\begin{equation}
\sa = \sa(\theta_0, \phi_0, E_0)
\end{equation}
Clearly, the sensitive area of the instrument is also dependent on the 
instrument mode.

\subsection{Energy Dispersion}
Finally, the analysis must consider energy dispersion.  The energy 
dispersion function gives the distribution of measured energy for a 
given true energy.  The measured energy varies from the true energy 
because of noise in the photomultipliers, fluctuations in the shower 
leakage from the calorimeter, and incomplete correction for energy 
losses elsewhere in the instrument.
\begin{equation}
\ed = \ed(E; E_0, \theta_{0}, \phi_{0})
\end{equation}

Taken together, these three functions yield the point-spread width 
approximation in \eq{intro:containment} in the following way.  
We first notice from the 
calibration data that the point-spread function is roughly azimuthally 
symmetric, and that it does not vary widely with the true inclination
angle.  Then we can find
\begin{equation}
	\label{stats:psfeq}
\psf(\theta) = \frac{2 \pi}{N} \int_{E = E_{{min}}}^{E_{{max}}}
    \int_{E^{\prime} = 0}^{\infinity} {E^{\prime}}^{-\alpha} \psf(\theta, E^{\prime}) 
	\ed(E^{\prime}, E) dE^{\prime}dE
\end{equation}
where $E$ is the measured \gammaray\ energy, $E^{\prime}$ is the true 
\gammaray\ energy, and $N$ is a normalization factor.  The deviation 
between the apparent incident angle and the true incident angle is 
$\theta$.  The point-spread function $\psf(\theta)$ is the integral of 
the true-energy dependent point-spread function, weighted by the 
spectrum, integrated over the measured energy band from $E_{min}$ to 
$E_{max}$, and integrated over all true energies, weighted by the 
energy dispersion function.  This reflects the fact that there is 
some probability that a \gammaray\ of any given true energy will have 
a measured energy between $E_{min}$ and $E_{max}$.

This integral was done numerically for a number of energy bands, and 
a Gaussian fit to the results led to \eq{intro:containment} 
\cite{egretcalibrate93}.

\section{Likelihood Analysis}
\label{stats:likelihood}
The sparsity of astrophysical \gammarays\ and the complicated instrumental
response of the \egret\ instrument suggests statistical data analysis that
functions at the photon-by-photon level, taking into account backgrounds
and the instantaneous instrument state to extract the most information from
each photon.  Early analyses of \cosb\ data were based on a cross-correlation
method \cite{crosscor}.  However, this method could not easily handle the
highly structured background that is typical of high-energy \gammaray\
astrophysics.  Later, a maximum likelihood technique was brought to bear
on \cosb\ data with much greater success \cite{pollock}.  Based on this 
success, maximum likelihood techniques were adopted for use with 
\egret~\cite{like}.

The central idea of likelihood analysis is very simple.  Given a set of models,
we wish to find the model which is most likely to be responsible for the 
observed data.  The {\em likelihood} is defined as the probability of the observed
data, given a choice of model.  The likelihood is written as follows:
\begin{equation}
\like(D | M) \nonumber
\end{equation}
where $D$ represents the observed data, and $M$ the model.  
Quantities to the right of the $|$ sign are taken as given and fixed, making
the likelihood a conditional probability.  The {\em maximum
likelihood} method determines the best model by maximizing this likelihood
function.  A special, but very common, case is that of a parameterized model.
For example, consider that we wish to measure the flux of a \gammaray\ 
source.  We will imagine an idealized detector with 100\% efficiency and
an angular area of $\alpha$~sr from the source.  The number of \gammarays\
emitted in a unit time is Poisson distributed, so the likelihood of measuring
$n^\prime$ photons from a source with intensity $\mu$ is given by
\begin{equation}
\like(D | \mu) = \frac{\alpha}{4 \pi} \frac{e^{-\mu} \mu^n}{n!}
\end{equation}
where $n$ is the number of photons emitted from the source, and
$n^\prime = (\alpha/4\pi) n$.  To find our best estimate
of $\mu$, we maximize this likelihood.  In fact, we will find it more
convenient (and mathematically equivalent) to maximize the logarithm
of the likelihood. 
\begin{equation}
\label{intro:loglikeex}
\ln \like(D | \mu) = -\mu + n \ln \mu + \ln \left(\frac{\alpha}{4 \pi n!} \right)
\end{equation}
The last term is constant, and thus for maximization purposes can be ignored.
Setting the derivative of \eq{intro:loglikeex} to zero, we find
\begin{equation}
0 = -1 + n/\mu \Longrightarrow \mu = n = \frac{4\pi}{\alpha} n^\prime
\end{equation}\index{$\Longrightarrow$}
Unsurprisingly, in this simple example, we find that the number of photons
measured per unit time divided by the subtended angle fraction is the best
estimate of the flux.

One other example is illustrative.  Let us assume we have some Gaussian 
process with a constant mean and a known variance $\sigma^2$.  
A series of observations $\myvec{x}$ is made,
and the most likely mean value $\mu$ is desired.  The likelihood function is
\begin{equation}
\like(\myvec{x} | \mu) = \prod_i e^{- \frac{(x_i - \mu)^2}{2 \sigma^2}}
\end{equation}
Taking the logarithm, we have
\begin{equation}
\ln \like(\myvec{x} | \mu) = -1/2 \sum_i \frac{(x_i - \mu)^2}{\sigma^2}
\end{equation}
We notice that $-2 \ln \like$ is formally equivalent to \chisq.  In fact,
this is an example of a general result:  in the limit that all distributions
involved are Gaussian, the maximum likelihood result is the same as the
\chisq\ minimizing result.  This is an example of a more general result
known as Wilks' Theorem~\cite{wilks}; it will be described more thoroughly
in \sect{stats:hypotest}.

\paragraph{Maximum Likelihood Confidence Regions.}
\label{stats:likeconf}
Maximum likelihood methods also yield confidence regions.
Following Eadie, \etal\ \cite{eadie}, we first consider the case of a Gaussian
distribution with unit variance and unknown mean $\mu$.  The likelihood is

\begin{eqnarray}
\like(\bar{x} | \mu) & = & (2 \pi)^{-N/2} \exp
 \left[- \frac{1}{2} \sum_{i=1}^N (x_i - \mu) ^2 \right] \nonumber \\
& = & (2 \pi)^{-N/2} \exp \left[ - \frac{N}{2} (\bar{x} - \mu)^2 \right]
\exp \left[ - \frac{1}{2} \sum_{i=1}^N (x_i - \bar{x})^2 \right]
\end{eqnarray}
The $\ln \like$ is thus a parabola in $\mu$ of the form $- \frac{N}{2} 
(\mu - \bar{x})^2$.  In the case that $N =1 $, let $\ln \like \geq 
-1/2$.  This corresponds to the interval $-1 \leq \mu - \bar{x} \leq 
+1$.  From the properties of the normal distribution, we know that 
this must contain 68.3\% of the distribution.  
Similarly, the interval corresponding to $\ln 
\like \geq -2$ contains 95.5\% of the distribution.

This would be merely a curiosity if not for the following.  Suppose that the
likelihood function is a continuous function of $\mu$, with only one maximum.
In that case, we may reparameterize our observed variable in terms of a new
variable $g(\mu)$ such that the likelihood as a function of $g$ {\em is}
parabolic.  We may now find the confidence region in $g$ as we did above.  
Given that region, we may invert the reparameterization to find a confidence
interval in $\mu$.  Furthermore, we notice that the function is the 
same, whether it is parameterized by $\mu$ or $g$.  That is,
\begin{equation}
\ln \like(\bar{x} | \mu) = \ln \like(\bar{x} | \mu(g))
\end{equation}
Thus, we can find the confidence region in $\mu$ directly by determining the
point at which the $\ln \like$ has decreased by 1/2 for 68\% or 2 for 95.5\%,
without ever finding the reparameterization $g$.  
However, note that the interval is central in $g$ since the likelihood 
as a function of $g$ is a 
Gaussian, but it is not necessarily central in $\mu$.

\section{Applying Likelihood Analysis to \egret}
\label{stats:egretlike}
We can see now how to proceed in \egret\ data analysis.  Take all the 
data accumulated in some time period.  The likelihood of that data is 
the product over differential elements of angle and energy of the 
Poisson probability density of detecting the photons, given the rate of photons 
times the probability of detecting each photon.  The rate $\mu$ can be 
expressed as
\begin{eqnarray}
\label{stats:rate}
	\lefteqn{\mu(\myell, b, E; \myell_{0}, b_{0}) = } \\
& &	\int_{E=0}^{\infinity}
	\left[ I(\myell_{0}, b_{0}) \psf(\myell, b, E^{\prime}; \myell_{0}, b_{0}) +
	B(\myell, b) \right] 
	\sa(\myell, b, E^{\prime}) \ed(E^{\prime}, E)  dE^{\prime} \nonumber
\end{eqnarray}
The rate is a function of the measured energy and position on the 
sky.  The source intensity $I$ depends only on the true source 
position ($\myell_{0}, b_{0})$.  The background $B$ is assumed to have
already been convolved with the point-spread function.  The three instrument
functions depend on the instrument mode $m$ as well; this dependence will be
suppressed for clarity.

The likelihood is then the product over all differential parameter 
elements of the Poisson probability of measuring (or not measuring, as 
was the case) a photon in that element, given the rate in that element 
from \eq{stats:rate}. The appearance of derivatives of products encourages us 
to use the logarithm of the likelihood.  Denoting the integrated rate 
over measured energies as $\bar{\mu} = \int_{E_{min}}^{E_{max}} 
\mu(\myell, b, E)\,dE$, the log likelihood becomes
\begin{equation}
\ln \like(\myell_{0}, b_{0}) = \int_{\myell} \int_{b} 
\left[ -\bar{\mu}(\myell, b; \myell_{0}, b_{0}) + 
     n \ln \bar{\mu}(\myell, b; \myell_{0}, b_{0})  \right]  d\myell\, db
\end{equation}
where $n$ is the number of photons observed in a differential element 
$d\myell\, db$.  Since the element is differential, this must be either 1 
or 0.  The integral thus divides into an integral and a sum:
\begin{equation}
	\label{stats:likeeq1}
	\ln \like(\myell_{0}, b_{0}) = - \int_{\myell} \int_{b} \bar{\mu}(\myell, b; 
	\myell_{0}, b_{0}) d\myell\, db + \sum_{i=1}^N \ln \bar{\mu}(\myell_i, b_i; 
	\myell_{0}, b_{0})
\end{equation}
where the sum is evaluated for the parameters of each photon.
While this looks fairly simple, $\bar{\mu}$ is quite a complicated 
object, implicitly containing four integrals.  We would like to 
maximize \eq{stats:likeeq1} over $\myell_{0}$ and $b_{0}$, the source 
position.  Computationally, this is a herculean task, requiring 
evaluation of a six-dimensional integral at each trial point ($\myell_0, b_0$).  
Clearly, some simplification is necessary.

\subsection{Binned Likelihood}
The most obvious simplification is to give up on individual photons, 
and create binned maps.  While binning is always undesirable, as it 
loses information in the data, in this case binning is minimally 
undesirable.  Our derivation above considered differential elements 
in the parameters; essentially, we let the bin size go to zero.  It 
has been shown that using a finite but small bin size speeds 
computation dramatically at very little expense of accuracy 
\cite{billthesis}.

\begin{figure}
\centering
\includegraphics[width = 5.5 in]{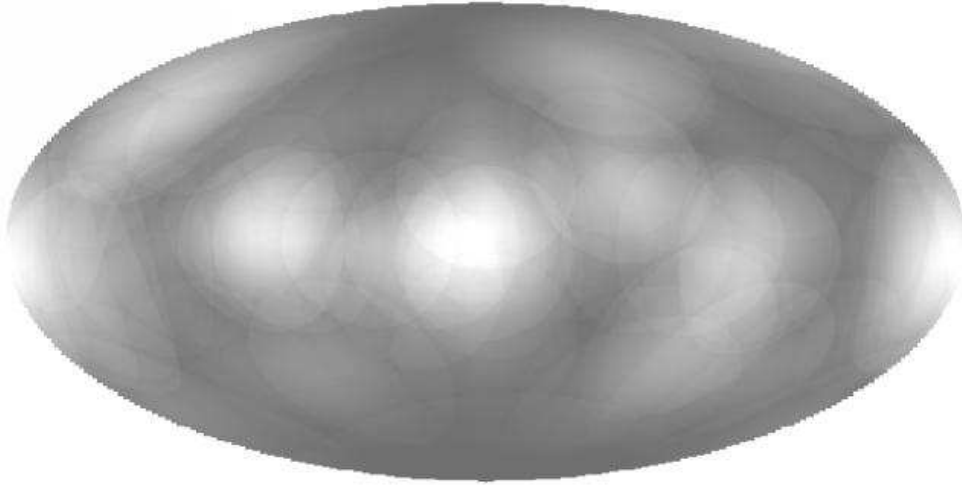}
\caption[Full sky exposure map]{\label{stats:allsky_exposure}
Combined \egret\ exposure in phase I, II, and III\@.  Phase I was an all-sky
survey, with roughly even exposure.  Phases II and III consisted of pointed 
observations for sources of interest.  Lighter shade corresponds to more
exposure.}
\end{figure}

A standard analysis program called \likeprog\ was developed for the 
map-based likelihood analysis of \egret\ data \cite{like}.  \likeprog\ 
considers the total expected \gammaray\ rate in each pixel, typically 
0\fdg5 \by 0\fdg5 on the sky.\footnote{ This scale is somewhat 
arbitrary.  It was based largely on the scale to which the Galactic 
hydrogen has been mapped, as well as to ensure sufficient photons in 
each bin.  It was not chosen to optimize the likelihood method.  
Nevertheless, for most energies the bin size is much smaller than the 
instrument point-spread width.} This rate is the sum of the expected 
rates from the isotropic background, the galactic background, and a 
point source.  In order to estimate rates for binned maps, the photons 
from any desired observation interval are binned as in 
\fig{stats:allsky_counts}.  In addition, the instrument exposure to 
the sky must be calculated.  This is a function of the amount of 
observing time and the sensitive area during that time.  Let 
$T(\theta, \phi, m)$ be the amount of observing ``livetime,'' (that 
is, elapsed viewing time that the instrument was active, excluding 
occultations and instrument dead time) for a location $(\theta, \phi)$ 
on the sky in an instrument observing mode $m$ between time $t_1$ and 
$t_2$.  Only the total observing time in each mode is relevant; it 
need not be contiguous.  Then the exposure ${\cal E}(\Delta E; \theta, 
\phi)$\index{${\cal E}$} for a given measured energy range $\Delta E$ 
to a point $(\theta, \phi)$ on the sky is
\begin{equation}
{\cal E}(\Delta E; \theta, \phi, t_1, t_2) = \sum_m T(\theta, \phi, m; t_1, t_2) 
\bar{A}(\Delta E; \theta, \phi, m)
\end{equation}
where
\begin{equation}
\bar{A}(\Delta E; \theta, \phi, m) =
\int_{E = E_{min}}^{E_{max}} \int_{E^{\prime} = 0}^{\infinity}
{E^{\prime}}^{- \alpha} \sa(E^{\prime}; \theta, \phi, m) \ed(E, E^{\prime})
dE^{\prime}\, dE
\end{equation}
This exposure explicitly depends on the spectral index of the source or
background, whichever is being observed.  This is the first example of a
continuing difficulty; the spectral index is required to calculate the exposure,
but it is not known until after the flux is determined.  In principle, the 
spectral index should be allowed to vary everywhere during the maximization 
process.  An approximation would be to iterate.  However, since most spectral
indices are very nearly 2.0, and varying the index requires recalculating the 
exposure maps (a computationally expensive prospect), a constant index is 
assumed.  The exposure can be calculated for each pixel in the standard map;
an example is shown in \fig{stats:allsky_exposure}.

To allow for the degrees of freedom in the background maps described above,
the background is assumed to be a linear combination of isotropic and 
galactic background whose coefficients will be optimized.  The isotropic
diffuse count estimate for a given bin is $g_b {\cal E}_{ij}$\index{${\cal E}$}, 
where $g_b$ (``g-bias'') is the coefficient to be fit in units of
photons \perareasecsr, and 
${\cal E}_{ij}$\index{${\cal E}$} is the instrument exposure to pixel $i, j$ 
in units of cm$^2$~s~sr. 

The Galactic component of the rate will depend on both the diffuse map 
and the instrument point-spread function, as well as the exposure.  
Denoting the binned radio diffuse gas map as $f_{ij}$, the rate due to 
the Galactic diffuse background is \index{${\cal E}$} 
\begin{equation} 
\label{stats:gasmapeq}
G_{ij} = \frac{\sum_{kl} f_{kl} {\cal E}_{kl} \psf(\theta_{ij,kl})} 
{\sum_{kl} \psf(\theta_{ij,kl})}
\end{equation} 
where $\theta_{ij,kl}$ is the angular distance between pixel $i,j$ and 
pixel $k,l$.  The denominator is a normalization factor, necessary since
the sums over $k$ and $l$ may not be over the entire sky.  Pixels far 
from the point of interest will contribute negligibly to the estimation 
of source flux and location, but it is critical to good flux measurements
to have a normalized point-spread function.  Therefore, we allow analysis
of only the pixels within some radius of analysis \ranal, but renormalize
the point-spread function.  In addition, to allow for the unknown cosmic
ray flux and the CO/H$_2$ ratio, we will allow a constant multiplier
$g_m$ (``g-mult'') to be optimized as well.

So, then, given $k$ sources in the field of view, the expected number of
counts in bin $i, j$ is
\index{${\cal E}$} \begin{equation}
\label{stats:genrate}
\mu_{ij} = g_m G_{ij} + g_b {\cal E}_{ij} + \sum_k c_k \psf(\theta_{ij, k})
\end{equation}
where $\theta_{ij,k}$ is the angular distance between the position of source 
$k$ and pixel $i,j$.  This count estimate is a function of $3 k+2$ parameters:
the $k$ source strengths, latitudes, and longitudes; $g_m$, and $g_b$.  
The fit values of $g_m$ are 
consistently in the range 0.92--1.08.  The fit level of $g_b$ is usually
around 2 \by 10$^{-5}$ photons \perareasecsr\ \cite{joethesis}.

\subsection{Maximizing the Likelihood with \likeprog}
\begin{figure}
\centering
\includegraphics[width=5.5 in]{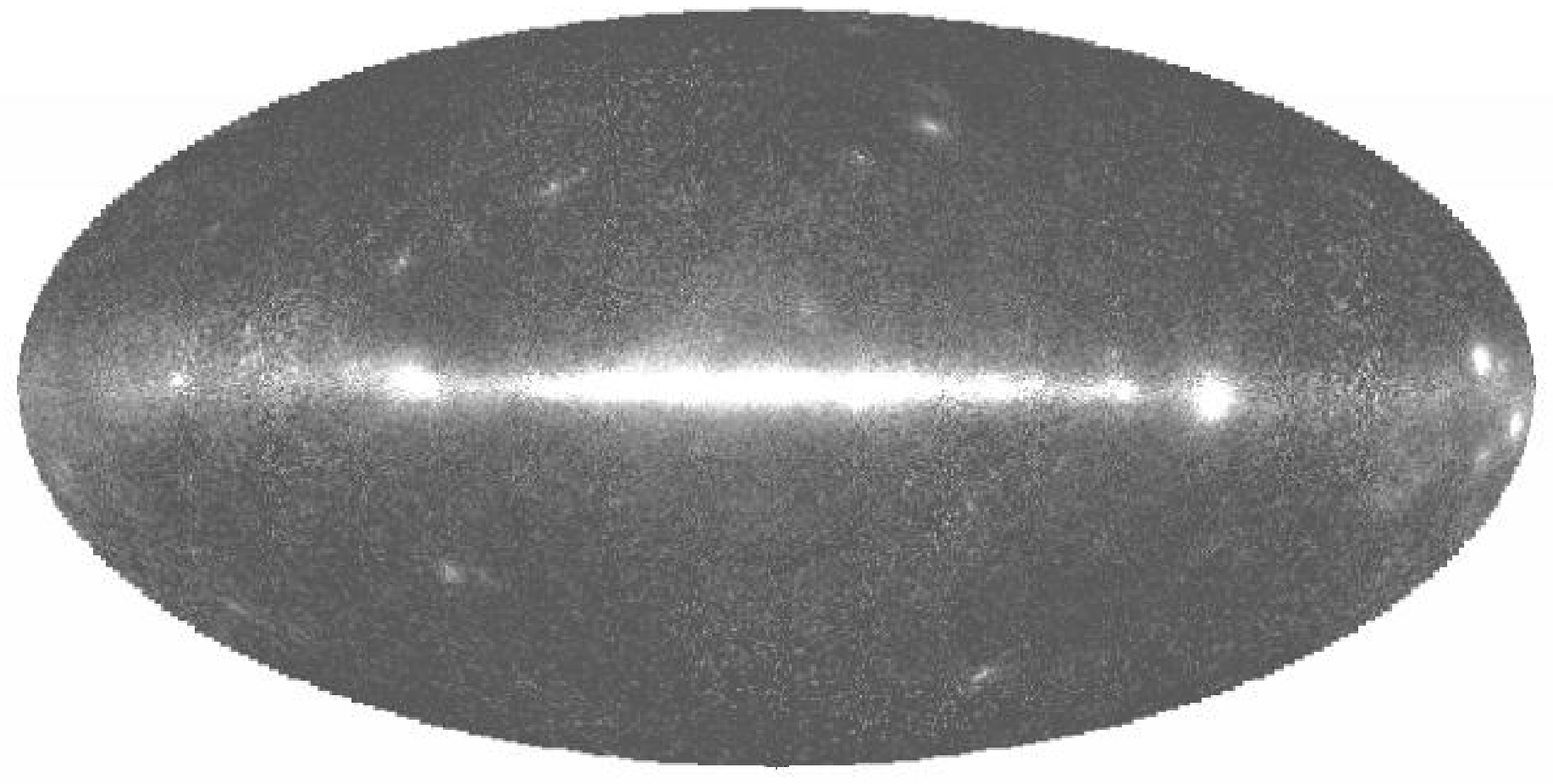}
\caption[Full sky intensity map]{\label{stats:allsky_intens}
\Gammaray\ intensity as measured by \egret.  The counts map in 
\fig{stats:allsky_counts} is divided by the exposure map in 
\fig{stats:allsky_exposure} to yield the \gammaray\ intensity. }
\end{figure}

Just as in the exact case, the distribution of photon counts in each 
bin is Poisson.  The likelihood of a given map, then, is:
\begin{equation}
\like(D | c_k, \myell_k, b_k, g_m, g_b) = 
\prod_{ij} \frac{\mu_{ij}^{n_{ij}} e^{-\mu_{ij}}}{n_{ij}!}
\end{equation}
with $\mu_{ij}$ given by \eq{stats:genrate}.  The maximum likelihood 
estimates of $c_k, \myell_k, b_k, g_m,$ and $g_b$ are simultaneously
solving the the set of equations
 \index{$_{\hat{\mu}}$}\begin{eqnarray}
\left. \dbyd{c_k} \ln \like(\mu) \right|_{\mu = \hat{\mu}} & = & 0 \\
\left. \dbyd{\myell_k} \ln \like(\mu) \right|_{\mu = \hat{\mu}} & = & 0 \\
\left. \dbyd{b_k} \ln \like(\mu) \right|_{\mu = \hat{\mu}} & = & 0 \\
\left. \dbyd{g_m} \ln \like(\mu) \right|_{\mu = \hat{\mu}} & = & 0 \\
\left. \dbyd{g_b} \ln \like(\mu) \right|_{\mu = \hat{\mu}} & = & 0 
\end{eqnarray}
where $\hat{\mu}$ is optimized to satisfy all these conditions.
This is just the multi-dimensional maximization of the likelihood
function---there are $3 k + 2$ equations to solve simultaneously.  For this reason,
it is usually computationally much faster (although less accurate) to find
the brightest source first, then fix its parameters and fit the next source.
Once the locations of the sources are established, \eq{stats:genrate} divided
by the exposure map yields an intensity map of the sky in 
\fig{stats:allsky_intens}; equivalent to the
optical image observed simply by peering through the eyepiece of an 
optical telescope. 

\section{Parameter Estimation vs. Hypothesis Testing}
\label{stats:hypotest}
The foregoing discussion concerned the estimation of the 
position and flux of a source which was assumed to exist.  The maximum likelihood
technique not only allows for parameter estimation as described above
but also for hypothesis testing.  Hypothesis testing estimates the
significance of the source detection.  Highly significant sources 
are reliable; low significance sources may be statistical fluctuations.

Consider two hypotheses, $M_0$ and $M_1$.  For concreteness, let us 
assume that $M_0$, the null hypothesis, states that there is no source 
in the field of view.  Further, we assume that $M_1$ states that there 
is a source with non-zero flux in the field of view.  If there are $n$ 
degrees of freedom in $M_0$, and $m$ degrees of freedom in $M_1$, then 
Wilks' Theorem \cite{wilks} states that $-2$ times the logarithm of 
the likelihood ratio is asymptotically distributed as 
$\chi^2_{(m-n)}$:
\begin{equation}
\label{stats:wilks}
-2 \ln \frac{\like(D | M_0)}{\like(D | M_1)}\; \mysim\; \chi^2_{(m-n)}
\end{equation}
 In the case of a source whose position is known from other 
experiments, there are 3 degrees of freedom in $M_1$
 ($c_k, g_m, g_b$) and 2 degrees of freedom in $M_0$ ($g_m, g_b$).  
Thus, the likelihood is distributed as $\chi^2_1$.  
We define the \egret\ test statistic 
$\TS \equiv\hbox{\index{$\equiv$}} -2 (\ln \like_0 - \ln \like_1)$.
From the definition of the $\chi^2$ distribution, this implies that the
significance of a detection is given by $\sqrt{\TS} \sigma$ in the standard
nomenclature.  Thus, a $\TS = 16$ source is detected at
the 4$\sigma$ level.   Of course, that is for a single trial, and in fact
many sources observed by other telescopes have been searched for with \egret.

The case of previously unknown sources is a little more complicated.  $M_1$ now 
has 5 degrees of freedom: $c_k, \myell_k, b_k, g_m, g_b$.  But the 
null hypothesis is degenerate in the position---the likelihood is 
independent of position when the flux is zero.  In that case, Wilks' 
Theorem does not hold, so the distribution is not known.  Some 
theoretical work has been done to determine this distribution 
\cite{bill_chi2}.  Based on Monte Carlo simulations, \egret\ sources 
are accepted as detections if, for sources at Galactic latitudes 
$|b|\, \gt\, 10\deg$, $\sqrt{\TS}\, \gt\, 4$, and for sources at 
Galactic latitudes $|b|\, \lt\, 10\deg$, $\sqrt{\TS} \,\gt\, 5$.  
Monte Carlo simulations have shown that for a perfect background 
model, roughly one spurious excess with $\TS\, \gt\, 16$ will be 
detected in an analysis of the entire sky \cite{like,joethesis}.

\section{Bayesian Methods}
\label{stats:bayes}
Despite their formal similarity, conceptual differences between maximum
likelihood analysis and Bayesian methods have kept the latter relegated,
for the most part, to the statistical backwaters of astrophysics.  Some
recent work using Bayesian methods has yielded useful 
results~\cite{bayes_blocks,loredo90,loredo92};
since Bayesian methods will be an appropriate alternative framework for 
later chapters, a brief overview will be presented here.

Traditional, or {\em frequentist}, statistics are predicated on the notion
that extreme values of some function of the data given a null hypothesis indicate that
the null hypothesis is probably wrong, and therefore the test hypothesis is 
probably true.  Much of the confusion of the general population about ``Statistics''
is due to the two twists involved in frequentist analysis: first of all, 
the goal is to disprove the thing we suspect false, rather than find evidence for
what we suspect true; and second, this is done
by considering all possible outcomes from an ensemble of data sets.  In the
case of likelihood statistics, we calculate the statistic ($-2 \ln \like$)
and compare it to the distribution expected from the ensemble of data sets that
might be generated if the null hypothesis were true.  If the measured value of
the likelihood is extreme, according to this distribution, we claim that the
null hypothesis has been excluded to some confidence level.

\subsection{Bayes' Theorem}
In contrast, the Bayesian method demands that we stay at all times in the realm
of probability: specifically, the probability that the test hypothesis is true.
To develop the mathematics, let us begin with what Scargle \cite{bayes_blocks}
calls the ``obviously true'' form of Bayes' Theorem:
\begin{equation}
\label{stats:obtrue}
P(M | D) P(D) = P(D | M) P(M)
\end{equation}
Formally, this follows from the definition of conditional probabilities.  The
probability of some statement $M$ being true, given that another statement
$D$ is true is the joint probability of $M$ and $D$.  Any joint probability
may be expressed as the probability of one statement times the conditional 
probability of the other.  Or stated another way, the probability of $A$ and $B$
equals the probability of $B$ and $A$. Thus Bayes' Theorem is proved.

Clearly, the notation in \eq{stats:obtrue} is suggestive.  As we have identified
above, the probability of the data, given a model, is known as the likelihood of
the data.  \Eq{stats:obtrue} points out that while we have been calculating
likelihoods, what we really desire is $P(M|D)$; that is, given the data that
we have, what is the probability that a given model is true?  This is the 
question which Bayesian methods set out to answer.  To that end, we rearrange
\eq{stats:obtrue} into a more useful form.
\begin{equation}
\label{stats:bayesthm}
P(M | D) = \frac{\like(D | M) P(M)}{P(D)}
\end{equation}
\Eq{stats:bayesthm} gives us exactly what we want: the probability of a model 
being true, given the data that we have observed.  We do not resort to any
hypothetical ensembles of data, and more importantly, we make direct statements
about the model in question.

Analogously to our likelihood calculations in \sect{stats:likelihood}, the model
in question may be one of a discrete series, or it may be parameterized.  In the
case that the model is a function of a  continuous parameter, then the 
left side of \eq{stats:bayesthm}, known as the {\em posterior} probability, becomes
a function $P(M(\theta) | D)$.  It is interpreted as the probability that the 
true value of the parameter $\theta_0$ is given by $\theta$.

Given the importance of the posterior probability, it behooves us to understand
the right side of \eq{stats:bayesthm}.  The likelihood has been fully discussed.
$P(M)$ is known as the {\em prior} probability of the model.  Except (apparently)
in the case of quantum mechanics, probabilities are used to represent ignorance.
We do not know the true value of a parameter, or which model is correct, so we
assign probabilities to represent the knowledge that we do have.  $P(M)$ represents
the knowledge we have about the system before we receive the data $D$.  A common
example is that of a parameterized model in which we know that the true value
of the parameter $\theta$ lies somewhere between $\theta_{min}$ 
and $\theta_{max}$\footnote{It will often be possible to take the limit of our
results as $\theta_{min} \rightarrow -\infinity$\index{$\rightarrow$} 
and $\theta_{max} \rightarrow \infinity$\index{$\rightarrow$}.
}\index{\footnotemark[2]}.
The prior is then flat over the interval $[\theta_{min}, \theta_{max}]$ and
zero elsewhere.  In other situations, it may be more appropriate to take a 
scale-invariant prior, which would have a logarithmic form.  
Such a prior is generally referred to as a ``least informative prior.''
It reflects only information about the structure of the experiment, and does
not favor any specific outcome.  Of course, if there
is specific information about the true parameter value, the prior should 
reflect that information.

Finally, we must address the denominator of \eq{stats:bayesthm}.  This term, the
probability of the data, serves as a normalization.  It expresses the probability
of measuring the data regardless of which model is actually true.  If this can 
be rigorously calculated, then the left side of \eq{stats:bayesthm} will be a
well-normalized probability distribution.  In practice, it is often more
practical to sidestep the issue by forming the odds ratio.

\subsection{The Odds Ratio}
Usually it is the case that we compare two discrete models, or that we compare
a discrete model with a class of models characterized by a finite number of 
continuous parameters.  In that case, the odds ratio conveniently handles our
normalization issues.  The odds ratio is formed by comparing the posterior
probabilities of the different models.  Consider two discrete models, $M_0$ and
$M_1$.  For concreteness and comparison with \sect{stats:hypotest}, let us
assume that $M_0$ states that there is no source, and that $M_1$ states that
there is a source of a given flux at a given position.  Then the odds ratio
is
\begin{equation}
\frac{P(M_0 | D)}{P(M_1 | D)} = \frac{\like(D | M_0) P(M_0)}{\like(D | M_1) P(M_1)}
\end{equation}
This gives the  probability that there is no source relative to the probability
that there is a source.  Note that it only allows those two possibilities.  A
value of \onethird, for example, would mean it was three times more likely that
there was a source than that there was no source.  We know nothing about the 
probability that there were two sources.

Of course, the example in \sect{stats:hypotest} was more complicated than this.
The model $M_1 = M_1(\mu)$ was a class of models, parameterized by the source
strength.  The odds ratio in that case is also a function of the parameter:
\begin{equation}
O(\mu) = \frac{P(M_1(\mu) | D) P(M_1(\mu))}{P(M_0 | D) P(M_0)}
\end{equation}
The function $O(\mu)$ gives the odds that there is a source of strength $\mu$
at a given position versus that there is no source.  In most cases, we will
be interested in the total odds ratio; that is, the odds that there is a 
source regardless of its strength.  This is akin to the detection significance
calculated in \sect{stats:hypotest}. We may calculate this ratio using the
procedure known as marginalization.

\subsection{Marginalization and Confidence Regions}
\label{stats:marginalize}
We may eliminate uninteresting parameters by {\em marginalization}.  This
process acquired its odd name through a historical accident, when ``integrations''
were carried out numerically by adding columns of numbers into the margins
of the page.  In essence, the method is mathematically simple.  Given any
conditional probability, and the probability of the condition, we may integrate
to eliminate the condition:
\begin{equation}
P(A) = \int_{B=B_{min}}^{B_{max}} P(A | B) P(B) dB
\end{equation}
The application to the odds ratio is immediately clear.  We simply integrate
the numerator over all possible $\mu$ to find the total odds ratio.  

A very similar process may be used to find confidence intervals.  Consider
the situation when a source is known to exist.  We then wish to find the 
best estimate of its flux, and a confidence interval for that estimate.
Since the flux $\mu$ must take on some positive value, we may evaluate
the denominator of \eq{stats:bayesthm}:
\begin{equation}
P(D) = \int_{0}^{\infinity} \like(D | M(\mu)\,) P(\mu) d\mu
\end{equation}
Then the probability that the true value of $\mu$ is between $\mu_-$ and
$\mu_+$ is
\begin{equation}
P(\mu_- \le \mu_0 \le \mu_+) = \frac{
\int_{\mu_-}^{\mu_+} \like(D | M(\mu)\,) P(\mu) d\mu}
{ \int_{0}^{\infinity} \like(D | M(\mu)\,) P(\mu) d\mu}
\end{equation}
For a given confidence level, there are an infinite number of choices 
of confidence intervals, as there are with frequentist statistics.  
There is no requirement that confidence intervals be contiguous.  
However, useful intervals can often be found by requiring $\mu_- = 
\hat{\mu} - \delta\mu$ and $\mu_+ = \hat{\mu} + \delta\mu$, where 
$\hat{\mu}$ is the most likely value of $\mu$~\cite{eadie}.  The 
resulting interval is central by definition, and in the limit that the 
probability density is symmetric about the maximum and is smaller 
everywhere outside the interval than it is inside the interval, it is 
minimal.

\subsection{Advantages and Disadvantages}
Advocates of frequentist and Bayesian methods have unfortunately been 
polarized into two extreme camps, with only the most acrimonious 
communication between them.  
In fact, both methods are rather like a 
powerful hunting rifle.  Used properly, they are efficient and 
successful at doing their job.  Improper use may result in permanent 
catastrophic injury and/or death. 
We will briefly examine the 
objections to both methods, and in doing so find that the 
disagreements are actually objections to using the methods improperly.

The major objection to the Bayesian method is the use of a prior.  It is
said that the prior subjectivizes what should be an objective procedure, and
therefore reduces the results to a sort of ``modified best guess'' of the
experimenter.  The Bayesian responds that that is exactly true; indeed, if
we use probabilities to express our ignorance, then we should hope that
the results of an experiment reduce our ignorance.  Furthermore, the Bayesian
claims that all assumptions and prior knowledge are made explicit under
the Bayesian formulation.  It is certainly clear that the prior is the
Achilles' heel of the Bayesian method.  There is no objective, prescribed method
for obtaining a prior.  However, there is also no objective, prescribed method
for choosing a traditional statistic.  The use of \chisq, for example, 
 implicitly assumes 
that the errors in the measurements are Gaussian, which may or may not be the
case.

The primary objection to traditional statistics is the {\em ad hoc} procedure
of selecting a statistic.  Statistics are chosen based on their power and
appropriateness, to be sure, but the choice is also largely guided by 
experience.  The justification for the choice of statistics is generally
its past success, rather than any {\em a priori} reasoning.  In contrast,
the Bayesian method prescribes the calculation for any experiment: form the
likelihood and weight by the prior.  

A secondary objection is philosophical
in nature.  Bayesians prefer to treat the true value of the parameter as a 
random variable, with the experimental data as the fixed and unchanging
measure of reality.  The parameter then
has some probability of falling within the confidence region.  The traditional
approach treats the true value of the parameter as fixed and unchanging, and
the data as only one possible outcome in an imagined ensemble, a shadow or
projection of the true reality.  The confidence
region that we calculate from this instantiation of the data then has some
probability of covering the true value of the 
parameter.
\footnote{The philosopher will note a certain correspondence to
historically important epistemological viewpoints. Sartre may debate Plato 
on the true nature of reality, but both are crushed by the rock
of Sisyphus.}

Despite these objections, the two methods are actually compatible, and under
the right circumstances, equivalent.  In most circumstances, the maximum
likelihood method is equivalent (for parameter estimation) to the Bayesian
result with a flat prior.  Whenever the implicit assumptions of a traditional
analysis are matched by the explicit assumptions of a Bayesian analysis, the
results will be the same. 

\section{Calculating Upper Limits}
\label{stats:uplim}
A significant positive detection of a source is the goal of all telescopes.
However, null results can also be useful 
\cite{michelson_morley2,tompkins_crab}.  It is often valuable to set an 
upper limit on the \gammaray\ flux of a source known from X-ray, optical, 
or radio observations.  Determining the value of the upper limit is a 
statistical endeavor that requires a very careful definition of the goal.
It has been noted that ``the question of how to calculate an upper limit in 
the vicinity of a physical boundary is one of the most divisive in high-energy
physics'' \cite{ppdb}.  Unfortunately, this is precisely the limit in which
we find ourselves.  A negative source flux is unphysical; nevertheless, a
measurement of a weak (or non-existent) source in the presence of a large
background may easily result in a data set best fit with a negative source 
flux.  In order to combine such a result with other results, the flux must 
be reported as negative, with the confidence region found as in
 \sect{stats:likeconf} under the maximum likelihood method \cite{ppdb}.
Otherwise, any combination of results from different observations or
different experiments would be biased.

Once we have agreed upon a point estimate of the parameter, we must
consider the confidence region that we wish to comprise the upper limit.
In analogy to the confidence regions around point estimates, we will 
define the ``$1\sigma$ upper limit'' (in frequentist terms) to be the
top of an interval which will, when constructed from an ensemble of data sets,
contain the true value of the flux 68.3\% of the time.
In Bayesian parlance, this means that the integral of the posterior
probability distribution from zero to the $1\sigma$ upper limit will be 68.3\%.

\subsection{Upper Limits from \likeprog}
The upper limits generated by \likeprog\ do not fulfill this definition.
There are three situations for which \likeprog\ must generate an upper limit.
The first is when the flux measurement is positive, but the confidence region
extends to negative flux; e.g., a measurement of $5 \mypm 10$.
The second is when the flux measurement is negative, but the confidence region
extends into positive territory.  The third is when the flux measurement and the 
entire confidence interval are negative.

\likeprog\ handles these situations in the following way.  
An upper limit is always quoted, and the flux measurement is quoted
only if certain conditions are met.
If $\TS\, \gt \, 1$ and the flux is positive, the flux and confidence
regions are quoted.  The upper limit is the top end of the confidence
interval.
If $\TS\,\lt\,1$ or the flux is negative, only an upper limit is quoted.
(Note that this immediately introduces the bias discussed above.)
If the flux is positive, the upper limit is the top end of the confidence
interval.
If the flux is negative, \likeprog\ finds the width of the confidence
region, then shifts it so that it is centered on zero.  The upper limit
is then the top end of the shifted confidence interval.

The upper limit calculated for strong sources with positive flux
clearly does not fulfill our definition.  The confidence interval
on the point estimate will contain the true value 68.3\% of the time.
The interval between zero and the upper limit is larger than the
confidence interval, and so must contain the true value at least
68.3\% of the time.  

The upper limit for sources with positive flux whose confidence 
intervals extend to negative fluxes also does not fulfill our
definition.  The 68.3\% upper limit is found from the one-sided
confidence interval; that is, the value $\sigma_{u.l.}$ such that
\begin{equation}
\int_{-\infinity}^{\sigma_{u.l.}} f(\hat{\mu} | \mu_0) d\hat{\mu} = 68.3\%
\end{equation}
where $f(\hat{\mu} | \mu_0)$ is the probability of measuring $\hat{\mu}$
given a true flux of $\mu_0$.  In the case of a Gaussian function $f$,
the confidence of the integral evaluated to $\sigma_{u.l.} = \hat{\mu} + \sigma$
may be evaluated. In general, the confidence of an interval extending 
to an upper limit of $\hat{\mu} + \sigma$ depends strongly on the
shape of $f$.

The most difficult situation is when the maximum likelihood source
strength is negative.  Barrett, \etal\ \cite{ppdb} suggest ``lifting
up'' the measured flux to zero, evaluating the likelihood function and
taking the upper limit.  Instead, \likeprog\ finds the width of the
confidence interval about the measured (negative) flux, then centers
that confidence interval about zero.  It is very difficult to estimate
the probability that a confidence interval obtained in this way would
cover the true source strength; one must consider, for each possible
value of the true source strength, all possible data sets.  Any one of
those data sets could fall into any of the categories we have outlined.

\subsection{Calculating More Accurate Upper Limits}
The upper limits calculated in this way are not only confusing and non-intuitive,
they also do not (in general) fulfill our requirement: a confidence interval
from zero to the upper limit should cover the true value 68.3\% (or some other
specified fraction) of the time.  A Bayesian approach is in this case more
intuitive and fulfills our requirements.

\begin{figure}
\centering
\includegraphics[width=3.5 in]{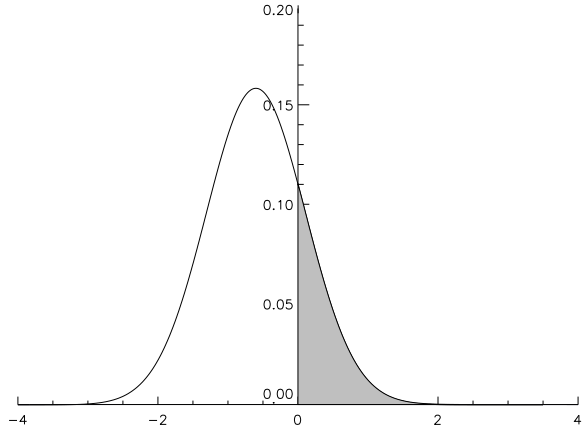}
\caption[Bayesian upper limits]
{\label{stats:bayesuplim}
Upper limits calculated by Bayesian
methods.  This hypothetical likelihood function represents a measurement
that gives a most likely value of some parameter to be less than zero.  If
the parameter is constrained by the physics to be positive, the Bayesian 
formalism suggests forming a prior which is zero for negative  parameters 
values, and constant for positive parameter values.  The posterior distribution
would then look like the shaded portion of the curve.}
\end{figure}

Instead of considering the ensemble of data sets that might produce a
given flux measurement, we start directly with the data, and consider
the range of true flux values that might produce the data.  The requirement
that all fluxes must be positive is easily fulfilled; we form a prior that is
flat for $\mu\,\ge\,0$ and zero for negative $\mu$.  Our posterior probability
distribution is identical to the likelihood, except that it is zero for all
negative values of $\mu$, and renormalized so that the integral over all values is unity.
 (\fig{stats:bayesuplim}).  The interpretation is then straightforward.
The most likely value of the flux is zero; the 68.3\% upper limit is found
by integrating the posterior distribution from zero to some value $\sigma_{u.l.}$
where the integral equals 68.3\%.

\subsection{Implications for Extant Conclusions}
There are two sorts of conclusions typically drawn from \egret\ upper limits.
The first is generally qualitative.  A known X-ray or radio source is not
detected in the \egret\ data.  A ``low'' upper limit is taken as evidence that
\gammaray\ emission is small or non-existent.  A ``high'' upper limit is 
taken as weak evidence that there is little \gammaray\ emission, but that
there was not sufficient data to draw a conclusion.  For these sorts of
conclusions, the \likeprog\ upper limits are adequate 
(\eg,~\cite{linSeyfert,seyfertrev}).

The second sort of conclusion is generally quantitative and statistical
in nature.  Often, possible source variability is examined in a number
of observations.  Some \egret\ observations of a source (usually
an AGN) yield significant detections, and some yield only upper limits.
Given flux measurements and confidence regions, it is a simple matter
to formulate a \chisq\ or likelihood test to determine if a constant
flux model is compatible with the data.  It is critical for such a test
that upper limits have well-defined statistical properties.  The upper limits
generated by \likeprog\ and quoted in the \egret\ catalogs do not have these
properties \cite{cat2,cat2sup,cat3}.  Unfortunately, catalogs of variability
have been compiled based precisely on these upper limits \cite{mclaughlin}. 
All variability conclusions which involve upper limits are suspect, and should
be treated as qualitative suggestions rather than quantitative results.  
Work by W. F. Tompkins is in progress to compile variability catalogs 
which have been calculated with statistically meaningful 
upper limits \cite{billthesis}; these should be used as soon as they are available.

%% file: GRB.tex
\chapter{\Gammaray\ Bursts}
\label{grbchap}

In the midst of the Cold War, in the late 1960s, a number of 
satellites were launched carrying \gammaray\ detectors.  Sensitive to 
\gammarays\ between \mysim 200 keV and \mysim 1.4 MeV, the Vela 
satellites were designed to detect the testing or use of nuclear 
weapons.  Between July of 1969 and July of 1972, four {\em Vela\/} satellites,
equally spaced in the same circular orbit, detected 16 bursts of 
\gammaray\ energy.  
Comparisons of the \gammaray\ arrival times in different satellites 
determined that the origin of the bursts was more than 10 orbital 
diameters away.  Thus, the first theory of \gammaray\ bursts, Soviet 
nuclear testing, was ruled out.  Although national security concerns 
delayed the publication of these intriguing results by Klebesadel, 
Strong, and Olson~\cite{firstburst} for a number of years, they would 
mark the beginning of the longest-standing mystery in astrophysics 
since the Shapley-Curtis debates.

A number of other bursts were observed over the next two decades 
\cite{metzger74,white78}.  A consensus emerged fairly early that the 
source of the mysterious \gammaray\ bursts was Galactic in origin 
\cite{gilman80}.\footnote{This work will deal only with the 
so-called ``classical \gammaray\ bursts.''  Another class of transient 
\gammaray\ sources, the soft gamma repeaters, are clearly a different 
type of object.}  Preliminary detections of 
flux over 1 MeV strengthened this conclusion~\cite{oldpat}; 
Schmidt ``showed'' that 
detection of emission over 1 MeV required a Galactic origin, since the 
source luminosity required for bursts at more than \mysim 100~pc would 
imply an energy density that would result in $\gamma$--$\gamma$ pair 
production \cite{schmidt78}---the optical depth to Earth for 1~MeV 
\gammarays\ would be $\gt 1$.  While there was never significant 
evidence as expected that the bursts were preferentially located in 
the Galactic disk, it was presumed that the \batse\ experiment on 
board \cgro\ would clear up any remaining ignorance about \gammaray\ 
bursts.

Despite the general consensus on the location of \gammaray\ bursts, 
the field attracted a wide variety of specific theories.  Nemiroff 
identifies 99 distinct \gammaray\ burst theories put forth between 
1968 and the end of 1991~\cite{nemiroff}.  Early theories placed 
bursts both locally and cosmologically distant, with energy generation 
mechanisms ranging from relativistic dust grains in the solar 
neighborhood to cometary collisions with neutron stars to white hole 
emission to vibrations of cosmic strings at $z = 1000$~\cite{cosmicstring}.
Nevertheless, by 1981, most models involved neutron stars 
in the Galactic plane.  (Note, however, that \pacz~\cite{bohdan86} put 
forth a theory of cosmological bursts with an optically thick \el\pos\ 
plasma outflow in 1986.  However, \pacz\ also put forth a number of 
other unrelated theories of \gammaray\ bursts; thus it is unclear 
whether his apparent success is due to prescience or judicious 
covering of theory space.)

In September 1991, Gerald Fishman, representing the \batse\ team
at the first Compton Symposium, 
summarized the results of the \gammaray\ burst search since the launch 
of \cgro\ in April of that year.  Everything that had been believed 
about the origin of bursts was shaken.  The data, published first by 
Charles Meegan, Fishman, and others in {\em Nature}~\cite{firstbatse}, 
showed an isotropic distribution of \gammaray\ bursts across the sky.  
Furthermore, there was already evidence of the so-called ``edge'' of 
the \gammaray\ burst distribution.  That is, there were fewer low-flux 
bursts than would be expected if bursts were standard candles 
uniformly distributed in Euclidean space.  The distribution was 
incompatible with a galactic disk population.  

\begin{figure}
\centering
\includegraphics[height = 4 in, angle=-90]{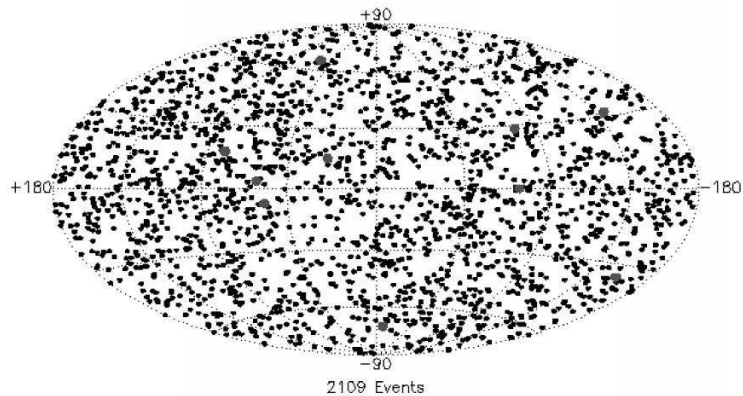}
\caption[\Gammaray\ bursts observed by \batse]{\label{grb:batsedist}
Locations of \gammaray\ bursts observed by \batse\ as of July 3, 1998, in
Galactic coordinates~\cite{currentbatse}.}
\end{figure}

Old theories die hard, and much effort went into searches for the 
expected spatial and temporal correlations between bursts in the 
\batse\ catalogs.  While it appeared in the first \batse\ catalog that 
some correlations might exist, additional bursts in the second catalog 
made isotropy much more likely~\cite{efron&pet95}.  Such statistical 
tests must be performed very carefully, due to the biases acquired 
from variable thresholds and exposures~\cite{vahe93,pet&lee96,lee&pet96}.  
Similarly, only very marginal evidence could be found for repeating \gammaray\ 
bursts~\cite{pet&efron95}.  The distribution
of all bursts observed to date (\fig{grb:batsedist}) is compatible
with a uniform distribution across the sky~\cite{currentbatse}.

The \batse\ data quickly winnowed away many \gammaray\ burst theories.  
But like opportunistic species after an ecological disaster, new 
theories quickly sprung up to fill the void (\eg,~\cite{tavani}).  The 
new theories were classified primarily by the location of the bursts: 
locally (in the Oort cloud), in the hypothesized large Galactic halo, 
or at cosmological distances.  The first of these was essentially 
abandoned when no suitable energy generation mechanisms could be 
envisioned.
\footnote{C. D. Dermer proposed a mechanism at a conference in Santa Barbara, 
California in 1995 involving antimatter comets annihilating 
with normal comets in the Oort cloud.  This theory was apparently 
never published.}
Halo models generally involved interactions with old 
neutron stars that had been ejected from the plane into the halo.  
While the presence of a dark Galactic halo has been inferred from the 
rotation curves of other galaxies, its extent has not been 
conclusively measured~\cite{brianugrad}.  The isotropy of the 
\gammaray\ burst distribution suggested that the radius of the halo 
would need to be many times the distance between the Earth and the 
Galactic center.  Similarly, the halo of M31 (Andromeda) would be 
evident in the distribution once \batse\ detected a large enough 
sample of bursts.  The isotropy of the first 1122 \gammaray\ bursts
detected by \batse~\cite{thirdbatse}
suggests if the bursts were in the Galactic halo, 
the halo would have a radius greater than 100~kpc. 
To be consistent with the lack of any excess toward Andromeda, the 
scale would have to be less than 400~kpc; a rather tight constraint.
If the bursts were at cosmological distances, 
evolution effects conveniently explain the observed ``edge'' of the 
distribution.  The energy required to produce the observed flux from 
such distances (\mysim$10^{51}$--$10^{54}$~ergs, modulo any beaming 
factor) is a few percent of the binding energy of a neutron star.  
Such an energy scale leads naturally to catastrophic theories 
involving neutron stars.

It had been realized quite early~\cite{schmidt78} that the large 
energy release required for distant \gammaray\ bursts would produce an 
environment that was optically thick to $\gamma$--$\gamma$ pair 
production.  An assumption of cosmological origin of the \gammaray\ 
bursts led to a bifurcation of theories into those describing the 
energy generation mechanism and those describing the propagation of a 
large amount of compact energy, regardless of its source.  The 
relevance of the latter theories for \egret\ observation will be 
discussed in \sect{grb:theory}.

\section{Recent Observations}
In 1981, Fichtel \& Trombka speculated that ``identification of the 
[\gammaray\ burst] objects with observations at other wavelengths will 
probably be required before significant progress can be made in 
determining their origin''~\cite{fichtel81}.  The day for which they 
were waiting arrived February 28, 1997, when the Italian satellite 
\sax\ discovered an X-ray afterglow to a \gammaray\ burst 
\cite{costa_iauc,costa_beppo}.  The \sax\ satellite has a collection 
of instruments, including a \gammaray\ burst monitor, a wide-field 
imaging X-ray camera with a positional resolution of about 
3~arcminutes, and a narrow-field X-ray camera with a position 
resolution of about 1~arcminute.  Within 8 hours of the detection of 
the burst and its imaging with the wide-field camera, the narrow-field 
camera observed a source, consistent with the burst error circle, at 
$\alpha_{2000}=05\hr01\mn57\mysc, 
\delta_{2000}=11\deg46\arcmin42\arcsec$ \cite{frontera97}.  Optical 
and radio transients at the same position were quickly 
found~\cite{galama97huntsville,paradijs97}.  Some time later the 
optical transient was also seen by the Hubble Space Telescope 
\cite{sahu97}.  The optical, radio, and X-ray sources were observed to 
decay as a power law consistent with the fireball blast wave theories 
of \meszaros\ and Rees~\cite{mes&rees97} and Wijers \etal\ 
\cite{wijers97}.

In the following months several other bursts would be found to be 
consistent with transient sources in other wavelengths.  Another 
revelation came from GRB 970508, observed May 8, 1997 by the 
wide-field camera on \sax\ and by its narrow-field camera 5.7~hours 
later.  An optical counterpart was quickly found by 
Bond~\cite{bondiau} and a radio counterpart found by Frail 
\etal~\cite{frail97} consistent with the narrow-field \sax\ position.  
Metzger and his colleagues at Caltech identified a set of absorption 
features implying a redshift of $\ge 0.835$~\cite{metzgeriau1}.  By 
early June, they detected emission lines as well, at the same 
redshift~\cite{metzgeriau2}.  For the first time since their discovery 
30 years before, there was direct experimental evidence that 
\gammaray\ bursts were cosmological~\cite{970508redshift}.

Another notable burst, GRB 971214~\cite{find971214}, was detected at a 
number of different wavelengths: optical~\cite{halpern}, 
infrared~\cite{gorosabel}, and ultraviolet~\cite{971214euve}, as well 
as being detected by a number of \gammaray\ and X-ray instruments 
including \batse, \xte, and \ulysses~\cite{971214xte}.  
Absorption and emission lines have been detected from this burst as
well, yielding a redshift measurement of 3.43---firmly establishing 
the cosmological nature of the \gammaray\ bursts~\cite{kulkarni98}.
Most recently, the redshift of the host galaxy of GRB~980703 has
been measured at the Keck Observatory.  Absorption and emission lines
yield a redshift of $z =0.966$~\cite{djorgovski98}.  A good review
of the current state of observational affairs is given by Wijers~\cite{wijers}.

Now that the observational mechanism for establishing \gammaray\ burst
counterparts has been established, the field is changing rapidly.
As of this writing, 11 \gammaray\ bursts have been observed in radio
and optical wavelengths~\cite{leonardreview}.  These observations and 
others which will certainly be made in the near future have strongly 
affected, and will continue to
affect, the leading theories of \gammaray\ bursts.

\section{\Gammaray\ Burst Models}
\label{grb:theory}
Until the launch of \batse, \gammaray\ burst theories outnumbered the 
bursts themselves.  In light of the data provided by \batse, \sax, and 
other satellites, it is worthwhile to examine the leading theories 
with an eye to understanding the consequences observable by \egret.  
We will first look at the theories of the energy source powering 
\gammaray\ bursts, and then examine some of the theories of the 
\gammaray\ generating shock waves created by the bursting source.

\subsection{Energy production mechanisms}
Gigantic explosions hold fascination for the theoretical physicist as 
much as for the schoolchild.  Cosmological \gammaray\ bursts require 
the largest explosions known, and thus attract theorists in droves
(\eg,~\cite{cannonball}).  
As long ago as 1975, Malvin Ruderman noted~\cite{rudermanquote}, ``For 
theorists who may wish to enter this broad and growing field, I should 
point out that there are a considerable number of combinations, for 
example, comets of antimatter falling onto white holes, not yet 
claimed.''  Mercifully, the considerable experimental data amassed 
since 1975 has largely narrowed the field to two general mechanisms: neutron 
star--neutron star mergers and massive stellar collapses known as 
``collapsars'' or ``hypernov\aeind.''  Nevertheless, at this stage in 
our understanding of \gammaray\ bursts it is unreasonable to think that
all theories could be placed in one of only two categories; these represent
only the most prominent of the theories.

\paragraph{Neutron star mergers.} 
The most popular theory at the time of this writing is the neutron 
star merger model~\cite{narayanNSmerge}.  The basic premise is very 
simple.  Neutron star binary systems slowly lose energy as a result of 
gravitational radiation.  Eventually, the neutron stars will spiral 
into each other, presumably resulting in a large release of energy.  A 
variant of the model replaces one of the neutron stars with a black 
hole.  Narayan~\cite{narayanNSmerge} cites several advantages of such 
a model, as well as a number of issues to be resolved.  First, the 
source population is known to exist.  Neutron star
binaries have been observed, and their energy loss corresponds
with that expected from gravitational radiation to better 
than 1\%~\cite{taylor89}.  
Second, the energetics are of the right order.  The energy release 
from such a merger would exceed $10^{53}$~ergs in \mysim1~ms within 
100~km.  Finally, the frequency of such mergers may be approximately 
right.  Estimates of the merger rate include the range from 
$10^{-6}$~to~$10^{-4}$~yr$^{-1}$ per standard galaxy, which matches 
the observed burst rate under a cosmological scenario and isotropic 
emission from the energy source.

An early objection to the neutron star merger model was the so-called ``no host 
galaxy'' problem.  Optical observations of \gammaray\ burst error 
circles before the launch of \sax\ failed to find the galaxies expected as the hosts 
of the colliding neutron stars.  However, it was realized that 
binaries can often acquire substantial recoil velocities as a result 
of their two supernova explosions.  For reasonable parameters, 
binaries may travel 1--1000~kpc before merging~\cite{tutukov94}.  
Therefore, by the time a binary neutron star system becomes a 
\gammaray\ burst, it may have left its progenitor galaxy.  The 
diverse character of  
observed \gammaray\ bursts has been pointed out as a problem for the 
merger model.  Neutron stars are expected to have a very 
narrow mass range.  Merging binaries should almost always consist of 
about 3 $M_{\sun}$\index{$M_{\sun}$} collapsing in a very clean 
gravitational system.  Such a homogeneous energy 
generation mechanism should result in a fairly homogenous population 
of \gammaray\ bursts, though beaming effects, magnetic fields, and an 
inhomogenous environment may explain the observed differences.

\paragraph{Hypernov\aeind.}
A much newer idea is the hypernova model due to 
\pacz~\cite{hypernova}.  The idea is similar to Woosley's ``failed 
supernova'' model~\cite{woosley93}.  A very massive star undergoes 
core collapse to form a \mysim10~$M_{\sun}$\index{$M_{\sun}$} black 
hole.  If the star is rapidly rotating, then the angular momentum of 
the star requires the formation of a rotating dense torus of material 
around the Kerr black hole.  Any previously existing magnetic field 
will be significantly affected by the collapse; most possibilities 
strengthen the local magnetic field.  \pacz\ estimates that the 
rotational energy of the black hole should be of order $10^{54}$~ergs.  
He also finds that the maximum rate of energy extraction is
\begin{equation}\index{$\approx$}\index{$M_{\sun}$}
L_{max} \approx 10^{51} \hbox{ergs s$^{-1}$}
\left( \frac{B}{10^{15}\, \hbox{G}} \right)^2
\left( \frac{M_{\hbox{\scriptsize BH}}}{10\, M_{\sun}} \right)^2
\end{equation}
Achieving the required energy release requires fields on the order
of $10^{15}$ G.  No mechanism for generating such fields is offered,
although many theorists are currently working on the details of 
the effects of the core collapse on the ambient magnetic fields.

The observed spectrum from either model of \gammaray\ bursts depends 
more on the details of the fireball model discussed in the next 
section than on the energy generation mechanism.  The primary 
difference between these mechanisms is their location.  While neutron 
star mergers are often expected far from galaxies, hypernov\aeind\ are 
found in star-forming regions, since their progenitors are rapidly 
evolving massive stars.  The lack of optical observation of GRB 
970828, despite fairly deep searches, combined with the reported large 
hydrogen column density \cite{murakami98} has led \pacz\ to conclude 
that the optical emission was not observed due to extinction by dust.

A substantial number of detections of \gammaray\ bursts at multiple 
wavelengths should illuminate this question in the very near future.

\subsection{Blast wave theories}
While very early experiments seemed to detect thermal radiation from 
\gammaray\ bursts, better instruments soon observed a characteristic 
power law emission in the X-ray and \gammaray\ regime.  Theoretical 
explanations have centered on the blast-wave model, where the 
injection of a large amount of energy into a very small volume leads 
to an expanding fireball, optically thick to pair production, 
expanding into some surrounding medium.  The simplest model---also due 
to \pacz~\cite{bohdan86}---assumes the spherical expansion of an 
optically thick \el\pos\ plasma with no surrounding medium.  This 
model results in a blackbody spectrum.  A series of more and more 
complicated models were based on this simplistic one, finally 
culminating in the model of \meszaros, Rees, and 
Papathanassiou~\cite{mrp94}, which calculated spectra expected for a 
variety of magnetic field configurations and particle acceleration 
efficiencies, including the shock fronts and reverse shocks which 
arise from the expansion of the burst ejecta into the surrounding 
medium.

Shock fronts in blast waves have become the baseline from which 
various energy generation mechanisms diverge.  Their popularity stems 
from their success: they can naturally explain how energy can be 
reconverted into \gammarays\ from the particles that must emerge from 
the fireball, they can naturally explain the observed burst 
time scales, and, assuming the medium into which the fireball expands 
varies slightly from burst to burst, they can explain the wide variety 
of burst profiles and time scales 
observed~\cite{batsespectral,waxtimescales}.  
The observed afterglows may be explained as emission from the 
slowing, cooling shock~\cite{waxafterglow,vietri}.
However, several 
theoretical issues remain.  The mechanism of particle acceleration in 
relativistic shocks is not well understood, though it is widely 
assumed that such acceleration results in a power law energy 
spectrum~\cite{bednarz}.  
The magnitude and nature of the magnetic field is unknown.  
Pre-existing magnetic fields, if sufficiently strong, will provide 
significant synchrotron radiation in the particle blast wave.  In 
addition, turbulent mixing in the shock can generate significant 
fields due to charge separation.  Finally, the coupling between 
electrons, baryons, and the magnetic field is not well understood.  
Energy radiation from elections is very efficient, but most of the 
energy is in the baryons, or perhaps the magnetic field, which do not 
radiate their energy so readily.  The nature 
of their coupling affects the total radiated power as well as the 
time scale of the radiation \cite{chiang&dermer98,rees&mes98}.

While a detailed investigation of the various blast-wave models is 
beyond the scope of this work, a general overview is given, based on 
that found in Chiang \& Dermer \cite{chiang&dermer98}.  The burst 
begins with the deposition of $10^{51}$--$10^{55}$~ergs of energy in a 
radius of about 100~km.  The nature of this energy is not known.  It 
is assumed that most of this energy will be transformed into kinetic 
energy of baryons, which expand adiabatically.  The baryons soon 
become cold; that is, the baryon speeds in the comoving frame of the 
bulk flow become sub-relativistic.  The bulk Lorentz factor is then 
given by $\Gamma_0 \simeq E_0/M_0 c^2$\index{$\simeq$} where $M_0$ is 
the rest mass of baryons.

As this sphere freely expands into the surrounding medium, it sweeps 
up material.  A shock front begins to form.  At some ``deceleration 
radius'' $r_d$ the shell can no longer be approximated as freely 
expanding, and the bulk kinetic energy of expansion begins to be 
reconverted into internal energy of the baryons.  This radius is the 
point at which the integrated momentum impulse of the swept-up matter 
equals the original baryonic rest mass: $r^3_d \approx (3/4 \pi \rho 
\Gamma_0) M_0$\index{$\approx$} where $\rho$ is the density of the 
surrounding material.

Predictions of what happens at this deceleration radius form the core 
of most blast-wave models.  A function $\Gamma(r)$ is derived which 
expresses essentially the rate at which baryonic kinetic energy is 
converted to radiation.  Since cosmological scenarios require high 
initial $\Gamma_0$ of order $10^2$--$10^3$, much of the energy release 
in all scenarios is compressed into the first few tens of seconds in 
the observer's frame.  Nevertheless, models of the blast wave at the 
deceleration radius and beyond make predictions about the observed 
spectrum at various times as well as the observed burst time scales.

Given the recent evidence that \gammaray\ bursts really are 
cosmological in origin, the energy required for the fluences observed 
on Earth make some sort of fireball almost inevitable.  Blast-wave 
models have qualitatively reproduced some aspects of \gammaray\ burst 
observations, and appear to be a promising theoretical road to pursue.  
However, it should be noted that blast waves as they have been 
described cannot be the whole story.  In either the neutron star 
merger model or the hypernova model, there is a natural symmetry 
axis associated with the source.  A symmetric blast wave then seems 
rather unlikely.  Asymmetry in the blast wave will lead to beaming on 
some scale.  Beaming will of course reduce the energy required per 
burst, but increase the rate at which the bursts must occur.  It is 
not clear how beaming would affect the radiation produced by shock 
waves.  The radiation reaching the Earth has already been beamed into 
a cone of order 1/$\Gamma_0^2$.  This tight beaming means that all of 
the \gammarays\ observed at the Earth are emitted from a very small 
piece of the blast wave; therefore, the amount of large-scale 
symmetry in the wave may be irrelevant to the spectrum observed.

\subsection{Observable consequences}
Without the compass of observations, we are doomed to drown in an 
ever-deepening sea of theories.  Fortunately, the tools of the 
experimental astrophysicist are becoming more and more powerful.  A 
number of X-ray satellites are currently operating, with more planned, 
which produce error boxes small enough to allow radio astronomers to 
bring their substantial instruments to bear.  \glast\ will 
revolutionize high-energy \gammaray\ astronomy in the next decade just 
as \egret\ did in this decade.  And new types of astronomy, previously 
relegated to science fiction or crackpot dreams, are beginning to 
provide useful data.  Air \cerenkov\ detectors can localize 
high-energy \gammarays\ from a few hundred GeV up to many TeV to much 
less than a degree.  B\"{o}ttcher\index{\"{o}} \& Dermer calculate the 
high-energy emission from proton synchrotron radiation and 
photopion-induced pair cascades, and find that future high-sensitivity 
\cerenkov\ telescopes with low energy cutoffs (or, it should be noted, 
good very-high energy response from \glast; see \chapt{glast:tng}) 
could measure the level of the infrared background 
radiation~\cite{bottcher98},  since high-energy \gammarays\ will 
interact with the infrared background to produce electron-positron pairs.
While no current neutrino detectors have 
sufficient sensitivity, a measurement of the neutrino flux from a 
\gammaray\ burst could shed light on the energy generation mechanism 
\cite{waxneutrino}.  Additionally, an intense gravitational collapse 
will produce gravitational waves, which could be detectable
with the detectors currently under construction~\cite{ligo}.



\section{\egret\ observations}
While \batse\ has enjoyed the spotlight for most of the contributions 
of \cgro\ to the \gammaray\ burst problem, \egret\ has made some 
important observations as well.  At least 16 bursts occurred outside the 
field of view of the spark chamber, but were nevertheless detected in 
the calorimeter~\cite{catelli,egret910601,egret940301}.  
Five \gammaray\ bursts have been detected in the \egret\ spark 
chamber~\cite{dingus}.  Each has characteristics worthy of some examination.  

The first burst detected in the \egret\ spark chamber occurred on May 
3, 1991.  Six photons were detected in the spark chamber in two 
seconds, coincident with the signal received from 
\batse~\cite{egret910503}.  Measurements of the background before the 
burst suggested that 0.18 photons per two seconds were expected in the 
spark chamber.  Additionally, the burst was evident in the TASC 
calorimeter data, as well as the anticoincidence trigger rate.  The 
anticoincidence dome is sensitive to photons above about 20~keV via 
Compton scattering, while 
the TASC has four triggering levels, corresponding approximately to 
energies above 1.0~MeV, 2.5~MeV, 7.0~MeV, and 20.0~MeV. Firm detection 
of this event with all three systems, coincident with the \batse\ 
trigger not only demonstrated \gammaray\ burst detection with \egret\ 
but also measured the first \gammaray\ burst emission above 100~MeV: 
one of the spark chamber photons had an energy of approximately 230~MeV.

The next burst was detected a month later, on June 1, 
1991~\cite{egret910601}.  This burst was also seen in the TASC and 
anticoincidence dome, and four photons were detected in the spark 
chamber when 1.5 background photons were expected.  The detection of 
high-energy photons from these bursts began to shift the evidence 
toward power-law emission instead of thermal emission.  Furthermore, 
the last photon to arrive in the spark chamber was nearly 70 seconds 
after the initial \batse\ trigger.

Buffalo Bills quarterback Jim Kelly remembers January 31, 1993 as the 
day the Bills lost their third consecutive Super Bowl, this time to 
the Dallas Cowboys by a score of 52--17.  High-energy astrophysicists 
worldwide remember that date for the detection of what still stands as 
the most intense \batse\ burst ever seen~\cite{kouvSuperBowl}.  
\batse\ count rates exceeded $2 \by 10^6$ counts s$^{-1}$.  
Significant structure to the burst was observed at scales below 10~ms, 
and the first intense spike lasted 200~ms.  Fortunately, the burst was 
in the field of view of the \egret\ spark chamber, which detected 16 
\gammarays\ (compared to 0.04 expected by chance) in 25 
seconds~\cite{egret930131}.  Two of the \gammarays\ had energies of 
nearly 1~GeV. \egret\ measured a power-law photon spectrum with index 
\mysim$-2$.  The hard photon spectrum lent further credence to shock 
acceleration models, although the high-energy photons were taken as 
evidence that the burst was closer than \mysim50~pc.  A total fluence 
measurement could not be made by \egret, since the dead time per 
spark chamber event is approximately 100~ms.  It is likely that a number of 
high-energy \gammarays\ passed though the spark chamber in the first 
100~ms.

It would be more than a year before \egret\ would detect another 
burst, but that burst would provide plenty of fuel for the raging 
\gammaray\ burst debate.  The burst, which arrived on February 17, 1994
and whose initial pulse lasted for 
180~s, was detected by \comptel~\cite{comptel940217}, \ulysses, 
\batse, and \egret.  Of primary theoretical importance was a single 
\egret\ photon, detected some 4500~s after the initial burst, with an 
energy of 18~GeV~\cite{egret940217}.  The probability of this photon 
originating in the background was $5 \by 10^{-6}$.  In fact, \egret\ 
detected a total of 28 photons from GRB 940217.  Ten photons were 
detected during the first 180~s, concurrent with the \batse\ 
detection, including a 4~GeV photon and a 3.5~GeV photon.  The 
remaining 18 photons were detected over the next 1.5 hours.  Hurley 
\etal~\cite{egret940217} point out a number of conclusions which can 
be immediately drawn.  While the universe is optically 
thin at \gammaray\ energies, the cosmic microwave background becomes 
an efficient medium for $\gamma$--$\gamma$ pair production for 
sufficiently high \gammaray\ energies.  For a 25~GeV photon, this 
consideration limits the source to $z \lt 55$.  To avoid attenuation 
from the intergalactic infrared background, the source distance is 
further limited to $z \lt 2.5$, ruling out early universe 
theories~\cite{cosmicstring}.  Finally, they note that if the 
spectrum of the first 180~s of GRB 940217 is extrapolated to 1~TeV, 
the predicted fluxes would be detected by air \cerenkov\ telescopes and 
the Milagro air-shower array.  Unfortunately, bursts with such 
high-energy emission are evidently somewhat rare, and the fields of 
view of air \cerenkov\ detectors are quite small.  Nevertheless, the 
current network of fast burst position detections may allow 
observation of delayed emission like that of the February 17 burst.  
The very long delay of emission has also been the source of much 
theoretical speculation.  The existence of such delayed emission is a 
natural consequence of blast-wave models.  While the initial 
\gammaray\ burst is caused by the formation of a shock with the 
surrounding medium, the blast front will continue to radiate as it 
decelerates.

The most recent burst detected in the \egret\ spark chamber arrived
only a few weeks later, on March 1, 1994.  This burst was similar
to the first two which had been detected, with 7 photons of maximum
energy 160~MeV, arriving within 20~s~\cite{egret940301}.  The spectrum
of this burst was softer (\mysim$-2.5$) than the February 17, 1994 
burst.

It is unlikely that any more bursts will be observed in the \egret\ 
spark chamber, as instrumental lifetime concerns have reduced \egret\ 
to Target of Opportunity observation only.  Nevertheless, the 
observations made by \egret\ create significant constraints on the 
models used to explain \gammaray\ bursts.  The \glast\ instrument, 
with its very wide field of view (\sect{bt:baselinesect}) and large 
sensitive area up to very high energies (300~GeV), should further 
constrain the high energy behavior of \gammaray\ bursts.

\section{Possible \egret-only Bursts}
Observations of \gammaray\ bursts in the last decade have shown 
conclusively that the high-energy emission has a power-law spectrum.  
If, for some bursts, the burst has a harder spectrum than the 
background, the signal-to-noise ratio should increase at higher 
energies.  Some blast wave theories predict that a spectral break may 
occur at or just below \egret\ energies \cite{dcb98}.  In such models, 
the total peak spectral power early in the burst is emitted above 
100~MeV.

It is thus possible that some \gammaray\ bursts will be detectable by 
\egret, but not by lower-energy instruments such as \batse.  The 
evidence of such bursts would therefore be lurking in the \egret\ 
photon database.  As was discussed in \sect{stats:likelihood}, the 
best way to deal with the photon-by-photon nature of \egret\ data is 
through the use of likelihoods.  Buccheri \etal~\cite{buccheri} 
developed a likelihood method to search \egret\ data for \gammaray\
bursts.  While they did find the known bursts in the data they 
searched, they apparently did not perform a comprehensive search of 
the \egret\ database.  The difficulty with such a search is that the 
exposure for each photon has to be calculated individually.  This 
process is very time-consuming, due to the the extended nature of the 
\egret\ calibration files.  Therefore, a comprehensive search required 
some adaptation of statistical methods~\cite{bbj}.  Those methods, and 
their results, will be described below.

\subsection{Statistical Methods}
The \gammaray\ burst search algorithm was designed to be fast and 
efficient at finding bursts.  From a statistical standpoint, a burst 
may be defined as a time interval with a measured rate which is, to 
some specified confidence, incompatible with Poisson fluctuations.  
The method, then, is fairly straightforward: first, find the 
background rate; second, find any time intervals with sufficiently 
high rates to rule out Poisson fluctuations at the given confidence 
level; and finally, verify that the spatial distribution of the 
photons is consistent with a point source.  The sequential nature of 
this method makes it fast compared with a full likelihood analysis; 
however, it also involves some binning, and as we will see, 
complicates our estimates of the significance of detections.

To further speed the algorithm, we search for significant intervals in 
two steps.  The first compares photon arrival rates only.  The rate is 
defined as the number of photons from some area on the sky per second.  
It requires only the amount of instrument live time and a photon 
count.  The second step will measure the photon fluxes (photons 
\perareasec).  This is certainly the more physically relevant quantity, 
but it is also computationally more expensive to determine the state 
of the \egret\ instrument for the entire interval.  Therefore, 
promising candidate intervals will be found by comparing rates, and 
then checked for significance by comparing fluxes.

To establish a background rate, the field of view is binned into 
$5\deg\,\times 5\deg$ squares.  All photons in the viewing period with 
energies over 100~MeV and zenith angles less than $100\fdg5$ are 
sorted into these bins.  The total instrument live time to the center 
of each square is calculated.  This yields an array of approximate 
background count rates.  The critical number of photons required for a 
significant interval, $N_{crit}$, is given implicitly by the Poisson 
formula
\begin{equation}
\label{grb:poisson}
\alpha = \sum_{n=0}^{N_{crit} -1} \frac{e^{-\mu t} (\mu t)^n}{n!} 
\end{equation}
where $\mu$ is the average rate, $t$ is the time interval 
being searched, and $\alpha$ is the confidence level.  The probability 
of observing $N_{obs} \gt N_{crit}$ is $1 - \alpha$.

To acquire a set of candidate events while avoiding specific time 
binning, each photon is considered to be the start of an interval.  
All photons arriving within the standard interval length (either 1 
hour, 30 minutes, 10 minutes, or 3 minutes) and within $5\deg$ of the 
initial photon are considered part of the same event and counted.  If 
the number of observed photons exceeds $N_{crit}$ for that area of the 
sky, the candidate event is accepted for further evaluation.

Because the number of candidate events which pass the first cut
is relatively small, it 
is practical to calculate instrument exposure and compare the 
candidate flux to the expected background flux.  The single trial 
probability of such an event, $P_{S}$, is calculated as in 
\eq{grb:poisson}, with the time $t$ replaced by the exposure ${\cal E}$
and the rate $\mu$ replaced with the flux.

Unfortunately, $P_{S}$ cannot be interpreted as the probability that 
the flux in a given interval is not a Poisson fluctuation.  In fact, 
we have searched many different intervals, and the appropriate 
confidence level must take into account the number of trials.  If the 
intervals had been fixed and preselected so as not to overlap, then 
each interval would be statistically independent.  The total 
probability $P_{N}$ that a given flux could not be attributed to 
Poisson fluctuations would be given by the complement of the product 
of the chances that it would not be seen in each of any one interval:
\begin{equation}
\label{grb:pn}
P_{N} = 1 - P_{S}^{N}
\end{equation}
where $N$ is the number of independent trials.

However, preselecting non-overlapping time bins would not be a good 
way to search for \gammaray\ bursts.  If a burst did not happen to 
fall entirely within one bin, but instead split its flux between two 
bins, it would probably not be significant in either bin.  The sliding
interval described above was designed to avoid this problem.  
Nevertheless, the
sliding interval introduces its own problems: namely, that the 
intervals are no longer independent.

\begin{figure}[t]
\centering
\includegraphics[angle=90,width = 3.75 in]{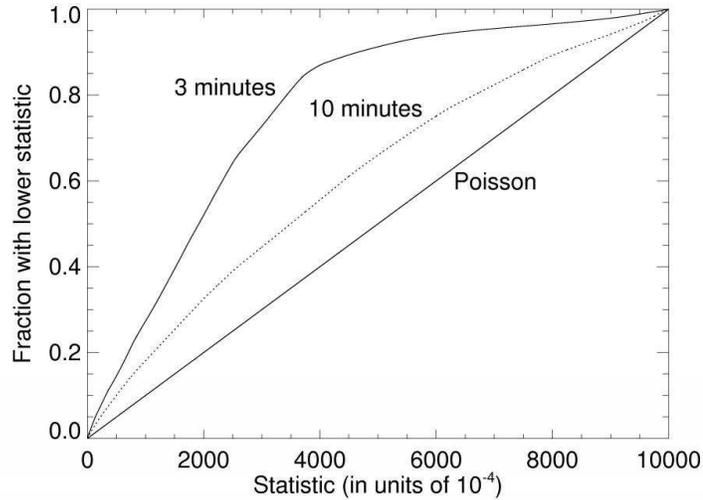}
\caption[Overall \gammaray\ burst significance from the Monte Carlo 
simulations.]{\label{grb:mcresults}Overall significance $P_{T}$
 versus raw probability $P_{N}$ 
from the Monte Carlo simulations.  $P_{N}$ is calculated for each
Monte Carlo event from \eq{grb:poisson} and \eq{grb:pn}.  A cumulative
count of the Monte Carlo events yielding such a $P_{N}$ yields the
cumulative probability of that events, $P_{T}$, given by the vertical
axis.}
\end{figure}

In order to evaluate the actual significance of an interval $P_{T}$ as 
a function of $P_{N}$, a Monte Carlo simulation was performed.  It was 
expected (and verified) that the significance of an interval should
be monotonic in $P_{N}$ (\fig{grb:mcresults}).  
Approximately $5.7 \by 10^{8}$ simulated photons were 
generated for the Monte Carlo data set; the actual data set searched 
contained $1.2 \by 10^{6}$ photons.
Each Monte Carlo photon was assigned an arrival time, drawn from a 
Poisson distribution with a given background rate.  Each photon was 
also assigned a uniformly distributed $x$ and $y$ coordinate between 
$0\deg$ and $40\deg$, simulating \egret\ field of view.  The data 
were simulated using actual \egret\ exposure information calculated 
for a typical point on the sky over many viewing periods, yielding an 
exposure set representative of the windowing and non-continuous 
exposure actually obtained with \egret.  This exposure set was used 
as many times as necessary to generate the entire Monte Carlo data 
set.  To determine appropriate background rates, 104 actual fluxes 
were measured from random points in differing viewing periods.  These 
fluxes were sorted, and the highest and lowest 20\% were discounted 
in order to exclude nonrepresentative outliers.  The range of this 
tightened distribution of fluxes was used as the range for the 
uniformly distributed fluxes in the Monte Carlo simulation.  This was 
done to ensure the use of typical, though not rigorously 
representative, background fluxes.  Since all probabilities are found 
from the difference between measured and expected fluxes, they depend 
only weakly on the precise values of the background fluxes.  The 
range of backgrounds found was $7.0 \by 10^{-7}$--$4.8 \by 10^{-6}$
photons \perareasec\ for a $5\deg$ circle on the sky.

In some cases, the search algorithm may have detected the same event 
more than once; that is, the interval following the first photon in 
the event yielded a significant rate increase, and the interval 
following the second photon also yielded a significant rate increase.  
In these cases, the two probabilities must be independently combined, 
because the Monte Carlo distribution was found by counting all 
significant events, regardless of whether or not they were part of the 
``same'' event.

In addition to exhibiting a flux increase, \gammaray\ burst photons should be 
statistically consistent with a point source origin.  Candidate bursts found in the time 
series analysis were examined spatially using a likelihood analysis 
technique \cite{eadie}.  Likelihoods were calculated for all photons 
within approximately $20\deg$ of the first photon of the interval.  A 
null model of smooth background was compared to a model with a 
variable position point source plus a background rate taken to be the 
average background found above.  To simplify computation, errors in 
photon position were taken from the width of the energy dependent best 
fit Gaussian containing 68\% of the \egret\ point-spread function
given by \eq{intro:containment}.
The usual \egret\ test statistic is then defined~\cite{like} as
in \sect{stats:hypotest}: 
$\TS \equiv\hbox{\index{$\equiv$}} {-2} (\ln \like_s - \ln \like_n)$,
where the $\like_s$ and $\like_n$ are the likelihoods of the source 
and null models, respectively.  The spatial analysis required 
likelihood calculation for each individual photon.  The standard \egret\ 
likelihood software, \likeprog~\cite{like}, is designed to evaluate 
likelihoods based on maps of photon counts.  Separate likelihood 
software was thus developed for this study.  Simplifications in this 
implementation make direct comparison of these values of TS with those 
from LIKE inexact.  For the purposes of source detection in \egret\ 
analyses, a TS of 16 is considered to be a significant ($4\sigma$) 
detection.  This is based on the measured distribution of TS. However, 
the likelihood statistic calculated here suffers from a very low count 
rate; often the likelihoods are computed from less than 10 photons.  
The statistics of TS with very few counts are not well characterized.  
Furthermore, the background rate across the whole $20\deg$ is 
taken to be the same as that found for the central $5\deg$ circle.  
In regions where the background is spatially varying, this may distort 
the measured TS.

Nevertheless, the TS measurement adds valuable information to the 
evaluation of burst candidates.  Candidates with a very low TS, 
corresponding to little or no point-like structure, should be 
discarded.  A sharp variation in photon arrival rate across a large 
area of sky is probably not due to an astrophysical process.  It 
should be remembered that there is already a selection effect of 
candidate events due to the fact that the time series analysis 
considers photons confined to a five degree circle.  This selection 
effect as well as the low count statistics make translation of TS into 
a confidence level problematic.  Note also that the spatial and time 
series probabilities are not completely independent.

\subsection{Results}

\begin{table}
\centering
\scriptsize
\begin{tabular}{llccccccc} 
 &  & & & &{\em Number} & {\em Number} & {\em Max Energy} & \\
\multicolumn{1}{c}{\em Date}  & 
\multicolumn{1}{c}{\em Time} &  $\myell$ & $b$ & TS &
{\em Expected} &
{\em Observed} & 
(MeV) & 
{ $P_T$} \\ \hline
1993 Jan 31 & 18:57:12 & 287 & 51 & 33.1 & 0.048  &6 &1240 &   $\gt 99.994\%$ \\
1994 Feb 17 & 23:03:05 & 152 & -55 & 18.8 & 0.074 & 5 & 3382 & 98.8\% \\
1994 Apr 27 & 01:31:01 & 121 & -0.7 & 26.7 & 0.120 & 6 & 680 &  95.4\% \\ 
1993 Mar 17 & 06:40:42 & 65  & 21 & 17.1 & 0.009 & 3 & 361 &  52.2\% \\
1992 Oct 8 & 04:35:04 & 201 & 31 & 15.7 & 0.047 & 4 & 508  & 32.9\% \\
1991 May 4 & 05:16:33 & 204 & 13 & 18.7 & 0.011 &3 &960 &     29.0\% \\
\end{tabular}
\caption[Three minute burst time scale results]
{\label{grb:3min}Three minute time scale results.  Burst candidates are
listed with their times (UT), galactic coordinates $\ell$ and $b$ of the
maximum likelihood position, spatial TS, expected and observed numbers
of photons, highest photon energy, and significance, as derived from
the Monte Carlo simulations. }
\end{table}

\begin{table}
\centering
\scriptsize
\begin{tabular}{llccccccc} 
 &  & & & & {\em Number} & {\em Number} & {\em Max Energy} & \\
\multicolumn{1}{c}{\em Date}  & 
\multicolumn{1}{c}{\em Time} &  $\myell$ & $b$ & TS &
{\em Expected} &
{\em Observed} & 
(MeV) & 
{$P_T$} \\ \hline
1993 Jan 31 & 18:57:12 & 287 & 51 & 33.1 & 0.160 & 6 & 1240 &  99.85\% \\
1993 Jul 20 & 07:07:55 & 121 & 42 & 23.2 & 0.097 & 5 & 878 &   74.0\% \\ 
1993 Feb 22 & 06:46:44 & 20 & -52 & 21.8 & 0.099 & 5 & 615 &   72.2\% \\  
1992 Mar 22 & 08:25:50 & 333 & 0.4 & 29.6 & 0.623 & 8 & 11100 & 28.1\% \\   
\end{tabular}
\caption[Ten minute burst time rcale results]
{\label{grb:10min}Ten minute time scale results, as in \tbl{grb:3min}.}
\end{table}

\begin{table}[t]
\centering
\scriptsize
\begin{tabular}{llccccccc} 
 &  & & & & {\em Number} & {\em Number} & {\em Max Energy} & \\
\multicolumn{1}{c}{\em Date}  & 
\multicolumn{1}{c}{\em Time} &  $\myell$ & $b$ & TS &
{\em Expected} &
{\em Observed} & 
(MeV) & 
{$P_T$} \\ \hline
1993 Jan 31 & 18:57:12 & 287 & 51 & 33.1 & 0.387 & 8 & 1240 & ~97\% \\ 
1993 Jul 20 & 06:50:45 & 120 & 42 & 28.8 & 0.205 & 6 & 880 &   ~77\% \\ 
\end{tabular}
\caption[Thirty minute time scale results]
{\label{grb:30min}Thirty minute time scale results, as in \tbl{grb:3min}.}
\end{table}

\begin{table}[t]
\centering
\scriptsize
\begin{tabular}{llccccccc} 
 &  &  & & & {\em Number} & {\em Number} & {\em Max Energy} & \\
\multicolumn{1}{c}{\em Date}  & 
\multicolumn{1}{c}{\em Time} &  $\myell$ & $b$ & TS &
{\em Expected} &
{\em Observed} & 
(MeV) & 
{$P_T$} \\ \hline
1993 Jan 31 & 18:57:12 & 287 & 51 & 33.1 & 0.387 & 8  & 1240 & ~97\% \\ 
1994 Apr 27 & 21:34:18 & 169 & 3 & 26.0 & 1.146 & 10  & 522 & ~70\% \\ 
\end{tabular}
\caption[One hour burst time scale results]
{\label{grb:1hour}One hour time scale results, as in \tbl{grb:3min}.}
\end{table}

We analyzed 182 viewing periods using these
methods, corresponding to observations between 1991 April 22 and 
1996 March 21.  Viewing periods after this time used new instrument 
modes to compensate for various instrument malfunctions and 
narrow-field viewing.  This data was not searched, since the field 
of view was significantly smaller, and the systematics of the new 
modes are not as well understood. Each viewing 
period was searched on four time scales: one hour, 30 minutes, 10
minutes, and 3 minutes.  Results are presented in Tables 
\ref{grb:3min}---\ref{grb:1hour}.
In each table, $P_T$ is the probability that no such
events would be detected in the entire \egret\ data set for that time
scale in the null hypothesis.
All fluxes quoted below are found by dividing the number of photons by
the total exposure in the scale time, even if the photons evidently
arrived in a much shorter time.  For events with very few photons, the
flux is not very well defined, since an arbitrary change in cutoff
time may reduce the exposure without changing the number of photons.
Thus, the flux for the purposes below is always found by considering
all of the exposure in the period.

Two of the bursts triggered by BATSE, the Super Bowl Burst of
1993 January 31 and the 1994 February 17 burst,  were also 
independently detected with this algorithm.
The other three \egret-detected BATSE-triggered bursts were not 
strong enough to be independently detected. 
For comparison, these bursts and their characteristics are listed in
\tbl{grb:triggers}.  Several independent detections were made, the
most significant occurring on 1994 April 27.

\begin{table}[t]
\centering
\begin{tabular}{lrcccc} 
& & {\em Number} & {\em Number} &   & \\
\multicolumn{1}{c}{\em Date} &
\multicolumn{1}{c}{\em Scale} &
{\em Expected} &
{\em Observed} &
{\em $P_S$} &
{$P_T$} \\ \hline
1991 May 3 & 3 min & 0.202 & 3 & 99.88\% & $<10^{-300}$ \\
& 10 min & 0.672 & 5 & 99.93\% & $<10^{-300}$ \\
& 30 min & 1.288 & 5 & 98.97\% & $<10^{-300}$ \\
& 1 hour & 1.288 & 5 & 98.97\% & $<10^{-300}$ \\
1991 June 1 & 3 min & 0.396 & 4 & 99.93\% & $<10^{-300}$ \\
& 10 min & 0.633 & 5 & 99.95\% & $<10^{-300}$ \\
& 30 min & 0.633 & 5 & 99.95\%  & $<10^{-300}$ \\
& 1 hour & 2.064 & 9 & 99.97\%  & $<10^{-300}$ \\
1994 March 1 & 3 min & 0.102 & 3 & 99.98\% & $<10^{-300}$ \\
&10 min& 0.333 & 4 & 99.96\%  & $<10^{-300}$ \\
& 30 min & 0.569 & 4 & 99.72\%  & $<10^{-300}$ \\
& 60 min & 0.569 & 4 & 99.72\%  & $<10^{-300}$ \\
\end{tabular}
\caption[Triggered bursts not independently detected]
{\label{grb:triggers}Known \batse-triggered \gammaray\ bursts
not detected independently by \egret\ in each time scale.  The
expected and observed numbers of photons are given, along with
the single trial probability $P_S$ and the significance $P_T$
as determined by the Monte Carlo.}
\end{table}

\subsection{Discussion}

Each $P_T$ is the chance that
measured photon arrival times are {\em not} due to random Poisson
noise in the entire \egret\ data set for that particular time scale.
Thus, an event $P_T=90\%$ in the one hour time scale would be a
spurious detection in 10\% of the ensemble of possible \egret\ data sets
searched on a one hour time scale.  However, four time scales have
been searched.  If each time scale were 
independent, this would be counted as four trials.  But the same data
were searched in each time scale, so the trials are not
independent.  Rigorous evaluation of the overall number of independent
trials is problematic, and is not attempted for fear of producing meaningless
results.

The calculated probabilities also do not take into account
the photon energies.  However, high energy photons are rarer than lower energy
photons, so the arrival of a similar number of high energy photons
would constitute a more significant event.  If the detected bursts
exhibited a plethora or paucity of high energy photons,
the event energy spectrum would be expected to 
differ appreciably from that of the background.  Because each event has
too few photons to meaningfully determine a spectrum, a hardness ratio was
calculated.  All photons from all bursts in each time scale were
collected, and a 
hardness ratio for the set was calculated.  Even such a
collection suffers from poor counting statistics, as evidenced by the
large errors (\tbl{grb:hardness}).  The hardness ratio
is defined as the number of photons with energies greater than 300 MeV
divided by the number with energies between 100 and 300 MeV. 
A typical background spectral index of 2.0 
 corresponds to a hardness ratio of 0.5.  The measured hardness
ratios are consistent with the background hardness ratios, implying
that the $P_T$ would not be significantly modified
if energy information were taken into account.

All probabilities are dependent on the statistical distribution found
from the Monte Carlo simulation.  Thus, any possible factors which
might make the Monte Carlo an imperfect simulation of the real data
must be examined.  Edge effects may slightly affect the final
distribution, since the 5$\deg$ circles drawn around photons near the
edge of the region will reach past the edge.  However, a short
simulation was done in a $20\deg$ by $20\deg$ square, and the resultant
distribution did not differ significantly from that calculated with a
larger area,
indicating that edge effects play a small part.  

The Monte Carlo simulation also assumed spatially uniform exposure and
background rate in order to make computation time reasonable.  These
assumptions will be poor only when the exposure or rate varies
significantly over the 5$\deg$ circle.  Exposure generally varies
smoothly, so this is probably a good approximation.  The background
rate can vary extensively if a strong point source is nearby, but the diffuse
background does not vary rapidly on this scale.  Thus, we may need to worry
about the distribution if events in the real data are detected near
steady point sources; otherwise, the constant background approximation
should be relatively good.

\begin{table}[t]
\centering
\begin{tabular}{lr} 
{\em Time Scale} & {\em Hardness Ratio} \\ \hline
3 min & $0.723 \mypm 0.338$ \\
10 min & $0.385 \mypm 0.202$ \\
30 min & $0.500 \mypm 0.433$ \\
1 hour & $0.500 \mypm 0.306$ \\
\end{tabular}
\caption[Hardness ratios for burst candidates]
{\label{grb:hardness}Hardness ratios for each time scale.  The hardness
ratio is defined as the number of photons above 300~MeV divided by the
number of photons between 100--300~MeV.}
\end{table}

Spatial correlations must also be considered in the evaluation of
burst candidates.  Unfortunately, rigorous treatment of spatial
correlations is fraught with difficulty.  The maximum likelihood
distribution is not well characterized for low counts.  Furthermore,
the $5^{\circ}$ search radius used will select for spatially
correlated events.  It was hoped that spatial analysis would add 
significantly to our understanding of these events; unfortunately, 
because of these biases and correlations the spatial analysis has
yielded little additional insight.

The most interesting candidate event is clearly the 1994~April~27~01:31
event in the 3 minute time scale.  The significance of this event,
while much lower than the \batse-independent detection of the Super
Bowl Burst, is almost as high as the \batse-independent detection of
the 1994 February 17 burst.  
This detection occurred while \egret\ was in its
most common, largest effective area mode.  The Earth zenith angle to
the event was $\sim40 \deg$, well away from the horizon.  It was
observed $27 \deg$ off the instrument axis.  
It thus seems unlikely
that it is a spurious 
detection caused by pushing the operating envelope of the instrument.
In addition to the six photons within three minutes of elapsed time
classified as good events, there were
seven additional events which were rejected 
by the standard \egret\ data analysis for a variety of reasons.  One of
these generated too few sparks in the spark chamber, one had its vertex
in the wall of the chamber, and the rest failed to produce an acceptable
time of flight measurement.  Estimated
trajectories of these events suggest that they originated from the
same area on the sky as the good events.  Although a precise
calculation of the significance of these events is impossible, there
appear to be more such events from that direction on the sky than
would otherwise be expected.
%

The time interval corresponding to the 1994 April 27 candidate was
examined in the 30 - 100 MeV range as well.  No photons were detected
during that period.  However, \egret\ sensitive area to this energy
range is comparatively small, so that only 0.046 background photons
were expected (compared to 0.120 in the 100 MeV and above range.)
The TASC and anti-coincidence dome rates did not show a significant
change in rate.
With such small numbers, it is difficult to assess the significance of
the lack of low energy photons.

Furthermore, a check was done of BATSE detections nearby.  No trigger
occurred at that time, although BATSE triggering was enabled.  The
last previous trigger occurred some 20 hours earlier.

No other bursts were detected so significantly.  However, there were
several less significant detections, making it worthwhile to consider
the probability that at least one \gammaray\ burst has been detected.
In the three
minute time scale, the April 27, 1994 burst is fairly significant by itself.
However, it is interesting to calculate the joint probability that all
the events in each time scale are due to 
random Poisson noise.  We will exclude the Super Bowl Burst
and the February 17 Burst, since they are
already known to be a \gammaray\ bursts.  
Table~\ref{grb:jointconfs} shows the joint probabilities in each
time scale from the arrival time data.  There is apparently
some evidence for burst occurrence in the short time scales,
with no significant evidence for occurrence in the half-hour and hour long time
scales.

The detected candidate events were compared to the Second BATSE Burst
Catalog to see if any coincident detections were made.  With the
exception of the bursts actually triggered by BATSE, that is, the
Super Bowl and February 17 bursts, no coincidences were found.

\begin{table}
\centering
\begin{tabular}{l|r}
{\em Time Scale} & {Probability} \\ \hline
3 min & 99.0\% \\
10 min & 94.8\% \\
30 min & 77\% \\
1 hour & 70\% \\
\end{tabular}
\caption[Joint burst probabilities]
{\label{grb:jointconfs}Joint probabilities of detecting all the marginal
candidates in each time scale.  While systematic errors could be 
responsible, there is apparently evidence for burst activity at or below
the three minute time scale.}
\end{table}


While the Poisson error analysis may be encouraging, it is important
to consider the contribution of systematic errors before coming to any
conclusions.  The search for \gammaray\ bursts by the method used is
most sensitive to 
systematic errors in the determination of background rates. 
 If the background is consistently underestimated
relative to the burst photons, many apparently significant events are
not actually significant.

The most likely source of systematic errors in the background rate is
the error in the estimate 
of the instrument exposure.  A subtle point should be explored here.
\egret\ switches observing modes every few minutes.  Each mode has a different
amount of sensitive area to a given position on the sky, as well as a different
field of view.  Approximately 80\% of the time \egret\ is in the single mode with
the largest field of view and sensitive area.
In the limit that the instrument is always in the same observing mode,
any exposure errors will cancel each other, since the error induced
in the background flux will be exactly the same as the error induced
in the burst flux.  However, if the exposure error is different for
each instrument mode, then a burst candidate detected in a rare mode
might exhibit considerable systematic error.

The sensitivity of \egret\ diminishes as the gas in the spark chamber
ages.  This effect has been partially characterized
 and compensated for in the \egret\ exposure data.  The time
scale for this drift is typically several months, much longer than the
burst search scales and also at least twice as long as the longest
background averaging time.

The nondetection of the 1994 April 27 candidate burst by BATSE must be
considered in light of current models for \gammaray\ burst production.
The relativistic fireball model of \meszaros, Rees, and Papathanassiou
\cite{mrp94} suggests the
possibility of a high energy flat spectrum tail.  If the spectrum is
flat down to the BATSE energy range, the signal would be lost in the
soft-gamma background.  However, that model
does not predict a flat spectrum at BATSE wavelengths.
Significant redshift effects might move the break of the fireball
model spectrum to below BATSE energies, but then x-ray detection would
be expected.  Nevertheless, other models \cite{dcb98} suggest 
that at the earliest times after the burst, the spectral break
could be at \egret\ energies.  

\section{Conclusions}

The mystery of the \gammaray\ bursts has recently experienced its third 
major revelation.  The first occurred in the late 1960s and early
1970s, when the energetic bursts were discovered.  The next occurred
in 1992, when \batse\ made clear the isotropic, non-homogeneous 
distribution of the bursts---pointing to the almost inconceivable
conclusion that the bursts were of cosmological origin.  The third
was the association, first by \sax, and then by many others of a 
few bursts with counterpart afterglows in many wavelengths, resulting
in the measurement of cosmological redshifts.

One of the outstanding questions about \gammaray\ bursts has thus been
answered.  Nevertheless, we are far from a complete understanding
of the burst mechanism.  The central engine
is probably either a 
merger of neutron stars or a hypernova. 
Energetic concerns suggest fairly strongly
that the initial energy release goes into baryonic kinetic energy,
and that the observed burst is radiation from a shock front created
when the outflowing baryons sweep up surrounding material.  The 
existence of shock fronts suggests high-energy power law emission,
but the exact mechanism is still unknown.

Given the uncertainty of the field, it was judged prudent
to examine the \egret\ data for the presence of \gammaray\ 
bursts not detected in other wavelengths.
One possible detection of a high-energy \gammaray\
burst event has been found in \egret\ data independent of a trigger
from \batse.  Two previously detected \gammaray\ bursts were
independently detected, verifying the search method.
The new event, occurring on 1994 April 27, was detected
with a statistical significance of 95.4\% on a 3 minute time scale 
although this does not include systematic errors.
Pointlike structure and the presence of several additional
photons which converted in the walls of the \egret\ instrument lend
qualitative support to the detection.  

 Taken together, several less
significant events suggest burst activity 
on the three minute time scale with probability 99.0\%.  Activity is
suggested at a somewhat lower probability on the ten minute time
scale, and at significantly lower probability on thirty minute and one
hour time scales.

%% file: Timevar.tex
\chapter{Periodic Time-Series Analysis}
\label{timevarchap}

The full-sky map of the \gammaray\ sky that \egret\ made in its first 
year of operation provided unprecedented discovery and localization of 
\gammaray\ sources (\tbl{intro:compare}).  However, \egret\ also 
provides excellent arrival time information about individual photons.  
Combined with good directional information, this timing data can be 
used to look for coherent periodic variation.  Six pulsars have been 
positively identified in \egret\ data, and the precise rotation of 
pulsars makes them an obvious candidate for periodic analysis.  X-ray 
binary systems also exhibit regular periodic behavior in the X-ray 
bands; it may be hoped that some of these could be detected by \egret\ 
as well.

Since there have been extensive \egret\ observations of pulsars, we 
will begin by looking at pulsar models for which \egret\ (or \glast) 
may have discriminating power, as well as the contributions made to 
pulsar understanding by \egret\ so far.  We will develop the necessary 
statistical apparatus to search for periodicity in a \gammaray\ 
signal.  We will then see how to apply likelihood statistics to pulsar 
analysis, and the results of that analysis.  Finally, we will see how 
the statistical tools for periodic time-series analysis may be applied 
to X-ray binary searches.

\section{Pulsars}

Two years before the Vela satellites would begin to secretly detect 
\gammaray\ bursts, Jocelyn Bell and Anthony Hewish detected a periodic 
radio signal that pulsed every 1.3~s~\cite{firstpulsar}.  The signal 
remained coherent for days at a time.  However, the mysterious source 
of the signal would be divined before the \gammaray\ bursts would be 
declassified.  The discovery of the Vela pulsar, with a period of 
89~ms~\cite{vela_discovery}, and the Crab pulsar, with a period of 
33~ms~\cite{crab_discovery}, would narrow the field of prospective 
sources to one: a rapidly rotating neutron star.  Such an object is 
the only astrophysical object capable of radiating the power observed 
at the pulse frequencies observed without flying apart under its own 
rotation.

The basic description of the pulsar was worked out over the next few 
years.  It had long been realized~\cite{chandrasekar} that a 
sufficiently massive body would overcome its electron degeneracy 
pressure by self-gravitation.  This situation can come about when a 
massive star ($\gt M_{\sun}$\index{$M_{\sun}$}) exhausts its nuclear 
fuel.  Without the energy released by the fusion reactions in the 
core, the star collapses, the gravitational potential condenses the 
electrons and protons, and a central core of neutrons develops.  The 
remaining outer material is blown off in a catastrophic explosion: the 
supernova.  Conservation of some fraction of the original stellar 
angular momentum results in the central neutron star rotating with a 
frequency of 0.1--100~Hz.  Conservation of the magnetic flux in the 
star (assuming it remains a perfect conductor) results in a magnetic 
field at the neutron star surface of order $10^{12}$~G. The observed 
flux from pulsars, then, is the result of a rapidly rotating magnetic 
dipole~(\eg,~\cite{joethesis}).

Some basic pulsar parameters may be inferred from elementary 
mechanics.  If we make the assumption that all radiated pulsar power 
comes from the rotational energy, then the power is given by the time 
derivative of the rotational energy $\frac{1}{2} I \Omega^2$, where 
$I$ is the moment of inertia and $\Omega$ is the rotation frequency.  
Converting this to the pulsar period and period derivative, we have
\begin{equation}
\dot{E}= (2 \pi)^2 I \dot{p}/p^3
\end{equation}
Making some basic assumptions about a pulsar as a rotating dipole, 
Ostriker \& Gunn~\cite{ostriker} find that the radiated power can be 
related
to the magnetic field:
\begin{equation}
\dot{E} \simeq \frac{2}{3} \; \frac{B^2 a^6 \Omega^4}{c^3}
\end{equation}
where $B$ is the surface magnetic field and $a$ is the radius of the
pulsar (typically 10-15~km).
Combining these two, we arrive at an estimate of the field strength in
terms of the period and period derivative
\begin{equation}
	\label{tv:magfield}
B^2 = \frac{2}{3} (2 \pi)^2 \left( \frac{c^3}{a^6} \right) I p \dot{p}
\end{equation}
Typical values of pulsar parameters are given in \tbl{tv:typpul}.  For 
any given pulsar, we assume that its radius and moment are constant.  
If we assume also that its magnetic field remains constant, we arrive 
at the differential equation $p \dot{p} = k$ governing the pulsar's 
age.  This equation is integrable, yielding
\begin{equation}
\frac{1}{2} (p^2 - p_0^2) = k \tau
\end{equation}
where $p_0$ was the pulsar period at time zero, and $k$ is a constant 
formed from the parameters in \eq{tv:magfield}.  By convention, we 
assume that the final period is much larger than the initial period, 
and, solving for $\tau$ above, designate the pulsar's characteristic 
age by
\begin{equation}
\label{tv:ageeqn}
\tau = \frac{1}{2} p/\dot{p}
\end{equation}
The characteristic age seems to be a reasonable approximation to the 
actual age for the only pulsar we can verify.  The Crab pulsar was 
born in a supernova which was observed at Earth on July 4, 
1054~\cite{ancientcrab,ancientcrabII}.  The Crab rotates with a period 
of \mysim 33.39~ms, and has a period derivative of $\mysim 4.21 \times 
10^{-13}$, yielding a characteristic age of about 1280~yrs.  This is 
correct to within 30\%, but the assumption of linear spindown is 
probably least accurate at very short times after the supernova 
explosion.  This effect will diminish with pulsar age, while glitches 
in the pulsar period and its derivative will add error to our age 
estimate.  One possible mechanism is gravitational wave emission due 
to mass multipoles~\cite{ostriker}.  Ostriker \& Gunn find that to 
fully account for the age difference, approximately $1/6$ of the 
Crab's current luminosity would be in gravitational waves.

\begin{table}
\centering
\begin{tabular}{lr}
Parameter & Typical Values \\ \hline
Period $p$ & 1.6~ms--6~s \\
Period derivative $\dot{p}$ & $10^{-15}$--$10^{-13}$ s/s \\
Magnetic Field & $10^{12}$--$10^{14}$ G \\
Characteristic Age & $10^3$--$10^6$ yr \\
Moment of Inertia & $\mysim 10^{45}$ g cm$^2$ \\
Radiated Power $\dot{E}$ & $\mysim 10^{31}$--$10^{34}$ erg s$^{-1}$ \\
\end{tabular}
\caption[Typical pulsar parameters]{\label{tv:typpul}
Typical pulsar parameters for \gammaray\ pulsars.}
\end{table}

Some elementary physics gives a reasonable first approximation to a 
pulsar model.  Nevertheless, physics is a fractal subject, and the 
number of questions still to be answered is seemingly independent of 
the number of questions already answered.  Fortunately for \egret, 
\gammarays\ turn out to be an excellent window on pulsar 
energy-generation mechanisms.
  
\subsection{\egret\ Contributions}
The \egret\ instrument along with the other instruments on board the 
\cgro\ has revolutionized our understanding of at least one class of
pulsars.  Extensive reviews of those 
contributions~\cite{joethesis,fierro_pulsarsII,nel_pulsarsIII,
nolan_survey,thompsonreview,thompson_pulsarsI} have been published 
elsewhere; we will only review their highlights.

Six pulsars have been unambiguously discovered in high-energy 
\gammarays.  All six share some common features.  First of all, they 
tend to have double-peaked light curves, with the peaks separated by 
somewhat less than 180\deg.  \Gammaray\ pulsars tend to have harder 
spectra than most sources (spectral indices range from $-1.4$ to 
$-1.8$, compared to the average \gammaray\ source spectral index of 
$\mysim 2.3$).  Five of the six have among the highest measured values 
of $\dot{E}/D^{2}$, the radiated power divided by the pulsar distance 
squared.  $\dot{E}/D^{2}$ should be proportional to the observed power 
at Earth.  It may be calculated by assuming that all of the kinetic 
energy lost from the pulsar spin-down is radiated isotropically, and 
that the radio dispersion measure gives a good distance estimate to 
the pulsar.

The Crab pulsar was detected by both \sas~\cite{sascrab} and 
\cosb~\cite{cosbcrab,cosbcrab2} in \gammarays\ some ten years after 
its initial radio discovery~\cite{crab_discovery}.  It was strongly 
detected in many different \egret\ viewing periods, yielding high 
quality light curves and 
spectra~\cite{nolan_crab,ulmer_crab,joethesis,fierro_phaseresolved}.  
All \gammaray\ pulsars are observed to have very steady luminosities; 
in fact, they have been used as constant calibration sources for some 
variability studies~\cite{mclaughlin}.  Early observers suggested, 
however, that the ratio of the luminosity in the two \gammaray\ light 
curve peaks of the Crab pulsar varied with a 14-year 
cycle~\cite{wills14year,ozel14year}.  However, further observations 
ruled out significant variation on timescales of order 14 
years~\cite{tompkins_crab}.

When it was first discovered in radio 
observations~\cite{vela_discovery}, no one expected the Vela pulsar to 
be the brightest object in the \gammaray\ sky.  Nevertheless, it was 
the strongest source detected by \sas~\cite{sas_vela}, and remains the 
brightest steady source in the \egret\ catalog~\cite{cat3}.  Although 
there is significant timing noise (that is, the pulsar period 
fluctuates on short time scales), Vela is easily detected in a number 
of \egret\ viewing periods~\cite{kanbach_vela}.

The most fascinating \gammaray\ pulsar is Geminga.  The third 
brightest object in the \gammaray\ sky after Vela and the Crab, it was 
unique when discovered by \sas\ in 1975 in that it had no obvious 
counterpart in other 
wavelengths~\cite{gem_discoveryI,gem_discoveryII}.  Most notably, it 
had no radio counterpart---a prominent feature of all other known 
pulsars.  It would be 17 years before pulsation was discovered in the 
X-ray emission~\cite{gem_pulsation}.  Using the X-ray ephemeris, 
pulsation was quickly detected in \egret~\cite{bertsch_geminga}, 
\sas~\cite{sas_gem}, and \cosb~\cite{cosb_gem} data.  Indeed, the 
\gammaray\ signal is so strong that various period searching 
techniques would have independently discovered the pulsation in 
Geminga without the aid of an X-ray ephemeris 
(\sect{tv:detect_geminga},~\cite{fouriersearch}).  Recently, there has 
been an unconfirmed report of pulsation detected in 
radio~\cite{radio_gem}; whether or not this detection can be 
confirmed, the radio emission is clearly very weak.

The remaining three \gammaray\ pulsars are less well-known, but no 
less important to our understanding (or lack thereof) of \gammaray\ 
pulsar physics.  \stos\ was only discovered as a radio pulsar in 
1992~\cite{radio_1706} and soon thereafter in the \egret\ 
data~\cite{thompson_1706a}.  Unlike the other five \gammaray\ pulsars, 
which have two peaks separated by 140\deg--180\deg\ in phase, \stos\ 
exhibits a single peak in its \gammaray\ light curve.  Unexpected in 
its own way is \tff.  It was discovered as a \gammaray\ pulsar in 
\egret\ data in 1993~\cite{fierro_1055}, although it ranks 
approximately 29th in known $\dot{E}/D^2$~\cite{joethesis}.  Why \tff\ 
is observed when other, apparently more luminous pulsars are not 
remains a mystery.  As a final confusion, \ntfo---fourth in 
$\dot{E}/D^2$---was not observed as a pulsed source in \egret\ 
Phase~I. Further pointed observations in Phase~III resulted in a weak 
pulsed signal~\cite{murthy_1951}.

\egret\ observations of \gammaray\ pulsars pose a unique challenge for 
theorists.  The small number of pulsars and the wide variety of 
observed behavior makes it difficult for any one model to match all 
the observations while remaining simple enough to make concrete 
predictions.  Nevertheless, two main classes of models dominate the 
thinking of most pulsar theorists today.

\subsection{Models}
The theory and ramifications of \gammaray\ pulsar models are 
extensively discussed 
elsewhere~\cite{lynebook,joethesis,thompsonreview}; we will review the 
highlights of the two major models.  Most current \gammaray\ pulsar 
models may be categorized as either ``polar-cap'' or ``outer-gap'' 
models.  These two classes differ mainly in the region where the 
energy is emitted.

\subsubsection{Polar-cap models}
The polar-cap models build on the early pulsar models of Goldreich \& 
Julian~\cite{standardpulsar}, which defined the standard pulsar 
framework.  Although many researchers have extended their work, 
Daugherty and Harding are largely responsible for developing the model 
for \gammaray\ emission~\cite{polarcap,polarcapII}.  The ``polar cap'' 
is a region at the magnetic pole of the pulsar, which in general is 
not aligned with the rotation axis.  Some magnetic field lines emerge 
from one pole and re-enter the pulsar at the other pole.  However, 
since the pulsar is rapidly rotating, there is some radius (given by 
$r = \Omega/c$) at which the field lines can no longer corotate with 
the pulsar without exceeding the speed of light.  This radius defines 
the light cylinder.  The closer the point from which the field line 
emanates is to the magnetic pole, the closer the field lines will 
approach the light cylinder.  Field lines which emanate from within 
some critical radius will not return to the pulsar; instead, they will 
close at infinity.  The polar cap is that region around the magnetic 
pole inside the critical radius.

The polar cap is important because free charges placed on open field 
lines will escape to infinity.  Potential differences across the polar 
cap can rip charge off of the pulsar surface.  This charge then flows 
along the curved field lines, emitting curvature radiation.  At the 
lowest altitudes above the pulsar surface, this curvature radiation is 
so energetic that it pair-produces in the magnetic field.  Eventually, 
the charged particles have lost enough energy that the photons 
(\gammarays) no longer pair produce.  These \gammarays\ are observed 
by \egret.  The details of the field shapes and the emission regions 
determine the \gammaray\ and radio light curves.  Currently, it seems 
as if many \gammaray\ pulsar characteristics can be explained with the 
polar cap model.  However, it has been suggested that 
general-relativistic frame-dragging effects may more strongly 
influence the \gammaray\ emission than the precise shape of the 
magnetic field~\cite{muslimov,mus&harding}.

\subsubsection{Outer-gap models}
A substantially different model of energy emission is the 
``outer-gap'' model, first proposed by Cheng, Ho, \& 
Ruderman~\cite{outergap,outergapII}.  In these models, vacuum gaps may 
arise along the last closed field lines in the pulsar magnetosphere.
The gaps separate charge of different sign on 
opposite sides of the last closed field line, causing significant 
electric potentials.  Charged particles may be accelerated across 
these potentials, following the magnetic field and emitting curvature 
radiation.  The details of the shapes of the emitting regions and the 
resultant light curves are not immediately obvious, but extensive work 
has been done to calculate the impact of the magnetosphere geometry on 
the \gammaray\ observations~\cite{romanimodel,romani}.  The outer-gap 
models also seem to describe many of the qualitative features observed 
in \gammaray\ pulsars.

\subsubsection{Remaining Questions}
It is still not clear which model of pulsar emission more accurately 
describes \gammaray\ pulsars.  In general, \gammarays\ emitted from 
outer-gap regions will be emitted in a more fan-like pattern, while 
\gammarays\ emitted from polar-caps will be more tightly beams.  
However, both models can reproduce a subset of the pulsar 
characteristics observed in \egret\ 
data~\cite{nolan_survey,thompson_pulsarsI,fierro_phaseresolved,joethesis}.

Unfortunately, the sample of observed \gammaray\ pulsars is still 
small, and it is difficult to determine which observed characteristics 
are generally representative and which are particular to the observed 
pulsars.  It is therefore critical to increase the number of observed 
\gammaray\ pulsars.  The \glast\ mission will bring a superior 
\gammaray\ telescope to bear on the problem (\chapt{glast:tng}).  
However, it is possible that there are additional Geminga-like 
radio-quiet pulsars which have already been observed by \egret.  Our 
task is to find likely pulsar candidates, and determine their 
pulsation parameters.

\section{Statistical Methods}
Searching for pulsation in \egret\  sources poses a unique set of
problems to the pulsar researcher.  Dispersion, the scourge of 
radio pulsar searches, is non-existent in \gammarays.  However,
since the flux is quantized into \gammarays, \egret\ might detect
a photon from a pulsar only once per thousand rotational periods.
Long integrations of coherent pulsar emission are required to 
resolve the periodic behavior.
Furthermore, the particular structure of \egret\  data, organized by
individual photon information, is significantly different than the
flux information found in other wavelengths.  Low count rates mean
that the instrument state can vary significantly between the arrival
of individual photons.  The low rates also require that as much
information as possible be extracted from each photon.

For these reasons, special statistical care must be taken in 
analyzing \egret\ data for pulsar signals.  Extensive coherent 
pulsation analysis has been already been done, and it will be 
instructive to examine the challenges specific to \egret\ previously 
overcome (\sect{tv:prevmet}).  First, however, we must approach some 
difficulties basic to all pulsation analysis.

In order to be sensitive to pulsed emission from the fastest pulsing 
sources known (a few milliseconds), we must have timing accuracy to a 
fraction of the pulse period.  \egret\ records photon arrival times
to an absolute accuracy of 100~\us, two and a half orders of magnitude
smaller than the period of the Crab.
However, the radius of the Earth's 
orbit is 8 light-minutes.  Photons from source observations separated 
by 
six months time will have light travel times differing by hundreds of 
thousands of pulse periods.  In order to compensate for the motion of 
the Earth, we will translate all measured arrival times into Solar 
System Barycenter (SSB) times.  The SSB time is the time when a 
photon would have arrived at the solar system barycenter if none of 
the mass of the solar system were present.  The conversion between 
measured time ($t_{\utc}$) and barycenter time $t_{b}$ has been 
worked out in detail elsewhere~\cite{joethesis,taylor89}.  The 
expression will be motivated and examined in \app{timedelay}; the 
schematic result is:
\begin{equation}
\label{tv:schemdelay}
	t_{b} = t_{\utc} 
	+  \Delta_{\hbox{\scriptsize convention}}
	+ \Delta_{\hbox{\scriptsize location}}
	+ \Delta_{\hbox{\scriptsize Einstein}}
	+ \Delta_{\hbox{\scriptsize Shapiro}}
\end{equation}
where adjustments to Universal Time are made to correct for 
bookkeeping conventions such as leap seconds, for the current
location of the spacecraft with respect to the barycenter,
for the gravitational redshift (``Einstein delay'') effects, and
for the delay due to the gravitational potential (``Shapiro delay'').  The
full result is given by \eq{delay:full}.

\subsection{Previous Methods}
	\label{tv:prevmet}
Two methods have been used to examine coherent periodicity in \egret\ 
data.  The first one, used for analysis of pulsars with known 
ephemerides, was implemented by Joe Fierro in a program called 
\pulsar~\cite{pulsardef}.  \pulsar\ selects photons within an 
energy-dependent cone about the source position.  The acceptance 
radius, chosen to maximize the signal-to-noise ratio, is given by 
\eq{intro:containment}.  The pulsar period, along with the first and 
second period derivatives, are used to assign to each photon a {\em 
phase}, corresponding to the fraction of a rotation through which the 
pulsar has turned since a reference time $t_{0}$.  Recall that a 
\gammaray\ will arrive, on average, only once per hundred or thousand 
rotations of the pulsar; by retaining only the fractional 
rotation, the emission from many rotations is added together 
coherently.  This process is known as {\em epoch folding}.  Plotting 
the resulting photon rate as a function of phase yields a light 
curve---the average profile of emission through a rotation.  Photons 
can then be selected as a function of phase for further analysis.

The phase of a photon is determined from the pulsation period and 
its derivatives, as measured from some reference epoch:
\begin{equation}
\label{tv:phase}
\phi(t) = \phi(t_0) + \nu(t - t_0) + 1/2 \dot{\nu}(t-t_0)^2 
+1/6 \ddot{\nu}(t-t_0)^3 + \dots
\end{equation}
where $\nu = 1/p$ is the pulsation frequency and $p$ the period.
In practice, it is only rarely that the second-derivative term
of the Taylor expansion is important; higher orders are negligible.
The constant term in the expansion is kept only to allow \gammaray\ 
light curves to be compared with radio light curves.

Of primary importance in such an analysis is the determination of the 
presence of intensity modulation at the known frequency.  Ephemerides 
for radio pulsars have been collected into a database stored at 
Princeton University~\cite{ptondbase}.  \Gammaray\ photon phases 
based on 
these ephemerides may be tested for periodicity.

\subsubsection{The $\chi^2$ test}
The simplest test is the $\chi^{2}$ test.  Photons are sorted into $m$ 
bins, based on their phase.  Since photon count statistics are 
distributed as a Poisson, the variance in the number of photons in 
each bin is $N_m$, the number of photons in the bin.  The usual 
$\chi^{2}$ statistic is calculated, and the probability of the 
observation under the null model is computed.  A sufficiently low 
probability of observation under the null model is taken as evidence 
for variability.  While this method is simple, it has a number of 
undesirable aspects.  First, the significance of the detection depends 
strongly on accidents of bin boundaries---a problem inherent to 
binning schemes.  Second, the $\chi^{2}$ is invariant under 
permutations of bins.  We expect pulsar light curves to smoothly vary, 
with one or two main peaks, and would like our statistical method to 
have some power to identify such light curves.

\subsubsection{The $Z_m^2$ test}
The $Z_m^2$ test was developed to try to address some of these 
difficulties~\cite{beran,buccheri83,joethesis}.  The basis of this 
test is the realization that the light curve is well represented by 
the phase density $f(\phi)$, which is the fraction of the integrated 
intensity in each $d\phi$.  In its simplest form, corresponding to 
$m=1$ and known as the Rayleigh statistic~\cite{mardia}, this phase
density will be compared to a sinusoid of unknown amplitude and 
phase.  The general form uses more terms of the Fourier expansion.
The best estimate of the phase density 
function is a series of $\delta$-functions:
\begin{equation}
f(\phi) = \frac{1}{N} \sum_{i=1}^N \delta(\phi_i)
\end{equation}
We will define $f_m(\phi)$ as the function composed of the first $m$ 
Fourier components of $f(\phi)$.  To calculate the $Z_m^2$ statistic, 
we determine a comparison (or model) phase density function 
$\tilde{f}_m(\phi)$.  
We then find the squared difference from $f_m(\phi)$ and 
$\tilde{f}_m(\phi)$
\begin{equation}
Z_m^2 \equiv 2 \pi N \int_0^{2 \pi} \left[ f_{m}(\phi) - 
\tilde{f}_m(\phi) \right]^2 d\phi
\end{equation}\index{$\equiv$}
If we take the comparison density function to be constant, this is 
simply computed; it can be shown to be equal to $2 N \sum_{k=1}^m 
(\alpha_k^2 + \beta_k^2)$, where $\alpha_k$ and $\beta_k$ are the 
even 
and odd Fourier coefficients from $f_m(\phi)$.  $Z_m^2$ is 
asymptotically distributed as $\chi^2$ with $2 m$ degrees of 
freedom~\cite{bendat}.

This approach clearly mitigates some of the problems with the 
$\chi^2$ test.  Primarily, photon binning has been eliminated.  In 
addition, the Fourier decomposition expects a periodic signal---the 
degeneracy to permutation of flux peak location is removed.  
Unfortunately, the $Z_m^2$ statistic has shortcomings of its own.  
The 
significance of a detection is a strong function of the number of 
harmonics chosen.  It is not difficult to see why this should be the 
case.  Retaining only $m$ harmonics of a sum of $\delta$-functions 
amounts to smoothing the observed phase density function.  Too much 
smoothing (a very small $m$) washes out any signal that may be 
present.  Too little smoothing (a very large $m$) results in the 
signal being swamped by noise from the higher harmonics.  To 
ameliorate this difficulty, another statistic was invented.

\subsubsection{The $H$-test}
In the quest to obtain more and more significant results, it naturally 
occurred to astrophysicists to try a variety of values of $m$, and see 
which one results in the most significant value of $Z_m^2$.  
Fortunately, de Jager \etal\ realized that the set of $Z_m^2$ obtained 
in this way would no longer be distributed as $\chi^2_{2 m}$.  At this 
point, there is no further guiding insight; an {\em ad hoc\ } limit of 
20 is set on $m$, and the distribution of $H \equiv \max_{1 \le m \le 
20} Z_m^2 - 4 m + 4$ is found from a Monte Carlo simulation.  The 
inclusion of the term involving $m$ favors models with fewer Fourier 
components, and the constant ensures that the statistic will be 
positive.  This test has become the primary tool to search for a 
pulsed signal in \egret\ data~\cite{joethesis}.

\subsubsection{Limitations of these statistics}
\label{tv:htestoops}
Statistics for analyzing pulsar data have evolved, with each 
generation of statistics alleviating some of the problems of the 
previous generation, but creating problems of its own.  The discussion 
of likelihood statistics in \sect{stats:likelihood} sheds some light 
on why these difficulties arise.  A full discussion of the application 
of likelihood statistics to this problem will be given in 
\sect{tv:like}; here we will quickly examine the above statistical 
measures.

Since the $H$-test is the most sophisticated measure, we will focus on 
it.  If we permit ourselves to use some Bayesian language, we see that 
the goal of the $H$-test is to eliminate the nuisance parameter $m$.  
The method is really to take $H$ as a function of $m$, and maximize 
this $H(m)$ with respect to $m$.
Each $Z_m^2$ has the form of the squared difference between the null 
model and a representation of the data comprised of $m$ 
harmonics\footnote{Note that each representation, labeled by $m$, is 
derived from the data, which is not permitted in Bayesian analysis.  
Therefore, the likelihood formed is not ``the likelihood of the data, 
given the model,'' but ``the likelihood of the representation, given 
the model.''  Since the representation is a (lossy) smoothing of the 
data, we have no reason to expect a one-to-one relationship between 
the likelihoods.  Nevertheless, the qualitative argument is still 
valid---this only means that multiplying the $Z_m^2$ by a prior and 
integrating over $m$ would still give incorrect numerical results.}.  
This is just the logarithm of the likelihood of the ``data 
representation,'' assuming a null model and Gaussian errors of 
variance $N_m$.  The constant term only insures positivity; clearly it 
does not affect the distribution of the statistic.  The only remaining 
term contains an $m$.  To see its significance, let us write the 
exponentiation of $H(m)$---that is, instead of the logarithm of the 
likelihood, the likelihood itself.
\begin{equation}
	\exp H(m) = \exp (Z_{m}^{2} - 4 m + 4) = 
	(e^4 e^{- 4 m}) \exp \int \left[ f_{m}(\phi) - \tilde{f_{m}}(\phi) 
\right]^{2} 
	d\phi
\end{equation}
The integral is the likelihood; it is the exponential of a sum of 
squared differences, which of course is a Gaussian.  The first term, 
$e^{-4m}$, is readily identified as the prior!  The Bayesian would now 
integrate\footnote{if only the $Z_m^{2}$ were a well-formed 
likelihood.} $H(m)$ over $m$ to eliminate the unnecessary parameter 
$m$ as described is \sect{stats:marginalize}.  However, if $H(m)$ is 
at all peaked as a function of $m$, then $e^{H(m)}$ is even more 
peaked, and the integral of $e^{H(m)}$ over $m$ is very nearly equal 
to the maximum.  The $H$-test, then, is an approximation to the 
integral over $m$ of the likelihood times a prior of $e^{-4m}$.

This example makes the comparison of the frequentist and Bayesian 
schools quite explicit.  The Bayesian makes arbitrary, subjective 
choices of priors.  The frequentist makes arbitrary, subjective 
choices in the method of smoothing---for example, it is not clear that 
the basis of sines and cosines is the optimal smoothing for the 
$H$-test.  The advantage to the Bayesian method is that it makes 
explicit when the subjective choices are to be made, and completely 
prescribes the rest of the analysis process.

As we have seen, the further disadvantage is that the scientist 
creating new statistics cannot always consider all the ramifications 
of his choices.  Unfortunately, the $H$ statistic cannot be easily 
adapted to a Bayesian analysis; for while $Z_{m}^{2}$ may appear to be 
a true likelihood, it is not.  The likelihood must represent the 
probability of the data.  $Z_{m}^{2}$ represents the probability of 
$f_{m}(\phi)$, which in general, does not have the same probability 
density as the actual data.  Furthermore, it is not at all clear that 
$e^{-4 m}$ is the most suitable prior.  If de Jager \etal\ had claimed 
to use such a prior, they would have had to spend some time justifying 
their choice.  However, under the guise of creating an {\em ad hoc} 
statistic, they have surreptitiously inserted that prior into the 
analysis---probably fooling even themselves.

Is the $H$-test then worthless, or worse yet, misleading?  No.  Within 
the constraints of the simulations done to quantify its significance, 
and within the assumptions that it implicitly makes, it is completely 
valid.  However, the likelihood approach offers a method which is both 
simpler to understand, and at least as powerful.  The correct 
generalization of the $H$-test will be discussed in \sect{tv:unbin}.

\subsubsection{Fourier Transforms}
The statistical tests described above have been used extensively in 
analyzing \egret\ data, but have only had success when a pulsar 
ephemeris from radio or X-ray observations was used to find the photon 
phases $\phi_i$.  The Geminga pulsar is the first known 
``radio-quiet'' pulsar, with little or no radio emission despite a 
strong pulsed signal in X-ray and \gammaray\ 
observations~\cite{bertsch_geminga,hans_geminga}.  It is natural to 
search for other such radio-quiet pulsars in the \egret\ data.  A 
likelihood method of pulsar searching will be discussed in 
\sect{tv:pulsarsearch}; here we will briefly touch on a Fourier 
transform method.

The traditional method of searching for periodic behavior at an 
unknown frequency is the fast Fourier transform (FFT).  The main 
advantage of the FFT is its speed.  Indeed, such an algorithm has been 
implemented to search for periodic signals from pulsar 
candidates~\cite{fouriersearch}.  It has been shown to be successful 
in detecting Geminga---although as we will later see, the signal from 
Geminga is so strong that this is not a definitive test.  The 
drawbacks to the FFT are also significant.  First of all, the FFT 
cannot detect a period derivative.  Trial period derivatives must be 
removed from the photon phases, and a new FFT performed for each 
period derivative.  The FFT also cannot naturally deal with 
noncontiguous variable exposure and backgrounds.  In order to detect 
the weakest pulsar signals, we will need to identify, to the extent 
possible, the photons most likely to be source photons, and account 
for the instrument exposure on a photon-by-photon basis.

	\subsection{Maximum Likelihood}
\label{tv:like}
The ideal statistical method for pulsed-signal analysis would be 
sensitive to variations in pulsed flux from the candidate source, 
rather than pulsed count rates in a region around the source.  Given 
the success of the maximum likelihood method for spatial analysis 
(\sect{stats:egretlike}), it seems natural to apply likelihood to the 
problem of periodic signal analysis.  Again, we will find that the 
mathematics may be interpreted in a maximum likelihood framework or a 
Bayesian framework.  Following a suggestion by Gregory \& 
Loredo~\cite{gl}, we will generalize the binned $\chi^2$ method to a 
maximum likelihood method.  In \sect{tv:unbin} we will examine the 
possibility of generalizing an unbinned method.

The first thing we must do is to construct the likelihood---the 
probability of observing our data, given our model and the state of 
the instrument.  Since the state of the instrument and the background 
are different for every photon, we first divide our time interval and 
spatial intervals into sufficiently small subintervals that there is 
either zero or one photon in each subinterval.  We then have an array 
in parameter space of volume elements $\Delta t \, \Delta \myell \, 
\Delta b$.  The probability that a given element has no photons in it 
will be the Poisson probability
 of detecting zero photons during a
time $\Delta t$ from a direction element $\Delta \myell \, \Delta b$
with rate $r(t, \myell, b)$, which depends in general on time and
observed position.  The probability of detecting one 
photon follows similarly:
\begin{eqnarray}
\label{tv:12probs}
p_0(t, \myell, b) &=& e^{-r(t, \myell, b) \Delta t \Delta \myell 
\Delta b} \nonumber \\
p_1(t, \myell, b) &=& r(t, \myell, b) \,\Delta t \Delta \myell \Delta 
b \,e^{-r(t, \myell, b) \Delta t \Delta \myell \Delta b}
\end{eqnarray}
We identify each element in $t, \myell,$ and $b$ with a sequential, 
unique label.  Furthermore, let us suppose that $N$ of these 
parameter-space volume elements contain one photon, and $Q$ of them 
contain no photons.  $N+Q$ is then the total number of parameter-space 
volume elements, and $N$ is the total number of photons.  We may find 
the likelihood of this configuration by multiplying the probabilities 
of each element:
\begin{equation}\index{$\in$}
\like=\prod_{i \in \alpha_1} p_1(t_i, \myell_i, b_i) \prod_{j \in 
\alpha_0}  p_0(t_j, \myell_j, b_j)
\end{equation}
where $\alpha_0$ is the set of $Q$ volume elements with zero photons,
and $\alpha_1$ is the set of $N$ volume elements with one photon.
Plugging in our expressions from \eq{tv:12probs}, we may factor out 
the
exponential and arrive at
\begin{equation}\index{$\in$}
\like=(\Delta t \Delta \myell \Delta b)^N \prod_{i \in \alpha_1} 
r(t_i, \myell_i, b_i) \exp \left[- \sum_{j=1}^{N+Q} r(t_j, \myell_j, 
b_j) \,\Delta t \Delta \myell \Delta b \right]
\end{equation}
The sum in the exponential is over all volume elements.  It is simpler
to separate this sum into three independent summations, one for each 
parameter.
In the limit that $\Delta t$, $\Delta \myell$, and  $\Delta b$ are 
very
small, these sums become integrals.
Now, the $i$ are just arbitrary labels.  We renumber the volume
elements so that elements 0 through $N$ contain one photon, and we 
have
\begin{equation}
\label{L(r)}
\like=(\Delta t \Delta \myell \Delta b)^N \prod_{i=1}^{N} r(t_i, 
\myell_i, b_i) \exp \left[-
\int_t \int_\myell \int_b
r(t, \myell, b) \,dt \,d\myell \,db \right]
\end{equation}
Note that nowhere have we assumed that the $\Delta t$ are contiguous.
Thus, this likelihood expression is valid even if there are gaps in
the data.

The rate $r(t_i, \myell_i, b_i)$ is a function of both the source
strength and background.  We assume that the background is constant in
time, and varies in space according to the standard \egret\ gas
map~\cite{bertsch_diffuse,hunter_diffuse}. 
 The positional dependence of the source rate
will be taken from the instrument point-spread function.
We will thus take the rate to be
\begin{equation}
r(t_i, \myell_i, b_i) = A(t_i, \myell_i, b_i) \tilde{f}(t_i) + B(t_i, 
\myell_i,
b_i)
\end{equation}
where $\tilde{f}$ is the time-dependent model source flux as before, 
$A(t_i, \myell_i, b_i)$
represents the instrument response function, and $B(t_i, \myell_i, 
b_i)$
is the background flux.  We will worry about the detailed forms of $A$
and $B$ below; however, we note here that the time dependence of $A$
and $B$ is a result of the time dependence 
of the instrument sensitive area.  

To simplify the products and exponents, we take the logarithm of the 
likelihood:
\begin{equation}
\ln \like = N \ln (\Delta t \Delta \myell \Delta b) + \sum_{i=1}^N
\ln (A_i \tilde{f}(t_i) + B_i) - \int_\myell \int_b \int_t (A_i 
\tilde{f}(t_i) +
B_i)\,d\myell\,db\,dt
\end{equation}
where $A_i$ and $B_i$ are shorthand for $A(t_i, \myell_i, b_i)$ and
$B(t_i, \myell_i, b_i)$.

{\samepage
Now, eventually we will either maximize the likelihood, or find a 
ratio of likelihoods.  Thus, constant additive terms in $\ln \like$ 
may be dropped.  We drop the first term and separate the integral to 
get \pagebreak
\begin{eqnarray}
\ln \like & = & \sum_{i=1}^N \ln (A_i \tilde{f}(t_i) + B_i) - 
\nonumber \\ \nopagebreak
& & \int_\myell \int_b \int_t A(t, \myell, b) 
\tilde{f}(t)\,d\myell\,db\,dt - \int_t \int_\myell \int_b
B(t, \myell, b)\,dt\,d\myell\,db
\end{eqnarray}
}
Although the last term is (implicitly) a function of source position,
it will be constant for our purposes, as we take source position to be
given.  Since it is constant, we drop it.  In the second term,
we may split $A$ into $\psd(\myell,b) \sa(t)$, the point-spread 
density
and the sensitive area.  The
point-spread function is normalized, 
by construction, for all times, and $\sa(t)$ depends only on the
source position, not the photon positions.  Therefore we may split the
integral into
\begin{equation}
- \int_\myell \int_b A'(\myell, b)\,d\myell\,db \int_t \tilde{f}(t) 
\sa(t) dt
\end{equation}
The spatial integrals are normalized, and the resulting likelihood 
function
is given by
\begin{equation}
\label{tv:finallike}
\ln \like = \sum_{i=1}^N \ln (A_i \tilde{f}(t_i) + B_i) - \int_t 
\tilde{f}(t) \sa(t) dt
\end{equation}
Remember that $\tilde{f}$ is the {\em model} source flux.  It is this 
model that we will vary to find the model most likely to have produced 
our data.  The data itself appears only implicitly in 
\eq{tv:finallike}, in the $A_i$ and $B_i$, and most importantly in the 
$t_i$ at which the model flux is sampled.

	\subsection{Application to \egret}
We have not yet selected a functional form for $\tilde{f}$.  It is 
tempting to follow the lead of the $Z_m^2$ statistic, and let 
$\tilde{f}$ be a function composed of some number of harmonics, 
perhaps with parameters to be determined.  We will delay such an 
attempt to \sect{tv:unbin}, and first develop a method based on 
$\chi^2$ which we will find to be less computationally demanding.

The functional form of $\tilde{f}$ that corresponds to $\chi^2$ 
binning is stepwise constant, with $m$ different steps.  The immediate 
benefit of such a choice comes in the simplification of the second 
term of \eq{tv:finallike}:
\begin{equation} \hbox{\index{$\longrightarrow$}}
\int_t \tilde{f}(t) \sa(t) dt \longrightarrow
\sum_{j=1}^m \tilde{f}_j \tau_j
\end{equation}
where $\tau_j$ is the sensitive area integrated over the livetime in 
bin $j$; that is, the total exposure in bin $j$.  Now 
\eq{tv:finallike} may be stated as a sum over bins
\begin{equation}\index{$\in$}
\label{tv:binlike}
\ln \like = \sum_{j=1}^m \left[ 
\sum_{i \in \hbox{\scriptsize bin\ } j}  \ln (A_i \tilde{f_j} + B_i) 
- \tilde{f_j} \tau_j
\right]
\end{equation}
Thus the maximum of $\ln \like$ may be found by independently 
maximizing the $\tilde{f}_j$ for each bin, instead of simultaneously 
maximizing all the parameters of $\tilde{f}$---an enormous 
computational savings.

The analysis is thereafter quite simple, and quite slow.  A candidate 
pulsar is selected for analysis.  All photons within some radius, 
usually $15\deg$, are collected.  Each photon with sufficient energy 
($\gt 100$~MeV) which arrives from a position sufficiently far from 
the Earth's limb ($\lt 95\deg$ from the instrument axis) during a 
valid instrument mode is accepted for analysis.  For each of these 
photons, the arrival time is corrected to the Solar System barycenter, 
and the phase-independent likelihood factors $A_i$ and $B_i$ are 
calculated.  For periods less than a minute, it may reasonably be 
assumed that the instrument exposure will be evenly distributed across 
the bins.  For longer periods, such as X-ray binary orbital periods, 
all instrument exposure during the observation is calculated as a 
function of SSB time.

To ameliorate the primary objection to binning, namely, that the 
arbitrary placement of the bin boundaries may obscure a signal, we may 
allow an offset to the bin boundary.  In practice, this means 
rebinning the data for a number of different boundary offsets.  Of 
course, such an operation increases the number of trials.

\subsubsection{Sampling Density}
\label{tv:density}
A subset of parameter space is identified to be searched for pulsed 
signals.  Some boundaries of the range of period and period derivative 
are set by the physics of the system being examined; those will be 
motivated in \sect{tv:pulsarsearch} and \sect{tv:xrb}.  The sampling 
size required is determined by the statistics.  To determine the 
number of steps required to sample the period space, we consider the 
difference in phase of the last photon in the observation as a 
function of the change in trial periods.  If two trial periods differ 
by so little that the change in phase of the last photon as computed 
by \eq{tv:phase} is less than the width of a bin, then the most likely 
$f_j$'s computed by maximizing \eq{tv:binlike} will be unchanged.  So, 
we choose the minimum spacing so that the phase of the last photon 
will change by an amount of order the bin width, and calculate the 
number of steps to take in period across the desired range.
\begin{equation}
N_p = (\nu_{max} - \nu_{min}) \, m \, \delta t
\end{equation}
where $\delta t$ is the length of the observation.  Note that a longer
observation means more photons will be observed, increasing the
signal-to-noise ratio, but the period space must be searched more 
densely.

Similarly, we can find the minimum period derivative we expect could
make a difference.  In this case, our condition is that 
$\phi(\delta t) - \nu \delta t$ is an appreciable fraction of a bin.
Neglecting higher derivatives, this happens when 
$\frac{1}{2} \dot{\nu} \delta t \, \mysim \, p/m$.  Since the 
frequency derivative
$\dot{\nu} = \dot{p}/p^2$, we have
\begin{equation}
\label{tv:pdotmin}
\dot{p}_{min} = \frac{2 p^3}{m (\delta t)^2}
\end{equation}
The maximum period derivative must be determined by the physics.

With the range of period derivatives for a given period in hand, we
may compute the number of samplings in period derivative required.
Again, we find the step size by equating the difference in 
computed phase with the bin size:
\begin{equation}
\frac{1}{2} \dot{\nu}(\delta t)^2 - \frac{1}{2} (\dot{\nu} + \delta 
\dot{\nu})(\delta t)^2
= 1/m
\end{equation}
and thus the number of steps required in the period derivative:
\begin{equation}
N_{\dot{p}} = \frac{1}{2} (\dot{\nu}_{max} - \dot{\nu}_{min}) \, m \, 
(\delta t)^2
\end{equation}

\subsubsection{Significance Estimates}
For each point in parameter space, the flux in each bin is estimated 
by maximizing the likelihood.  Since the sampling of parameter space 
was chosen to make each point roughly independent, the number of 
points sampled in parameter space should be a good estimate of the 
number of independent trials.  The significance is then the 
probability of measuring no likelihoods greater than or equal to that 
observed in the number of trials.

\subsubsection{Automated analysis with \timevar}
These methods have been automated into a tool for searching for 
periodicity in \egret\ data called \timevar~\cite{tvmanual}.  Capable 
of running either interactively or in batch mode, \timevar\ searches a 
given region of parameter space, for a given number of bins.  It 
maximizes the likelihood for each period and period derivative, and 
reports the significance of the most likely period and period 
derivative.  In addition, it yields a light curve, both as the most 
likely flux in each bin, and for comparison with \pulsar, the number 
of photons within the 68\% containment radius in each bin.

\begin{figure}[t]
\centering
\includegraphics[width = 4.5 in]{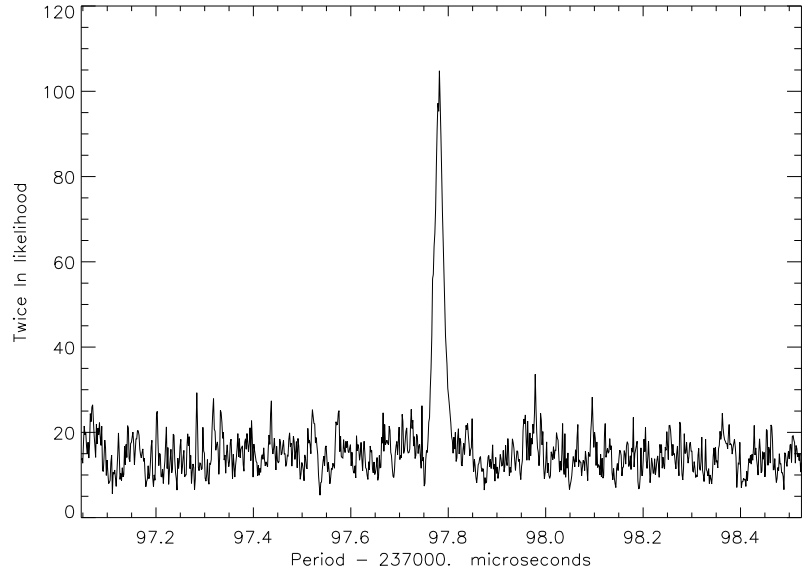}
\caption[Geminga likelihoods]{\label{tv:gemlike}The likelihood of 
periodic modulation from Geminga, as a function of period, for a 
30-bin light curve.  Plotted on the $y$-axis is $2 \ln \like$, which 
is distributed as $\chi^2_{29}$, since the average intensity in the 
null model is a free parameter.  The probability of the peak in the 
null model is less than $10^{-25}$.  The maximum likelihood value of 
the period is 0.237097785 s, assuming a period derivative of $1.09744 
\times 10^{-14}$ s/s and a truncated Julian day epoch 8750.0.}
\end{figure}

\begin{figure}[t]
\centering
\includegraphics[width = 2.5 in]{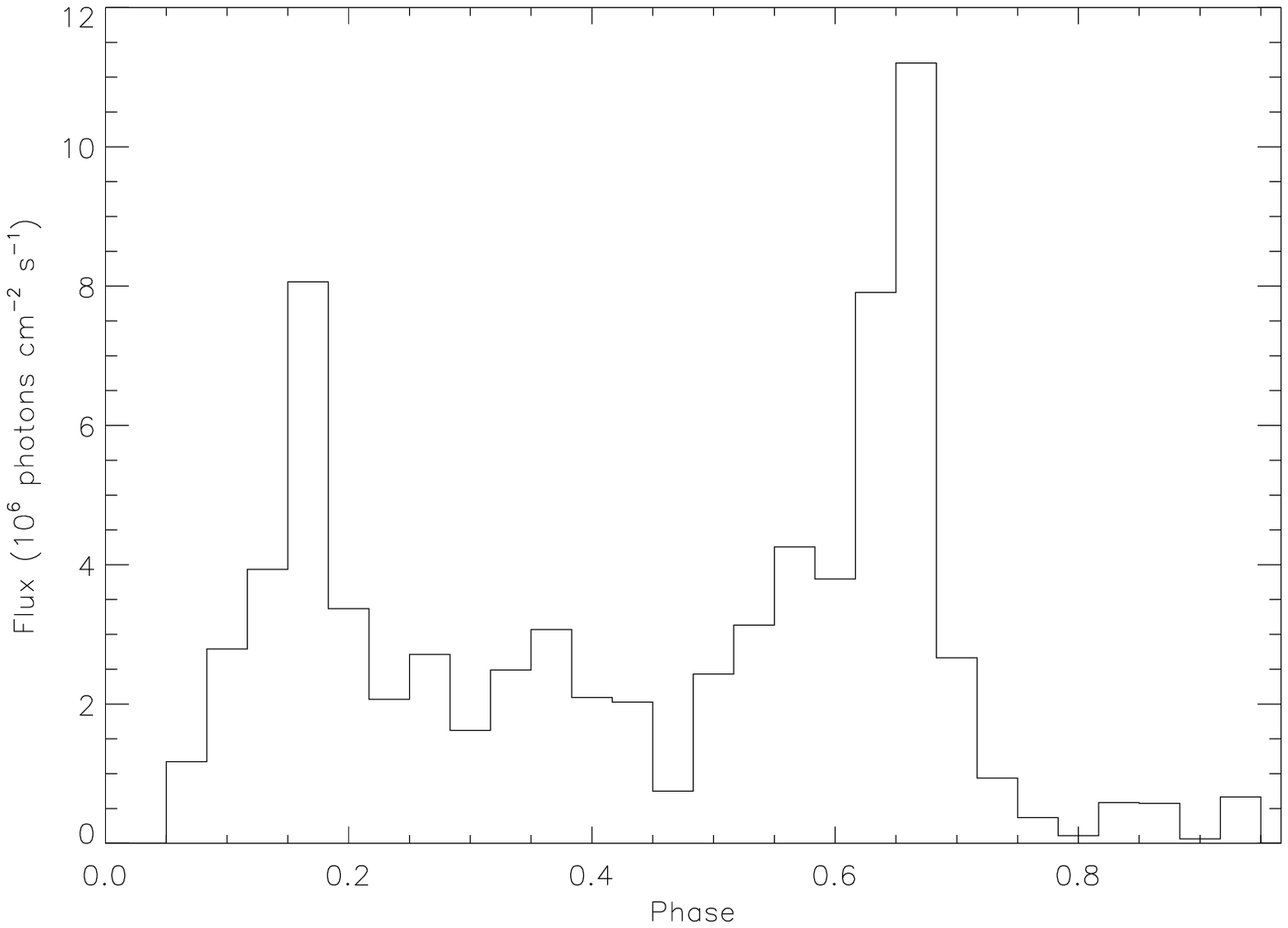}
\includegraphics[width = 2.5 in]{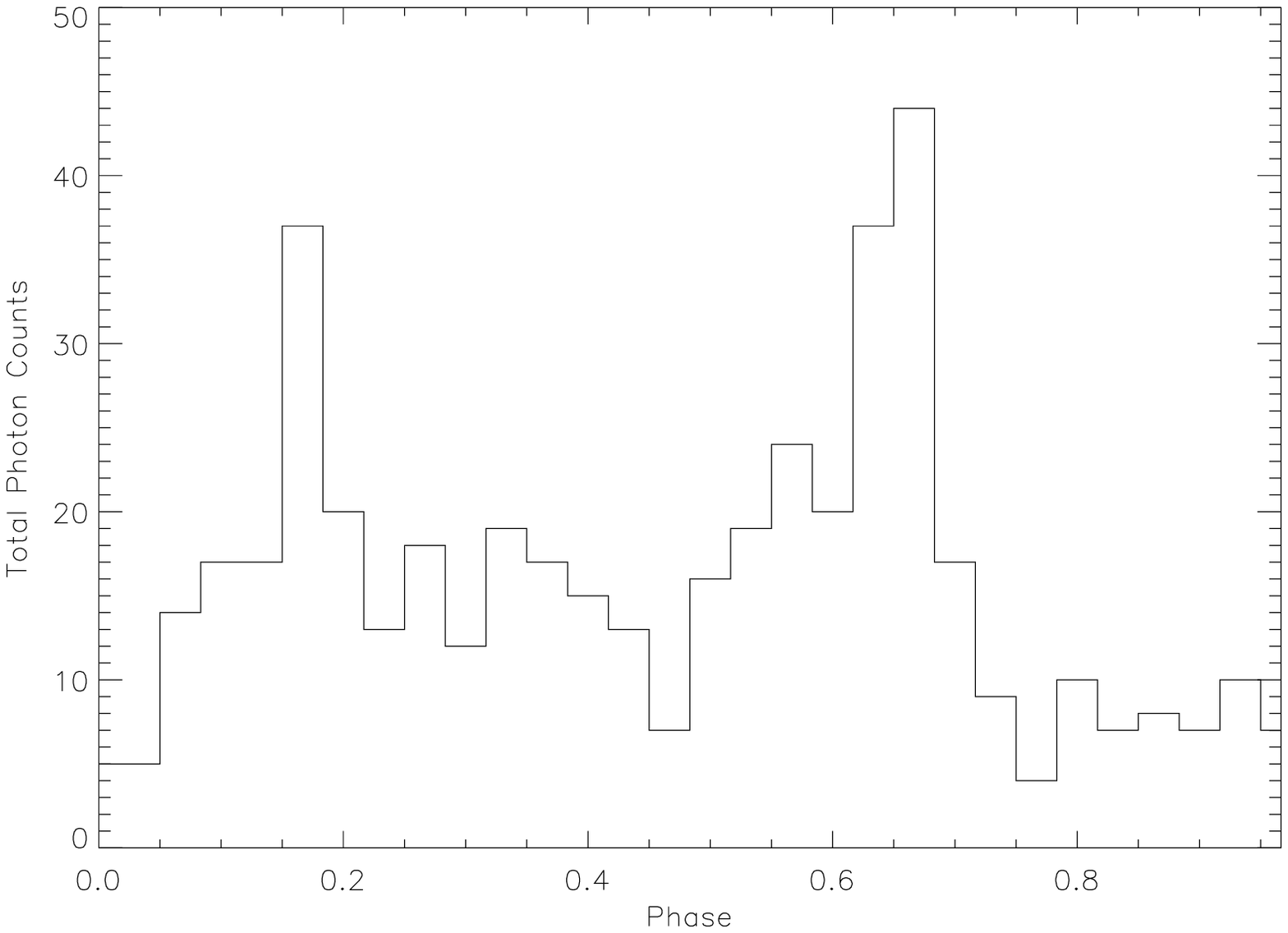}
\caption[Geminga light curves]{\label{tv:gemlight}Light curves of the 
most likely period as found in \fig{tv:gemlike}.  The light curve on 
the left is the most likely flux in each bin, after accounting for the 
instrument point-spread function and sensitive area and the 
background.  The light curve on the right is the raw photon count 
inside the 68\% containment radius, as given by 
\eq{intro:containment}.  This is the same as the light curve used by 
\pulsar.}
\end{figure}

\label{tv:detect_geminga}
To demonstrate the capabilities of \timevar, we examine the Geminga 
pulsar.  We will check viewing period 413.0, corresponding to 
observations between 7~March~1995 16:44:00 and 21~March~95 14:08:39.  
A small range around the known period was searched, and the results 
are presented in \fig{tv:gemlike}.  The light curves are shown in 
\fig{tv:gemlight}.  The probability of the observation in the null 
model is less than $10^{-25}$.

The great advantage of \timevar\ over previous analysis methods is 
that it combines period and period derivative searching ability with a 
proper likelihood treatment of the source and background as perceived 
through the instrument point-spread function and sensitive area.  
\pulsar\ has been useful to examine known pulsars with radio 
ephemerides, although it does not optimally use the point-spread and 
background information about each photon.  The disadvantage to 
\timevar\ is its computational expense.  Examining a single pulsar 
candidate over a reasonable range of period and period derivatives 
currently takes on the order of months of continuous running on a Sun 
Microsystems Sparc~10.  Mattox \etal~\cite{fouriersearch} have 
sidestepped this problem first by turning to a Fourier 
algorithm---which is much faster, but cannot properly account for 
backgrounds and point-spread functions---and second by using a 
massively parallel computer.  Further optimization may be possible to 
speed \timevar, and certainly additional computing power will reduce 
the required search time.  A similar program designed for \glast\ will 
be slowed by the additional order of magnitude in the number of 
photons to process, but since the sensitive area will be larger, the 
elapsed time required to observe a sufficient number of photons will 
be shorter, allowing period space to be sampled less densely.  In 
addition, faster computers will be available to do the calculations.  
The net effect will be that period searching with the \glast\ 
equivalent to \timevar\ will be quite feasible (\tbl{tv2:glast_search}).

	\subsection{Bayesian Inference}
A further advantage of the likelihood method employed by \timevar\ is 
the ease by which results can be evaluated under a Bayesian framework.  
Strictly speaking, the plot shown in \fig{tv:gemlike} is meaningless 
in a maximum likelihood framework---only the maximum value, and not 
the entire likelihood function, is relevant.  However, intuitively we 
expect the rest of the function to have some meaning.  Local maxima 
should represent frequencies more likely to be present in the data.  
We would expect harmonics of the main pulse frequency to be present.  
As we saw in \sect{stats:bayes}, the Bayesian formulation includes all 
this information into the posterior probability distribution.  We 
simply multiply the likelihood function given in \fig{tv:gemlike} by 
the prior probability distribution.  Then we may integrate the 
posterior distribution over interesting intervals, or simply present 
it as it is.

\subsubsection{Appropriate Priors}
The least-informative prior probability distribution for the pulsation 
period is scale-invariant.  That is, it should not depend on the units 
used for measuring period, and it should be the same whether the 
search is done in period or frequency space.  The appropriate solution 
is flat in the logarithm of the period.  Therefore, we have 
$P_{\hbox{\scriptsize prior}}(p) \propto 1/p$\index{$\propto$}.  The 
upper and lower bounds of the period search define the cutoff, so that 
the prior can be normalized:
\begin{equation}
\label{tv:priorp}
P_{\hbox{\scriptsize prior}}(p) = \left( \ln \frac{p_{min}}{p_{max}} 
\right)\; \frac{1}{p}
\end{equation}

The least informative prior for the period derivative is uniform.  The 
period derivative is a dimensionless number, and thus is the most 
natural unit in which to work.  The last parameter of interest is the 
phase offset of the bin boundaries.  This is also dimensionless, and 
thus the least informative prior is flat.

Gregory \& Loredo~\cite{gl} point out that marginalization over $m$ 
yields the best light curve independent of the number of bins.  This 
is exactly the spirit of the $H$-test.  However, this is 
computationally impossible for a period search given the present state 
of the art.  It is interesting to note here that the proper treatment 
of this problem is very difficult using frequentist statistics.  Two 
models with different $m$ have a different number of free parameters.  
The Bayesian framework takes care of this naturally; each free 
parameter (that is, each of the $m$ fluxes) has its own prior; thus 
the complete prior for all the fluxes (assuming a flat prior between 
zero and the large cutoff) will be $P_{\hbox{\scriptsize 
prior}}(\tilde{f}_m) = f_{max}^{-m}$.  This prior is a quantization of 
Ockham's Razor: models with more free parameters (that is, larger $m$) 
are discouraged.  The frequentist method compares each model to the 
null model, but cannot directly compare the two.

\begin{figure}
\centering
\includegraphics[width=3 in]{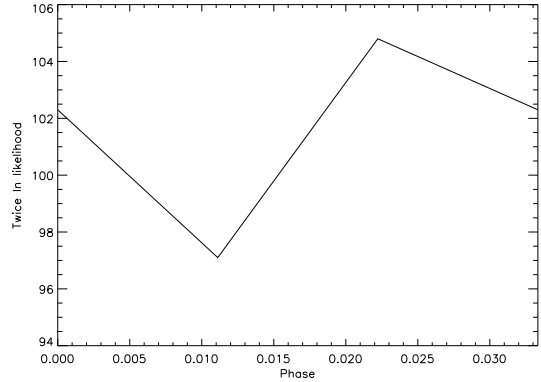}
\caption[Geminga phase offsets]{\label{tv:gemoffset}
The likelihood of pulsation in Geminga as a function of the phase 
offset of the bin boundaries, for fixed period.  The period chosen is 
that of the maximum likelihood for any offset, and corresponds to the 
peak seen in \fig{tv:gemlike}.  The width of the plot corresponds to 
one bin; therefore, the leftmost point and the rightmost point are the 
same.}
\end{figure}

\begin{figure}[t]
\centering
\includegraphics[width=4.5 in]{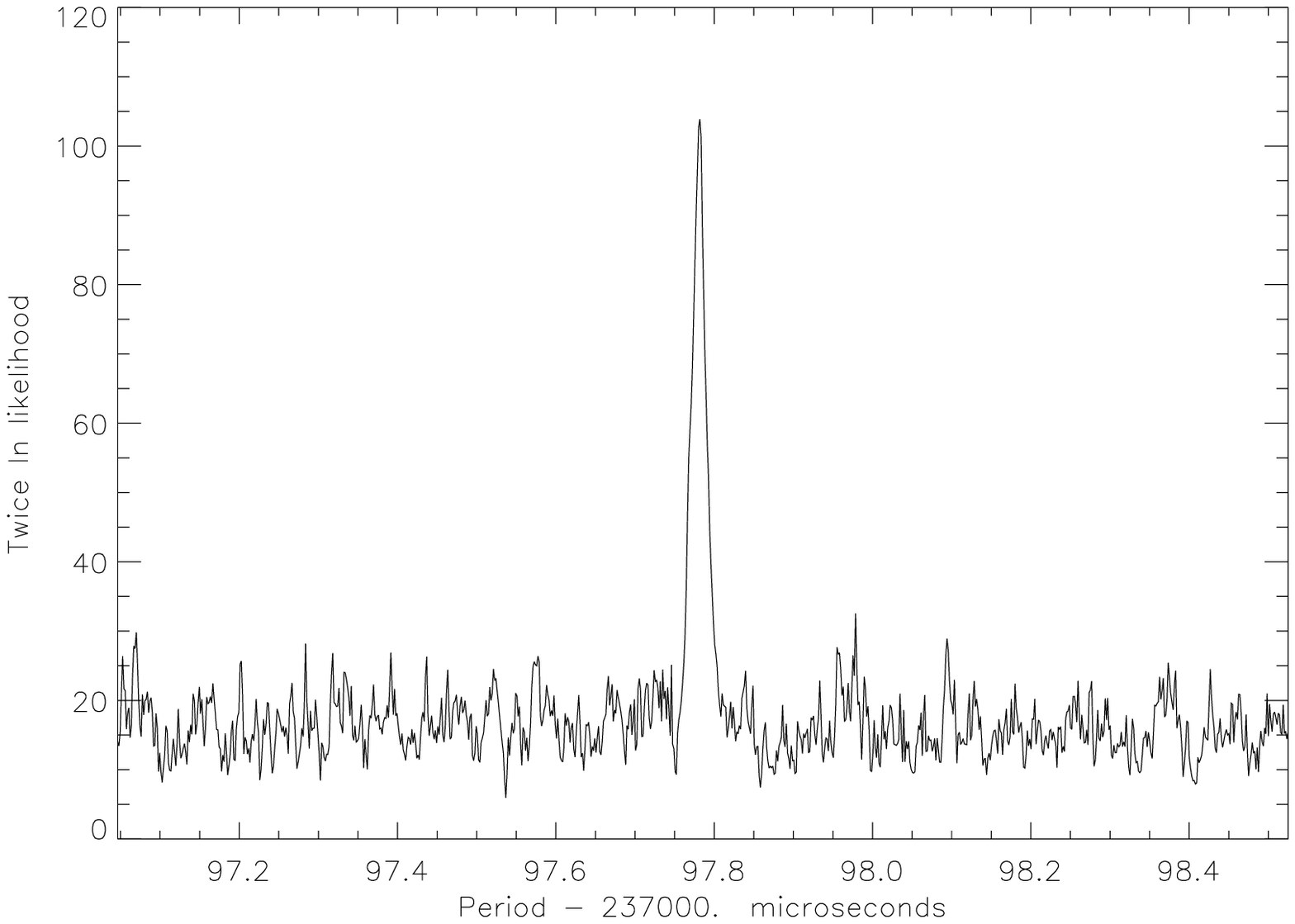}
\caption[Marginalized geminga likelihood]{\label{tv:gemlike2}
The likelihood of pulsation in Geminga.  This likelihood function
is similar to that shown in \fig{tv:gemlike}, except that the 
likelihood
has been marginalized over the phase offset.}
\end{figure}

\fig{tv:gemlike} is actually one of three different likelihood 
functions measured for different phase offsets.  It is the slice of 
the likelihood corresponding to the most likely offset.  The 
perpendicular slice through the likelihood as a function of offset 
fixed at the best period is shown in \fig{tv:gemoffset}.  The phase 
offset is a classic nuisance parameter; it arose by the arbitrary 
choice of $t_0$ in \eq{tv:phase}.  We therefore marginalize over the 
offset and arrive at the likelihood function shown in 
\fig{tv:gemlike2}.

To find the overall significance of a detection\footnote{In our case, 
this will be the overall significance of a detection for all models 
with $m$ bins.  In principle, however, the parameter $m$ could be 
marginalized over as well, yielding the true overall detection 
significance.} in the Bayesian framework, we multiply the likelihood 
function shown in \fig{tv:gemlike2} by the period prior given in 
\eq{tv:priorp}.  The ratio of this probability with the probability of 
the null model ($m$=1) gives the odds ratio.  In this case, the 
probability of periodic modulation is completely dominated by the 
peak, and the significance of the detection is $1 - 7 \times 
10^{-46}$.

The Geminga pulsar is clearly detected by either method.  The 
detections are so strong that even if a very large region of phase 
space had been searched, the detections would still be significant.  
Finding the significance that would have resulted from such a search 
may be calculated by assuming that the peak we found would be the 
maximum likelihood over the entire search, and that the number of 
trials would be the number of steps found in \sect{tv:density}.  A 
thorough search of viewing period 413.0 from 50 \us\ to 500 \us, over 
the appropriate period derivative ranges would require approximately 
$7 \times 10^8$ trials.  The significance of the detection in 
\fig{tv:gemlike} would still be $1 - 10^{-20}$.  However, it is clear 
that weaker detections will not survive such a harsh attack.  It is 
therefore also useful to develop some formalism for upper limits and 
detection thresholds.

	\subsection{Upper Limits and Thresholds}
\label{tv:thresholds}
The problem of upper limits was briefly touched upon in 
\sect{stats:uplim}.  There we noted that a desirable quality of an 
upper limit is that it be statistically well-behaved; that is, that a 
``$1\sigma$ upper limit'' be greater than the true value in 68\% of 
the ensemble of possible data sets.  The value of this upper limit 
depends on the details of the data.

In contrast, in some cases it may be more useful to calculate a 
threshold.  We define a threshold to be the intensity (of a source, or 
of pulsation) that would be detected by our criteria in some fraction 
of the possible data that might be measured from that source.  This 
may be quantified: an $x$\% detection threshold is that true parameter 
value for which $x$\% of possible data sets would result in detection.  
We make a statement about the outcome of a hypothesis test (is there 
pulsation?)  instead of a statement about a point-estimate (is the 
pulsed fraction less than some value?).

There are three advantages to this.  First of all, it more closely 
reflects the intuitive ``upper limit'' concept---that is, it is now 
reasonable to say, ``The threshold for detection is $f$.  Since we do 
not detect, then the actual value is probably less than $f$.''  We 
will quantify this below.  Second, we may calculate thresholds from 
the underlying distributions of the data; upper limits can only be 
calculated with the specific data observed.  This means that 
thresholds can be precalculated.  This is a great advantage, 
especially for instruments which are still being designed or built.  
Finally, since we can use distributions, we can calculated thresholds 
(at least partially) analytically.

For the purposes of pulsation analysis, we will define the ``pulsed 
fraction threshold'' as the fraction of the flux (background plus 
source) that must pulse such that a fraction $\xi$ of data sets 
observed with true sinusoidal variation with threshold amplitude will 
result in detection.  We will assume that any data set with 
significance greater than $\alpha$ will be a detection and the 
variation will be sinusoidal.

For simplicity, we define $C \equiv 2 \ln \like$\index{$\equiv$}.  We 
will take $\xi = 0.5$, so that we may find the $\alpha$-point of the 
integral $\chi^2$ distribution.  That is, we seek $C$ such that 
$\int_0^{\expct{C}} \chi^2(d, t)\,dx = \alpha$, where $d$ is the 
number of degrees of freedom, and $t$ is the number of 
trials.\footnote{Assuming that the mean of $C$ is a good approximation 
to the median, so that roughly half of the realized values of $C$ are 
greater than $\expct{C}$, and half are less.}  This follows from 
Wilks' theorem (\sect{stats:hypotest},~\cite{wilks}), which states 
that $C$ is distributed under the null model as $\chi^2$ with the 
number of degrees of freedom equal to the number of free parameters in 
the model.  To find the mean value of $C$, we integrate over all data 
sets, weighted by their true probability:
\begin{eqnarray}
\label{tv:mainavg}
\expct{C} & = & \int_D \like_T(D) C(D) dD \nonumber \\
&  = & \int_D \like_T(D)\, 2 \left[ \ln \like_{BF}(D) - 
\ln \like_{BN}(D) \right] dD
\end{eqnarray}
where $\like_T(D)$ is the true likelihood of data $D$, $\like_{BF}(D)$ 
is the likelihood of the best model fit to data $D$, and 
$\like_{BN}(D)$ is the likelihood of the best null model fit to data 
$D$.

To proceed, we turn once again to Wilks' theorem.  If the null model 
is true, then from \eq{stats:wilks} we have $2(\ln \like_{BF} - \ln 
\like_{null}) \; \mysim \; \chi^2_d$.  From the definition of the 
expectation value, we know that
\begin{equation}
\expct{A(D)} = \int_D P(D) A(D)\,dD
\end{equation}
for any function $A(D)$.  
The expectation value of $\chi^2_d$ is $d$~\cite{eadie}, and the 
probability
of the data is the likelihood of the data under the (unknown) true 
model 
$\like_T(D)$, so
\begin{equation}
\label{tv:cheat}
\expct{2 \left[ \ln \like_{BF}(D) - \ln \like_T(D) \right]} = d  
\end{equation}
Therefore, we conclude that 
\begin{equation}
\int_D \like_T(D) 2 \left[ \ln \like_{BF}(D) - \ln \like_T(D) 
\right]\, dD = d
\end{equation}
Substituting into \eq{tv:mainavg}, we have
\begin{equation}
\label{tv:intermedavg}
\expct{C} = 2 \int_D \like_T(D) \ln \like_T(D) \, dD + d - 
2 \int_D \like_T(D) \ln \like_{BN}(D) \, dD 
\end{equation}

We now make some further assumptions about the pulsation.  As before, 
we will let $f_m(\phi)$ to be the model flux as a function of $\phi$ 
with $m$ parameters.  In particular, we will define the null model 
$f_1(\phi)= f_1$, with one parameter, the mean flux.  We will further 
assume that the number of photons in each bin is large enough to 
approximate the Poisson distribution as a Gaussian (that is, $n_i$ is 
large).  Then in each bin $i$ we may write
\begin{equation}
z_i = \frac{n_i - f_m(\phi_i)}{\sqrt{f_m(\phi)}}
\end{equation}
where $n_i$ is the number of photons in bin $i$, and $f_m(\phi_i)$ is 
the 
model number of photons in that bin.  Then 
$\like_T(D) = \prod_i \frac{e^{-z_i^2/2}}{\sqrt{2 \pi}}$.
For each bin, the first term of \eq{tv:intermedavg} becomes
\begin{eqnarray}
\int_{D_i} \like_T(D) \ln \like_T(D) \, dD & = &
\int_{-\infinity}^{\infinity} \frac{1}{\sqrt{2 \pi}} e^{-z_i^2/2} 
\left( -\frac{z_i^2}{2} - 
\ln \sqrt{2 \pi} \right) \, dz_i \nonumber \\
& = & - \ln \sqrt{2 \pi} - 1/2
\end{eqnarray}
This does not depend on the model $f_m$ since for each bin, the best 
fit model should describe the data in approximately the same way.  
That is to say, the statistical deviation from the best fit model 
should not depend on the details of that model.  Using this result, 
\eq{tv:intermedavg} becomes
\begin{equation}
\expct{C} = -2 m \ln \sqrt{2 \pi} - m + d - 
\int_D \like_T(D) \ln \like_{BN}(D) \, dD
\end{equation}
The last term contains the best fit null model.  This must be related 
to the 
true null model in the same manner as \eq{tv:cheat}.
\begin{eqnarray}
\int_D \like_T(D) \ln \like_{BN}(D) \, dD & = & 
\int_D \like_T(D) \left[ \ln \like_{TN}(D) + 1/2 \right] \, dD 
\nonumber \\
& = & 1/2 + \int_D \like_T(D) \ln \like_{TN}(D) \, dD
\end{eqnarray}
Again assuming Gaussian fluctuations, $\like_{TN}(D) = \prod_i 
\frac{1}{\sqrt{2 \pi}}
\exp \left[ - \left( \frac{n_i - f_1}{\sqrt{f_1}} \right)^2 \right]$, 
yielding
\begin{eqnarray}
\label{tv:penultimateavg}
\expct{C} &= &- 2 m \ln \sqrt{\pi} - m + d - m +2 m \ln \sqrt{2 \pi} 
\nonumber \\
& & + \sum_{i=1}^m \int_{D_i} \frac{1}{\sqrt{2 \pi}}
\left( \frac{n_i - f_1}{\sqrt{f_i}} \right)^2
e^{-z_i^2/2} \, dD_i
\end{eqnarray}
The number of degrees of freedom $d$ is $m-1$.
To facilitate the integral, we write $n_i$ in terms of $z_i$.  The 
integral over
the data $D_i$ is properly normalized as an integral over $z_i$.
We assume that $n_i$ is large enough
that expanding the bounds of integration from $-\infinity$ to 
$\infinity$ will
introduce only a small error.
\begin{equation}
\label{tv:finalavg}
\expct{C}= -m -1 + \sum_{i=1}^m \int_{-\infinity}^{\infinity} 
\frac{1}{\sqrt{2 \pi}}
e^{-z_i^2/2} \left( \frac{z_i \sqrt{f_m} + f_m - f_1}{\sqrt{f_1}} 
\right)^2 \, dz_i
\end{equation}
The threshold is then given implicitly as a function of the 
parameters of $f$
through $z_i$, as the value of the parameters such that 
$\int_0^{\expct{C}} \chi^2(d, t) = \alpha$.

For example, one form of $f$ might be a sinusoidal modulation plus a 
constant background: $f(\phi) = B + A \sin 2 \pi \phi_i$.  The average 
value over $\phi$ is $B$, so $f_1 = B$.  We may now ask the question: 
for a given background $B$, what is the threshold for detecting a 
modulation of strength $A < B$?  Comparisons of the threshold with the 
modulation expected on physical grounds can put null experiments into 
perspective (\sect{tv:xrb},~\cite{bbjxrb}), as well as help identify 
potential candidate sources for pulsation detection.

	\subsection{Bin-free Maximum Likelihood}
	\label{tv:unbin}
The application of maximum likelihood methods to pulsation searches 
binned photons to simplify and speed calculations.  It is important to 
remember that this binning is not an essential element of likelihood 
methods.  Recall from \eq{tv:finallike} that the likelihood function 
depends on a generic model function $\tilde{f}$.  We may construct the 
maximum likelihood analogue of the $Z_m^2$ by taking
\begin{equation}
\tilde{f}_m(\phi) = \alpha_0 + \sum_{k=1}^m ( \alpha_k \sin 2 \pi k 
\phi
+ \beta_k \cos 2 \pi k \phi)
\end{equation}
This function actually has $2m+1$ degrees of freedom.  Given this, we 
can calculate \eq{tv:finallike}, and maximize it with respect to the 
$\alpha_k$ and $\beta_k$.  In contrast with the binned case, we must 
now maximize the likelihood simultaneously for the model parameters.  
Instead of $m$ one-dimensional maximizations, we must perform one 
($2m+1$)-dimensional maximization\footnote{Reducing the number of 
Fourier components retained would not help this problem.  Optimization 
with $n$ Fourier components still requires an $n$-dimensional 
maximization, as opposed to $n$ one-dimensional maximizations in the 
binned case.} Since the $A_i$ and the $B_i$ are taken from numeric 
tables and the functions involved are rather complex, it is unlikely 
that it will be possible to maximize the functions analytically.  
Numerical methods for multidimensional maximizations are notoriously 
slow and prone to finding local minima~\cite{numrec}.  Nevertheless, 
all the principles discussed above remain valid.  The analogue to the 
$H$-test would be found by marginalizing over $m$ with a suitable 
prior.

Sinusoids are familiar, and so are an obvious choice of basis.  
However, any set of basis functions may be used in a parameterization 
of $\tilde{f}$.  Generalizing in a different way, we may imagine $k$ 
bins with arbitrary boundaries, so that the bins may not be the same 
size.  This would be represented by a stepwise function
\begin{equation}
f(\phi) = f_j,\; \left\{ \begin{array}{l}
	 k_j \le \phi \lt k_{j+1} \\
	j \in [0, k) \end{array}
\right.
\end{equation}
where $k_j$ indicates the beginning of bin $j$, $f_j$ is the flux in 
bin $j$, and there are $k$ bins.  Again, however, if the $k_j$ are 
allowed to vary, the likelihood function is no longer separable, and 
all the model parameters must be simultaneously optimized.

%% file: Timevar2.tex
\section{Searching for Pulsars}
\label{tv:pulsarsearch}

Now that we have a statistical framework in which to search for pulsed flux
in \egret\ data, it is natural to use it to examine unidentified sources.
A number of unidentified \egret\ sources are positionally coincident (within
large error regions) with known radio 
pulsars~\cite{joethesis,fouriersearch,michelson_pulsarupperlimits}.  
Other \egret\ sources
are concentrated in the Galactic plane.  Since there are no other known strong
Galactic \gammaray\ sources, and the Geminga pulsar is known to be radio-quiet,
it is widely assumed~\cite{mukherjee_unid,pohlpulsar,merck_unid,fouriersearch}
that some or all of these sources are pulsars.  It may be possible to detect 
pulsation from these sources without the use of ephemerides from other wavelengths.

\begin{figure}
\centering
\includegraphics[width = 3.5 in]{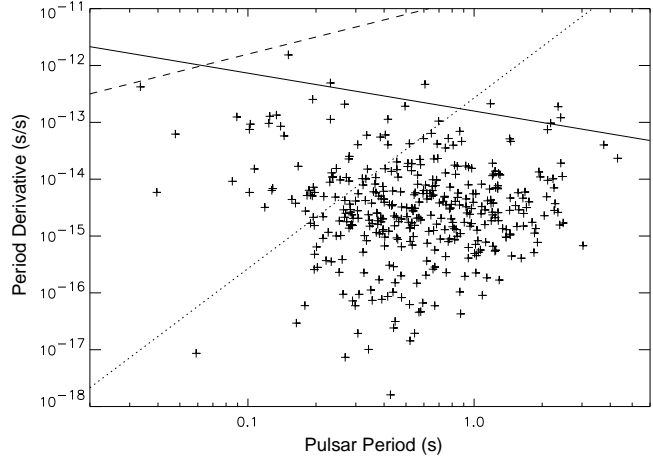}
\caption[Pulsar periods vs.\ period derivatives]{\label{tv2:deathline}
Periods vs. period derivatives for pulsars in the Princeton 
database~\cite{ptondbase}.  The solid line is the high period-derivative cut-off 
line; almost all pulsars lie below and to the left of this line.
The dotted line is the sampling limit; below it, period derivatives may be
neglected for pulsar searches. This limit depends inversely on the square
of the length of the observation; 10 days is assumed here.  
The dashed line corresponds to a characteristic
age of 1000 years; no younger pulsars are expected.}
\end{figure}

One of the difficulties of such an analysis is the choice of the range of 
parameter space to search.  The fastest millisecond pulsar has a period of 1.5~ms,
and the slowest pulsars have periods of several seconds.  The longest
period pulsar observed in \gammarays\ is Geminga, with a period of about 237~ms.
A reasonable search strategy would cover as much of this space as possible.
The range of period derivative to search depends strongly on the period being
searched.  \Eq{tv:pdotmin} gives the minimum period derivative. 
Observed pulsars fall below an empirically determined cut-off in period derivative~(\fig{tv2:deathline}).
All known \gammaray\ pulsars have period derivatives smaller than the cutoff.
 The known pulsar population occupies a relatively
compact area on a plot of period versus period derivative, allowing the appropriate
regions of parameter space to be searched for periodic signals.

Unfortunately, according to \eq{tv:pdotmin}, the minimum period derivative 
to search decreases
with decreasing period, and according to \fig{tv2:deathline}, the maximum period
derivative increases with decreasing period.  Therefore, as we search shorter and
shorter periods, the number of period derivative steps required increases rapidly.
However, for periods larger than that where the sampling limit given by
\eq{tv:pdotmin} crosses the
period-derivative cut-off, we may ignore the period derivative.
The time required to search a given region of parameter space thus depends weakly
on the maximum period to be searched, and strongly on the minimum period to be 
searched.  A minimum period of 50~ms and a maximum period of 500~ms would contain 
all of the known \gammaray\ pulsars except the Crab.  This range of period, and
the appropriate range of period derivative for each period, is taken as the 
canonical search range.

	\subsection{Measurement of Known Pulsars}
To verify that \timevar\ is both statistically accurate and bug-free, several
known pulsars were examined as if they had been the subject of a search.  The
results of those analyses are summarized in \tbl{tv2:known}.  Each known
\gammaray\ pulsar was examined in one viewing period, chosen for good exposure to
the source.  A periodicity search was carried out for a small range around the
known period, and the significance of the detection is given as if a single
period had been tried.  To find the total detection probability, the number of
trials is found by calculating the number of parameter space samplings required
for the canonical search range for that observation.  Unfortunately, we will find 
that this range is too ambitious to search with current technology.

\begin{table}
\begin{minipage}{3 in}
\footnotesize
\begin{tabular}{lc r@{.}l r@{.}l ccc}
Pulsar & VP & \multicolumn{2}{c}{True Period} & \multicolumn{2}{c}{\timevar} &  One-trial    & Trials    & Final \timevar \\
       &    & \multicolumn{2}{c}{(ms)}        & \multicolumn{2}{c}{period}   &  probability & Needed\footnote{
\scriptsize Searching a period range from 50~ms to 500~ms}  & significance \\ \hline
Geminga &  310.0 & 237&097785  & 237&097784 & $4.8 \times 10^{-47}$ & $5.5 \times 10^{12}$ & $1 - 2.6 \times 10^{-34}$ \\
Crab &     001.0 & 33&38981606 & 33&38981541 & $9.5 \times 10^{-146}$ & $5.1 \times 10^{16}$\footnote{\scriptsize Searching from 10~ms to 500~ms} & $1 - 4.8 \times 10^{-129}$ \\
Vela &     301.0 & 89&29607605 & 89&2960604808 & $6.7 \times 10^{-76}$ & $8.7 \times 10^{12}$ & $1 - 5.8 \times 10^{-63}$ \\
B1706-44 & 232.0 & 102&4544637 & 102&24544626 & $2.6 \times 10^{-14}$ & $8.8 \times 10^{12}$ & 0.7712 \\
B1055-52 & 014.0 & 197&1102023 & 197&110146607 & $4.3 \times 10^{-4}$ & $8.4 \times 10^{12}$ & 0 \\
1951+32  & 203.0 & 39&530908008 & 39&5308978685 & $2.0 \times 10^{-4}$ & $1.7 \times 10^{17}$ $\,^b$ & 0 \\
\end{tabular}
\end{minipage}
\caption[Detection of known \gammaray\ pulsars by \timevar]
{\label{tv2:known}
Detection of known \gammaray\ pulsars by \timevar.  
The one-trial probability is the chance of observing the maximum likelihood value 
in a single trial.  The number of trials is determined from the required sampling
(\sect{tv:density}) to search a period range from 50~ms to 500~ms, with the 
period derivative range appropriate for each period.  For the Crab pulsar 
and PSR 1951+32
the search range is 10~ms to 500~ms.  The total significance is the chance that
no such likelihood would be observed in the given number of trials under the 
null model.}
\end{table}

\begin{figure}[p]
\centering
\includegraphics[width =4 in]{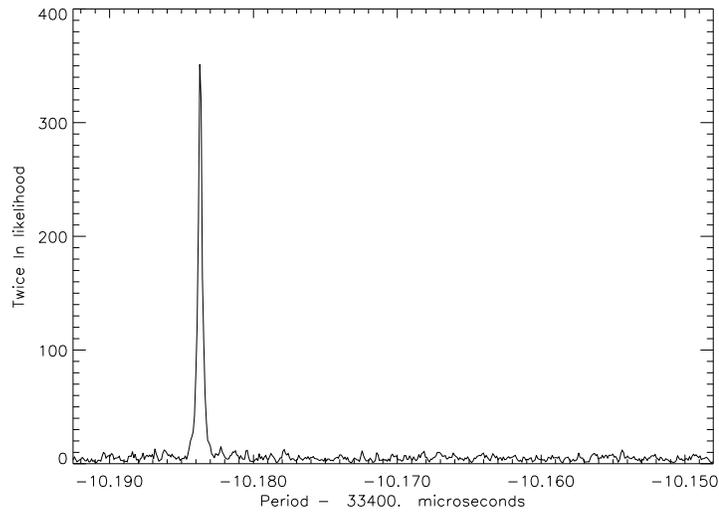}
\caption[Likelihood of flux modulation from Crab]
{\label{tv2:crablike}Twice the likelihood ($C$) of flux modulation from 
Crab as a function of trial period for a 10 bin light curve.  $C$ is distributed
as $\chi^2_9$ in the null model.  The maximum likelihood period is 33.38981541~ms.}
\end{figure}

\begin{figure}[p]
\centering
\includegraphics[width = 2.5 in]{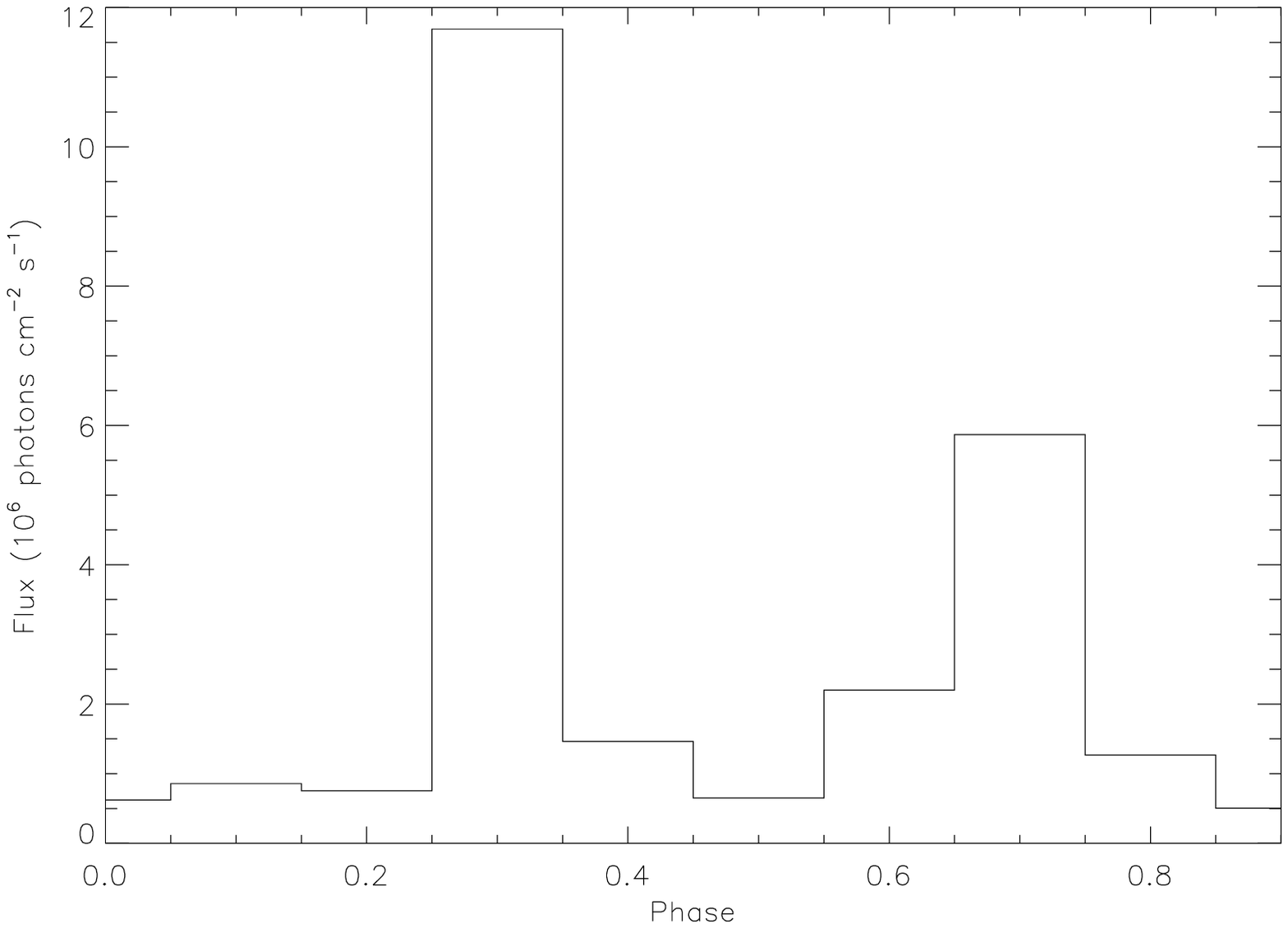}
\includegraphics[width = 2.5 in]{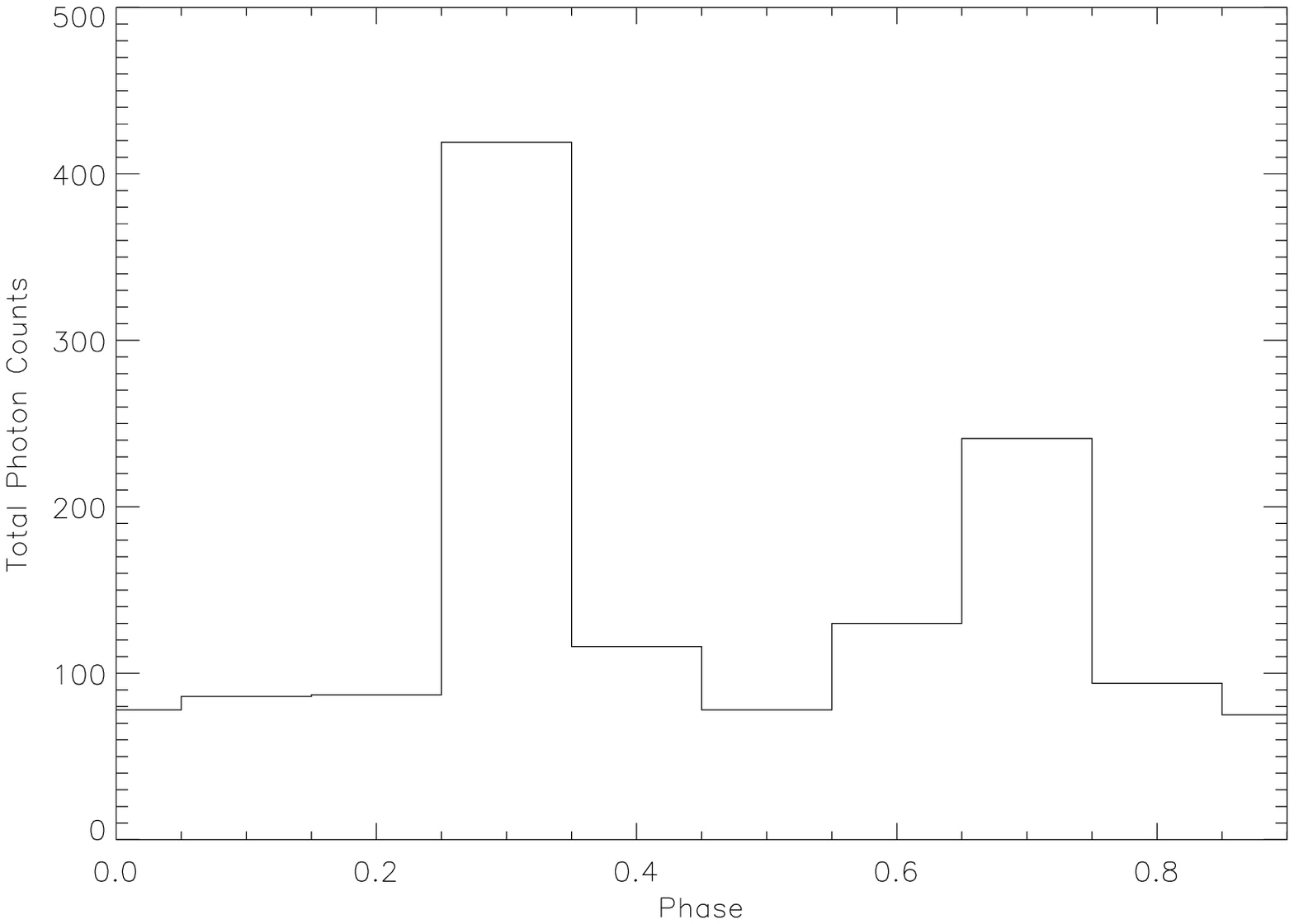}
\caption[Light curves of Crab]{\label{tv2:crablight}
Light curves of Crab at the
maximum likelihood period.  
The left light curve is the maximum likelihood flux
in each bin.  The right light curve is the raw photon count within the energy-dependent
68\% containment radius.}
\end{figure}

The three brightest steady-state \egret\ sources are Vela, Geminga, and the Crab.
All three of these objects would have been easily discovered by \timevar\ in a
search of \egret\ data with no prior information about their periods.  Their
significance is definitive; there would be no doubt about the validity of the
detection.  A light curve from Geminga is shown in \fig{tv:gemlight}.  A plot
of likelihood versus period for the Crab is given in \fig{tv2:crablike}, and
the light curves associated with the most likely period are given in
\fig{tv2:crablight}.  The likelihood peak in \fig{tv2:crablike} is very sharp.
Note, however, that the probabilities given in \tbl{tv2:known} are taken only 
from the peak value of the likelihood.  The Bayesian significance would be
found by multiplying the likelihood function by the prior, and integrating over
the width of the 
peak.\footnote{In principle, one would integrate over the entire
range.  Since \fig{tv2:crablike} represents $\ln \like$, the contribution of all
points not in the peak is negligible.}  In this case, the choice of prior is
largely irrelevant, since the likelihood is so sharply peaked.  \fig{tv2:crablight}
shows the maximum likelihood light curve as the flux measured in each bin, as
well as the raw photon count within the energy-dependent 68\% containment radius 
as given by \eq{intro:containment} for the same pulsar period.  

\begin{figure}
\centering
\includegraphics[width=4 in]{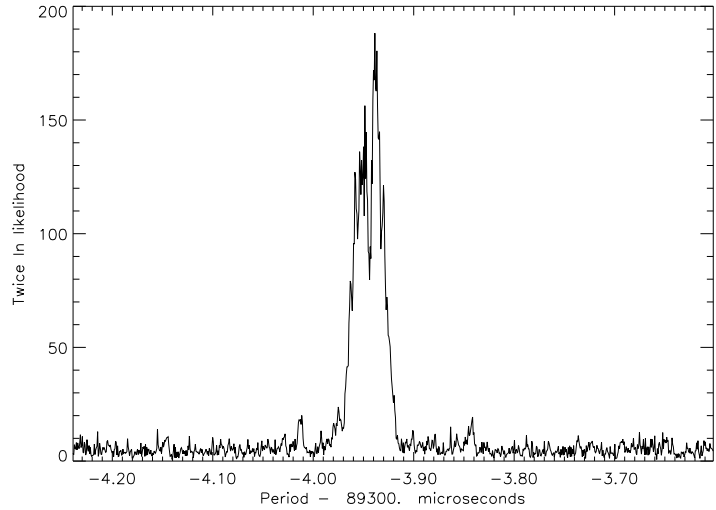}
\caption[Likelihood of flux modulation from Vela]
{\label{tv2:velalike}Twice the likelihood of flux modulation in Vela
as a function of trial period for 10 bin light curves.  The maximum likelihood
period was 89.2960605~ms.}
\end{figure}

The likelihood peak for Vela in \fig{tv2:velalike} is not nearly as sharp as that
for the Crab.  The viewing period analyzed began on 22~August~1991 and ended on
5~September~1991, soon after a pulsar glitch on 20~July~1991~\cite{velaglitch}.
It is likely that Vela's period was not very constant during this observation,
causing the broad likelihood peak observed.

\begin{figure}
\centering
\includegraphics[width=2.5 in]{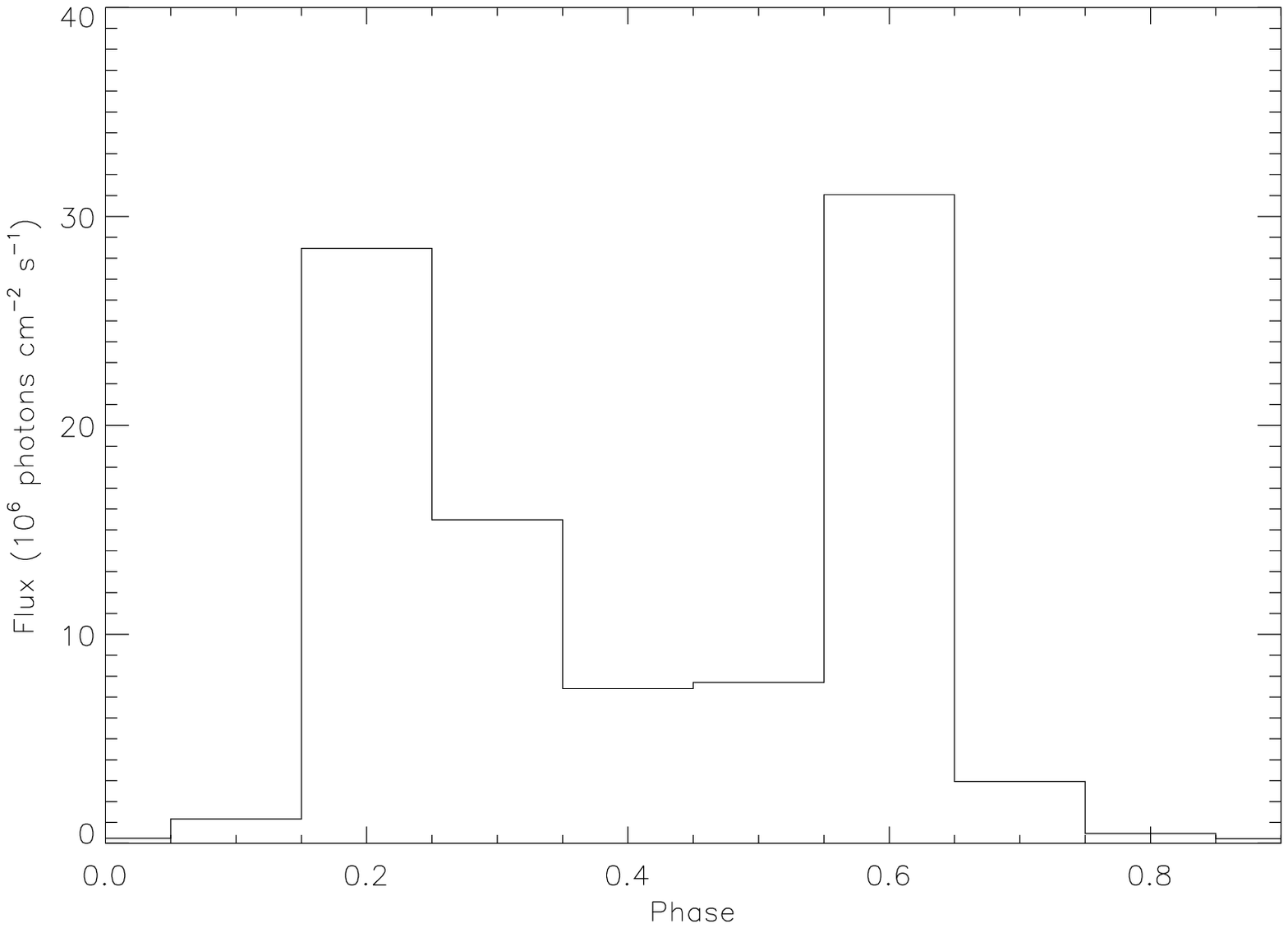}
\includegraphics[width=2.5 in]{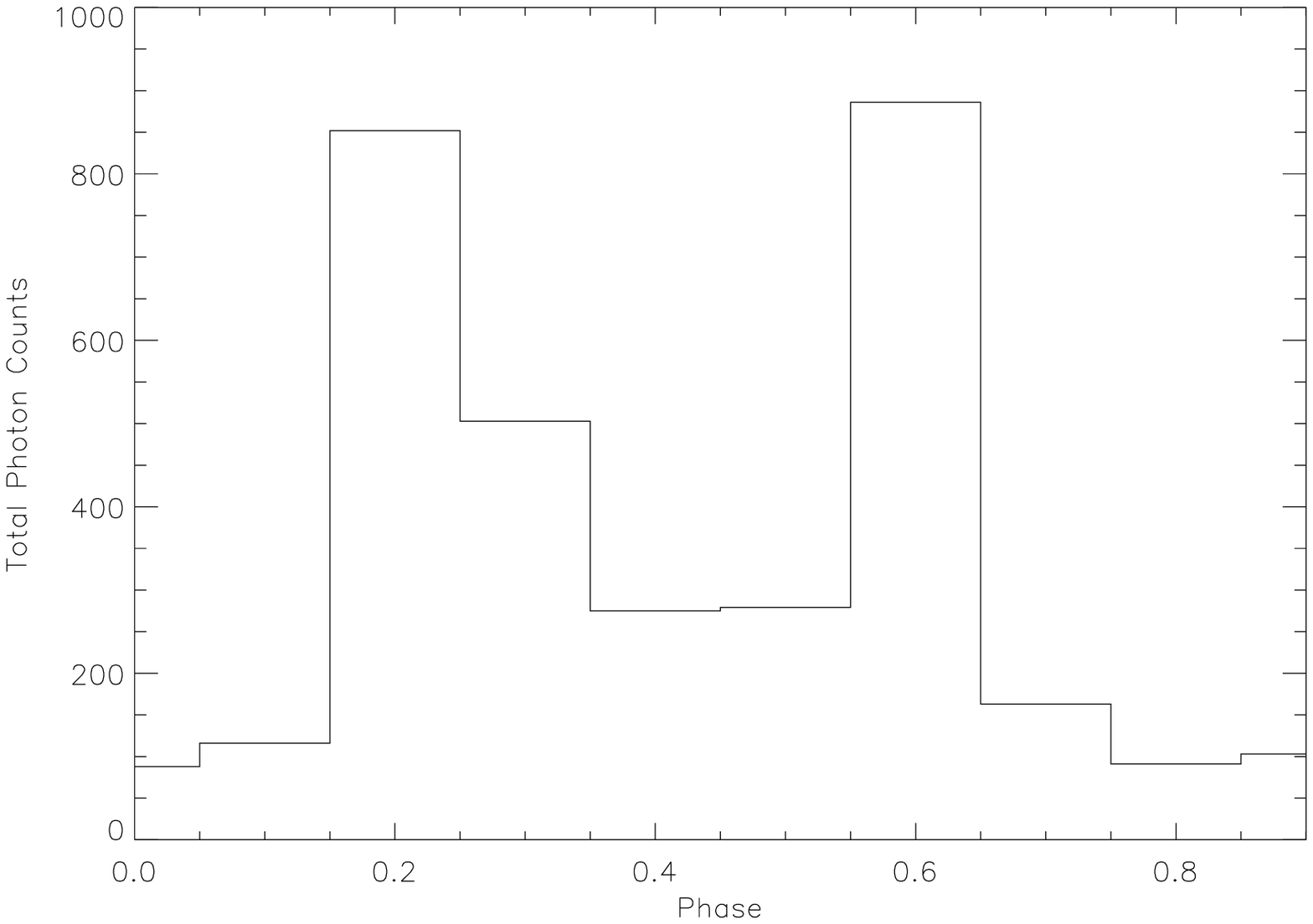}
\caption[Light curves of Vela]{\label{tv2:velalight}
Light curves of Vela at the maximum likelihood period.
The left light curve is the maximum likelihood flux
in each bin.  The right light curve is the raw photon count within the energy-dependent
68\% containment radius.}

\end{figure}

\begin{figure}
\centering
\includegraphics[width=4 in]{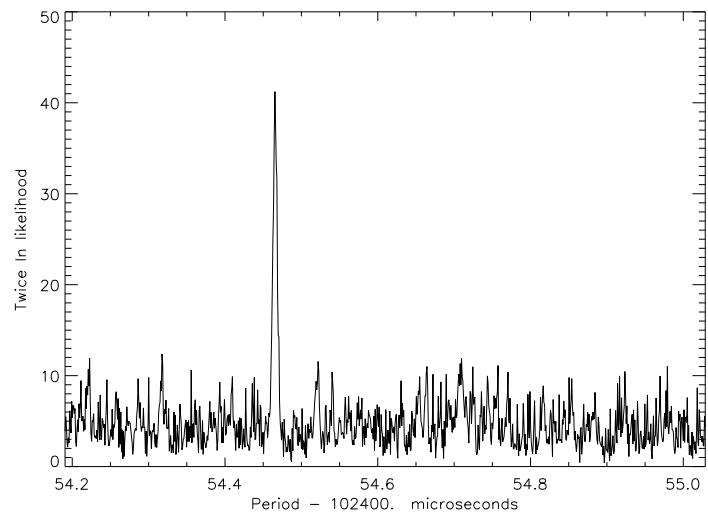}
\caption[Likelihood of flux modulation in \stos]{\label{tv2:1706like}
Twice the likelihood ($C$) of flux modulation from \stos\ as a function
of trial period for a 10 bin light curve. The maximum likelihood period was
102.24544626~ms}
\end{figure}

\begin{figure}
\centering
\includegraphics[width=2.5 in]{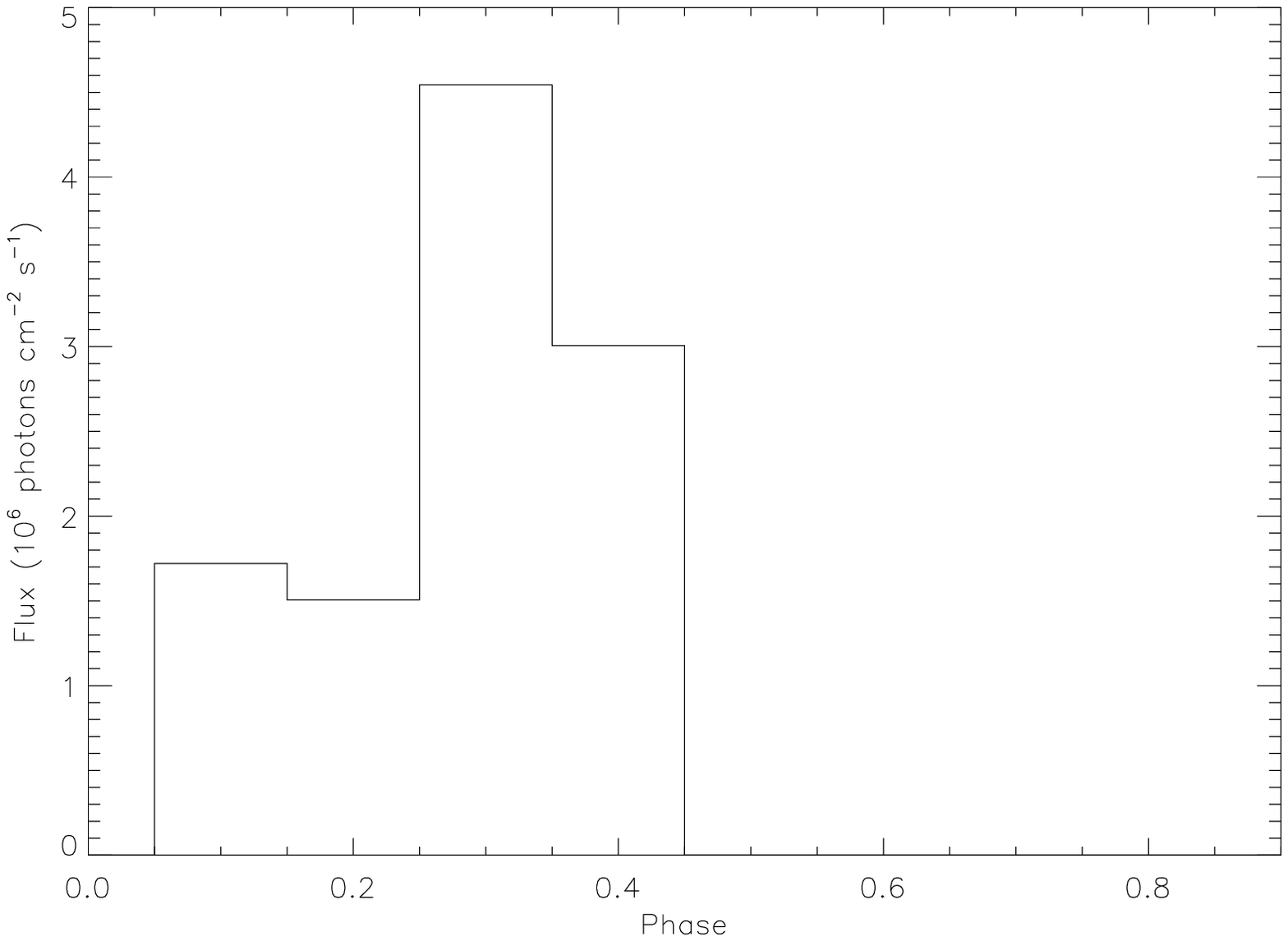}
\includegraphics[width=2.5 in]{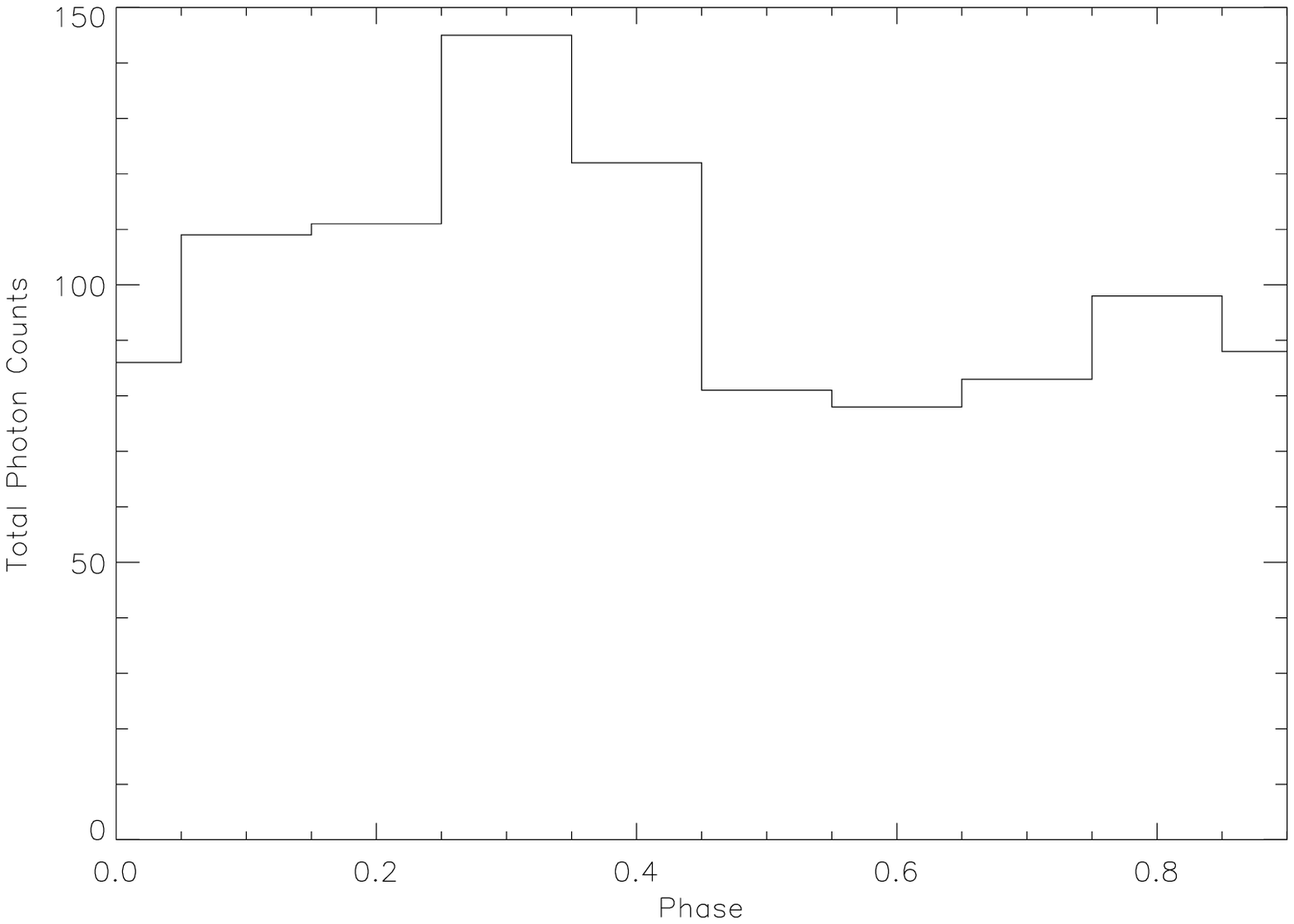}
\caption[Light curves of \stos]{\label{tv2:1706light}
Light curves of \stos\ at the maximum likelihood period.
The left light curve is the maximum likelihood flux
in each bin.  The right light curve is the raw photon count within the energy-dependent
68\% containment radius.}
\end{figure}

In addition, \stos\ would have been detected with a significance of 
\mysim77\% (\fig{tv2:1706like} and \ref{tv2:1706light}).  
While this is not high enough to conclude that the emission is
truly pulsed, it is based on the data from a single viewing period.  Such detections
should be rare enough that a small search in a different viewing period would
be feasible to confirm the detection.

The remaining two known \gammaray\ pulsars, 1055-52 and 1951+32, were not
detected significantly.  

	\subsection{Searches for Geminga-like Pulsars}
Of course, given the presence of Geminga, largely invisible to radio observations, 
as well as the large number of unidentified \egret\ sources near the plane, it
is natural to suspect that some of these sources are Geminga-like pulsars 
(\eg,~\cite{mukherjee_unid,merck_unid,pohlpulsar}).  To begin the gargantuan
task of examining the unidentified \egret\ sources, we separate those which 
are likely to be good pulsar candidates.  The criteria for determining good
candidates are steady emission, a hard spectrum, and close proximity to the
Galactic plane~\cite{joethesis,fouriersearch}.

One promising candidate is 2CG075+00, which is also known as 
GRO~J2019+3719, and hereafter will be referred to as 2CG075.
It lies very close ($ \lt 0.5\deg$) to the galactic
plane, and has a very hard photon spectrum ($\alpha=-1.40 \pm 0.14$)
with a break occurring at about 1~GeV.  Similar behavior is seen in Vela,
Geminga, \stos, and perhaps \tff~\cite{joethesis}.
There are no known radio pulsars inside EGRET's 95\% error
contours of 2CG075. 

The brightest unidentified \gammaray\ source is known as GRO~J1745-28, and
is positionally coincident with the Galactic center.  While it is unlikely 
that the Galactic center is occupied by a pulsar, it is quite possible that 
a pulsar may be nearby.  In addition, a significant spectral break has made
it a popular choice as a pulsar candidate~\cite{fouriersearch}.

The \egret\ error box on 2EGS~J1418-6049 is smaller than the field of view
of one of the X-ray cameras on the \asca\ satellite.  Analysis of the \asca\ 
data in the \egret\ error box has identified a potential X-ray 
counterpart~\cite{malloryI}.  The \egret\ data was searched for periodicity 
over a small region of parameter space in coordination with the X-ray effort.


Before too much computer time is spent examining candidate pulsars for fluctuation,
it is worthwhile to consider the thresholds for pulsation detection.
We can find the threshold for pulsed signal detection, assuming that all of the
source flux is pulsed, and using the background in the direction of the source. 
The thresholds and actual flux levels for some sources of interest are given
in \tbl{tv2:threshs}.  In accordance with what we have already seen, Vela, Geminga,
and Crab significantly exceed their threshold flux.  \stos\ is somewhat below 
threshold; in fact, in the absence of independent data to confirm our detection, we
would not have considered \stos\ to be detected in a \gammaray\ period search.

\begin{table}
\centering
\begin{tabular}{lrrrrrc}
Source & VP & $\expct{B}$ & $I_{\hbox{\scriptsize TH}}$  & $I_{\hbox{\scriptsize TH,trials}}$  
& $\expct{I}$ & Detection Possible? \\ \hline
Geminga & 310.0 & 443 & 66 & 131 & 562 & Yes  \\
Vela & 8.0 & 1937.18 & 138 & 277 & 3699 & Yes \\
Crab & 310.0 & 1638 & 127 & 279 & 218 & Possible \\
1706-44 & 232.0 & 2162 & 146 & 293 & 170 & Unlikely \\
2CG075 & 203.0 & 2736 & 184 & 367 & 314 & Possible \\
J1745-28 & 5.0 & 2134 & 162 & 377 & 158 & Unlikely \\
J1418-61 & 314.0 & 1507 & 136 & 316 & 184 & Unlikely \\
\end{tabular}
\caption[Thresholds for pulsed sources]{\label{tv2:threshs}
Thresholds for pulsed sources.  For each source, a viewing period is selected
for observation.  $\expct{B}$ is the expected number of background photons within
5\fdg85 of the source. $I_{\hbox{\scriptsize TH}}$ is the minimum number of
source counts needed to detect pulsation in a single trial, and 
$I_{\hbox{\scriptsize TH, trials}}$ is the number of counts needed to detect
pulsation in a period search.  $\expct{I}$ is the expected number
of source counts, based on the source flux and exposure in the viewing period. }
\end{table}

\begin{figure}
\centering
\includegraphics[height = 3.0 in]{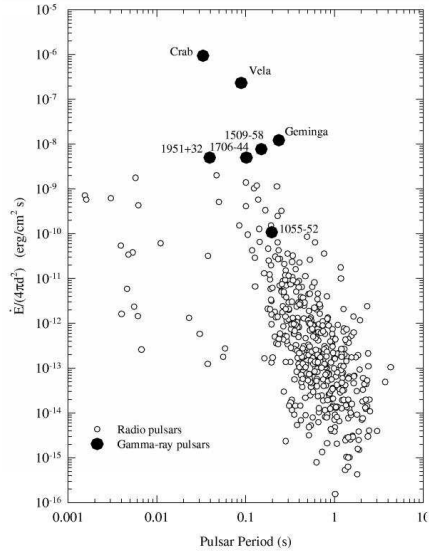}
\caption[$\dot{E}/D^2$ vs.\ pulsar period]{\label{tv2:edot}
$\dot{E}/D^2$ vs.\ pulsar period.  Pulsars most likely to be identified 
have the highest apparent luminosities.  The \egret\ threshold is 
approximately $5 \times 10^{-11}$~\cite{thompsonreview}.}
\end{figure}

Another useful application of threshold analysis is the estimation of the
number of radio-quiet, Geminga-like pulsars that may be observed by \glast.
There have been many estimates of the Galactic population of such sources,
based on the predictions of various pulsar 
models~\cite{romani,mukherjee_unid,merck_unid,ion_unid}.  Here we will
estimate the instrumental threshold.  As we have seen, the threshold depends
on the amount of exposure to the source.  Assuming we do not already know
the pulse period, then a longer exposure means that we must search period
space more densely (\sect{tv:density}).  This requires more trials, and thus,
more computation time.  An example is shown in \tbl{tv2:glast_search}.  
The \egret\ gas map is integrated over a hypothetical
\glast\ 68\% containment radius of 1\fdg5 to obtain a background 
flux ($1.9 \times 10^{-7}$ \perareasec).  The detection threshold for
a period search from 50~ms to 500~ms is given.  The threshold continues
to decline for longer exposures, while the number of trials required increases.
The threshold includes the statistical effects of the number of trials, so the
only limiting factor is computation time.

\begin{table}[t]
\centering
\begin{tabular}{ccccc}
Elapsed Time & Exposure  & Threshold  & Trials & Threshold  \\
 (Days)  & ($10^8$ cm$^2$ s) & (known period) &  & (period search) \\  \hline
1.7 & 1.5 & 11.1 & $1.7 \times 10^{10} $ & 20.7 \\
3.4 & 3.0 & 7.9 &  $1.3 \times 10^{11} $ & 15.0 \\
5.1 & 4.5 & 6.4 &  $4.3 \times 10^{11} $ & 12.4 \\
6.8 & 6.0 & 5.6 &  $1.0 \times 10^{12} $ & 10.9 \\
\end{tabular}
\caption[\glast\ pulsed signal thresholds]{\label{tv2:glast_search}
\glast\ thresholds in units of $10^{-8}$\perareasec\ for the detection of 
pulsed sources near the Galactic plane for known ephemerides, and for
period searches.  The number of trials is that required to search
from 50~ms to 500~ms, with appropriate period derivatives.}
\end{table}

While the exact value of the detection threshold will depend on the details of
the structured background at the point of interest, and the details of the 
instrument exposure, we may take $1.5 \times 10^7$ to be a reasonable threshold
for the detection of pulsation in radio-quiet pulsars in a period search.
Yadigaroglu \& Romani \cite{ion_unid} list 35 unidentified \egret\ sources near the plane.
Thirty of these have fluxes well above our threshold; three more are close
to the threshold, and three are below it.  While \glast\ will detect many more
sources than \egret, it is unlikely to detect many new bright sources.
It seems likely, then, that \glast\ will be capable of definitively searching
25--30 radio-quiet pulsar candidates, either detecting pulsation or setting
stringent upper limits, effectively eliminating pulsars as an identification
of these sources.

\subsection{Results}
Searching \egret\ data for unknown-period pulsations is a very slow process.
2CG075 was analyzed with an array of approximately five Sparc~5 and Sparc~10
processors.  Part of one viewing period (203.0) was searched from 137.8~ms
to 500~ms in period and the requisite range of period derivative (\sect{tv:density}).  
The analysis required approximately three months of continuous computation.
The most significant pulse period and period derivative, while unlikely in
a single trial ($P = 5.6 \times 10^{-7}$), was not significant given the 
number of trials ($4.6 \times 10^{8}$).  To make sure that the detection
was a statistical anomaly, a different viewing period (2.0) was searched 
in a very small window around the best ephemeris.  There was no significant
pulsing in the independent data.

J1418--604 was searched for periodic modulation in viewing period 314.0.  No
significant pulsation was seen.  J1745-28 was searched in viewing period 5.0.  
While a very low significance detection was made, examination of the candidate
pulse period in viewing period 223.0 revealed no significant modulation.

\begin{table}
\centering
\begin{tabular}{lc r@{--}l r@{.}l c}
Candidate & VP & \multicolumn{2}{c}{Period Range} & \multicolumn{2}{c}{Best Period} & Significance \\ \hline
2CG075 & 203.0 & 137.8&500~ms & 140&950161 & $1.19\times 10^{-34}$ \\
J1418--604 & 314.0 & 226.5&300~ms  & 271&870713  &  $0.55$\% \\
J1745--28 & 5.0 & 235.3&400~ms & 277&760942 & $87.7$\% \\
\end{tabular}
\caption[Results of period search with \timevar.]{\label{tv2:notfound}
Results of period search with \timevar.  The infinitesimal significance
of the 2CG075 detection stems from that fact that the single-trial significance
of the most significant period would be expected to be observed about 260 times
in the number of trials performed.}
\end{table}

\section{X-Ray Binaries}
\label{tv:xrb}
X-ray binary systems offer an excellent example of the advantages of 
coherent pulsation analysis with the methods used in \timevar. 
 Most orbital periods
are between tens of minutes and several hours.  This is comparable to,
or longer than, the typical time scale of exposure changes in \egret; 
thus any attempt to resolve orbital flux modulation must take
explicit account of exposure changes.  Meanwhile, \gammaray\ fluxes of
X-ray binaries are low enough that direct observation of a single period of
modulation is impossible.  As with pulsar analyses, epoch folding
offers a way to improve the signal-to-noise ratio and increase
chances for the detection of orbital flux modulation.  However, not
only must the arrival times of photons be epoch folded, but also the
changing exposure must be folded at the same period and phase to find
accurate fluxes.~\cite{bbjxrb}

X-ray binaries consist of a neutron star or black hole accreting material
from a normal star.  They constitute the brightest sources in the X-ray
sky.  X-ray binaries are divided into ``low-mass'' and ``high-mass'' systems,
in reference to the normal companion.  Low-mass X-ray binaries have an older
star as a companion that generates little or no stellar wind.  In order to 
emit significant X-ray power, the companion star must overflow its Roche
lobe.  The overflowing material then accretes onto the neutron star or black hole.
If the system contains a neutron star with a large magnetic field, the material
will flow along the field lines and accrete onto the polar cap.  Otherwise, the
material will form a thick accretion disk.  X-ray emission can come from
either the polar caps~\cite{mitsuda} or the inner edge of the accretion 
disk~\cite{shakura}.  High-mass X-ray binaries have O or B stars as companions, 
which generate substantial stellar winds, depositing 
$10^{-6}$--$10^{-10}\, M_{\sun}$
per year onto the compact object.  This wind is sufficient to generate the 
observed X-ray luminosities~\cite{xrb_book}.

Two major features are observed in X-rays over the binary orbit period.  The
first is a simple eclipse of the compact object, resulting in a rapid drop
in the X-ray luminosity.  The second is a dip in the X-ray luminosity, presumably
caused by the obscuration of the compact object by the thick accretion 
disk~\cite{xrb_book}.  Individual X-ray binaries may exhibit one or both of
these features.  The periods of the observed X-ray binary orbits range from
11.4 minutes to 398 hours.

	\subsection{Thresholds and Searches}
The low-mass
X-ray binaries listed in \tbl{tv2:lmxb} and the high-mass X-ray binaries
listed in \tbl{tv2:hmxb}, taken from the lists in White, Nagase, \& 
Parmar~\cite{xrb_book}, were found to have a non-zero average 
flux with a significance of greater than $1\sigma$.   
However, without a
detection of pulsation at the X-ray period it is not justified to claim
an association of the \gammaray\ excesses with the X-ray binaries.
Only one source,
Cygnus~X-3, was detected at the $5\sigma$ level in \egret\ data.  

Thresholds for detection of periodicity were calculated using the method 
described in \sect{tv:thresholds} for an extensive list of X-ray binary 
candidate sources.  It was assumed that the source variation was sinusoidal.
For any given values of source and background strengths, the detection
threshold is a strong function of the duty cycle of the source.  While it may
be hoped that some X-ray binaries have duty cycles as short as 20\%, it is
more likely that \gammaray\ emission is fairly constant for most of the orbit,
then drops sharply but briefly during eclipse.  It was further assumed that
the number of photons in any bin was large enough so that the Poisson distribution
is well approximated by a Gaussian.  This required a rather coarse division
into five phase bins.

X-ray binaries which would be near the \egret\ detection threshold if all their
flux were modulated at the orbital period were examined with \timevar.  All photons
from the source were epoch folded for a small range of periods around the known
X-ray orbital period.  They were assigned likelihood values with regard to the
source and background strengths.  The instrument exposure to the source was also
epoch folded, and the exposure for each bin was obtained.  The likelihood of
source fluctuation could then be calculated as in \sect{tv:detect_geminga}.

\begin{table}
\small
\centering
\begin{tabular}{lrrrr}
Name & VP & Photons/bin & Max Modulation & Threshold \\ \hline
X0543--682 = Cal 83 & 329.0 & 8  & 14.4\%  & $>$80\% \\
X0547--711 = Cal 87 & 224.0 & 13 & 33.2\%  & 75\% \\
	           & 17.0  & 57 & 13.6\%  & 45\% \\
X1124--685 = N'Mus 91 & 230.0 & 22 & 14.6\% & 65\% \\
X1323--619 = 4U1323-62 & 23.0 & 63 & 4.7\% & 40\% \\
X1455--314 = Cen X-4 & 217.0 & 14 & 15.3\% & 75\% \\
X1625--490 	& 23.0 & 94 & 7.2\% & 35\% \\
		& 529.5 & 76 & 9.0\% & 40\% \\
XB1636--536	& 27.0 & 135 & 1.8\% & 30\% \\
X1656+354 = Her X-1 & 9.2 & 37 & 6.7\% & 50\% \\
XB1658--298 	& 232.0 & 226 & 2.0\% & 25\% \\
		& 5.0 & 438 & 1.2\% & $\sim$10\% \\
X1659--487 = GX339-04 & 336.5 & 98 & 2.6\% & 35\% \\
		& 270.0 & 136 & 5.3\% & 30\% \\
		& 226.0 & 127 & 2.1\% & 30\% \\
		& 323.0 & 178 & 2.1\% & 27\% \\
		& 210.0 & 31 & 4.4\% & 60\% \\
		& 214.0 & 38 & 7.6\% & 52\% \\
		& 5.0 &  246 & 2.2\% & 22\% \\
		& 219.0 & 12 & 18.8\% & 75\% \\
		& 302.3 & 68 & 2.2\% & 40\% \\
		& 423.0 & 54 & 1.7\% & 45\% \\
X1735--444	& 226.0 & 183 & 0.4\% & 27\% \\
X1755--338	& 226.0 & 183 & 0.7\% & 27\% \\
		& 229.0 & 18 & 4.9\% & 68\% \\
X1820--303	& 323.0 & 260 & 0.4\% & 17\% \\
X1822--371	& 508.0 & 47 & 2.5\%  & 47\% \\
		& 529.5 & 46 & 3.1\% & 47\% \\
X1908+005 = Aql x-1 & 43.0 & 24 & 14.7\% & 60\% \\
X1957+115 = 4U1957+11 & 331.5 & 42 & 3.1\% & 50\% \\
X2023+338 = V404 Cyg & 303.2 & 62 & 13.4\% & 42\% \\
X2127+119 = AC211 & 19.0 & 46 & 1.6\% & 47\% \\
\end{tabular}
\caption[Low mass X-ray binaries]
{\label{tv2:lmxb}Low mass X-ray binaries.  For each source, the 
viewing period of the observation is listed, along with the average number 
of photons in each phase bin, the modulation fraction that would be measured
if all the source flux were modulated, and the threshold modulation fraction
that would yield a 99\% significance detection in half of all possible data sets.}
\end{table}

\begin{table}
\centering
\begin{minipage}{5.5 in}
\small
\begin{tabular}{lrrrr}
Name & VP & Photons/bin & Max Modulation & Threshold \\ \hline
X0532--664 = LMC X-4	& 6.0  & 44 & 16.1\% & 50\% \\
			& 17.0 & 52 & 13.6\% & 45\% \\
			&224.0 & 12 & 37.3\% & 80\% \\
X0538--641 = LMC X-3	& 6.0  & 42 &  7.9\% & 50\% \\
X0540--697 = LMC X-1	& 17.0 & 55 & 24.0\% & 43\% \\
			&  6.0 & 46 & 10.9\% & 47\% \\
X1119--603 = Cen X-3	& 14.0 & 164 & 3.7\% & 27\% \\
			& 402.5 & 31 & 18.6\% & 57\% \\
			& 402.0 & 27 & 19.3\% & 57\% \\
			& 208.0 & 19 & 21.5\% & 70\% \\
			& 215.0 &  8 & 22.3\% & $>$80\% \\
X1538--522 = QV Nor	& 516.1 & 20 & 1.2\% & 70\% \\
X1700--377 = HD153919	& 508.0 & 36 & 5.4\% & 54\% \\
X1956+350 = Cyg X-1	& 318.1 & 60 & 7.3\% & 44\% \\
			& 601.1 & 36 & 15.8\% & 54\% \\
X2030+407 = Cyg X-3	& 203.0 & 476 & 13.3\% & 15\% \\
			&   2.0 & 296 & 14.4\% & 21\% \\
			&   7.1 & 133 & 10.8\% & 30\% \\
			& 212.0 & 243 & 14.8\% & 23\% \\
			& 303.2 & 77  & 16.1\% & 38\% \\
			& 328.0 & 67  & 27.9\% & 42\% \\
			& 331.0 & 30  & 26.4\% & 57\% \\
			& 331.5 & 48  & 15.9\% & 47\% \\
			& 333.0 & 64  & 3.6\%  & 42\% \\
			& 601.1 & 32  & 26.1\% & 57\% \\
			&  34.0 & 54  & 5.1\% & 47\% \\
			& p12\footnote{\scriptsize Combined data from Phases 1 and 2.}  
 & 1039 & 10.6\% & 12\% \\
\end{tabular}
\end{minipage}
\caption[High mass X-ray binaries]
{\label{tv2:hmxb}High mass X-ray binaries.  For each source, the viewing 
period of the observation is listed, along with the average number of photons
 in each phase bin, the modulation fraction that would be measured if all the
 source flux were modulated, and the threshold modulation fraction that would
 yield a 99\% significance detection in half of all possible data sets.}
\end{table}

	\subsection{Results}
Most X-ray binaries \cite{xrb_book} have source-to-background ratios so
low that even if their orbital modulation fraction were 100\%, the modulation
would be undetectable.  It would still be possible to detect
modulation in such sources if the duty cycle were sufficiently short,
but under the standard model of X-ray binary emission, this is unlikely.
For candidate sources that
have parameters near threshold, a maximum likelihood period
search is done.  The sinusoidal assumption is dropped in favor of a
five independent bin light curve model.  The choice of five bins was
made to maximize flexibility in the model while retaining sufficient
numbers of photons in each bin.  Photons are barycenter corrected and
epoch folded with trial periods in a small range ($\pm$10-20\%) of the known X-ray
orbital periods.   In contrast to the pulsar searches, the longer periods
of X-ray binaries made it necessary to epoch fold the instrument exposure as well.
While the exposure was fairly evenly distributed among bins in the shortest
period binaries, it could be quite uneven for longer period X-ray binaries.
Some sources were observed in more than one viewing period; 
Fourteen promising low-mass X-ray
binaries (\tbl{tv2:lmxbtv}) and four promising high-mass X-ray
binaries (\tbl{tv2:hmxbtv}) yielded no periodic signal detections significant at
the 99\% level.  

\begin{table}
\centering
\begin{tabular}{l r@{.}l r@{.}l}
Source Name & \multicolumn{2}{c}{X-ray Orbital}  & \multicolumn{2}{c}{Significance of} \\
 	    & \multicolumn{2}{c}{Period (hr)}    & \multicolumn{2}{c}{\gammaray\ modulation} \\ \hline
X0543-682	&  \hspace*{0.25 in}25&0  &  \hspace*{0.4 in}0&23\% \\
X0547-711	&   10&6  &  56&0\% \\
X1124-685	& 10&4  &  0&7\% \\
4U1323-62	& 2&93  &  0&25\% \\
Cen X-4		& 15&1  &  54&4\% \\
X1624-490	& 21&0	&  54&2\% \\
XB1636-536	&  3&8	&  8&8\% \\
Her X-1		& 40&8 	&  5&0\% \\
XB1658-298	&  7&1	& 49&7\% \\
GX339-04	& 14&8	& 60&4\% \\
X1755-338	&  4&46	&	1&\% \\
4U1957+11	&  9&3	&	0&25\% \\
X2023+338	&  5&7  & 	94&0\% \\
X2127+119	& 17&1  &	0&65\% \\
\end{tabular}
\caption[Low-mass X-ray binary results]{\label{tv2:lmxbtv}
Low-mass X-ray binaries that were searched with \timevar, their orbital
periods, and the significance of \gammaray\ flux modulation as found
by \timevar.}
\end{table}

\begin{table}
\centering
\begin{tabular}{l r@{.}l r@{.}l}
Source Name & \multicolumn{2}{c}{X-ray Orbital}  & \multicolumn{2}{c}{Significance of} \\
 	    & \multicolumn{2}{c}{Period (hr)}    & \multicolumn{2}{c}{\gammaray\ modulation} \\ \hline
LMC X-4		&  \hspace*{0.25in}33&6	&	\hspace*{0.4in}77&5\% \\
LMC X-3		&  40&8	&	10&1\% \\
LMC X-1		& 101&28	&	54&5\% \\
Cyg X-3		&  4&8  &	17&8\% \\
\end{tabular}
\caption[High-mass X-ray binary results]{\label{tv2:hmxbtv}
High-mass X-ray binaries that were searched with \timevar, their orbital
periods, and the significance of \gammaray\ flux modulation as found
by \timevar.}
\end{table}

Several sources were analyzed despite a low signal-to-noise ratio, in case
 they were to display variation with a very short duty cycle.
Only Cyg X-3, which has been extensively studied in gamma rays 
\cite{lamb_cygX3,michelson_cygX3,mori_cygX3}, was bright enough to
have a non-negligible chance of being detected, assuming a sinusoidal 
light curve.  No evidence for variation was found in any of the sources.

%% file: Conclusions.tex
\chapter{Conclusions}

Most astrophysicists consider the study of statistics and statistical methods 
only slightly more interesting than taxonomy and speeches by university presidents.
While this feeling is by no means unique to astrophysics, it is particularly 
unfortunate in a field where statistics play such a pivotal role in our
understanding of the scientific data.

Astrophysics may be differentiated from other specialties by the unique nature
of experimentation.  In fact, we have a separate word to describe experimental
astrophysics: observation.  The choice of language highlights the fact that 
in astrophysics more than any other realm of physical study we are most often
passive collectors of data, rather than active experimenters manipulating 
controlled environments.  This is not to say that astrophysicists are lazy; 
indeed, enormous amounts of work have gone into the design, construction, and 
analysis of all kinds of observatories.  Concentrating on high-energy
astrophysics, we have seen the great contribution of \sas, \cosb, and \egret.
We anticipate further advances from \glast.

Unfortunately, the institutional disinterest in statistics has put some of
this tremendous effort to waste.  \Gammaray\ instruments are carefully 
designed to have great sensitivity to individual photons.  The scarce quanta
are jealously collected, for they each contain a great deal of information.
Indeed, the entire theoretical field of \gammaray\ burst mechanisms has 
been forced to account for a single \egret\ photon~\cite{egret940217}.
It seems puzzling, then, that after all the quality efforts made to 
improve the capabilities of each of the previous \gammaray\ telescopes,
the statistical analysis methods were developed as an afterthought.
The result~\cite{like} was that the capabilities of the \egret\ instrument
were not fulfilled until very late in the mission~\cite{cat3}.  Since \likeprog\ does not fully use all the information
in the data, various attempts have been made to cajole the standard analysis
programs into being more efficient with the data.  

An example will help clarify the point.  \likeprog\ applies the average
photon point-spread function to all photons in the data set.  Of course,
the width of the \egret\ point-spread function depends strongly on the
photon energy.  This causes \likeprog\ to understate the information in
the location of high-energy photons, and overstate the information in
the location of low-energy photons.  The net result is that the error estimates
of point-source locations is significantly larger than it needs to be.
The data set may be restricted to a smaller energy range~\cite{over1GeV},
thereby making the average point-spread function a better estimate of the
actual point-spread function for more of the photons in the data set.  However,
this improvement comes at the expense of cutting drastically the number of
photons available.  The efforts of the instrument designers to capture more
photons have been wasted.

There are many factors that go into the design of statistical tools to 
analyze \gammaray\ data.  Simplicity of design and computational speed
are often taken as the driving considerations.  A better approach, however,
is to begin with a correct statistical implementation that uses all of the
information in the data.  This theoretical implementation can then be
simplified to achieve speed and simplicity requirements.  The advantage of 
the ``top-down'' approach is that the approximations and simplifications
can be made rationally, fully weighing the losses in accuracy against the
gains made in other areas.  This is the spirit in which \timevar\ has
been designed for periodic signal analysis.  Tompkins~\cite{billthesis} has 
developed a
successor to \likeprog\ along these lines.  The result is a statistical
method which is no more complicated than \likeprog\ that computes in 
reasonable speed and offers much better position estimates and much smaller
error regions.  Such an approach avoids the pitfalls of an empirical
design of statistical methods, in which {\em ad hoc} adaptations may
have unforeseen consequences (\sect{tv:htestoops}).

We stand now on the brink of the next generation of \gammaray\ astronomy.
The efforts of scores of scientists at dozens of institutions throughout
the world are producing the design of a \gammaray\ telescope that will
revolutionize high-energy astrophysics.  While experts make extensive
computer simulations of the instrument to optimize the design, and engineers
carefully design the electronic, mechanical, and thermal structures of
\glast, relatively little effort is going into the design of the statistical
apparatus to analyze the wealth of data that will someday pour forth.
Unfortunately, this is not a project which can profitably be left until
the design stage is over.  Realistic interpretations of computer simulated 
data rely on high-quality data analysis software.  The design of that
software depends (\sect{stats:insteffects}) on the instrument design.

The final goal of any telescope is to make observations which lead to 
increased understanding of astrophysical objects.  In the case of \gammaray\ 
telescopes, this is achieved by the design of an instrument that 
simultaneously optimizes various scientific goals (sensitive area, point-spread
function, energy range and sensitivity, photon timing) with various
spacecraft limits (power consumption, heat production, telemetry limits).  
It does so by optimizing different instrument subsystems:  the \el\pos\ 
tracking system, the calorimeter, and the triggering system, among others.
A well designed instrument would consider the data analysis methods and software
to be a separate subsystem, just as important to the success of the instrument
as the other systems.  In fact, instrument design choices must take into consideration
the impact on data analysis.  Various instrument design choices, such as pointing
modes, zenith cuts, and instrument modes can have significant impact on the 
data analysis process, and in some cases, can limit the precision of the 
results.

The purpose of this work is to elucidate the statistical methods best 
suited to the analysis of periodic flux modulation in \gammaray\ data.
These methods should be appropriately modified for use with \glast.  In 
addition, the calculation of thresholds for various \glast\ configurations
and modes can be used to concentrate analysis efforts on objects that 
seem most likely to be detected.  It is likely that some simplification
of the methods will be necessary to enable the analysis to be performed
in a useful and timely way; but these simplifications can now be made
from a basis of a well-developed statistical method, rather than in
an {\em ad hoc} manner which may unnecessarily degrade the quality of 
\glast\ results.




Statistical techniques are the double-edged sword of high-energy 
astrophysics.  Without them, the field would not exist.  The existence
of high-energy emission from some \gammaray\ bursts would be unknown. 
Our understanding of pulsar energy-generation mechanisms would be 
strikingly curtailed.  Nevertheless, their misuse has caused large
segments of the astrophysical community to doubt their validity.  
So called ``$4\sigma$'' results are believed with a confidence of about
90\%---far from the nominal 99.9937\% which they claim.  This lack 
of confidence comes from the misuse and misunderstanding of statistics
and statistical methods, which is in turn a direct result of the general
disinterest in statistical methods in astrophysics.  I hope that this 
work has inspired interest in developing useful and correct statistical
methods that dispel the indifference and improve the results of future
high-energy astrophysical experiments.

%% file: Beamtest.tex
\chapter{\glast: The Next Generation} 
\label{glast:tng}
The success of the \egret\ \gammaray\ telescope has answered many 
questions, but it has also given rise to new ones.  The bounty of 
unidentified \egret\ sources undoubtably holds the key to understanding a
wide variety of astrophysical systems.  Several of these sources at 
low Galactic latitude are likely to be Geminga-like pulsars
 \cite{romani,merck_unid,pohlpulsar}. 
High-latitude sources may be unobserved AGN, or may be a new class of 
sources not yet associated with \gammaray\ emission
\cite{ozel_highlatunid}. 
Furthermore, \egret\ has positively identified many \gammaray\ sources 
that deserve further study.  While a number of \gammaray\ pulsars have 
been extensively studied \cite{fierro_pulsarsII,nolan_survey},
additional high-quality \gammaray\ data would discriminate between 
competing models of energy-generation mechanisms 
\cite{polarcap,outergap,romanimodel}.  
Multiwavelength campaigns to simultaneously observe AGN from radio 
wavelengths to \gammarays\ have become an important tool in 
understanding energy generation in these distant yet powerful galaxies 
\cite{hartman_multiwavelength,montigny_multiwavelength}.  The recent 
discoveries of optical counterparts to \gammaray\ bursts 
\cite{costa_beppo} underscores the need for a large field-of-view, 
high-energy \gammaray\ detector.  In order to achieve these goals we 
require a \gammaray\ telescope with a large effective area, a narrow 
point-spread function, and good energy and timing resolution.

\begin{figure}
\centering
\includegraphics[width = 5in]{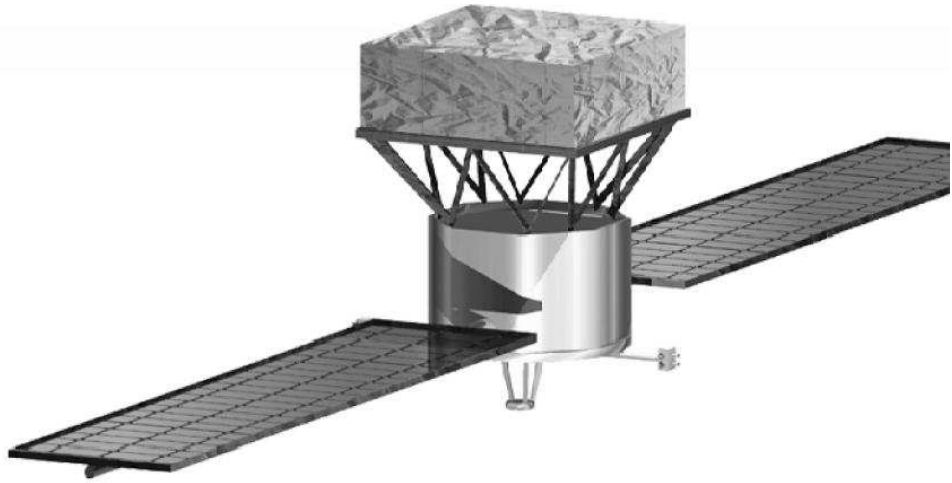}
\caption[The \glast\ satellite]{\label{glast:glastpic}
Artist's conception of the \glast\ satellite.  The square detector array
is held away from the spacecraft bus to minimize background.  Courtesy of
the \glast\ Facilities Science Team.}
\end{figure}

A proposed future telescope to that end is \glast, the Gamma-ray Large 
Area Space Telescope (\fig{glast:glastpic}).  
\glast\ will be based on solid-state silicon 
strip detector technology to provide high-quality \el\pos\ tracks from 
pair conversion events which can be reconstructed to give good 
directional information about the incident \gammaray.  A calorimeter 
will provide energy information and possibly some directional 
information as well~\cite{glast:proposal95,glast:atwood94,glast:bloom,glast:doe_proposal}.

\section{Potential Improvements}
Given the great success of \egret, is there any call to spend significant
time, energy, and resources to build the next generation \gammaray\ 
telescope?  Potential budget ramifications are beyond the scope of this
work; nevertheless, it is useful to look into the science gains that 
might be hoped for.  We will focus on the \gammaray\ bursts, pulsars, 
and X-ray binaries, the subject of \chapt{grbchap} and \chapt{timevarchap}.

\paragraph{\Gammaray\ Bursts.}
As we discussed in \chapt{grbchap}, only a handful of \gammaray\ bursts
were detected by \egret.  It seems likely that some bursts have a very soft
spectrum, or perhaps a spectral break, so that the \gammaray\ luminosity 
was far below the \egret\ threshold.  However, there are also a number of
instrumental limitations on \egret\ \gammaray\ burst observations.  The 
first of these is the instrument field-of-view.  \batse\ observes approximately
one burst per day in the entire sky.  The \egret\ field-of-view is somewhat 
less than 1.5~sr; an instrument with a field-of-view closer to $2 \pi$~sr 
would observe a bursts approximately every day as well.  At any given time, 
\egret\ has a sensitive area of 800--1000 cm$^2$.  Increases in the sensitive
area translate directly into increases in the number of photons observed from
each burst, yielding better information on burst flux, time profile, and 
position.  Perhaps most importantly, \egret\ estimates of \gammaray\ burst
flux were strongly constrained by dead-time considerations.  With the \egret\ 
instrument inactive for about 100~ms after every trigger, many \gammaray\ 
burst photons are probably missed in the initial spike of \gammaray\ flux.
There is no way to estimate the missed flux without recourse to another 
instrument, since the burst time scale is of the same order as the dead time.
Thus, the ideal high-energy \gammaray\ burst instrument would have 
a wide field-of-view, large sensitive area, good angular resolution at 
low energies, and very short dead times. 

\paragraph{Pulsars and X-ray binaries.}
The two primary difficulties in searching for flux modulation from pulsars
are the low signal-to-noise ratios, and the large region of parameter space
that must be searched.  The instrumental parameters which alleviate these
issues are increased sensitive area, and improved point-spread function.  
Increasing the sensitive area means that more photons are detected in
less elapsed time.  The sampling density required (\sect{tv:density}) depends
on the total observation time, so the increased sensitive area means 
that, for the same signal-to-noise ratio, period space may be searched 
less densely.  The increased sensitive area combines with an improved
point-spread function to improve the signal-to-noise ratio.  This lowers
the thresholds for both pulsar and X-ray binary pulsed flux detection.
It is likely that X-ray binaries other than Cygnus~X-3 will be detected
with \glast\ in steady state, and many of the X-ray binaries listed in
\tbl{tv2:lmxb} and \tbl{tv2:hmxb} will be above the threshold for 
the detection of flux modulation.  In addition, the sensitivity of 
\glast\ to photons down to \mysim20~MeV will greatly enhance the sensitivity
to soft sources like X-ray binaries.

The proposed \glast\ instrument will excel at all three of the desired
capabilities.  As described in the next section, the sensitive area
will be nearly an order of magnitude greater than that of \egret\ 
at high energies.  The point-spread function will be almost a factor
of two smaller in radius, and the dead time will be about 100  times less than
the \egret\ dead time.  These factors will combine to make \glast\ an
excellent instrument for observing \gammaray\ bursts, pulsars, and X-ray
binaries.
\section{The Baseline \glast\ Instrument}
\label{bt:baselinesect}
\egret\ has revealed the \gammaray\ sky to be a vast resource of 
astrophysical information.  As described fully in \sect{intro:egret}, 
the \egret\ instrument is composed of three sections: a calorimeter, 
an \el\pos\ tracker, and an anti-coincidence system.  \glast\ will 
follow the same paradigm, although the technologies used for each 
component have advanced significantly in the two decades since 
\egret\ was designed.

\begin{table}[t]
\centering
\begin{minipage}{5.5 in}
\centering
\begin{tabular}{r@{\hspace{0.5 in}}l}
\gammaray\ Telescope Characteristic & \glast\ Baseline Performance \\ \hline
Energy Range & 			20~MeV to 300 GeV \\ 
Energy Resolution & 		$\lt$ 25\%, 10 MeV to 300 GeV \\ 
&				$\lt$ 10\%, 100 MeV to 10 GeV \\ 
Effective Area &		8000 cm$^2$ above 1 GeV \\ 
&				4000 cm$^2$ at 50 MeV \\
Point-Spread Width  & 3.1\deg \by\ 100 MeV/$E$ \\
& (68\% containment) \\
Off-axis\footnote{where the sensitive area drops
to half of its on-axis value} width & 1.4 \by\ on-axis width \\
Field of View (FWHM) & 2.6 sr \\
Point Source Sensitivity & 3.5 \by\ $10^{-9}$ photons \perareasec \\
& (1 year, $E \gt$ 100 MeV, 5$\sigma$ significance)\\
Point Source Location & 30 arcsec--5 arcmin \\
Mission Life & 5 years (2 year minimum) \\
Mass & 3000 kg \\
Power & 600 W \\
Telemetry (average) & 100 kbps \\
Orbit & 600 km low-inclination \\
\end{tabular}
\end{minipage}
\caption[\glast\ baseline characteristics]{\label{bt:baseline}
Characteristics of the baseline \glast\ instrument.  Values found
by \glastsim\ simulation of the baseline instrument 
\protect\cite{glast:doe_proposal}.}
\end{table}

The instrumental requirements for \glast\ are driven by the scientific 
questions we wish to answer.  Locating point sources and separating nearby 
sources requires compact point-spread functions.  Observing \gammaray\ 
bursts requires large photon collection area and very short dead times 
at the lowest energies, where burst photons are more plentiful.  
Pulsar timing also requires large photon collection area, but the hard 
spectra of pulsars necessitates good sensitivity to higher energy 
photons.  Spectral measurements of all sources require good energy 
resolution.  Monitoring for transient events like AGN flares and 
\gammaray\ bursts requires a large field of view (FOV).  In 
additional, all of these goals must be achieved while maintaining 
excellent ($\gt\,1:10^5$) background rejection to eliminate cosmic ray 
contamination, while staying within the structural and 
power constraints dictated by the spacecraft design.

The \glast\ instrument is still in the planning stages, and as such 
many options for achieving these goals are being discussed.  Most of 
these discussions take place as variations on the baseline design 
as developed by an international collaboration of scientists from 
many institutions around the world (\tbl{bt:institutions}).
This design was proposed for funding to the Department of Energy
and NASA in 1998, as well as to other international agencies \cite{glast:doe_proposal}.  
The baseline 
design currently calls for a 5~\by~5 array of modular towers.  Each 
tower would consist of 17 trays arranged to hold 16 layers of silicon 
strip detectors.  Each layer would measure the location of passing 
charged particles in $x$- and $y$-projections.
plane.  In addition, each tower would contain 80 CsI(Tl) blocks of 
approximately 2.3~cm~\by~3~cm~\by~31~cm arranged horizontally in eight 
layers of ten blocks in alternating $x$--$y$ orientations to measure
the total energy of the \gammaray.  PIN diodes 
attached to each end of each block allow differencing of the light 
detection in order to determine a lateral displacement.  The baseline 
anticoincidence detector (ACD) covers the entire instrument to identify 
cosmic rays.  It 
consists of plastic scintillator tiles 1~cm thick and approximately 
the size of a single tower, arranged in two offset layers to cover all 
cracks.

\begin{table}
\small
\begin{tabular}{cc}
Aerostudi, S.r.l.		& Shibaura Institute of Technology \\
Boston University		& Sonoma State University	\\
Commissariat a L'Energie Atomique (CEA) & Stanford Linear Accelerator Center \\
\'Ecole\index{\'E} Polytechnique & Stanford University		\\
Hytec, Inc.			& Texas A\&M University--Kingsville \\
ICTP and INFN, Trieste		& U.S. Naval Research Laboratory \\
Kanagawa University		& University of California, Santa Cruz \\
Laboratory for High Energy Astrophysics  & University of Chicago	\\
Lockheed Martin			& University of Rome			\\
Max Planck Institut f\"ur Extraterrestrishe Physik & University of Tokyo\\
NASA Ames Research Center	& University of Utah			\\
NASA Goddard Space Flight Center & University of Washington		\\
\end{tabular}
\caption[Institutions in the \glast\ collaboration]{\label{bt:institutions}
Institutions in the \glast\ Collaboration}
\end{table}

The baseline tracker design consists of planes of silicon strip 
detectors (SSDs).  Although some alternative designs (gas microstrip 
detectors \cite{gasmicro} and scintillating fiber detectors) 
are being considered, the 
prototype instrument tested in the October 1997 \beamtest\ was based 
on silicon strip technology.  Therefore, we will concentrate on the 
silicon strip baseline.

Optimization studies have been performed \cite{glast:doe_proposal} 
using Monte Carlo simulations for a variety of tracker configurations.  Some of the 
considerations that go into such an optimization are discussed in 
\sect{bt:tracker}.  The best design found so far consists of 17 trays of 
detectors, with 3.5\% radiation length (\radlen) Pb radiators to convert \gammarays\ to 
electron--positron pairs (\sect{bt:pairproduction}), and 400 \micron\ 
thick SSDs with a 195 \micron\ pitch, or distance between strips.  Each 
tray has detectors on the top and bottom of a thick wafer of 
low-density material.  Thus the SSDs on the bottom of one tray are 
close to those on the top of the next.  Such a pair of SSD layers are 
treated as one logical ``plane'' for the purposes of \el\pos\ 
measurements.  Since there are no SSDs on either the top of the first 
tray or the bottom of the last tray, we have 16 detector planes in the 
tracker.  Each layer will consist of a 5~\by~5 array of SSDs, each one 
of which is 6.4~cm~\by~6.4~cm.  The SSDs will be connected into chains 
along their strip axis, resulting in an effective size of 
6.4~cm~\by~32~cm.  The signal-to-noise ratio in each strip is 
approximately 23:1 for a minimum ionizing particle.

\chapter{Testing the \glast\ Science Prototype}
In order to demonstrate the feasibility of the \glast\ project, as well as 
to confirm observational parameters measured in simulations of the 
instrument, a science prototype was constructed at the University of 
California, Santa Cruz, the Naval Research Laboratory, and Goddard 
Space Flight Center.  It was tested in the parasitic electron beam at 
the Stanford Linear Accelerator Center (SLAC) in October of 1997 
\cite{glastnim}.  Modeling instrument response, both analytically and
through Monte Carlo simulations, and verifying those results experimentally
has become a viable way to optimize instrument design while minimizing
costs \cite{wcjthesis}.  \Gammaray\ reconstruction software was
developed and tested with Monte Carlo simulations (\chapt{reconchap}), 
and then was used to analyze the experimental results.  Comparison of 
the actual \beamtest\ results with simulation confirmed that the 
simulations represent an accurate model of the \beamtest\ instrument, 
and by extension, of the baseline \glast\ design \cite{psfVSmc}.  
These results will be discussed in \chapt{responsechap}.


\section{The SLAC \el\ Beam}
\label{bt:beam}

\begin{figure}[t]
\centering
\includegraphics[height=3 in]{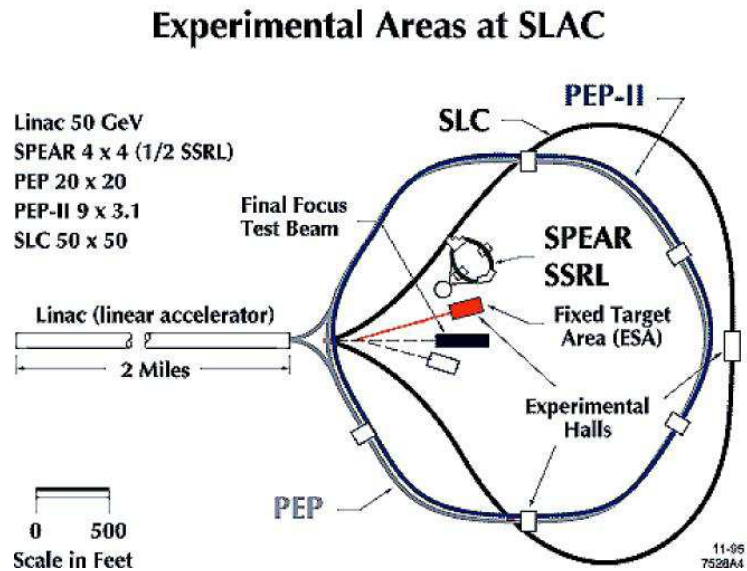}
\caption[Location of End Station A at SLAC]{\label{bt:slacpic} 
Location of End Station A at SLAC.  The October \beamtest\ took
place in End Station A (ESA), also known as the Fixed Target Area.
The linac was simultaneously providing beam for SLC.}
\end{figure}

The October 1997 \beamtest\ was conducted in End Station A at the 
Stanford Linear Accelerator Center \cite{glastnim}.  End Station A is 
located at the end of the main linear accelerator beam 
(\fig{bt:slacpic}).  This choice of testing location was a combination of design 
requirements and practical considerations.  For the purposes of our 
test, we desired a beam of incident electrons arriving approximately one at a 
time, which we could either directly measure with our apparatus, or 
first convert to high-energy \gammarays.  End Station A accommodates 
such a beam with the linac in {\em parasitic beam} operation 
\cite{parasitic}.

Parasitic mode means that the beam consists of particles which have 
been scraped away from the main beam.  The main experiment at SLAC 
during October 1997 was SLD, a search for $Z_0$ vector bosons.  SLD 
required rapid, well-focused bunches of approximately $10^{11}$ 
electrons or positrons for collision.  The main linac produces bunches 
of electrons or positrons at a rate of 120~Hz, and accelerates them to 
50~GeV. The SLD beam profile was defined with a small aperture to 
achieve a compact, nearly monoenergetic beam.  Electrons (and 
positrons) in the wings of the profile would stop in the massive 
shields around the aperture.  Bremsstrahlung \gammarays\ would 
continue forward, while the rest of the main electron beam was 
magnetically steered into the SLC arcs.  These \gammarays\ could then 
be converted back to electrons by passing through a high-Z foil.  The 
number of electrons per pulse could be limited by adjusting the size 
of the momentum acceptance in the transport line~\cite{daniel}.

This resulting beam of ``parasitic'' electrons then consisted of a mix 
of electrons and positrons with a broad range of energies.  Steering 
magnets selected electrons of the desired energy and delivered them to 
End Station A. The electron energy was tunable from approximately 
5~GeV to approximately 40~GeV. The number of electrons per bunch was 
tunable over a wide range from less than one to many tens of 
electrons.

Once the electron beam had been delivered to End Station A, we 
modified it appropriately for our own uses.  For several runs, we took 
the beam directly as it came; usually a 25~GeV electron beam.  This 
was useful to calibrate the calorimeter, to do backsplash studies with 
the anti-coincidence detector (ACD) and to look at straight tracks in 
the silicon tracker.  However, most of the data was taken with a thin 
Cu radiator (usually 3.5\% \radlen) inserted in the electron beam.  
Between the radiator and the instrument was a large magnet known as 
\BO. By adjusting the magnetic field in \BO, the electron beam could 
be steered aside and deposited in a hodoscopic calorimeter.  If the 
electron in the beam shed a bremsstrahlung photon in the Cu foil, then 
its energy was reduced, and its deflection in \BO\ was greater.  The 
hodoscopic calorimeter had 88 fingers arrayed horizontally to measure 
this deflection.  This allowed us to tag the energy of \gammarays\ 
incident on our instrument to an accuracy of about 250~MeV.

Events with multiple \gammarays\ in the tracker were quite 
undesirable.  It was likely that only one of the \gammarays\ would be 
detected in the tracker, while both would deposit energy in the 
calorimeter.  The apparent energy of such an event could be strikingly 
different than its true energy, and recognizing such events in the 
data would be very difficult.  Therefore, it was desirable to keep the 
number of \el\ per pulse small; for most of the runs, the momentum
slits were set to allow approximately one \el\ per pulse on 
average.  Of course, the number of \el\ in a pulse is a Poisson 
process, so there was exactly one electron in each pulse approximately 
one-third of the time.  When the Cu radiators were in the beam to 
produce \gammarays, the rate could be a little higher, since not all 
\el\ shed bremsstrahlung photons.

\section{The \BeamTest\ Instrument}
The \glast\ science prototype instrument was divided into three parts. 
The tracker consisted of 6 planes of silicon strip 
detectors, for precision measurement of the \el\pos\ tracks.  The 
calorimeter was composed of segmented blocks of CsI to measure 
deposited energy.  The anti-coincidence detector (ACD) was a set of 
plastic scintillators read out to photodiodes via wave-shifting 
fibers.  Each of these three components was connected to the data 
acquisition system of End Station A \cite{PerryZen}.

\begin{figure}
\centering
\includegraphics[width = 5 in]{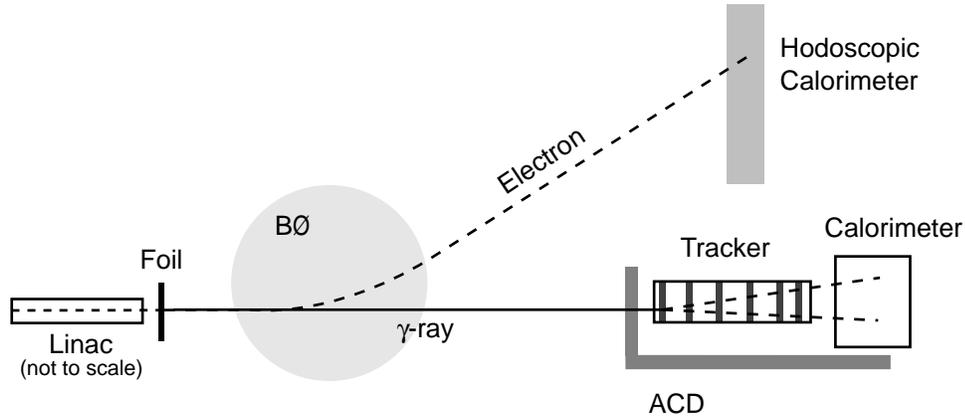}
\caption[\Beamtest\ experimental scheme]{\label{bt:instdiag}Diagram of the
\beamtest\ experimental scheme.  The 25~GeV electron beam entered from the
left, incident on a high-Z foil.  Some electrons emitted a bremsstrahlung
photon, which continued through the ACD, and into the silicon tracker.
Some \gammarays\ converted to \el\pos\ pairs in the tracker.  The
\gammaray\ energy could be measured with the calorimeter.  A magnet, \BO,
deflected the original electron into the hodoscopic calorimeter.  The
angular deflection caused by \BO\ was roughly proportional to the
bremsstrahlung \gammaray\ energy. }
\end{figure}

\subsection{Tracker}
\label{bt:tracker}
Competing physical effects lead to the adoption of a number of tracker 
design configurations.  Multiple scattering of the \el\pos\ pair is 
the dominant source of error in reconstructing the incident angle of 
low-energy \gammarays.  Multiple scattering (\sect{bt:mssect}) refers 
to the process by which an electron passing through a material is 
deflected by many small scatters; it is generally inversely 
proportional to the electron energy \cite{ppdb}.  However, at high 
energies the granularity of the strip pitch can cause significant 
errors in the angle estimations.  These two competing effects make two 
parameters relevant to the design of a silicon-strip \gammaray\ 
telescope: the ratio of the strip pitch to the gap between planes, and 
the amount of radiator inserted between planes to facilitate 
conversions.  Reducing the pitch-to-gap ratio improves the resolution 
of the instrument at high energies at the expense of increasing the 
number of channels---corresponding to greater instrument complexity 
and power usage---or of reducing the field of view.  Reducing the 
amount of radiating foil between planes decreases the amount of 
multiple scattering that each electron experiences, at the expense of 
fewer \gammaray\ pair conversions---corresponding to a reduced 
detection efficiency and thus less exposure.

\begin{figure}[t]
\centering
\includegraphics[height=2.5 in]{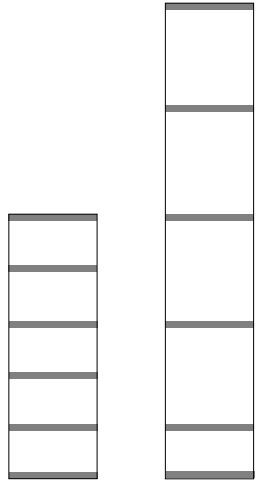}
\caption[Pancake and Stretch tracker configurations]{\label{bt:configs}
Pancake (left) and Stretch (right) tracker configurations, 
illustrated to scale.  The gray bars represent the
 locations of the planes.  The distance between the planes in pancake
configuration was 30.0 mm; in stretch, 60.0 mm, except for the last 
gap, which was 30.0 mm.}
\end{figure}

In order to explore this two-dimensional parameter space, the 
\beamtest\ instrument was built with adjustable spacing between 
planes, and with adjustable lead radiating foils between planes.  Each 
of the six cards built for the instrument had two silicon-strip 
detectors (SSDs) attached to it; one with strips in the $x$ direction 
and one with strips in the $y$ direction.  For consistency with 
\glast\ documentation, a single detector in either direction will be 
referred to as a {\em layer}, while an $x$-$y$ pair of detectors will 
be referred to as a {\em plane}.  The test box was built with ten 
slots on 3 cm centers to accommodate the cards, allowing us to vary 
the pitch-to-gap ratio by putting the cards in different slots.  The 
\beamtest\ SSDs were 5 cm by 5 cm square, with a strip pitch of 236 
\micron\ and a thickness of 500 \micron.  Each SSD had 192 
instrumented strips, corresponding to 6 readout chips responsible for 
32 strips each, and a total instrumented area of 4.6 cm by 5 cm 
\cite{glastnim}.

In addition, each slot could accommodate a {\em radiator card}, a 
special card with no SSDs, but instead with a thickness of lead (Pb) 
foil.  The distance between the lead radiators and the silicon 
detectors was approximately 2~mm.  Radiator cards were prepared 
with approximately 2\%~\radlen, 4\%~\radlen, and 6\%~\radlen\ to allow 
us to vary the total radiator in the instrument.

\begin{table}[b]
\centering
\begin{tabular}{c | c}
Pancake & Stretch \\  \hline
0.00\% & 0.00\% \\
1.71\% & 1.71\% \\
3.71\% & 3.71\% \\
 & 5.4\% \\
\end{tabular}
\caption{\label{bt:radlenstbl}Radiation lengths of Pb available}
\end{table}

Simulations before the \beamtest\ \cite{presimres} suggested two 
instrument configurations that were adopted for study.  The first, 
so-called ``pancake'' mode, consisted of 6 planes of silicon, each 
containing an $x$ and $y$ layer, separated by 3 cm.  This relatively 
compact configuration maximized the number of pair electrons contained 
within the tracker.  However, at high energies when multiple 
scattering is small, the squat aspect ratio of this configuration 
accentuated the measurement error.  The second mode, called 
``stretch,'' placed the planes as far apart as experimental conditions 
would allow.  The first five planes were spaced 6 cm apart, and the 
last one was spaced 3 cm apart.  This configuration allowed more 
low-energy pairs to escape the tracker, but minimized measurement 
error for the high-energy pairs.

Data was taken in the stretch configuration with 2\%~\radlen, 
4\%~\radlen, and 6\%~\radlen\ radiators, as well as with the radiator 
cards removed (``0\%~\radlen'').  In the pancake configuration, data 
was taken with no radiators, 2\%~\radlen, and 4\%~\radlen.  There was 
not enough time to take 6\%~\radlen\ data in pancake configuration.

In most cases, analysis was done in ten standard energy bands to 
improve the statistics.  The energy bands are defined in 
\tbl{bt:ebands}.

\begin{table}[t]
\centering
\begin{tabular}{c r@{--}l r@{~}l}
Band Number & \multicolumn{2}{l}{Energy Range} & 
\multicolumn{2}{c}{Center} \\ \hline
0  & 10&20~MeV & 14 & MeV \\
1  & 20&50 & 33 \\
2  & 50&100 & 72 \\
3  & 100&200 & 140 \\
4  & 200&500 & 330 \\
5  & 500&1000 & 720 \\
6  & 1&2~GeV & 1.4 & GeV \\
7  & 2&5 & 3.3 \\
8  & 5&10 & 7.2 \\
9  & 10&20 & 14.4 \\
\end{tabular}
\caption[Energy bands used for analysis]
{\label{bt:ebands}Energy bands used for analysis.  
The approximate geometric center of the energy band 
was often used as the average energy, implicitly 
assuming an $E^{-1}$ spectrum.}
\end{table}

\subsection{Calorimeter}
\begin{figure}[t]
\centering
\includegraphics[height=2.8 in]{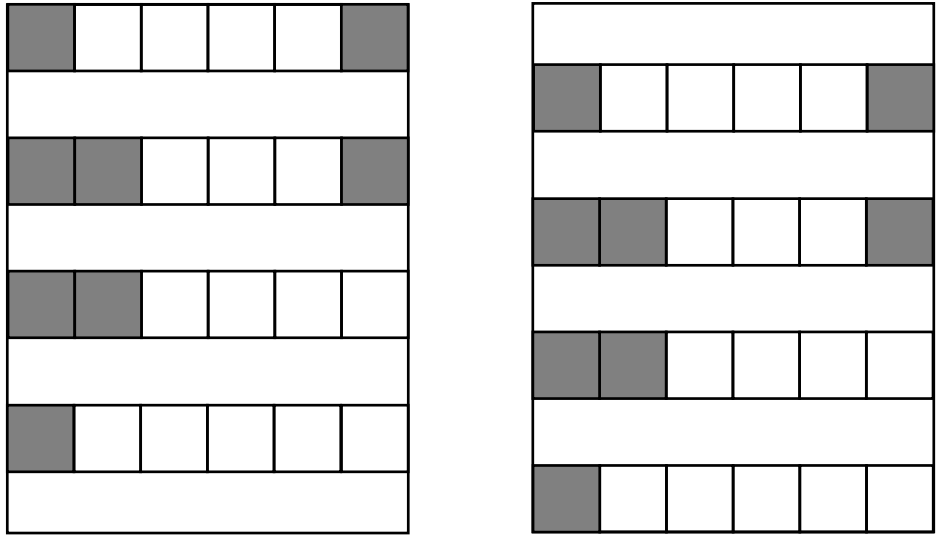}
\caption[CsI(Tl) Calorimeter]{\label{bt:cal}
CsI(Tl) Calorimeter in each projection.  Shaded blocks are non-instrumented Cu.  
Unshaded blocks are CsI(Tl) with PIN diodes on each end.  The beam was incident
from the top of the diagram.}
\end{figure}

Behind the silicon tracker was a prototype calorimeter made of 
segmented blocks of CsI(Tl).  For all of the tracker data runs, the 
calorimeter consisted of eight layers of 6 logs, each 3~cm by 3~cm by 
28~cm.  Thirty-two blocks were made of CsI(Tl), and fully 
instrumented with photodiodes on each end.  The other 16 blocks were 
made of Cu with holes bored into the material to simulate the 
equivalent number of radiation lengths of CsI(Tl).  A diagram of the 
locations of the instrumented blocks is given in \fig{bt:cal}.  The 
calorimeter, built at the Naval Research Laboratory, is fully 
described elsewhere \cite{nrlcal}.  For the purposes of the silicon 
tracker data analysis, the only calorimeter measurement used was the 
total energy deposited in all instrumented blocks.  This was found by 
summing the counts from the analog-to-digital counters (ADCs), 
subtracting the pedestals, and adjusting for the gain:
\begin{equation}
\label{bt:calgain}
E_{meas} = \alpha \left( \sum \mbox{ADC} - 1660 \right)
\end{equation}
where $\alpha$ is 1.6 for gain 1, 0.25 for gain 4, and 0.06 for gain 7.
The gain setting of the photodiodes used for measuring the energy deposition 
in the calorimeter blocks was recorded for each data run.

\subsection{Anti-Coincidence Detector}
\glast\ will be equipped with an anti-coincidence detector (ACD) to 
distinguish \gammarays\ from the charged-particle background.  While 
the ACD is designed to reject background charged particles, it is also 
very sensitive to the array of particles emitted from a \gammaray\ or 
other particle interaction in the calorimeter.  High energy 
\gammarays\ may ``self-veto'' as secondary particles created in the 
calorimeter activate the ACD. \egret\ was especially prone to this 
type of event due to its monolithic anti-coincidence scintillator, and 
in fact displayed a precipitous decline in sensitive area at high 
energies \cite{egretcalibrate93}.  The \glast\ ACD will be an 
arrangement of plastic scintillator tiles covering the instrument 
which will register the passage of charged particles.  The segmented 
tiles will provide independent measurements to facilitate background 
rejection while not reacting to backsplash events.  To test the 
scintillators and the associated electronics, an ACD was developed for 
the \beamtest\ \cite{glastnim}.

The system set up for the \beamtest\ consisted of 15 plastic 
scintillators.  Nine of these were arranged along the side of the 
silicon tracker and calorimeter, and six were placed in two layers in 
front of the silicon tracker.  
The segmented design allows the discrimination of true charged 
particle events from backsplash by lending position information.  The 
scintillators along the side of the \beamtest\ instrument were 
designed to verify this procedure.


All of the \beamtest\ scintillators were read out via waveshifting 
fibers to photomultiplier tubes.  These fibers allowed the 
scintillators to be placed very close to one another, minimizing 
cracks through which charged particles might penetrate undetected.

\section{Summary of Data}
Data was collected in runs of up to two hours.  The length of the run 
was limited by the amount of data that could be stored by the data 
acquisition system on one tape~\cite{PerryZen}.  Over 400 runs were 
made over 30 days in October 1997, running 24 hours each day, 7 days 
each week.  Approximately 30 collaboration members worked shifts to 
monitor the experiment and change instrumental or beam parameters as 
necessary.  The data acquisition system recorded $2.1 \by 10^{8}$ 
triggers, which required more than 200 gigabytes of tape.  Only the 
\gammaray\ runs were useful for the tracker study; electron runs were 
used for backsplash and calorimeter studies.  Useful \gammaray\ events 
were filtered from these triggers; this process will be described in 
\sect{br:cuts}.

\section{Simulations}
The October 1997 \beamtest\ was just as critical for the evaluation of 
\glast\ simulation techniques as it was for the evaluation of \glast\ 
technology.  The verification of simulation results for the \beamtest\ 
indicated to what extent simulations of the full \glast\ instrument 
could be trusted to accurately represent instrument performance in 
orbit.

Simulations of the \glast\ instrument have been successfully done 
using computer code called \glastsim.  The code is based on \gismo, a 
toolbox of routines that simulates the interaction physics for a large 
number of particles with a large number of materials \cite{gismo}.  
These particles and their interactions are taken from EGS, a 
highly-tuned analytical model of quantum electrodynamical interactions 
and transport established by the particle physics community for the 
simulation of high-energy physics experiments \cite{egs,chaputthesis}.

For the purposes of the science prototype \beamtest, we further 
modified \glastsim\ to simulate the prototype instrument that we would 
actually be using.  In the interest of realism, as much of the 
experimental apparatus was included in the simulation as possible.  
The SLAC electron beam is nearly monochromatic (to within a few 
percent in energy) because of the large steering magnets which are 
used to bring the beam to End Station A. The simulations thus assumed 
a monochromatic 25~GeV electron beam, directly incident on a 3.5\% \radlen\  
foil radiator.  The electrons would bremsstrahlung in the radiator 
according to the interaction cross-section.  A magnetic field then 
swept away the incident electron, allowing any bremsstrahlung photons 
to continue into the instrument.  Once the \gammaray\ entered the 
silicon tracker, it was allowed pair-produce using the standard EGS 
\gammaray\ interactor.  

Upon exiting (or missing) the tracker, the resulting particles were 
collected in a CsI(Tl) calorimeter.  Since tracking the particle 
shower in simulations of the calorimeter are complicated and thus 
quite slow, some simulations were done with monoenergetic incident 
\gammarays, without a calorimeter.  When a calorimeter was used, it 
was composed of 8 layers of 8 CsI(Tl) blocks, each 3~cm by 3~cm, for a 
total of 13 radiation lengths.  When the Monte Carlo data was read 
into the analysis code, energy deposition into blocks that were not 
instrumented was ignored.  Furthermore, since only a few percent of the 
electrons incident on the Cu foil shed a bremsstrahlung \gammaray, it was 
significantly faster to simply inject a bremsstrahlung spectrum of 
\gammarays\ directly.

\chapter{Reconstructing Events}
\label{reconchap}
All telescopes require a method of converting the raw data recorded by 
the instrument into relevant observational parameters.  For a 
\gammaray\ telescope, these parameters consist of information about 
individual photons.  A bright astrophysical \gammaray\ source might 
have an intensity of $10^{-6}$ \perareasec.  \egret\ might measure 30 
photons per hour from such a source.  Compare this to an ordinary 
light bulb, which emits something like $10^{20}$ photons per second.  
Of course, each optical photon is very much less energetic.  Nevertheless, 
if the total energy from our hypothetical \gammaray\ source had been 
emitted at optical frequencies instead of \gammarays, our telescope would 
receive around two billion photons per hour.

Clearly, the fact that photons arrive so rarely will profoundly 
influence the way we analyze our data.  ``Imaging'' must be done 
statistically, with long integration times.  Sometimes it may even be 
more advantageous to look at maps of some statistical measure, rather 
than directly at maps of intensity.  Likelihood techniques used in 
analyzing photon information to derive astrophysical information for 
\egret\ were discussed in \pt{part1}.  Here we will concentrate on the 
process of deriving photon information for the basic instrument 
response, beginning with the mechanism of \gammaray\ pair production.

	\section{Pair Production}
\label{bt:pairproduction}
High-energy \gammaray\ telescopes work on the principle of pair 
production.  According to the rules of quantum electrodynamics (QED), 
a photon passing through matter may convert into a electron-positron 
pair.

\begin{equation}
\gamma + \mbox{nucleus} \longrightarrow
 e^+ + e^- + \mbox{nucleus}
\end{equation}\index{$\longrightarrow$}

The probability of such a conversion taking place is roughly 
independent of the energy of the incident photon above 1~GeV, and 
falls off at lower energies.  However, not all interactions result in 
pair production.  At low energies, photons tend to Compton scatter 
more readily than pair produce.  At 20~MeV, approximately 70\% of 
interactions in Si result in pair production.  At 10~MeV, close to 
half of interactions in Si are Compton scatters \cite{ppdb}.  The 
total interaction cross section for all processes is fairly constant 
down to about 10-20~MeV. While the full pair-production cross section 
is quite a complex function of incident \gammaray\ energy, electron 
energy, positron energy, nuclear recoil energy, opening angle, 
azimuthal angle, and recoil angle \cite{motz}, several simplifying 
assumptions give simple estimates of bulk behavior~\cite{ppdb}.
%
%
 For a homogeneous material, the intensity of the incident
\gammaray\ beam falls off like
\begin{equation}
\label{bt:Idecay}
I = I_o \exp{ (- \frac{7}{9} t/X_o)}
\end{equation}
due to all interactions, 
where $t$ is the thickness of material and \radlen\ is the radiation
length of the material.  Therefore the probability of a particular
\gammaray\ interacting in the material is
\begin{equation}
\label{bt:probconv}
P(t) = 1 - \exp{ (- \frac{7}{9} t/X_o)}
\end{equation}

\paragraph{Pair Production \Gammaray\ Telescopes.}
\Gammarays\ that pair produce offer an opportunity for detection.  By 
tracking the resulting \pos\el\ pair, we can estimate the incident 
\gammaray\ energy and direction.  The reconstructed energy will be the 
sum of the \pos\ and \el\ energies, corrected for energy loss in the 
instrument, and the incident direction of the \gammaray\ must be the 
momentum-weighted average of the \pos\ and \el\ directions.  All but 
the lowest energy electrons detected by \glast\ will be relativistic, 
so we may use the energy-weighted average of their directions to 
calculate the incident \gammaray\ direction.

\label{bt:mssect}Accurately reconstructing the particle tracks is 
therefore of great importance.  Two effects hinder our efforts to do 
this.  The first is multiple scattering of the electrons.  (\pos\ and 
\el\ will be referred to collectively as ``electrons.'')  
 At large angles, it 
is not Gaussian; however, the core of the distribution (out to 
approximately $3\sigma$) is approximately Gaussian \cite{ppdb}, with a 
projected width of
\begin{equation}
\label{bt:mseqn}
\theta_o = \frac{13.6 \mbox{MeV}}{E} \sqrt{x/X_o} (1 + 0.038 \ln{ x/X_o)}
\end{equation}
for relativistic electrons, where $E$ is the electron energy and $x$ is
the thickness of material traversed.  In addition, there is some lateral
displacement of the electron from one side of the material to the other.
 The rms width of the displacement distribution
is given by
\begin{equation}
y_{rms} = \frac{1}{\sqrt{3}} x \theta_o
\end{equation}
Note that multiple scattering becomes smaller, on average, with 
increased electron energy, and with thinner radiating material.

The second effect which complicates track reconstruction is 
measurement error.  Most technologies proposed or used for \gammaray\ 
telescopes are based on wires or strips made of various materials.  
When an electron passes near the strip, it ``fires'' or records a hit.  
The strip pitch clearly affects the resolution of the telescope.  The 
strip pitch divided by the gap between planes roughly determines the 
minimum angle that the telescope can resolve.

Given a set of strip addresses which have been hit, we must 
reconstruct the electron tracks and determine the parameters of the 
incident \gammaray.  There may be noise hits, spurious tracks, missing 
hits, or ambiguous tracks.  We are limited by measurement error, and 
by energy-dependent multiple scatter.  Even if we have two 
well-defined tracks, we may not know the energy in each electron, only 
the combined energy deposited in the calorimeter.  Furthermore, the 
$x$ and $y$ projections of the instrument are read out separately.  
Given a track in the $x$ projection, the question of which $y$ track 
corresponds to it is ambiguous.  Clearly, a good method of finding and 
fitting electron tracks will be critical to the accurate estimation of 
the incident \gammaray\ direction.

	\section{Track Reconstruction}
The problem of establishing the most likely electron tracks falls 
naturally into two steps: finding and fitting.  The first step 
consists of choosing which hits in the tracker are part of the track 
in question.  Designing good algorithms to do this is an art; in fact, 
it is similar in some ways to the pattern recognition problems being 
worked on by computer scientists.  The second step consists of making 
the best estimate of the track of the electron that caused those hits.  
The latter is a science---an optimization problem---and is by far the 
more tractable problem.  We will address the simpler problem first.

\begin{figure}
\centering
\includegraphics[height=5 in]{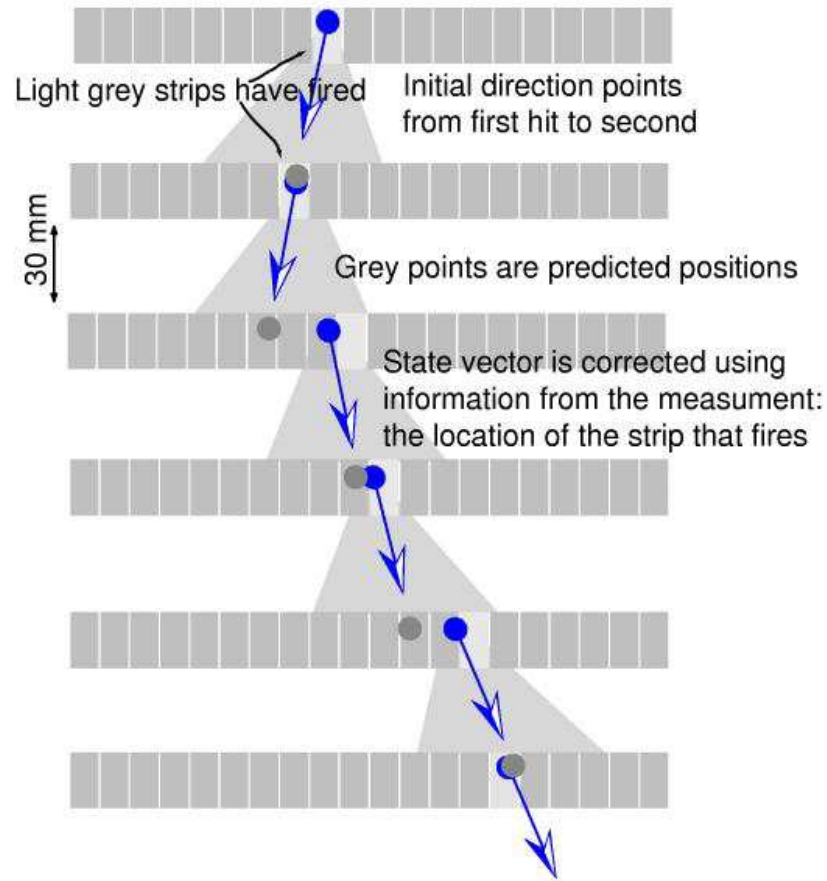}
\caption[Kalman filtering]{\label{bt:KalCar2}
The Kalman Filtering process.  A passing electron causes a strip to 
fire in each plane (light grey).  The Kalman filter uses the current 
track direction (black arrow) to predict the hit location on the next 
plane (grey circle).  The light grey area represents the range of likely 
directions after multiple scattering.  The location of the strip which 
fires is used to correct the track location, and predict the track 
direction to the following plane.}
\end{figure}

\paragraph{Least-squared Methods.}
The simplest method of track fitting is the linear 
least-squares fit.  We simply fit a straight line to all of the hits 
in the track.  Since we expect that the total angle scattered should 
increase as the track proceeds through additional layers, we assume the 
uncertainty in the measurement to grow increasingly larger as we travel down 
the tracker.  This method has the advantage of being very simple and 
fast.  At high energies, the track should be very nearly straight, so 
the least-squared fit line will be a good approximation to the real 
track.  However, at energies where multiple scattering is significant, a 
straight line is a poor approximation to the actual track.  
A line fit in this way will have reasonable information 
about the incident track direction, but its estimation of the final 
track position and direction may be quite poor.  If we wish to 
extrapolate the track to the calorimeter below, we will require better 
estimates of the track parameters at the bottom of the silicon 
tracker.  Furthermore, a linear least-squares fit rolls the multiple 
scattering and measurement error into one general error.  As we have 
seen, these errors behave very differently in different energy limits.

\begin{figure}
\centering
\includegraphics[height=5 in]{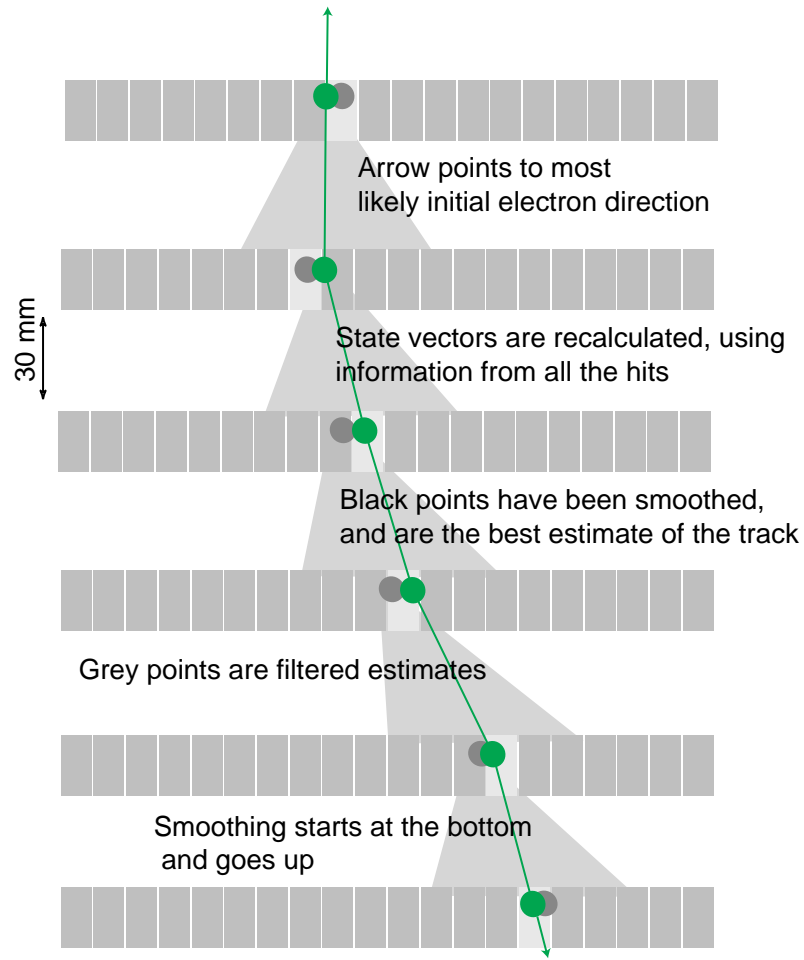}
\caption[Kalman smoothing]{\label{bt:KalCar3} 
The Kalman Smoothing process.  Starting at the bottom and working
up, the track estimate (grey circles)
is corrected further (black track), based on information
about the track below the plane in question.}
\end{figure}

\paragraph{Kalman Filters.}
Given a set of hits that make up a track, the optimal linear fitting 
method is called Kalman filtering \cite{kalman, physicistsguide}.  
``Kalman filtering'' is really a two-step process, 
consisting of a ``filter'' and a ``smoother.''  The filter begins at 
the first hit of a track, and makes a prediction for the location of 
the next hit.  That prediction is refined in light of the measured hit 
location, and the error matrices are updated.  This process, called 
``filtering,'' continues to the end of the track (\fig{bt:KalCar2}).  
Once a track has been filtered, it is then ``smoothed.''  When the 
filtering process is finished, our estimation of the track in any 
given plane has no information about the locations of hits in 
subsequent planes.  Smoothing incorporates that information.  It steps 
back up the track from the bottom, further refining the track 
parameters at each step based on the information found further down 
the track (\fig{bt:KalCar3}).  Fr\"{u}hwirth\index{\"{u}} 
\cite{fruhwirth} developed a practical implementation of Kalman 
filters that is applicable to particle track fitting.

The filter must balance the competing effects of multiple scattering 
and measurement error.  The problem simplifies immensely if either one 
of these is negligible.  If the measurement error were negligible 
compared to the multiple scattering, as expected at low energies, the 
filter would simply ``connect the dots,'' making a track from one hit 
to the next.  Most of the information about the \gammaray\ direction 
would come from the first two hits, where the cumulative effects of 
multiple scattering are the smallest.  However, if the measurement error 
is significant and multiple scattering is negligible (as it will be 
for high energy photons), all hits have information, and we should 
essentially fit a straight line to the hits.  The Kalman filter 
balances these limits properly for all energies, and thus earns its 
title as the optimal linear filter; in the limit that all errors and 
multiple scattering are Gaussian, it is the optimal filter.  That 
means that if all is Gaussian, it is completely equivalent to both a 
$\chi^2$ fit and a maximum likelihood fit.
The Kalman Filter was chosen for analysis of the beam test data.  
Accordingly, we will examine the method and its implementation in some detail
in \sect{bt:kalmaneqns}.

\paragraph{Track finding.} 
Track finding is a more subjective problem.  The identification of 
which hits belong to a track is a pattern recognition problem which 
does not admit an analytic solution.  The basic algorithm we have 
adopted is based on the filtering procedure described above.  At each 
plane, we use a Kalman filter to predict the most likely location of 
the hit.  We then assume, in most cases, that the nearest hit to that 
predicted location is the one which belongs to the track.  This simple 
criterion is complicated by caveats that allow for tracks leaving the 
tracker and for tracks sharing the same hit.  The complete algorithm 
will be discussed in \sect{bt:algo}.

	\section{The Annotated Kalman Filtering Formulae}
\label{bt:kalmaneqns}
We base the Kalman filtering equations on those found in 
Fr\"{u}hwirth\index{\"{u}} \cite{fruhwirth}.  At each instrument 
layer, we have the projected state vector, the ``filtered'' state 
vector, and the ``smoothed'' state vector.  The Kalman filter will 
successively calculate these estimates of the track parameters.  The 
state vector will contain all parameters of interest about the track.  
These might include the lateral position in $x$, $y$, or both; the 
height $z$ in the instrument; the direction of the track as either 
angles or slopes; and the energy of the track.  The state vector may 
be chosen to contain information from only one projection, separating 
the problem into fitting the $x$ and $y$ projections separately, or it 
may contain all parameters for a simultaneous fit.  However, a state 
vector of length $n$ will require inversion of $n \by n$ matrices.  
Computationally, it may be advantageous to separate independent 
variables into separate state vectors, and fit them separately.  For 
each layer, then, we have a system equation of the form

\begin{equation}
\myvec{x}_k = {\bf F}_{k-1} \myvec{x}_{k-1} + \myvec{w}_{k-1}
\end{equation}
where $\myvec{x}_k$ is the state vector containing track parameters in 
plane $k$, the {\bf F} matrix is the propagator from one layer to the 
next, and the random variable $\myvec{w}_{k-1}$ is the multiple 
scattering.

 The {\bf F} matrix takes the state vector on one plane to the state 
 vector on the next plane.  In general, this means it will combine 
 directional information with position information to compute a new 
 position.  In the absence of a magnetic field, the direction will not 
 change.  The {\bf F} matrix is indexed since it implicitly contains 
 information about the gap between one plane and the next, which may 
 not be the same for all planes.
 
The Kalman filter must also consider the measurement process.  Given a 
true track position $\myvec{x}$, we will measure

\begin{equation}
\myvec{m}_k = {\bf H}_k \myvec{x}_k + \myvec{\epsilon}_k
\end{equation}
where $\myvec{m}_k$ is the measurement that we make, {\bf H} is the
measurement matrix, and the random variable $\myvec{\epsilon}_k$ is the
measurement error.  

Also note that the dimensions of $\myvec{x}$ and $\myvec{m}$ may not be 
the same.  If all properties of the track are not directly measured, 
{\bf H} will not be square.  Again, remember that we can have a 
separate ${\bf H}_k$ for every plane.  If silicon tracker data were 
combined with data from a sampling calorimeter, the calorimeter's {\bf 
H} matrix would include a term indicating an energy measurement.

The second part of the system and measurement equations is the random 
variables we use to represent multiple scattering ($\myvec{w}_k$) and 
measurement error ($\myvec{\epsilon}_k$).  If their expectation values 
are zero, it will be sufficient to consider their covariance matrices.  
By suitably defining our state vectors and other matrices, this 
condition can always be satisfied.  For notational convenience, 
following Fr\"{u}hwirth\index{\"{u}}, we define

\begin{eqnarray}
{\bf Q}_k =  \mbox{cov} \left\{  \myvec{w}_k \right\}  \\
{\bf V}_k = {\bf G}_k^{-1} = \mbox{cov} \left\{ \myvec{\epsilon}_k
\right\} \\
\end{eqnarray} 

It is helpful to make the semantic distinction between multiple {\em 
scattering} and measurement {\em error}.  Both appear as random 
variables in our equations, but it is important to remember that 
measurement error is a description of our imperfect measurement, while 
multiple scattering is a physical distribution.

The displacement in the detector plane of the track due to multiple 
scattering is correlated with the angle through which the track 
scatters \cite{ppdb}.  If $z$ is the thickness of the detector, and 
$\theta_o$ is the rms multiple scattering width given by 
\eq{bt:mseqn}, then the multiple scattering covariance matrix is
\begin{equation}
{\bf Q} = \left( \begin{array}{cc}
	z^2 \theta_o^2 / 3 & z \theta_o^2 / 2 \\
	z \theta_o ^2/2 & \theta_o^2 \\
	\end{array} \right)
\end{equation}
assuming $\myvec{w}$ has two components, horizontal position $x$ and 
track angle $\theta$.

Finally, we have the covariance matrix of our state vector:
\begin{equation}
{\bf C}_k = \mbox{cov} \left\{ \myvec{x}_k - \myvec{x}_{k, true} \right\}
\end{equation}
This will provide our estimate of the errors in our estimate of the 
track parameters; that is, the estimate of the point-spread function 
for each electron track.  The details of this estimation may be found 
in \sect{bt:psf}.

\subsection{The Filtering Equations}

Now, we can write down the prediction, filtering, and smoothing 
equations.  Note that the equations are only stated as found in 
Fr\"{u}hwirth\index{\"{u}}, not derived here. 
First, we predict the next position, using the propagator and the 
position on the previous plane:

\begin{equation}
\myvec{x}_{k, proj} = {\bf F}_{k-1} \myvec{x}_{k-1}
\end{equation}
and the next covariance matrix, found by adding to the predicted covariance
the effects of the multiple scattering that happens in plane $k-1$:

\begin{equation}
\label{bt:projC}
{\bf C}_{k, proj} = {\bf F}_{k-1} {\bf C}_{k-1} {\bf F}_{k-1}^T + 
{\bf Q}_{k-1}
\end{equation}

These equations express the propagation, according to {\bf F}, of the
position and errors, with the addition of the multiple scattering
covariance. 

The filtering process refines the predicted position by using 
information from the measurement on that plane.  We must first refine 
the covariance matrix:

\begin{equation}
\label{bt:filteredC}
{\bf C}_k = \left[ ({\bf C}_{k, proj})^{-1} + {\bf H}_k^T {\bf G}_k
{\bf H}_k \right] ^{-1}
\end{equation}

Recall that ${\bf G}_k$ is the inverse of the measurement error 
covariance matrix.  If measurement errors were huge, then the 
measurement would contribute very little information.  ${\bf G}_k$ 
would be nearly zero, and the error matrix would be just the previous 
error matrix propagated to the next plane.

This is a good place to consider a limiting case.  If all of our state
vectors were scalars, ${\bf C}$ would look like $\sigma^2$.  Then
(\ref{bt:filteredC}) would look like

\begin{equation}
\sigma_{filtered}^2 = \left( 1/\sigma_{predicted}^2 + 1/\sigma_{measurement}^2
\right) ^{-1}
\end{equation}
We are simply weighting our estimate of $\sigma$ by the quality of the
measurements.

Using our refined ${\bf C}_k$, we can calculate
\begin{equation}
\label{bt:filterX}
\myvec{x}_k = {\bf C}_k \left[ ({\bf C}_{k, proj})^{-1} \myvec{x}_{k, proj}
+ {\bf H}_k^T {\bf G}_k \myvec{m}_k \right]
\end{equation}
Again, the size of ${\bf G}_k$ controls how heavily the measurement is
weighted.

In our limiting case, we now weight our track estimate by the quality
of the measurement: (\ref{bt:filterX}) becomes

\begin{equation}
\frac{x_{filtered}}{\sigma_{filtered}^2} = 
\frac{x_{predicted}}{\sigma_{predicted}^2} +
\frac{x_{measured}}{\sigma_{measured}^2}
\end{equation}

\subsection{The Smoothing Equations}
So far, all the equations have only used information about the 
measurements taken either on the same plane, or ``upstream'' of the 
current plane.  Smoothing is the process of further refining each 
position estimate in light of the information from all the 
measurements, upstream and downstream of the current plane.  We first 
calculate an auxiliary matrix {\bf A}:

\begin{equation}
\label{bt:Aeqn}
{\bf A}_k = {\bf C}_k {\bf F}_k^T ({\bf C}_{k+1, proj})^{-1}
\end{equation}
Then the smoothed position and covariance estimates are
\begin{eqnarray}
\label{bt:smootheqn}
\myvec{x}_{k, smooth} = \myvec{x}_k + {\bf A}_k (\myvec{x}_{k+1, smooth} -
\myvec{x}_{k+1, proj}) \\
{\bf C}_{k, smooth} = {\bf C}_k + {\bf A}_k ({\bf C}_{k+1, smooth} -
{\bf C}_{k+1, proj}) {\bf A}_k^T
\end{eqnarray}
 The difference between the smoothed and projected versions of 
 $\myvec{x}_k$ and ${\bf C}_k$ may be thought of as the effect on the 
 state vector and covariance matrix of the data from planes $k$ and 
 below.  Now we need to know how to calculate the influence of that 
 data on the estimates for the current plane.  The key is ${\bf A}_k$.  
 First, consider the case with no multiple scattering, when ${\bf 
 Q}_k$ is zero.  Plugging into (\ref{bt:Aeqn}) from (\ref{bt:projC}) 
 after adjusting indices, we find

\begin{equation}
{\bf A}_k = {\bf C}_k {\bf F}_k^T ( ({\bf F}_k^T)^{-1} {\bf C}_k^{-1}
	{\bf F}_k^{-1}) = {\bf F}_k^{-1}
\end{equation}
In this case, ${\bf A}$ is just the back-propagator, taking 
$\myvec{x}_{k+1}$ to $\myvec{x}_k$.  Then (\ref{bt:smootheqn}) just 
propagates $\delta \myvec{x}_{k+1}$ back to plane $k$, exactly what we 
would expect with no multiple scattering to complicate the issue.

The limit that multiple scattering is large compared to the covariance 
matrix is more complicated.  Returning to the scalar case, the various 
${\bf C}$'s become $\sigma^2$, and {\bf F} becomes a scalar.  Then our 
{\bf A} becomes

\begin{equation}
a = \frac{f \sigma^2_{k, filtered}}{\sigma^2_{k+1, proj}}
\end{equation}
Again plugging in from the scalar equivalent of (\ref{bt:projC}), we end up with
\begin{equation}
\label{bt:scalarA}
a = \frac{f \sigma^2_{k, filtered}}{f^2 \sigma^2_{k, filtered} + 
	\sigma^2_{ms}}
\end{equation}
In the limit that $\sigma^2_{k, filtered} \ll \sigma^2_{ms}$, this simplifies to
\begin{equation}
a \approx \frac{f \sigma^2_{k, filtered}}{\sigma^2_{ms}}
\end{equation}
So, the bigger the expected multiple scattering, the more we discount the 
information in the planes below.  In fact, both extremes can be seen in 
the scalar version: as multiple scatter goes to zero in (\ref{bt:scalarA}), 
$a$ approaches unity---the information from later planes is not discounted 
at all.

Now we have calculated the smoothed state vectors and covariance 
matrices at all points along the track.  These revised estimates 
represent the optimal linear filter of the particle track.  If the 
measurement and multiple scattering errors were Gaussian, it would be 
the optimal estimate of the track, and equivalent to a maximum 
likelihood estimate.  The smoothed state vector for the first plane 
gives us the initial track direction, while the smoothed covariance 
matrix for the first plane gives us the ``point-spread width'' for 
that track.

		\subsection{Goodness of Fit}
		\label{bt:kalchisq}
If the Kalman filter, under some assumptions, is identical to both 
least-squared fitting and maximum likelihood, we should demand that it 
produce a \chisq\ value or a likelihood as a measure of its goodness 
of fit.  In fact, it does produce a ``running \chisq'' as it filters 
and smooths.

For each plane, we find the residual vector
\begin{equation}
\myvec{r}_k = \myvec{m}_k - {\bf H}_k \myvec{x}_k
\end{equation}
and the covariance matrix of the filtered residuals
\begin{equation}
{\bf R}_k = {\bf V}_k - {\bf H}_k {\bf C}_k {\bf H}_k^T
\end{equation}
The incremental \chisq\ is then
\begin{equation}
\chi^2_+ = \myvec{r}_k^{\,T} {\bf R}_k^{-1} \myvec{r}_k
\end{equation}
The total \chisq\ of the track is given by the sum of the \chisq\ 
contributions for each plane.

The smoothed incremental \chisq\ can be similarly calculated:
\begin{eqnarray}
\label{bt:chisqeqn}
\myvec{r}_{k,smooth} = \myvec{m}_{k,smooth} - {\bf H}_k \myvec{x}_{k,smooth} \\
{\bf R}_{k,smooth} = {\bf V}_k - {\bf H}_k {\bf C}_{k,smooth} {\bf H}_k^T \\ 
\chi^2_+ = \myvec{r}_{k,smooth}^{\,T} {\bf R}_{k,smooth}^{-1} \myvec{r}_{k,smooth}
\end{eqnarray}

The incremental \chisq\ for each plane is distributed as 
$\chi^2(m_k)$, where $m_k$ is the dimension of $\myvec{m}_k$ 
\cite{fruhwirth}.  In fact, it is precisely in this sense that we may 
call it \chisq.  Since it measures residuals due to multiple 
scattering as well as those due to measurement error, strictly 
speaking it is $-2 \ln \like$, which of course is distributed 
as \chisq~\cite{wilks,eadie}.

This test may be used as a way to identify track outliers.  A 
measurement with a \chisq\ value corresponding to the $(1-\alpha)$ 
quantile may be rejected as not belonging to the track.  This process 
will reject a measurement which actually is part of the track with 
probability $\alpha$.

The complete set of Kalman filtering equations is summarized in 
\app{app:KalEqn}.

%% file: Beamtest2.tex
	\subsection{Kalman Filter Implementation for the \BeamTest}
\label{bt:smallangle}
In order to analyze the data from the October 1997 \beamtest, the 
Kalman filtering equations above were implemented in a C++ program 
called \tjrecon.  This program combined track finding and track 
fitting into one package which could analyze both \beamtest\ and Monte 
Carlo data.  The track finding algorithms will be discussed in 
\sect{bt:algo}.  

To maintain generality, the implementation was designed to be as 
flexible as possible.  The $x$ and $y$ projections were fit 
separately, and their results combined after the fitting process.  The 
state vector was chosen to be:
\begin{equation}
\myvec{x}_k = \left( \begin{array}{c}  \mbox{horizontal position} \\
				     \mbox{track slope}  \\
				     \mbox{current energy} \\ \end{array} \right)
\end{equation}
The propagation matrix was then simply (aside from unit conversions)
\begin{equation}
{\bf F} = \left( \begin{array}{ccc} 1  & 1 & 0 \\
				   0 & 1 & 0 \\
				   0 & 0 & 1 \\ \end{array} \right)
\end{equation}

Recall from \eq{bt:mseqn} that the multiple scattering {\em angle} is 
approximately Gaussian distributed.  Since we are using the slope of 
the track, the (1,2) component of ${\bf F}$ should really be 
$\tan(\mbox{track slope)}$.  Of course, this destroys the linearity of 
the propagation equation, and with it the Kalman filter.  So, we 
assume that the track slope will be small so that $\tan{\theta} 
\approx \theta$.  Consequences of relaxing this requirement will be 
discussed in \sect{bt:ekf}.

We assumed a measurement vector $\myvec{m}_k$ with an energy component, 
to allow the possibility of extending the Kalman filter to the 
calorimeter.
\begin{equation}
\myvec{m}_k = \left( \begin{array}{c} \mbox{strip number} \\
				\mbox{current energy} \\ \end{array} \right)
\end{equation}
Since there was no energy measurement information in the tracker, 
our {\bf H} matrix was, for all planes,
 \begin{equation}
{\bf H}_k = \left( \begin{array}{ccc} 1 & 0 & 0 \\
				      0 & 0 & 0 \\ \end{array} \right)
\end{equation}

	\section{Track Finding Algorithm}	
	\label{bt:algo}
Track finding is significantly more difficult than track fitting in 
the following sense: it is impossible to rigorously prove that one 
track finding algorithm is better than another.  One must implement 
both algorithms and run them on the data in question.
		\subsection{The Exhaustive Search}
An exhaustive combinatorial search is the exception; it is guaranteed 
to find the best track.  It is also simple to implement: for every 
possible combination of hits for each of the two tracks, apply the 
Kalman filter and look at the \chisq.  The best tracks are the 
combination that give you the lowest total \chisq.  However, the 
method soon becomes computationally infeasible.  For a single track, 
the number of combinations $n$ to try is roughly
\begin{displaymath}
n = p^h  
\end{displaymath}
where $p$ is the number of planes and $h$ is the average number of 
hits.  For the \beamtest, $p$ was 6, and $h$ was typically between 
zero and 6.  This would imply several hundred possibilities for each 
track in each event; an achievement which would be possible, though 
probably not at real-time speed.  However, a single tower of \glast\ 
will have 16 planes and could have as many as 10 hits per plane.  An 
exhaustive search would require $10^{12}$ trials, which is clearly not 
feasible for the expected data rates.


 		\subsection{\BeamTest\ Algorithm}
		\label{bt:explainalgo}
In an attempt to avoid the computational complexity of an exhaustive 
search, we will try to make a good initial guess for the track, then 
vary that guess in order to optimize the goodness of fit.  The Kalman 
filter naturally suggests the outline of a track finding algorithm, 
based on the prediction of the track location in the next plane.  
However, there are myriad details of such an algorithm left 
unconstrained.  Unfortunately, there is no continuous metric to guide 
our choices---the only metric by which to compare different algorithms 
is to implement them, and see which one does the best.  The specific 
choices we made in designing an algorithm for the \beamtest\ were 
verified on Monte Carlo data to succeed rather well in reconstructing 
events.  However, we can make no claim that this algorithm is optimal, 
nor that it will be the basis for an algorithm for use in \glast.

To make an initial guess for the track location, we find the first 
plane with a hit.  Presumably, the \gammaray\ converted in the lead 
above this plane, or in the plane itself.  We assign the leftmost hit 
in the first plane to the track, and the leftmost hit in the 
second plane to be the second hit in the track.  These will serve as 
initial guesses, but we will explore other possibilities later.  Given 
these two hits, we may run the Kalman filter, predicting a state 
vector, and therefore a position, for the third plane.  We assume that 
the closest hit to the predicted position is part of the track.  
Continuing this procedure to the bottom of the tracker yields our 
initial guess.

The metric for determining a good track is the total \chisq\ 
calculated in \eq{bt:chisqeqn} in a slightly modified form.  Because 
of the existence of noise, double hits, and the possibility of tracks 
leaving the instrument, we construct penalties which we add to the 
total \chisq\ when a track makes an undesirable choice of hits.  This 
is perfectly acceptable in the language of likelihoods---we have 
additional information to suggest that such a track is unlikely, so we 
subtract from the $\ln \like$ of that track.  However, while 
the relative sizes of some penalties may be rigorously derived, their 
actual values must be determined empirically.

There are two special cases that must be considered when making our 
initial track guess.  The first is the possibility that the track 
projects to a horizontal location outside the tracker.  In that case, 
the code creates a ``virtual'' hit with a position outside the 
tracker.  The track must pay a substantial penalty to its \chisq\ in 
order to use this hit.

The second special case involves the so-called ``noisy strips.''  It 
was found (\sect{bt:noise}) that a number of strips were defective.  
They were either completely unresponsive, never recording a hit, or 
they were very noisy, recording a hit even when no electron passed 
through.  All of these strips were marked as ``noisy,'' and masked 
off.  Nevertheless, the possibility remained that an electron actually 
did pass through the noisy strip.  In order to allow that possibility, 
the algorithm checks for noisy strips near the projected track 
position.  It then creates virtual hits at each noisy strip within a 
small (10 strip) radius around the projected position.  The track must 
add a penalty to its \chisq\ in order to use a noisy hit.

It is assumed that each event contains an \el\pos\ pair.  Once the 
first track guess has been made, a second track is proposed, starting 
with the same hit in the first layer and the rightmost hit in the 
second layer.  This track guess is made in the same way as the first, 
yielding a pair of initial track guesses.

\paragraph{Track Optimization.}
Once an initial track is established for each electron, several steps 
are taken to optimize the choice of hits.  First, the two tracks are 
untangled so they do not cross, regardless of whether the respective 
$\chi^2$ values decrease by doing so.  Subsequent track optimization 
proceeds more smoothly if the tracks are initially untangled.  This 
subsequent optimization will allow tracks to cross if that is justified
by the data.

On each plane, there may be hits which are not a part of either of the 
tracks.  The algorithm next checks each of these to see if replacing 
any hit in either track with an unclaimed hit will improve the fit.  
The free hit is substituted, and following projected positions are 
re-evaluated, substituting nearby hits if appropriate.

Then the hits of the two tracks are swapped, plane by plane, to see if
swapping hits will increase the total $\chi^2$ of both tracks.

The last part of the track finding algorithm consists of checking the 
possibility that a track may have left the tracker.  Often, a track 
may take a sharp scatter to leave the tracker, with the result that 
the predicted position is inside the tracker.  The large penalty on 
leaving the tracker prevents tracks from choosing this virtual hit 
unjustifiably, but the even larger penalty on sharing hits prevents 
unrealistic swerves in the track to make a single hit.

Finally, the issue of terminating tracks must be addressed.  For 
example, if one track in the $x$-projection leaves the tracker, then 
one of the tracks in the $y$-projection must be terminated.  To 
determine which track has left, it is assumed that the track with the 
higher $\chi^2$ probably has swerved precipitously to incorporate hits 
that do not belong to it.  These hits reside on planes below that at 
which the track left the tracker.  Therefore, the track is terminated 
at that plane, and is then taken to have no further hits.

A summary of the track finding algorithm may be found in \app{app:find}.

	\section{Measurement Error and Multiple Scatter Estimates}
\label{bt:meandms}
\begin{figure}[t]
\centering
\includegraphics[height=2.5 in]{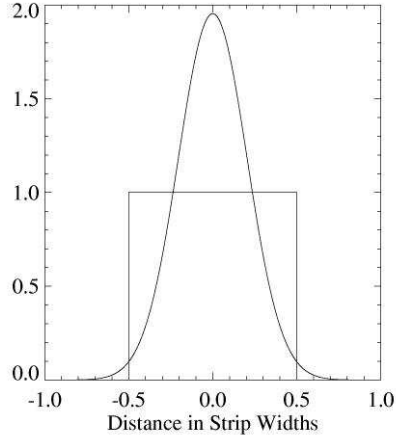}
\caption[Measurement error distributions]{\label{bt:sqrgau}Measurement 
error distributions.  The Gaussian approximation and square 
approximation to the measurement error distribution.  Both are 
normalized and have variance 1/12.}
\end{figure}

Of critical importance are the values used for the measurement error 
and multiple scattering.  For the measurement error, we assume that 
any electron coming within one-half the strip pitch on either side of 
the strip will cause it to fire.  This would result in a square 
distributions.  The Gaussian with the same mean and variance has a 
standard deviation of the box width over $\sqrt{12}$.  (This is easily 
shown: the variance of a distribution is its second moment, so we 
find $\int_{-1/2}^{+ 1/2} x^2 P(x) dx$.  For a square normalized 
distribution, $P(x) = 1$, and the integral equals 1/12; see 
\fig{bt:sqrgau}.)  There was no energy measurement, so its variance is 
arbitrary.
\begin{equation}
{\bf V} = \left( \begin{array}{ccc} 1/12  & 0 \\
				    0  & 1 \\ \end{array} \right)
\end{equation}
in units of strips \cite{ppdb}.

According to \eq{bt:mseqn}, multiple scattering depends on energy.  We 
must therefore establish the energy of the electron before we begin to 
fit its track.  A sampling calorimeter could, in principle, provide 
this information.  Otherwise, some approximation will have to be made.  
For the \beamtest, we assumed that the energy in each track was half 
the calorimeter measured energy.  The thickness of the SSDs was 500 
\micron, or about twice the strip pitch.  From the multiple scattering 
equations, it follows that the multiple scattering covariance matrix 
{\bf Q} should be approximately
\begin{equation}
\label{bt:Qmat}
{\bf Q} = \left( \begin{array}{ccc} 
\frac{1}{12} \theta_o^2  & \frac{1}{4}\theta_o^2 & 0 \\
\frac{1}{4}\theta_o^2 & \theta_o^2 & 0 \\
0 & 0  & 1 \\ \end{array} \right)
\end{equation}
If the state vector were to carry a meaningful energy estimate, the 
(3,3) component of {\bf Q} should be the variance of the energy loss 
by the electron as it passes through the detector.  However, the 
distribution of energy losses suffered by electrons passing through 
material is not at all well described by a Gaussian, and special care 
must be taken to account for these distributions \cite{stampfer}.  
Potential improvements to account for these effects are discussed in 
\sect{bt:better}.

	\section{Noise}
	\label{bt:noise}

Even before the \beamtest\ actually began, it became clear that there 
would be some dead strips that never fire, and some noisy strips that 
fire far too often.  At the time of the \beamtest, the \glast\ 
collaboration did not have the capability to wire bond the individual 
silicon strips to the front-end readout chips.  Therefore, a private 
company was contracted to perform the wire bonding.  When the 
detectors were returned from the wire bonder, several of the channels 
drew significantly more current than they had before the wire 
bonding.  Subsequent visual inspection revealed obvious mechanical 
damage to several of the strips in the form of scratches; in addition,
some of the bonds were of inferior quality.  Unfortunately, the detectors
were returned from the contractor only one week before the \beamtest\ 
began, and no repair was feasible.  We were thus left to incorporate 
the dead and noisy strips into our software analysis.

For our purposes, both dead and noisy strips were 
treated in the same way, and will be referred to as ``noisy.''  Such 
noisy strips will clearly confuse the track finding routines, and they 
must be dealt with.

\begin{figure}[t]
\centering
\includegraphics[width = 2.5 in]{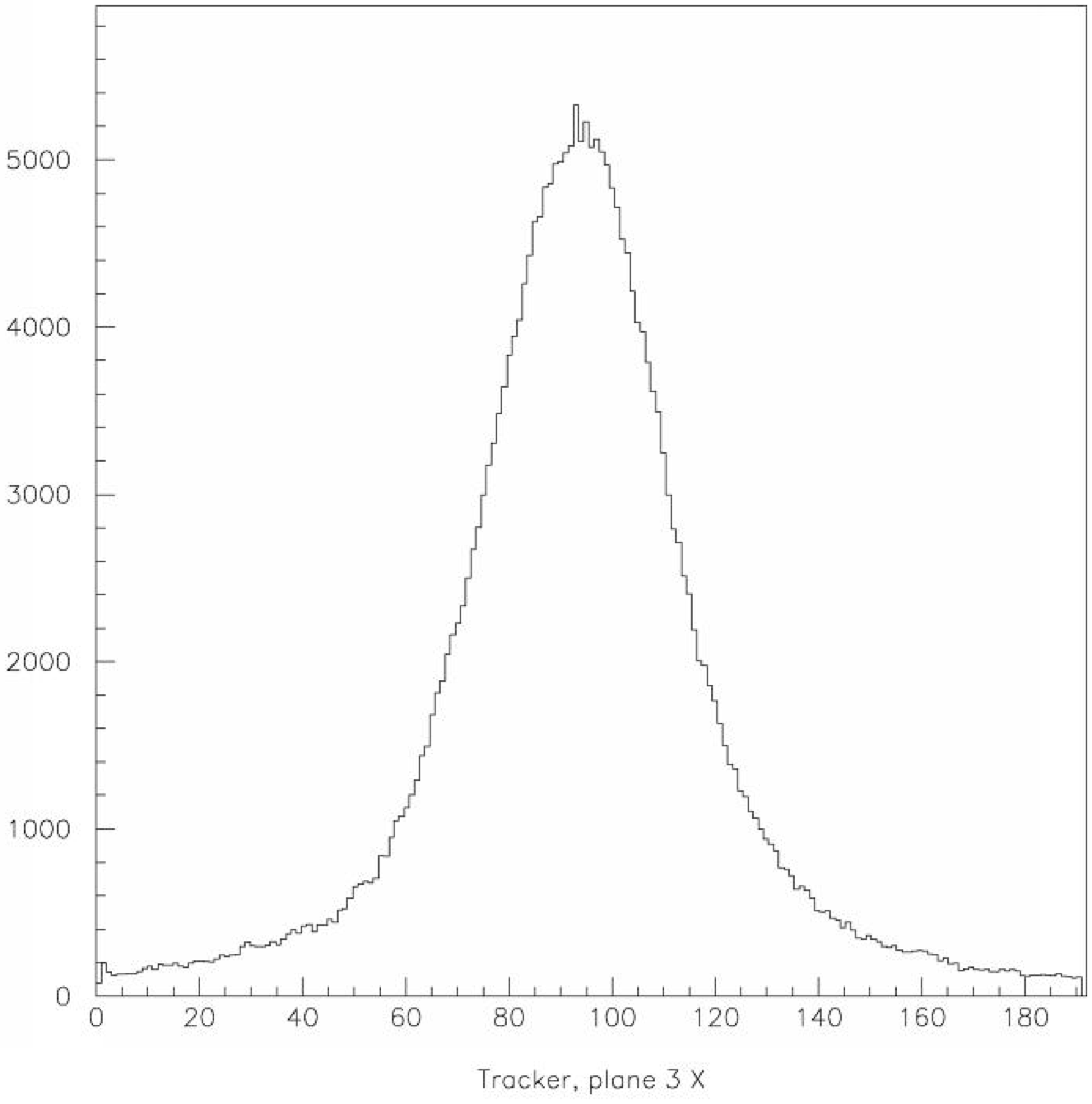}
\includegraphics[width = 2.5 in]{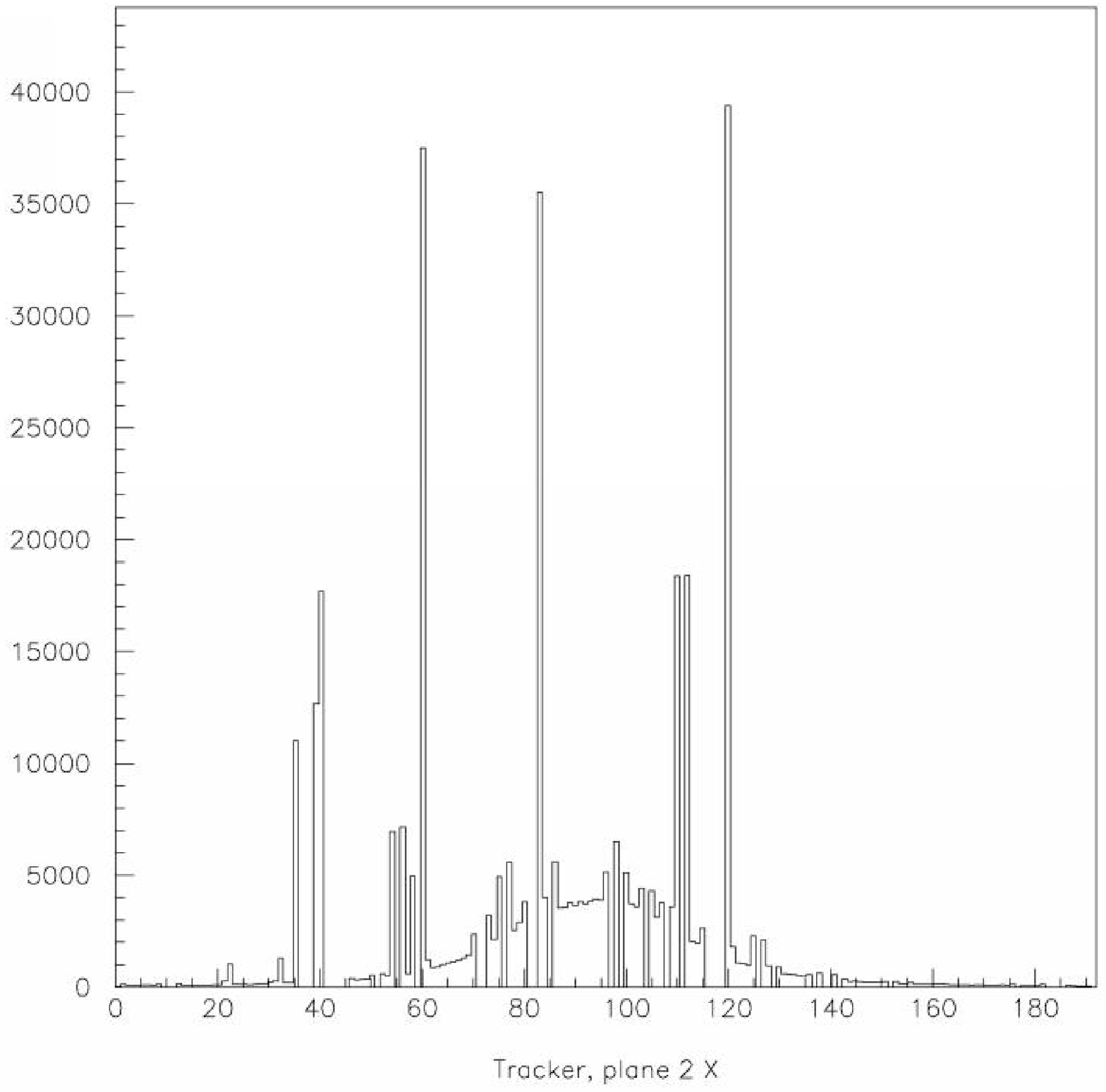}
\caption[Strip occupancies]{\label{bt:occupancy}
Occupancies in two planes during a typical run (Run 376).  The plane 
on the left is almost completely noise free; all recorded hits were 
due to electrons passing through the plane.  The plane on the right 
was mishandled by an outside wire-bonding contractor; 
it displayed excessive noise.  
The Gaussian shape of the real electron hits can be seen between the 
noisy channels.}
\end{figure}

 Initially it was thought by many that there would not be any noisy 
 strips at all in silicon strip detectors.  In fact, carefully 
 prepared silicon strip detectors do exhibit very few noisy strips.  
 However, the prototypical nature of the \beamtest\ instrument, and 
 especially the necessity of outsourcing some fabrication, led to 
 several types of noise (\fig{bt:occupancy}).

 First of all, about 2/3 of the planes exhibited one or a few strips 
 with high occupancy---typically higher than about 20\% of triggers.  
 In addition, there were also a fair number of strips which were dead.  
 The reconstruction code merely identified these strips and flagged 
 them for the reconstruction routines.

 In addition, there were two special planes to worry about.  The first 
 was a $y$-layer with a malfunctioning readout chip.  The silicon 
 strips connected to this chip were independently tested and found to 
 work correctly; however, the readout chip reported a set of 32 strips 
 to be either (almost) all on, or all off.  The strips numbers were 
 approximately 32 through 64.

The other special plane was the so-called noisy plane.  This was an 
$x$-plane that exhibited a large number of noisy strips.  The reasons 
for the large numbers of noisy strips has been traced to some quality 
assurance problems; they are detailed elsewhere~\cite{glastnim}.

 The first attempt at identifying noisy strips was done by hand.  
 Strips were noted as noisy when they appeared to have a higher 
 occupancy than expected based on other nearby strips.  This method 
 proved tedious in light of the number of strips involved, and the 
 fact the the noisy strips are not always constant between runs.  On a 
 gross level, the location of the noisy strips changed several times 
 during the \beamtest\ as a result of the physical rearrangement of 
 the planes within the tracker.  Clearly, noisy strips moved with the 
 planes.  In addition, several studies of the effect of different 
 discriminator thresholds changed the noise levels.

For these reasons, we decided to automate the noise finding procedure. 
The algorithm works as follows: first, all strips with an occupancy 
above 15\% were marked as ``dead.''  For our purposes, we flagged dead 
strips and excessively noisy strips in the same way.  Then we looked 
for the plane with the bad chip.  This was done by checking each 
$y$-plane in strips 32-34 and strip 60.  If they were the same to 
within a few percent, then that plane was assumed to hold the bad 
chip.  The entire range from 32-64 was marked as dead.

  Next we looked for noisy strips in the wings of the distribution.  
  Starting at strip 95 (approximately the center of the plane) we 
  moved out to the left until the number of counts had dropped by a 
  factor of 5.5 from its peak value, ignoring any dead strips.  We 
  then continued to the left and marked as dead any strip with more 
  than one fifth the number counts as the maximum channel.  The entire 
  process was independently repeated for the right side.

Finally, we fit a Gaussian to the core of the distribution.  The tails 
were not fit, since they were clearly non-Gaussian.  Any strip with  
more than 130\% of the hits predicted by the fit is marked as noisy.  
The Gaussian fit roughly represents the number of legitimate hits, and 
any additional hits are noise.  Thus, we throw away strips where more 
than 30\% of their hits are noise hits.

The resulting list of noisy strips is passed to the reconstruction
routines, which ignore all hits from those strips. 

	\section{Energy}
Energy resolution is an important part of any \gammaray\ telescope.  
Good measurements of the spectra of astrophysical sources and their 
cutoffs offer insight into their energy-generation mechanisms.  Good 
energy resolution is also important for good \gammaray\ 
reconstruction.  To conserve momentum, the \gammaray\ direction must 
be a energy-weighted average of the \el\ and \pos\ directions.  
Furthermore, the Kalman filter estimate of the electron track depends 
on the expected multiple scattering, inversely proportional to the 
energy of the electron.

		\subsection{Energy Splitting}
\label{bt:energysplit}
For these reasons, it is very useful to determine the energies of each 
electron.  A useful first approximation is to simply assume each track 
has half the total energy measured in the calorimeter.  This is a good 
enough approximation to allow reconstruction of the tracks.  One might 
hope that an iterative approach would allow better energy resolution 
and better track fitting.  We can imagine varying the fraction of 
energy in each track to maximize the likelihood.  The energy split which 
produces the maximum total likelihood would be the most likely energy split.

This method was implemented, but it was unstable.  In every case, the 
maximum likelihood energy splitting put all the energy in the 
straighter track, with no energy in the other track.  This instability 
is due to the non-Gaussian tails on the multiple scattering 
distribution, as well as the small angle approximation we made in the 
Kalman filter.  Tracks make large scatters much more frequently than 
would be expected from a Gaussian distribution.  In order to make 
these large scatters reasonably probable, the Kalman filter radically 
lowers the track energy.  This means that the track with a large 
scatter will have a grossly underestimated energy.  Some possible 
improvements on this method will be discussed in \sect{bt:better}.

		\subsection{Energy Dispersion}
Of course, the energy measured by the calorimeter is not the true 
energy.  In the case of the \beamtest, portions of the calorimeter 
volume were not even instrumented.  Therefore, we must expect some 
energy dispersion.  We used the simplest method of estimating total 
energy from the calorimeter.  Eric Grove of NRL \cite{nrlcal} provided 
us with a simple function (\eq{bt:calgain}) to translate raw 
analog-to-digital readout counts and gain settings to deposited 
energy.  We did not try to correct for shower leakage by fitting the 
shower profile, or to try to separate the energy deposited by each 
electron.

The hodoscopic calorimeter (\sect{bt:beam}), which measured the energy of the 
``leftover'' incident beam electron, allowed a rough calibration of 
the \gammaray\ energy.   However, this measurement was 
only good to an accuracy of about 250~MeV. The hodoscopic calorimeter 
did allow the detection of some events with multiple beam electrons in 
a single pulse.  These events could be identified when more than one 
leaded-glass block of the hodoscopic calorimeter recorded energy 
deposition.

\begin{figure}[t]
\centering
\includegraphics[width=4.5 in]{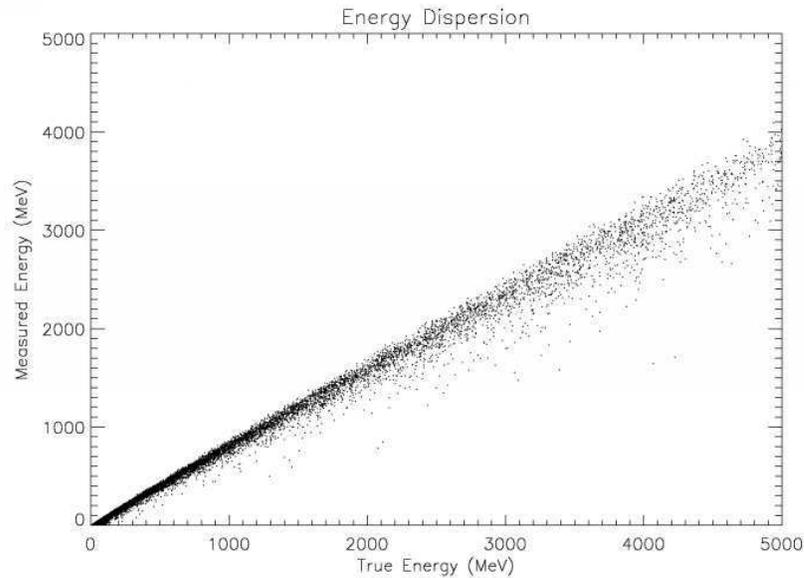}
\caption[Monte Carlo energy dispersion]{\label{bt:edplot}
Measured energy vs. true energy in the Monte Carlo calorimeter.
Measured energies are about 20\% lower than true energies, since
the calorimeter does not collect all of the \gammaray\ energy.
In addition, the width of the measured energy distribution at
any constant true energy is about another 20\%.}
\end{figure}

From 10~MeV to several GeV, we compared the energy deposited in the 
Monte Carlo simulation of the calorimeter with the true energies of 
the particles injected into the simulation (\fig{bt:edplot}).  Over 
most of the range, the calorimeter detected about 80\% of the incident 
energy.  The distribution was far from symmetric; as expected, many 
higher-energy photons were recorded as lower-energy photons, because 
of energy leakage out of the back and sides of the calorimeter as well 
as energy deposition in non-instrumented blocks.

	\section{Potential Improvements}
	\label{bt:better}
There are a number of potential improvements to the silicon tracker 
analysis.  Most of these would make very little difference for the 
results of the \beamtest---largely because they would affect both the 
actual \beamtest\ data and the Monte Carlo simulations in the same 
way.  However, they may be quite important for track finding and 
fitting of \glast\ data.

	\subsection{Track Fitting}

\paragraph{Improve estimate of measurement error distribution.}
The probability that a given strip will fire is not really a boxcar, 
of width one strip, centered on the strip.  In fact, it is a 
complicated function of the collection area geometry and bias voltage 
on the strip.  It is plausible, and empirically verified, that often 
two adjacent strips will fire.  This can occur when the electron 
passes between the strips, depositing enough energy into each of them 
to raise their potential over the threshold, or when the electron 
passes through the silicon at an angle, depositing energy in more than 
one strip.  Currently, the hit which is more easily incorporated into 
the track is selected.  Clearly, there is information in the fact that 
the other strip has also fired.  A careful analysis of the probability 
of each strip firing as a function of the electron position would 
allow an accurate characterization of the probability of the true 
electron position given that both strips fired.  The width of this 
distribution could then be used in the measurement error covariance 
matrix {\bf V} instead of the current $1/12$ times the strip 
pitch.

\paragraph{Transform multiple scattering distribution to Gaussian.}
The Kalman filtering formalism is fast and accurate because it assumes 
distributions are Gaussian.  However, nowhere does it require that the 
state vector be composed of physically meaningful values.  It may be 
possible to find a function that would transform the multiple 
scattering distribution to something more nearly Gaussian.  In that case, the 
state vector would contain the transformed variable instead of track 
slope.  This would result in better fits, and might allow energy 
estimation from the track shape.  The Kalman filter would find the 
most likely state vectors, yielding physically meaningful information 
via the inverse transformation.

\paragraph{Transform measurement error distribution to Gaussian.}
It is unlikely that a more accurate measurement error distribution 
like that described above would be Gaussian.  However, if a simple 
transformation would make it nearly Gaussian, we would achieve the 
same improvement as described for multiple scattering.

\paragraph{Account for total material traversed.}
Since both multiple scattering and energy loss depend on the amount of 
material that the electron traverses, electrons that cross the planes 
at a steep angle will be scattered more and will lose more energy.  
Unfortunately, this is an inherently non-linear phenomenon; large 
scattering leads to greater likelihood of crossing a plane at a steep 
angle, which leads to greater likelihood of large scattering.

\paragraph{Require agreement between calorimeter and tracker 
reconstructed directions.} By analyzing the shape of the 
electromagnetic shower in the calorimeter, it is possible to estimate 
an incident direction of the particle.  This information should be 
combined with the tracker reconstruction to find the best estimate of 
the incident \gammaray\ using all available information.

\subsection{Track Finding}
\paragraph{Exhaustively try the first two points in track.}
While a complete exhaustive search for the best track will be 
infeasible for \glast, trying all possibilities of the first two or 
three planes in the track would be possible.  Since the first few 
points are the most important in determining the rest of the track, 
this would increase the probability that the best track is actually 
found.

\paragraph{Require vertex in material.}
Pair production can only occur in the presence of nuclei in order to 
satisfy momentum and energy conservation.  Therefore we can expect the 
two electron tracks to project back to a common vertex which is in 
high-Z material.  Since the conversion may well happen in lead, we may 
not have a measurement of the electrons at the vertex.  However, we 
know the precise position of the lead foil, so the requirement that 
the vertex lie in high-Z material will improve the estimates of the 
tracks.

\paragraph{Account for tracks which start with a dead strip.}
It may happen that a \gammaray\ converts close to a dead strip.  In 
that case, the first plane will not record a measurement of the track.  
The algorithm designed for the \beamtest\ would assume that the 
conversion actually happened in the next plane.  However, it should be 
very rare that the conversion happens near a dead strip in both $x$ 
and $y$.  Furthermore, a fit track can be extended upwards to see if 
it could have converted in a dead strip above.  Therefore, it may be 
possible to identify situations in which the first hit of the track is 
missing.

\paragraph{Assign penalties to individual strips according to occupancy.}
Currently, dead strips and noisy strips are masked identically, and 
all are assigned the same penalty for track finding.  However, some of 
the noisy strips contain significant information.  For a strip with a 
relatively low occupancy, many or most of the hits recorded may be due 
to actual electron tracks.  A better penalty scheme would find the 
occupancy of each strip, and assign larger penalties to strips with 
higher occupancies.  The magnitude of the penalty would be 
proportional to the likelihood that a hit registered in that strip was 
actually noise.

\paragraph{Use empty triggers to find noisy strips.}
The occupancy of each strip could be measured by examining which 
strips are hit when there is nothing in the silicon tracker.  For the 
\beamtest, this could be done by looking at triggers in which the beam 
electron did not bremsstrahlung in the Cu foil.  For \glast, this can 
be done by reading out the tracker when there has not been a trigger.  
Careful establishment of the occupancies, remeasured frequently, would 
allow accurate penalties to be set for each strip.

	\subsection{Energy Estimations}
\paragraph{Take into account energy loss per plane.}
It would be very easy to include energy loss in each plane.  The 
energy loss in a material is well characterized.  The (3,3) component 
of the propagation matrix {\bf F} would then become the percentage of 
the electron's energy that it retains.  The multiple scattering 
covariance matrices of lower planes would then be calculated based on 
this reduced energy.  Unfortunately, the distribution of energy
losses is asymmetric.  A proper treatment of energy loss~\cite{stampfer}
requires this distribution to be taken into account in the track fitting.

\paragraph{Try to get energy estimates from tracks.}
If the Gaussian transformations described above could be found, then 
the tails of the multiple scattering distribution would no longer 
undermine our efforts to estimate the energy of the track from the 
track itself.  An iterative approach could then estimate the most 
likely energy, refit the tracks, and re-estimate the energy until it 
converged.  Then the energy splitting information could be used to 
make better estimates of the incident \gammaray\ direction.

\paragraph{Fit shower profiles in the calorimeter.}
Fitting the shower development profile in the calorimeter may lead to 
much better energy estimates.  Such fits allow leakage to be 
accurately estimated.  The \glast\ collaboration calorimeter team is 
implementing shower profile fits \cite{nrlcal}.
 
\paragraph{Measure individual electron energies in the calorimeter.}
A finely segmented calorimeter would enable the separation of the 
energy deposition from the individual electrons.  An accurate measure 
of the individual electron energies would increase the silicon tracker 
resolution by allowing better energy weighting of the electron track 
directions to find the incident \gammaray\ direction.  However, finer 
sampling of the calorimeter requires more electronics and thus more 
power, as well as more gaps between CsI blocks.  Sampling granularity 
small enough to resolve individual electron energies may not be 
feasible for \glast.

\paragraph{Use better energy splitting probabilities.}
Heitler \cite{heitler, motz} has calculated the probability 
distributions of the energy split between the electrons as a function 
of the total incident energy.  While it may not be possible to get 
good energy splitting estimates directly from the tracks, it is often 
possible to determine which track is more energetic.  The most likely 
energy split according to Heitler could be assumed, with the appropriate 
track receiving more energy.

\begin{figure}[t]
\centering
\includegraphics{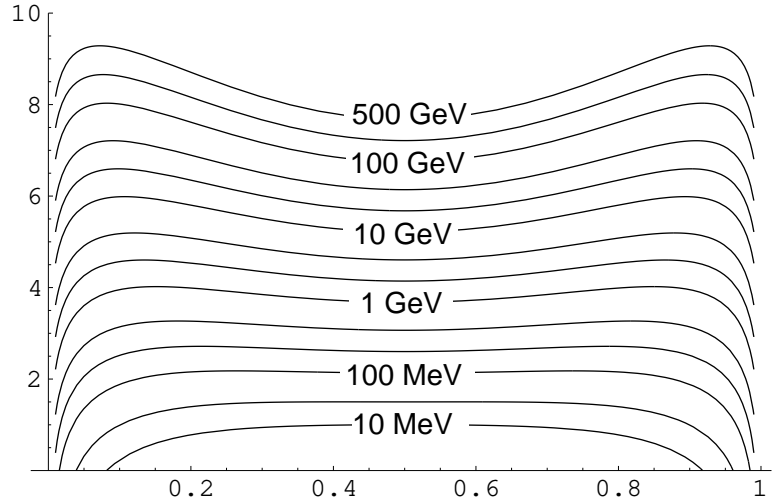}
\caption[Probability distributions of \el\pos\ energy 
split as a function of total incident energy]{\label{bt:Esplit}
Probability distributions of \el\pos\ energy 
split as a function of total incident energy.  The 
vertical scale is arbitrary; each
curve is normalized.  The vertical offset is for clarity.  Labels indicate
total incident \gammaray\ energy.  At the highest energies, asymmetric energy
splittings are relatively more likely; for most energies of interest to \glast,
the probability of all splittings except the most extreme ones is approximately
equal.}
\end{figure}

The relative probability of various energy splittings as a function of 
total \gammaray\ energy can be found from the total pair-production 
cross-section.  In the limit of an unscreened point nucleus, 
extreme-relativistic energies, and negligible nuclear recoil, in the 
first Born approximation, the cross-section as a function of energy 
splitting is \cite{motz}

\begin{equation}
\label{bt:dsigde}
\frac{\mbox{d} \sigma}{\mbox{d} f} = 4 \alpha Z^2 r_0^2
\left( \frac{4}{3} f^2 - \frac{4}{3} f + 1 \right)
\ln \left[2 f E (1 - f) \right]
\end{equation}
where $f$ is the fraction of energy in one electron, $E$ is the total 
\gammaray\ energy in MeV, $Z$ is the atomic number of the target nucleus (Si 
or Pb), and $r_0$ is the classical electron radius, $2.82 \by 
10^{-13}$ cm.  This function is plotted as relative probability versus 
energy split for a variety of total \gammaray\ energies in 
\fig{bt:Esplit}.  For the lowest energies, the probabilities appear to 
be negative for extreme values of the energy split; this is an 
indication that the assumptions noted above have broken down.

\section{\label{bt:psf}Calculating the Point-Spread Function}
The point-spread function of the \beamtest\ instrument could be 
estimated by the distribution of reconstructed \gammaray\ incident 
angles.  \glast\ may be calibrated in the same way either in a beam, 
or in space with a sufficiently bright \gammaray\ point source.  However, 
the Kalman filter offers us an opportunity to calculate the theoretical 
point-spread function under our Gaussian assumptions.  The variance in 
the estimate of the incident direction of each electron is given by 
the (2,2) component of the covariance matrix {\bf C}.  To calculate 
the point-spread function, we should look at the smoothed estimate of 
{\bf C} in the top layer.  We simply read off the variance of the 
estimate of the electron track direction.  To find the 
error in our estimate of the incident \gammaray\ direction, we need to 
combine our estimates of the two electron tracks.  If $E_{1}$ is the 
energy of one electron and $E_{2}$ is the energy of the other, our 
estimate of the incident \gammaray\ direction is
\begin{equation}
	\theta_{\gamma} = \frac{E_{1} \theta_{1} + E_{2} \theta_{2}}
	{E_{1} + E_{2}}
\end{equation}
Then the variance in $\theta_{\gamma}$ must be
\begin{equation}
	\sigma_{\theta_{\gamma}}^{2} = \frac{E_{1}^{2} \sigma_{\theta_{1}}^{2}
	+ E_{2}^{2} \sigma_{\theta_{2}}^{2}}{(E_{1} + E_{2})^{2}}
\end{equation}
The variance in our estimate of the incident \gammaray\ direction 
depends on the variances of the two electron track directions, and on 
the energy split between the electrons.  We can calculate the 
variances of the two electron tracks from the Kalman filtering 
equations.  Since our estimates of {\bf C} do not depend on the data, 
we can compute the variances ahead of time.  The variances will depend 
on the electron energy (through ${\bf Q}_k$, the multiple scattering
covariance matrix given by \eq{bt:Qmat}), the 
amount of radiator on each plane (also through ${\bf Q}_{k}$), the 
ratio of the strip pitch to the gap between planes (through the 
measurement error {\bf V} and the propagation matrix {\bf F}), and the 
number of planes with measurements.  Thus the variance depends on four 
variables, making it difficult to display in a single plot.  Given 
probability distributions for some of these variables, we may 
marginalize some or all of them in a Bayesian way (\sect{stats:marginalize}), 
leaving us with a 
representative ``average'' point-spread width.  \tbl{br:psftable} 
shows the calculated width of the point-spread function for several 
interesting configurations of the full \glast\ instrument.

It should be noted that all of these calculated point-spread widths 
assume that every quantity is normally distributed.  In particular, 
they ignore the large tails on the multiple scattering distribution, 
and they combine the estimates and variances of the two electron 
directions as if they were Gaussian distributed.  They ignore the 
effects of electrons leaving the tracker, which will shorten the 
average track length.  As such, they are useful for estimating 
instrument performance, but would not be suitable for likelihood 
analysis of \glast\ data.

\section{\label{bt:ekf}Extended Kalman Filters}
In \sect{bt:smallangle} we linearized the propagation matrix {\bf F}.  
That is, we chose to use the track slope in our state vector, and 
assumed that the additional slope from multiple scattering would be 
normally distributed, and would add to the previous slope value.  Of 
course, the multiple scattering {\em angle} is (roughly) normally 
distributed, and the additional scattering angle should be added to 
the previous track angle.  For small angles ($\lt\,30\deg$), this is a 
reasonable approximation.  Since all the \beamtest\ \gammarays\ were 
incident from the same direction (0\deg), this approximation was 
valid.  However, for \glast, the \gammarays\ will be incident from 
every direction, and a small-angle approximation will not be valid.  
There has been extensive work done \cite{julier1,julier2} to extend 
Kalman filters to non-linear propagators.  This would allow the 
electron track angle to be kept in the state vector.  A detailed 
discussion of such methods is beyond the scope of this work.

%% file: Beamresults.tex
\chapter{Instrument Response}
\label{responsechap}
Once the various parts of the \beamtest\ instrument had been built, 
and the software to analyze the data had been written, the instrument 
was placed in the beam line in End Station A. The data taken at SLAC 
during October 1997 yielded important insights into issues of
backsplash self-veto in the ACD 
as well as the establishment of the feasibility of pointing resolution 
using the calorimeter only \cite{glastnim}.  
In this chapter, we will focus on the silicon tracker: the specific 
adaptations necessary to analyze tracker data and the instrument 
parameters measured.

	\section{Alignment}
To avoid systematic errors in reconstructed particle direction, 
corrections were made to account for possible misalignment of the 
planes.  Machining errors on the scale of the strip pitch (236 
\micron) would make it appear that a track had scattered more (or 
less) than it actually had.  Smaller errors would influence track 
reconstruction in a statistically similar way.  In part to ameliorate 
this effect, data was taken with the Cu radiating foil removed, so 
that beam electrons were directly incident on the detector.  For 
25~GeV beam electrons, multiple scattering is negligible.  These 
tracks could thus be used to ``align'' the planes by finding the 
relative offset of each plane.  The offsets were used to correct the 
positions of the hits in software.

The straight electron tracks were fit with a line in each projection.
In each plane, the median value of the fit residual was taken as the
plane offset.  All strip positions in the plane were corrected, and
the process iterated until the median residuals converged to zero.  
Because the strip measurements were quantized, the linear fits were
subject to severe aliasing on the scale of the strip pitch.  To 
alleviate the effects of this aliasing, a small amount of random noise
was added to each strip position independently for each fit.  

The tracker was found to be nearly aligned as constructed.  The largest
offset was $\mysim250$~\micron.  The alignment procedure measured the strip
positions to an accuracy of $\mysim50$~\micron.

\section{Cuts}
\label{br:cuts}
The SLAC main electron beam runs nominally at 120 Hz.  The state of 
the \beamtest\ instrument, including all strips hit, all energy 
deposited in the calorimeter, and all ACD tiles hit were read out for 
each beam spill.  With an average of one electron per pulse, 
approximately 30\% of spills had no electrons in them at all.  Spills 
with one or more electrons shed bremsstrahlung photons with a 
probability dependent on the thickness of the Cu radiator 
foil (3.5\% \radlen, 5\% \radlen, or 10\% \radlen).  
Therefore only a fraction of the 
$2.1 \by 10^{8}$ triggers recorded on tape were useful.  A filtering 
program (not to be confused with Kalman filtering) was developed to 
extract the useful triggers.  The criteria for accepting a trigger as 
a useful event were the detection of hits in three successive tracker 
planes, or of more than 6~MeV for low gain or 160~MeV for high gain in 
the calorimeter.  These criteria were adopted so as to be sure to 
accept any particle that passed through the tracker (whether or not it 
hit the calorimeter) as well as any event that did not interact with 
the tracker, such as a \gammaray\ that did not convert until it 
entered the calorimeter.  The three-in-a-row requirement for the 
tracker was designed to ensure that random noise hits in the tracker 
would not pass the acceptance criteria.  When the Cu foils were in 
place to produce \gammarays\ from the main electron beam, 
approximately 20\% of the triggers were accepted.  When the foil was 
removed, and the electron beam was directly incident on the 
instrument, nearly all the triggers were retained.

These ``useful events'' were then analyzed by the reconstruction 
software.  Many of them had no interactions in the calorimeter and 
were thus thrown away.  In addition, any event with one or fewer hits 
in either projection was thrown away.  The remaining events were 
reconstructed, but not all of these were satisfactory to be included 
in the analysis.  The first requirement was that the hodoscopic 
calorimeter reported only one electron in the spill.  The presence of 
multiple beam electrons in the spill greatly increases the chance of 
multiple bremsstrahlung \gammarays\ entering the tracker at the same 
time.  Second, each anti-coincidence tile was required to have less 
than 1/4 MIP (minimum ionizing particle; see \cite{ppdb} for details) 
of energy deposited.  While the ACD is designed to detect charged 
particles, it represents  1\% \radlen\ of material, which can cause 
\gammarays\ to pair convert.  If a \gammaray\ pair converts in an ACD 
tile, each electron will deposit 1 MIP, times the fraction of the tile 
through which the electron passes.  For example, if the \gammaray\ 
converted halfway through the tile, then the two electrons would 
deposit $2 \by (1/2 \by 1 \mbox{MIP}) = $ 1 MIP. The 1/4 MIP threshold 
will therefore reject all events where a \gammaray\ converts in the 
top 7/8 of the thickness of the tile.  Of course, there are two layers 
of scintillator on the top of the instrument, so any conversions in 
the first layer will deposit 2 MIP in the second layer.  Thus this cut 
eliminates 15/16ths of the events which convert in the 
ACD, as well as any charged particle events.

In addition, cuts were made based on the characteristics of the tracks 
themselves.  All tracks were required to have at least three real 
hits, exclusive of ``virtual'' hits placed on noisy strips or outside 
of the tracker.  The total track \chisq\ was divided by the number of 
hits in the track, and this reduced \chisq\ was required to be less 
than 5.  The final cut demanded that all tracks start at least 4.7 mm 
(20 strips) from the edge of the active area of the tracker.  
Electrons from \gammarays\ that convert that close to the edge are 
likely to exit the tracker preferentially, and will introduce a bias 
to the distribution of reconstructed \gammaray\ directions.  The 
efficiency of each cut is given in Table \ref{bjwft:cutstable}.

\begin{table}[ht]
\centering
\begin{tabular}{l c }
\multicolumn{1}{ c }{Cut} &
 \multicolumn{1}{ c }{Triggers Kept} \\ \hline
Hodoscope & 42-55 \% \\
ACD & 57-84 \% \\
Three Hits & 77-87 \% \\
$\chi^2$ & 73-82 \% \\
Edge & 89-96 \% \\ \hline
All Cuts &  15-25 \% \\ 
\end{tabular}
\centering
\caption{\label{bjwft:cutstable}
Cut efficiencies} 
\end{table}

\paragraph{Monte Carlo cuts.}
In an effort to make the \beamtest\ data as directly comparable with 
Monte Carlo simulations as possible, the Monte Carlo data was 
subjected to very similar cuts.  The Monte Carlo included an 
anti-coincidence system, and a similar cut was made to reject events 
which converted in the plastic scintillator.  All of the cuts based on 
track parameters were made in the exact same way for both the Monte 
Carlo and the \beamtest\ data.  Since most of the Monte Carlo 
simulations were done with incident \gammarays\ drawn from a 
bremsstrahlung energy spectrum, there was no need to make cuts to 
ensure only one \gammaray\ in the tracker.

The particular cuts made were chosen to simplify analysis of the 
\beamtest\ data.  They are not meant to represent the types of cuts 
that will be made on for \glast.  \glast\ will be faced with a very 
different environment in space, replete with background particles, 
albedo \gammarays, and signal \gammarays\ incident from all 
directions.  It also has a very different geometry, which will require 
combining data from different towers and accounting for gaps and 
support structure.  The relevant feature of the cuts made here are 
that they are very nearly identical for the \beamtest\ instrument and 
the Monte Carlo simulations.  This will allow direct comparison of 
their results.  However, the instrumental parameters measured for the 
\beamtest\ instrument will not be directly scalable to \glast.

	\section{Expected \BeamTest\ Point-Spread Widths}
\label{br:kalpredict}.
Once the various cuts have been applied to the data sets, 
distributions of the reconstructed angle can be compared.  Before we 
compare these results, it may be useful to consider the theoretical 
widths we expect to find, based on the Kalman filter covariances.  In 
\sect{bt:psf} we saw that we could calculate the expected point-spread 
width of a given instrument configuration independent of the actual 
measured data, under the assumption that all distributions involved 
are Gaussian.  In order to estimate the point-spread widths we will 
measure with the \beamtest\ instrument, we must properly average the 
point-spread width over all the different classes of events we expect 
to measure.  Recall that the point-spread width $\sigma(E, f, p, d, x, 
N)$ is a function of the \gammaray\ energy $E$, the fraction $f$ of 
that energy in one electron, the silicon strip pitch $p$, the gap $d$ 
between adjacent planes, the total amount $x$ of material in the 
radiator and the detector in each plane, and the number $N$ of planes 
which measure the given event.  For any instrument configuration, 
strip pitch $p$, gap $d$ and radiator $x$ are fixed.

\begin{figure}
\centering
\scalebox{0.5}{\includegraphics{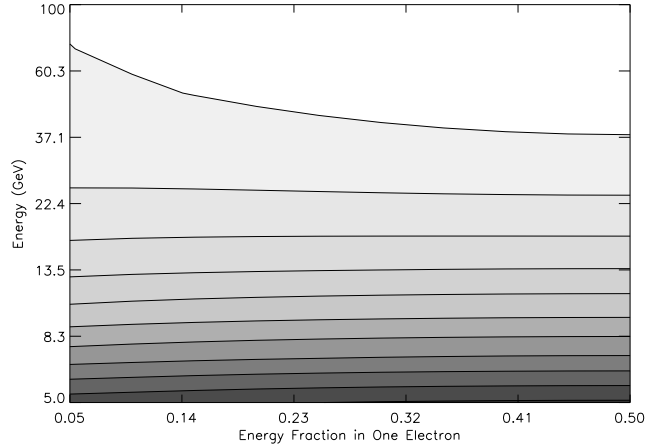}}
\caption[Point-spread width dependence on total energy and energy 
splitting fraction]{\label{br:highEvsf}
Point-spread width dependence on total energy and energy 
splitting fraction.  Darker shades indicate larger point-spread width.
At lower energies, the width is nearly constant as a function of the
splitting fraction.  At high energies, the point-spread function
is much larger for asymmetric splittings.  The \beamtest\ was
always in the low-energy limit for this purpose.}
\end{figure}

To facilitate comparison, we will look at events with approximately 
equal energies.  Therefore we must average over a range of electron 
energy splittings and numbers of planes traversed.  The point-spread 
width is an almost constant function of $f$ below 5~GeV. Above 5~GeV, 
the point-spread width is significantly smaller for even energy split 
(\fig{br:highEvsf}), although such a splitting is comparatively less 
common\footnote{The energy splitting is assumed here to be symmetric, 
as described by the Bethe-Heitler formula \cite{heitler}.  Below about 
3.3~MeV, asymmetries arise because the nucleus attracts \el\ while 
repelling \pos.  The asymmetric differential cross sections have been 
worked out by \overbo, Mork, and Olsen \cite{overbo}.}\index{\footnotemark[2]} 
(\fig{bt:Esplit}).  Therefore, we will assume that the point-spread 
width is approximately independent of the energy splitting fraction 
$f$.  The number of planes $N$ through which the \el\pos\ pair travel 
depends on the total number of planes in the instrument (6) and in which 
plane the \gammaray\ converted.  The probability that a \gammaray\ 
will convert in a given layer is given by \eq{bt:probconv} with 
$t/X_o$ equal to the thickness of the SSD plus the Pb radiator for the 
plane in question.  The intensity of the \gammaray\ beam at that plane 
is given by \eq{bt:Idecay}, with $t/X_o$ equal to the total number of 
radiation lengths of material above the plane in question.  This 
represents the beam attenuation from the planes above.  Therefore, the 
probability of a conversion in plane $n$ is
\begin{equation}
P(n) = I_o e^{ (- 7/9 (n-1) x)} (1 - e^{7/9 x})
\end{equation}
where $x$ is the total radiator in each plane.  To calculate the 
average point-spread width, we wish to weight the point-spread width 
for each number of planes by the probability that the conversion 
happened in that plane.  Recalling that we accept only events with at 
least three hits, we find the probability that a \gammaray\ converted 
in one of the top four planes:
\begin{equation}
P(n \le 4) = \sum_{n=1}^4 P(n)
\end{equation}
Therefore, the probability that any given pair-production event converted
in plane $n$ is
\begin{equation}
P(n | 1 \le n \le 4) = P(n)/P(n \le 4) \nonumber 
\end{equation}
Plugging in, we arrive at the weights we will apply to the variances
measured on each plane.
\begin{equation}
P(n) = \frac{ e^{-\frac{7}{9} (n-1) x} }{ \sum_{n=1}^4 
\exp{ \left( - \frac{7}{9} (n-1) x \right) }}
\end{equation}

It remains only to examine the point-spread width $\sigma$ as a 
function of total \gammaray\ energy $E$ for each instrument 
configuration.  The nature of the Kalman equations 
(\sect{bt:kalmaneqns}) makes the calculation of the point-spread width 
tedious.  However, it is a simple matter to compute electronically 
\cite{psfestimator}.


	\section{Conclusions}


While our theoretical calculations make some simplifying assumptions 
which may not be entirely accurate, we hope that the Monte Carlo 
simulations will more completely represent the actual data.  For 
example, the \gismo\ code has a sophisticated multiple scattering 
model which is much more accurate than the simplistic Gaussian 
assumption made above.  Furthermore, the Monte Carlo simulations 
included all the particular geometric elements of the \beamtest\ 
instrument.

\subsection{Comparison of \BeamTest\ and Monte Carlo Results}
\begin{figure}[t]
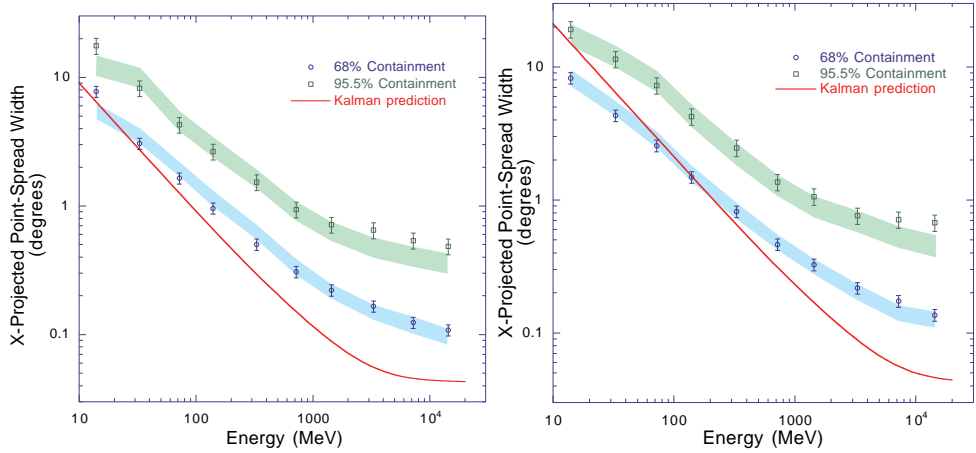

\centering
\includegraphics[width=2.5 in]{plots/p0x6895wKal.epsi}
\includegraphics[width=2.5 in]{plots/p4x6895wKal.epsi}
\caption[Pancake X-Projection PSF vs. Monte Carlo]{\label{br:pxpsf}
$x$-projected  point-spread widths for Pancake configuration
with no Pb radiators (left) and 4\% Pb radiators (right).  
Circles indicate the 68\% containment width, and
squares indicate the 95.5\% containment width.
Error bars are 2$\sigma$ statistical errors, and
shaded regions represent the 2$\sigma$ confidence regions of the Monte
Carlo estimates.  The line is an estimate of the point-spread
width from the Kalman filter.}
\end{figure}

\begin{figure}[t]
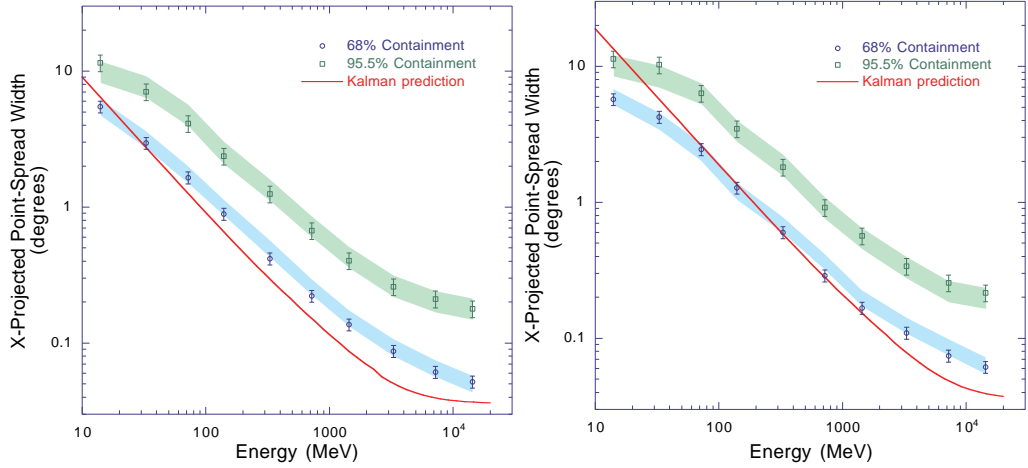

\centering
\scalebox{.45}{\includegraphics{plots/s0x6895wKal.epsi}}
\scalebox{.45}{\includegraphics{plots/s4x6895wKal.epsi}}
\caption[Stretch X-Projection PSF vs. Monte Carlo]{\label{br:sxpsf}
$x$-projected point-spread widths for Stretch configuration with no Pb 
radiators (left) and 4\% Pb radiators (right).  Circles indicate the 
68\% containment width, and squares indicate the 95.5\% containment 
width.  Error bars are 2$\sigma$ statistical errors, and shaded 
regions represent the 2$\sigma$ confidence regions of the Monte Carlo 
estimates.  The line is an estimate of the point-spread width from the 
Kalman filter.}
\end{figure}

\begin{figure}[t]
\centering
\includegraphics[width=2.4 in]{plots/atp4xe140MeV.epsi}
\hspace{.5 in}\includegraphics[width=2.4 in]{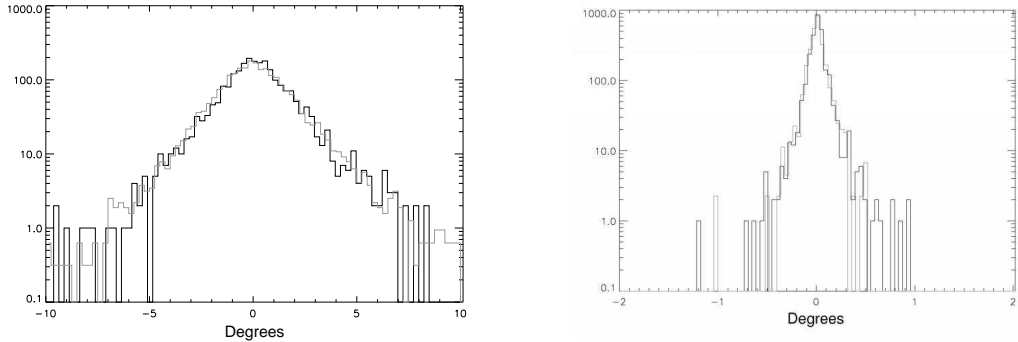}
\caption[Projected point-spread distributions]{\label{br:atplots}
Full distributions of the reconstructed projected incident \gammaray\ angle.
Left side is pancake configuration, 4\% Pb radiators, 100--200~MeV.  Right
side is stretch configuration, no Pb radiators, 5--10~GeV.  Vertical scale
is renormalized for comparison of distributions.}
\end{figure}

The measured point-spread widths for both the Monte Carlo simulations 
and the \beamtest\ data are shown for each instrument configuration in 
Figures \ref{br:pxpsf} and \ref{br:sxpsf}.  There is good agreement 
between the two out to the 95\% containment radius.  Two example 
distributions are shown in \fig{br:atplots}.

Three features of the Kalman filter estimate are relevant.  First of 
all, in general the Kalman estimate is lower than the measured widths.  
Second, the slope of the Kalman estimate with energy is steeper.  
Finally, the Kalman estimate reaches an asymptotic limit faster, and 
the limit is significantly lower.

The Kalman estimates are lower than the measured widths in general 
because the Kalman estimates are based on Gaussian errors. 
Furthermore, the point-spread widths estimated by the Kalman filter
are the standard deviations; that is, they correspond to the 68\% containment
radius of the point-spread function assuming it is Gaussian.  As we have 
noted, the multiple scattering distributions have significant tails, 
and the measurement error distributions are more square than Gaussian.  
These effects tend to degrade the fit quality, leading to larger 
point-spread widths.

The slope of the Kalman estimate is very nearly $1/E$ at low energies; 
this is expected since it is based on a Gaussian multiple scattering 
model with the width inversely proportional to the energy.  On the 
other hand, the actual \beamtest\ instrument is very narrow.  At low 
energies, this leads to a strong self-collimation effect.  \Gammarays\ 
whose electrons initially make large scatters would be reconstructed 
in a wide instrument with large apparent incident angles.  These 
events cannot be reconstructed with the limited data from the narrow 
instrument, and are thrown away.  As we expect, stretch configuration 
displays more self-collimation than pancake, and high-radiator 
configurations (with larger scattering) also display more 
self-collimation.

At high energies, the width of the point-spread function is dominated 
by the measurement error.  The assumptions made for the Kalman filter 
are not particularly good in this regime.  The Kalman filter assumes 
that the position measurement is continuous, with Gaussian errors.  In 
fact, the measurement is a discrete one, with roughly square errors.  
The effects of the square error distribution have been discussed in 
\sect{bt:meandms}.  The effects of the discrete measurement are more 
difficult to assess.  Since the tracker planes were very nearly 
aligned, and the \gammaray\ beam was very nearly aligned with the 
strip grid, the measured distributions are subject to aliasing.  If 
the planes are perfectly aligned, then the minimum non-zero half-angle 
which can be measured by the instrument in pancake mode is \onehalf\ 
\by\ 236 \micron /150.0 mm $\approx$ 0\fdg05.  That assumes that the 
\gammaray\ converts in the middle of a strip, and that the electron 
continues down the instrument, hitting the same strip in each plane 
until the last one, by which time it has drifted half the strip pitch 
and activates the next strip.  In fact, we see in \fig{br:pxpsf} that 
the Kalman estimate is well below 0\fdg05 above 5~GeV. Since the 
point-spread function for the \beamtest\ instrument and for the Monte 
Carlo simulations is quantized at that level, it is not surprising to find that 
its width is somewhat larger.

These aliasing effects will actually be helped by the slight offsets 
in the instrument planes acquired during the launch of \glast, 
assuming that they can be well measured using cosmic rays.  In 
addition, the additional planes will act to reduce the aliasing.  
Perhaps more importantly, \glast\ will be illuminated from a large 
fraction of the sky.  In scanning mode, the bearing to a source in 
instrument coordinates will be constantly changing, serving to smear 
the aliasing over all scales.  These effects should push the 
performance of \glast\ closer to the Kalman estimate.  In any case, 
since the Monte Carlo correctly models the aliasing, simulations of 
\glast\ from physically reasonable sources will make accurate 
predictions of the \glast\ point-spread width.

A final difference between the instrument as tested and simulated and 
the idealized Kalman assumptions concerns the geometry of the 
detectors.  The Kalman estimate assumes that the multiple scattering 
occurs in the same place as the the measurement, when in fact most of 
the multiple scattering occurs in the Pb radiators, which are a few 
millimeters away from the detectors.  Simulations of the baseline 
\glast\ instrument suggest that as a rule of thumb, the point-spread 
width is degraded by 3.5\% per millimeter of separation between the 
radiator and detector \cite{glast:doe_proposal}.  The separation in 
the beamtest instrument was about 2~mm, which would suggest a degradation 
of about 7\% from the theoretical Kalman estimate.

	\subsection{Implications for \glast}
The agreement between the Monte Carlo simulations and the \beamtest\ 
data is clearly encouraging to the \glast\ collaboration.  
The Kalman filtering formalism appears to succeed admirably at reconstructing
electron tracks.  Kalman estimates of the \glast\ point-spread width
for various configurations are given in \tbl{br:psftable}. 

\begin{table}[t]
\begin{tabular}{ccccc}
Strip pitch (\micron) & Radiator & 100 MeV PSF & 1 GeV PSF  & 10 GeV PSF \\ \hline
180 & 0.0\% & 0.91 & 0.11 & 0.023 \\
180 & 3.5\% & 2.0 & 0.22 & 0.035 \\
180 & 5.0\% & 2.3 & 0.25 & 0.038 \\
240 & 0.0\% & 0.91 & 0.12 & 0.029 \\
240 & 3.5\% & 2.0 & 0.22 & 0.040 \\
240 & 5.0\% & 2.3 & 0.26 & 0.043 \\
400 & 0.0\% & 0.93 & 0.13 & 0.044 \\
400 & 3.5\% & 2.0 & .24 & 0.051 \\
400 & 5.0\% & 2.3 & 0.27 & 0.054 \\
\end{tabular}
\caption[Calculated point-spread widths]{\label{br:psftable}
$1\sigma$ widths of the point-spread function as calculated from the
Kalman formalism, for different silicon strip pitches and amounts lead radiator
(given in radiation lengths), assuming 16 planes space 30.0 mm apart.  
The method used to generate these results
is given in \sect{bt:psf}.}
\end{table}

The equivalent Kalman estimates of the beamtest instrument point-spread
width proved to be a reasonable estimate of the instrument performance.
However, 
some care must be taken to interpret these results in light of their 
implications for \glast.  Primarily, it is important to remember that 
the point-spread widths measured by the beamtest may not be used as an 
estimate of the 
\glast\ point-spread width; they are two separate instruments with 
different characteristics.  Furthermore, the \beamtest\ instrument was 
measured in a controlled, low background beam environment, while 
\glast\ will be operating in space, bombarded by charged particles and 
albedo \gammarays.

However, this does not mean that these results are irrelevant to 
\glast.  The verification of the Monte Carlo code implies that 
simulations of \glast\ with \glastsim\ should yield accurate estimates 
about the final instrument parameters without building many expensive 
prototypes.  Promises made by \glastsim\ may be reasonably expected to 
be fulfilled by \glast.

%% file: timedelayapp.tex
\chapter{SSB Arrival Time Corrections}
\label{timedelay}

There are four components to the corrections to Solar System Barycenter (SSB) 
Time from UTC time in \eq{tv:schemdelay}~\cite{joethesis,lynebook}.  
The first is due entirely to timekeeping convention, and translates UTC
to a monotonic sequentially indexed time known as International Atomic Time:
\begin{equation}
\label{delay:conveq}
\Delta_{\hbox{convention}} = k + 32\fs 184
\end{equation}
where $k$ is the integral number of leap seconds since 1972.

The second correction is made for the position of the observatory.  Since the pulsar
is much further away from the barycenter than the observatory is, the difference
in path lengths is simple.  Dividing by the speed of light, we arrive at the 
correction to the arrival time:
\begin{equation}
\label{delay:locationeq}
\Delta_{\hbox{location}} = \frac{\hat{\myvec{n}} \vdot \myvec{r}_{bo}}{c}
\end{equation}
where $\hat{\myvec{n}}$ is a unit vector pointing at the pulsar, and $\myvec{r}_{bo}$
is the vector between the SSB and the observatory.  In practice, this is the
vector sum of the vector from the SSB to the center of the Sun, the vector from 
the Sun to the Earth, and the vector from the Earth to \cgro.  The position vectors
for the Earth and Sun are taken from the standard ephemeris published by the
Jet Propulsion Laboratory~\cite{standish90}.

The remaining two corrections are relativistic in nature.  The first 
is the Einstein delay $\Delta_{\hbox{Einstein}}$.  This is the 
manifestation of the old adage that ``heavy clocks run slowly.''  It 
depends only on the depth of the potential well in which the 
observatory sits, not the path that a \gammaray\ takes to arrive 
there.  Therefore, the Einstein delay is included in the JPL 
ephemeris.  There is an additional term that results from the 
ellipticity of the Earth's orbit of the form $\myvec{v}_{\earth} \vdot 
\myvec{r}_{bo} / c^2$.  The portion of this term due to the Earth's 
position is already included in International Atomic Time; we need 
only to add the part due to the satellite position:
\begin{equation}
\label{delay:einsteindelay}
\Delta_{\hbox{Einstein}} = \Delta_{\hbox{Einstein ({\sc JPL})}} + 
\frac{\myvec{v}_{\earth} \vdot \myvec{r}_{\earth o}}{c^2}
\end{equation}

The last correction is the Shapiro delay $\Delta_{\hbox{Shapiro}}$.  This
results from the delay induced by the gravitational potential
in the region of the Sun~\cite{shapiro64} and goes as
\begin{equation}
\label{delay:shapirodelay}
\Delta_{\hbox{Shapiro}} \simeq \frac{2 G M_{\sun}}{c^3} \ln (1 + \cos \theta)
\end{equation}\index{$\simeq$}
where $\theta$ is the angle between the vector from \cgro\ to the Sun and the
vector from \cgro\ to the pulsar.

Combining all these corrections, the arrival time in the SSB frame as a 
function of the measured UTC time is given by:
\begin{eqnarray}
\label{delay:full}
	t_{b}  & = & t_{\utc} + k + 32\fs 184 + (1/c) \hat{n} \vdot \myvec{r}_{O} +
\nonumber \\
& &	\Delta_{\hbox{Einstein}} + 
(1/c^{2})\myvec{v}_{\earth} \vdot \myvec{r}_{\earth O} + \nonumber \\
& & 	 (2 G M_{\sun} /c^{3}) \ln (1+ \cos \theta) 
\end{eqnarray}

The code to make these adjustments to \egret\ arrival times was written by
Joe Fierro~\cite{pulsardef}, based on calculations and data found 
in~\cite{brumberg,hunt71,shapiro64,standish90,ptondbase} and~\cite{taylor89}.

%% file: part2appen.tex
\chapter{Summary of Kalman Filtering Equations \label{app:KalEqn}}

\begin{singlespace}
\begin{tabbing}
Define the first state vector $\myvec{x}_0$ and covariance ${\bf C}_0$ \\
For each \= plane $k$ in the instrument: \\
\>	Project \= from the last plane: \\
\>		\> $ \myvec{x}_{k, proj} = {\bf F}_{k-1} \myvec{x}_{k-1} $ \\
\>		\> $ {\bf C}_{k, proj} = {\bf F}_{k-1} {\bf C}_{k-1} {\bf F}_{k-1}^T + {\bf Q}_{k-1} $ \\
\>	Filter \= the estimates: \\
\>		\> $ {\bf C}_k = \left[ ({\bf C}_{k, proj})^{-1} + {\bf H}_k^T {\bf G}_k {\bf H}_k \right] ^{-1} $ \\
\>		\> $ \myvec{x}_k = {\bf C}_k \left[ ({\bf C}_{k, proj})^{-1} \myvec{x}_{k, proj} + {\bf H}_k^T {\bf G}_k \myvec{m}_k \right] $ \\
\\
Starting \= at the second-to-last plane and working back up: \\
\>	Smooth \= the estimates: \\
\>		\> ${\bf A}_k = {\bf C}_k {\bf F}_k^T ({\bf C}_{k+1, proj})^{-1}$ \\
\>		\> $\myvec{x}_{k, smooth} = \myvec{x}_k + {\bf A}_k (\myvec{x}_{k+1, smooth} - \myvec{x}_{k+1, proj})$ \\
\>		\> ${\bf C}_{k, smooth} = {\bf C}_k + {\bf A}_k ({\bf C}_{k+1, smooth} - {\bf C}_{k+1, proj}) {\bf A}_k^T$ \\
\>		\> $\myvec{r}_{k,smooth} = \myvec{m}_{k,smooth} - {\bf H}_k \myvec{x}_{k,smooth}$ \\
\>		\> ${\bf R}_{k,smooth} = {\bf V}_k - {\bf H}_k {\bf C}_{k,smooth} {\bf H}_k^T$ \\
\>		\>$\chi^2_+ = \myvec{r}_{k,smooth}^T {\bf R}_{k,smooth}^{-1} \myvec{r}_{k,smooth}$ \\
\end{tabbing}
\end{singlespace}

\chapter{Track Finding Algorithm \label{app:find}}
\begin{singlespace}
\input{trackalg}
\end{singlespace}


%% file: trackalg.tex
While the Kalman filter offers a way to quickly and automatically find the best fit electron track to a given set of detector hits, identifying which hits belong to the track is not a science, but an art.  We have experimented with a number of algorithms, and have empirically fine-tuned this one.  This is the algorithm used by {\tt tjrecon} for finding tracks.

\begin{itemize}
\begin{singlespace}
\item\index{$\bullet$} Load the data from one event into the data structures.

\item\index{$\bullet$} For each projection:
	\begin{enumerate}
	\item Create a new track:
		\begin{itemize}
		\item Initialize all the matrices and arrays
		\item Assign first hit to be leftmost hit in first layer with hits
		\item Assign second hit to be leftmost hit in second layer with hits. 
		\item If there are $\leq 1$ hits in this projection, go to next projection
		\item Assign initial slope to first hit
		\item Assign half the calorimeter energy to the track
		\item Assign the multiple scattering matrix for that energy as the Covariance matrix.
		\end{itemize}
	\item Create a second track, using leftmost hit in first layer with hits, rightmost hit in second layer.  If there are no hits in the second layer, continue down until you find a layer with hits.
	\item Find the hit for each layer of the first track:
		\begin{itemize}
		\item Project state vector and C matrix to next layer
		\item Assign the hit for that layer:
			\begin{itemize}
			\item If the track projects out of the tracker, add a virtual hit outside the tracker, and assign that hit to the track.  
			\item If there is a dead strip within $\mypm 10$ strips, add a virtual hit on the dead strip. Don't assign it yet.
			\item Find the hit closest to the projected track
			\item If there is only one hit on that layer, assign it to the track.
			\item If the closest hit already belongs to another track, take the next closest hit.  Assign it to the track.
			\item If the closest hit is unclaimed, assign it to the track.		
			\end{itemize}
		\item Revise (``filter'') the projection of the state vector and C matrix in light of this measurement.
		\end{itemize}
	\item Smooth the first track:
		Starting at the bottom, apply the Kalman smoothing equations to the state vectors and C matrices

	\item Find the hit for each layer of the second track.
	\item Smooth the second track.

	\item Untangle the tracks by swapping hits from one track to the other so they do not cross.

	\item Check free hits for first track:
		\begin{itemize}
		\item For each level, starting at the top, for each unclaimed hit, swap the track's current hit with the unclaimed hit.
		\item Filter the track.  (giving a new set of predicted locations at each level below the current one)
		\item For each successive level, check to see if the new predictions suggest different hits in the track.  If so, and if they are real hits (not virtual ones), assign them to the track.  Filter.  Continue to the bottom of the tracker.

		\item Smooth the new track.
		\item Compare the $\chi^2$ of this new track with the original.  If it's lower, keep the new track.  Otherwise revert to the old track.
		\item Start the whole game over from scratch on the next level down.
		\end{itemize}
	\item If the track changed at all in the above step, check free hits again, this time starting at the bottom.

	\item Check free hits for the second track, starting at the top.
	\item If the second track changed, check free hits again, this time starting at the bottom.

	\item Starting at the bottom, check to see if swapping hits helps:
		\begin{itemize}
		\item For each level, swap the hits between the two tracks.  Filter and smooth both tracks.
		\item If the new $\chi^2$ is smaller, keep the new tracks.  Otherwise revert.
		\end{itemize}
	\item Make sure tracks don't cross between first and second planes.  If the two tracks do not share their first hit, it is likely that the tracks don't really cross.  Filter and smooth both tracks.

	\item Check to see if one of the tracks may have left the tracker:
		(If so, the two tracks may currently share hits on all layers below the exit point)
		\begin{itemize}
		\item Find the lowest level on which the tracks share hits.  If they don't share any, then quit.
		\item If the first shared hit is at least the third hit in the track, create a virtual hit outside the tracker in the direction of the current track.  Assign it to the track.
		\item Remove all successive hits from the track.
		\item Filter and smooth.
		\item Check free hits again, to see if, in this new configuration, the track would prefer different hits upstream.
		\item If the $\chi^2$ if smaller, keep the new tracks.  Otherwise revert.
		\end{itemize}
	\item Repeat all for the other projection.
	\end{enumerate}

\item\index{$\bullet$} If we don't have four good tracks, then throw the event away.

\item\index{$\bullet$} If at least one of the four tracks leaves the tracker:
	\begin{itemize}
	\item Find the track that leaves the tracker earliest.
	\item Find the track in the other projection with the larger $\chi^2$.  We will claim that this track is the other projection of the track that leaves.
	\item Remove all hits from that track below the exit point of it's opposite-projected track.
	\end{itemize}
\samepage{
\item\index{$\bullet$} For each projection:
	\begin{itemize}
	\item Vary the energy split between the two tracks.
	\item Filter and smooth with the new energy split.
	\item Find the split fraction that minimizes $\chi^2$.
	\item In practice, this always assigns all the energy to one track.  Assign 75\% of the energy to that track, 25\% to the other.
	\end{itemize}
\item\index{$\bullet$} Write tuple containing relevant information.
}
\end{singlespace}
\end{itemize}